\documentclass[twoside,12pt]{article}

\usepackage{amsmath,amsfonts}
\usepackage{amssymb,latexsym}
\usepackage{amsxtra}
\usepackage{upref}
\usepackage{exscale}
\usepackage[mathscr]{eucal}

\newcommand{\be}{\begin{equation}}
\newcommand{\ee}{\end{equation}}
\newcommand{\beq}{\begin{eqnarray}}
\newcommand{\eeq}{\end{eqnarray}}
\newcommand{\no}{\nonumber}
\newcommand{\bea}{\begin{array}}
\newcommand{\eea}{\end{array}}
\newcommand{\lb}{\label}

\newcommand{\mscr}{\mathscr}
\newcommand{\mfrak}{\mathfrak}
\newcommand{\ve}{\varepsilon}
\newcommand{\ds}{\displaystyle}
\newcommand{\ts}{\textstyle}
\newcommand{\pp}{\partial}
\newcommand{\im}{\imath}
\newcommand{\ppr}{^{\boldsymbol{\prime}}}

\newcommand{\bprime}{\boldsymbol{\prime}}
\newcommand{\wt}{\widetilde}
\newcommand{\ovv}{\overline}

\newcommand{\trmn}{\raisebox{-4pt}{$\mbox{tr}\atop {\scriptstyle \mu,\nu}$}}

\newcommand{\strab}{\raisebox{-4pt}{$\mbox{str}\atop {\scriptstyle \alpha,\beta}$}}
\newcommand{\STRAB}{\raisebox{-4pt}{$\mbox{STR}\atop {\scriptstyle a,\alpha;b,\beta}$}}
\newcommand{\ph}{\phantom}

\newcommand{\scz}{\scriptsize}

\numberwithin{equation}{section}
\allowdisplaybreaks[4]

\begin{document}

\begin{center}
{\large\bf Effective Sine(h)-Gordon-like equations for pair-condensates} \vspace*{0.1cm}\\
{\large\bf composed of bosonic or fermionic constituents} \vspace*{0.1cm}\\
{\bf Bernhard Mieck}\footnote{e-mail: "bjmeppstein@arcor.de";\newline freelance activity during 2007-2009;
current location : Zum Kohlwaldfeld 16, D-65817 Eppstein, Germany.}
\end{center}
\vspace*{-0.2cm}

\begin{abstract}
An effective coherent state path integral for super-symmetric pair condensates is investigated
with specification on the nontrivial coset integration measure. The non-Euclidean integration measure prevents
straightforward classical equations and solutions of the independent, anomalous field variables which
follow from variations of the actions in the exponential phase weight factors.
We examine a transformation with a suitable super-Jacobi matrix for the change of coset integration
measure to 'flat' Euclidean path integration fields of pair condensates.
The independent parameter fields of the super-symmetric anomalous terms are given by those of the
\(\mbox{Osp}(S,S|2L)/ \mbox{U}(L|S)\) coset super-manifold. The described, effective coherent state
path integral of pair condensates is obtained by a gradient expansion after a Hubbard-Stratonovich
transformation (HST) of the original path integral with super-fields of bosonic and fermionic atoms.
A modified HST of bosonic and fermionic super-fields converts the original path integral into one
with 'Nambu' doubled, super-symmetric self-energies. Due to the addition
of source fields, we consider a spontaneous symmetry breaking of the total \(\mbox{Osp}(S,S|2L)\)
super-group to the \(\mbox{Osp}(S,S|2L)/ \mbox{U}(L|S)\otimes \mbox{U}(L|S)\) coset decomposition
with the super-unitary \(\mbox{U}(L|S)\) group as the invariant subgroup of the background density field.
The nontrivial coset integration measure, determined by the square root
\((\,\mbox{SDET}(\hat{G}_{\mbox{{\scz Osp}}/ \mbox{{\scz U}}})\,)^{1/2}\) of the super-determinant
of the \(\mbox{Osp}(S,S|2L)/ \mbox{U}(L|S)\) coset
metric tensor \(\hat{G}_{\mbox{{\scz Osp}}/ \mbox{{\scz U}}}\), is eliminated
by the '{\it inverse square root}' of the coset metric tensor \((\hat{G}_{\mbox{{\scz Osp}}/ \mbox{{\scz U}}})^{-1/2}\)
as the appropriate super-Jacobi matrix; this results into Euclidean path integration variables
for the pair condensate fields. A diagonal construction of the coset metric
tensor \(\hat{G}_{\mbox{{\scz Osp}}/ \mbox{{\scz U}}}\)
allows a straightforward application of the super-Jacobi matrix \((\hat{G}_{\mbox{{\scz Osp}}/ \mbox{{\scz U}}})^{-1/2}\)
to the pair condensate fields of the \(\mbox{Osp}(S,S|2L)/ \mbox{U}(L|S)\) coset super-manifold
which also appears with this coset metric tensor in the gradients and kinetic energies
of the actions. Therefore, we acquire a considerable simplification of
the effective coherent state path integral in terms of anomalous, '{\it Euclidean path integration variables}'.
In analogy and in the sense of statistical thermodynamics, the particular property of being a '{\it state variable}'
is verified for the modulus of eigenvalues of the coset order-parameter matrix for anomalous fields.
On the contrary, the phase values of eigenvalues of the anomalous coset order-parameter matrix depend
on the chosen time contour path or previous history of transformation fields and therefore correspond
to the path dependent '{\it heat}' or '{\it work}' variables of thermodynamics in a transferred sense.
According to the transformation to Euclidean, anomalous path integration variables, first order variations of fields
can be performed for classical equations with inclusion of second and higher even order variations for universal
fluctuations determined by the coset metric tensor \(\hat{G}_{\mbox{{\scz Osp}}/ \mbox{{\scz U}}}\).
Furthermore, we mention how to extend finite order gradient expansions to infinite order by using
a suitable integral representation for the logarithm of an operator and similarly for its inverse.\newline
\end{abstract}

\newpage

\tableofcontents

\newpage

\section{Introduction} \lb{s1}

\subsection{Variation of classical actions in coherent state path integrals
with nontrivial integration measures}\lb{s11}

Generating functionals, as coherent state path integrals, allow various kinds
of approximations or even exact solutions apart from being representations
of many-particle quantum mechanics \cite{neg}-\cite{nag2}. Coherent state path integrals can be
examined by Monte Carlo methods according to appropriate importance sampling
and stationary phase filtering or even by exact solutions in the case of
integrable systems \cite{neg,ka}. At zero temperature they consist of a weight factor,
usually an exponential phase comprising a classical action, so that certain
classical field configurations can contribute a dominant part in the
weighting with the exponential. These dominant contributions are usually
obtained by the stationary phase or first order variation of the actions
in the exponent of the weight factor. In principle this variation can be
extended to second or even higher order variations around the solutions of
classical fields from the first order variation.

However, this process of variations for approximating by classical
solutions becomes nontrivial in the case of non-Euclidean integration
measures of the field variables. In this case one can apply the method of
steepest descent for a polynomial-like integration measure with the
exponential of classical actions \cite{Das1} or one transforms the
whole factor of the '{\it integration measure}' to its
'{\it exponential(logarithm(integration measure))}' form so that it
has an equivalent weight as the classical action terms in the exponents.
This method is straightforward; but, one can also
try to determine a more sophisticated transformation of the path
field variables so that the nontrivial integration measure is
eliminated for Euclidean integration variables by inclusion of an
additional Jacobi-determinant. The functional dependence of the
classical actions with the original fields is then altered by the
corresponding Jacobi-matrix of the new Euclidean integration
field variables. Both methods, the method of steepest descent (with
'{\it exponential(logarithm(integration measure))}' or the removal
of nontrivial integration measures by transformations with a suitable
Jacobi-matrix, are in general far from being equivalent.
In spite of the '{\it exponential(logarithm(integration measure))}' form,
the fields of the nontrivial integration measure in the variations of steepest descent method contribute
in a different manner than the fields of the actions weighted by exponentials.
In the case of the considered coherent state path integral \cite{mies1},
it is inevitable that the '{\it exponential(logarithm(integration measure))}' in the method
of steepest descent has its first contributions from second and
all higher even order variations of the fields whereas the main other actions have only
non-vanishing terms in odd-numbered order of variations with the independent fields
on the time contour. Therefore, the simple method of steepest descent causes
inconsistent treatment in the case of nontrivial path integration measures and
their variations on the time contour in comparison to the variations
on the time contour dependent fields in the actions of the exponentials.

We illustrate this problem in analogy to a multidimensional integral
\(Z[\vec{x}]\), (\(\vec{x}=\{x^{1},\ldots,x^{N}\}\)), with an action
\(\mscr{A}[\vec{x}]\)
\be\lb{s1m_1}
Z[\vec{x}]=\int d[\vec{x}]\;\sqrt{\mbox{det}\big(\hat{g}(\vec{x})\big)}\;\;
\exp\big\{\im\;\mscr{A}[\vec{x}]\big\}\;,
\ee
where the Euclidean integration measure \(d[\vec{x}]\) is modified by the
square root of a metric tensor \(\hat{g}_{ij}(\vec{x})\) as the nontrivial
integration measure
\be \lb{s1m_2}
\big(ds\big)^{2}=dx^{i}\;\hat{g}_{ij}(\vec{x})\;dx^{j}\;.
\ee
The transformation to Euclidean variables \(d[\vec{y}]\) is related to the inverse
square root of the metric tensor \(\hat{g}^{-1/2}(\vec{x})\)
as the appropriate Jacobi matrix where
the symmetry of the metric tensor allows a decomposition into orthogonal
matrices \(\hat{O}_{ij}(\vec{x})\) and real eigenvalues \(\hat{\lambda}^{k}(\vec{x})\)
\beq \lb{s1m_3}
\big(ds\big)^{2}&=&dx^{i}\;\hat{g}_{ij}(\vec{x})\;dx^{j}=
dx^{i}\;\hat{O}_{ik}^{T}(\vec{x})\;\hat{\lambda}^{k}(\vec{x})\;
\hat{O}_{kj}(\vec{x})\;dx^{j}=  \\ \no &=&
\underbrace{dx^{i}\;\Big(\hat{O}^{T}(\vec{x})\cdot\hat{\lambda}^{1/2}(\vec{x})\Big)_{i}^{\ph{i}k}}_{dy^{k}}\;
\underbrace{\Big(\hat{\lambda}^{1/2}(\vec{x})\cdot\hat{O}(\vec{x})\Big)_{kj}\;dx^{j}}_{dy_{k}}=
dy^{k}\;dy_{k}=dy^{k}\;dy^{k}\;;  \\ \lb{s1m_4}
dy^{j}&=& \Big(\hat{\lambda}^{1/2}(\vec{x})\cdot\hat{O}(\vec{x})\Big)^{j}_{\ph{j}i}\;dx^{i} \;;
 \\ \lb{s1m_5}
\hat{O}_{ji}(\vec{x})\;dx^{i}&=&  \Big(\hat{\lambda}^{-1/2}(\vec{x})\cdot d\vec{y}\Big)_{j} \;;
\\ \no
\;\;\Longrightarrow y^{j}&=&y^{j}(\vec{x})\;\;
\Longrightarrow x^{i}=x^{i}(\vec{y})\;.
\eeq
This yields with the additional Jacobi matrix
\(\hat{J}^{i}_{\ph{i}k}=(\pp x^{i}/\pp y^{k})=(\,\hat{O}^{T}(\vec{x})\cdot\hat{\lambda}^{-1/2}(\vec{x})\,)^{i}_{\ph{i}k}\)
Euclidean integration variables \(\vec{y}\) for \(Z[\vec{x}(\vec{y})]\)
\beq \lb{s1m_6}
\hat{J}^{i}_{\ph{i}k}&=&\frac{\pp x^{i}}{\pp y^{k}}=\Big(\hat{O}^{T}(\vec{x})\cdot
\hat{\lambda}^{-1/2}(\vec{x})\Big)^{i}_{\ph{i}k} \;; \\  \lb{s1m_7}
\mbox{det}\big(\hat{J}^{i}_{\ph{i}k}\big)&=&
\mbox{det}\Big[\big(\,\hat{O}^{T}(\vec{x})\cdot\hat{\lambda}^{-1/2}(\vec{x})\,\big)^{i}_{\ph{i}k}\Big]  =
\mbox{det}\big[\hat{g}^{-1/2}(\vec{x})\big]=\Big(\mbox{det}\big[\hat{g}(\vec{x})\big]\Big)^{-1/2} \;;  \\  \lb{s1m_8}
Z[\vec{x}(\vec{y})]&=&\int d[\vec{y}]\;
\underbrace{\mbox{det}\Big[\hat{g}^{\boldsymbol{-1/2}}(\vec{x})\Big]\;
\sqrt{\mbox{det}\big(\hat{g}(\vec{x})\big)}}_{\equiv 1}\;\exp\big\{\im\;\mscr{A}[\vec{x}(\vec{y})]\big\} \;; \\ \no &=&
Z\ppr[\vec{y}]=\int d[\vec{y}]\;\exp\{\im\;\mscr{A}\ppr[\vec{y}]\}\;; \\  \lb{s1m_9}
\mscr{A}\ppr[\vec{y}] &=& \mscr{A}[\vec{x}(\vec{y})]\;;\hspace*{0.5cm}
Z\ppr[\vec{y}]=Z[\vec{x}(\vec{y})]\;.
\eeq
The functional Taylor expansion of the action is then achieved straightforwardly
where 'classical equations' are determined in a transferred sense from the vanishing
of the first order variation which can be improved by Gaussian integrals of the
second order variation for fluctuations around the 'classical solutions'.
In order to obtain the transformation to Euclidean fields \(\vec{y}\),
it is of particular importance that the metric tensor \(\hat{g}_{ij}(\vec{x})\)
can be diagonalized to the eigenvalues \(\hat{\lambda}^{k}(\vec{x})\).
By taking the (inverse) square root of eigenvalues \([\hat{\lambda}^{k}(\vec{x})]^{\pm 1/2}\),
one acquires the (inverse) square root of the metric tensor
\([\hat{g}(\vec{x})]^{\pm 1/2}\) in combination with the
orthogonal matrix \(\hat{O}_{ji}(\vec{x})\) as the eigenvectors.

In this paper we investigate an analogous problem, but in the more involved context
of a super-symmetric coherent state path
integral \cite{mies1} with a nontrivial integration measure which also contains anti-commuting integration fields.
The measure is given as the square root of the super-determinant
\([\mbox{SDET}(\hat{G}_{\mbox{{\scz Osp}}/ \mbox{{\scz U}}})]^{1/2}\) of the metric tensor
\(\hat{G}_{\mbox{{\scz Osp}}/ \mbox{{\scz U}}}\) in a coset decomposition
\(\mbox{Osp}(S,S|2L)/ \mbox{U}(L|S) \otimes \mbox{U}(L|S)\) of
the ortho-symplectic super-group \(\mbox{Osp}(S,S|2L)\) with the super-unitary
\(\mbox{U}(L|S)\) group as subgroup \cite{cor}-\cite{alice1}. The independent field degrees of freedom
of the final effective actions are restricted to the anomalous
molecular- and BCS- pair condensates in a spontaneous symmetry breaking (SSB) \cite{gold,nambu}
with the coset decomposition \(\mbox{Osp}(S,S|2L)/ \mbox{U}(L|S)\otimes \mbox{U}(L|S)\).
We briefly describe the steepest descent
method by exponentiating and taking the logarithm of the coset integration
measure; furthermore, a detailed account is outlined for the transformation to a Euclidean
coherent state path integration measure with the inverse square root
of the metric tensor \((\hat{G}_{\mbox{{\scz Osp}}/ \mbox{{\scz U}}})^{-1/2}\) as the
appropriate super-Jacobi-matrix. The latter transformation completely
removes the coset integration measure and yields nontrivial classical
field dependence in the actions, but results in simple Euclidean
path integration measures of the independent fields. According
to the Euclidean path integration measure, the variations with the
classical fields in the actions of the exponents allow a
consistent treatment of the non-equilibrium time contour integrals
in the coherent state path integrals \cite{ke}-\cite{gil1}.

One may expect that all the transformations of the anomalous, coset fields only involve spatially
and time-like local expressions as one transforms to the 'flat' Euclidean path integration fields
with the super-Jacobi matrix given by the inverse square root \((\hat{G}_{\mbox{{\scz Osp}}/ \mbox{{\scz U}}})^{-1/2}\)
of the coset metric tensor \(\hat{G}_{\mbox{{\scz Osp}}/ \mbox{{\scz U}}}\). However, we verify that one has to take
into account previous values or time contour histories in the case of phase-valued
transformations of the eigenvalues of the coset order-parameter matrix;
this is in contrast to the {\it absolute values} of eigenvalues of the order-parameter matrix which only yield
local space-time expressions in the transformations. Therefore, the absolute values
of eigenvalues of the coset order-parameter are similar to '{\it state variables}' in the sense of thermodynamics;
on the contrary, the phase values of the eigenvalues of the coset order-parameter matrix require the detailed previous
time contour history in order to achieve the transformed, Euclidean path integration variables. In consequence,
one can compare the transformation of the phases of the coset eigenvalues with the path-dependent
'{\it work}' or '{\it heat}' variables of thermodynamics in a transferred sense. This observed property of our
transformations to Euclidean fields is in accordance with other models, as the transition from incoherent to coherent
laser light, where the phase of the laser light is treated separately (as e.g.\ in a phase diffusion model) or
in analogy to a second order phase transition for the laser threshold \cite{haken1}-\cite{scully}.

Section \ref{e1} is devoted to the issue of finite versus infinite order gradient expansion
of a (super-)determinant. Finite order gradient expansions have the advantage to be related
to known, integrable, classical Sine(h)-Gordon-like equations; however, as one only takes
into account gradually varying spatial gradients of coset matrices, it turns out that the 'inverse'
of these slowly altering gradients is inevitably involved yielding also strongly varying fields
in coordinate space. Therefore, we point out a suitable integral representation for the logarithm
and similarly for the inverse of an operator \cite{eng} so that infinite order gradients are considered in
a reliable manner \cite{inf_g1,bmdis2}.
In sections \ref{s12}, \ref{s13} general properties of super-matrices are reviewed
for the considered case of super-symmetric coherent state path integrals with
'Nambu' doubled super-matrices for the self-energy
(compare Refs.\ \cite{gold,nambu} for the doubling of fields and
see Refs.\ \cite{cor}-\cite{alice1} for more details
concerning super-groups with their super-algebras).
We define the underlying Hamiltonian with the combination of Bose- and Fermi-operators
and introduce symmetry breaking source fields for a coherent BEC-wavefunction
and for coherent molecular- and BCS- pair condensates (compare Refs.\ \cite{lipp}-\cite{dick}
for similar cases in many-body theory). In Ref.\ \cite{mies1} the various
steps and the analysis of super-symmetries \(\mbox{Osp}(S,S|2L)/ \mbox{U}(L|S)\) are
outlined for the transformation to a coherent state path integral with the 'Nambu'
doubled self-energy \(\delta\wt{\Sigma}_{\alpha\beta}^{ab}(\vec{x},t_{p})\;\wt{K}\)
taking values in the ortho-symplectic super-algebra \(\mbox{osp}(S,S|2L)\).
A gradient expansion, combined with a coset decomposition
\(\mbox{Osp}(S,S|2L)/ \mbox{U}(L|S) \otimes \mbox{U}(L|S)\),
reduces the independent angular momentum field degrees of freedom to anomalous terms
whose effective Sine(h)-Gordon-like actions are determined by a real, scalar self-energy
density as background field for effective coupling constants. Since the present paper
aims at the removal of the nontrivial coset integration measure
\((\mbox{SDET}(\hat{G}_{\mbox{{\scz Osp}}/ \mbox{{\scz U}}}))^{1/2}\) by a transformation with the
inverse square root of the coset metric tensor \((\hat{G}_{\mbox{{\scz Osp}}/ \mbox{{\scz U}}})^{-1/2}\),
we also trace out the detailed parametrization of the self-energy as an exact
element of the ortho-symplectic algebra \(\mbox{osp}(S,S|2L)\) in section \ref{s21}.
The \(\mbox{osp}(S,S|2L)\) self-energy generator is separated by a coset decomposition
into \(\mbox{u}(L|S)\) density terms as subalgebra and into \(\mbox{osp}(S,S|2L)/ \mbox{u}(L|S)\)
related anomalous molecular- and BCS- terms whose nontrivial integration measure is
briefly outlined in section \ref{s22}. Section \ref{s23} contains the effective
actions of the coset matrices for pair condensates following from the gradient
expansion with averaging of coupling parameters according to the background self-energy
density field. In section \ref{s24} we apply a scaling to dimensionless fields and
parameters of the actions in the exponentials of coherent state path integrals
with non-Euclidean path integration measure. After general symmetry considerations
in section \ref{s31}, sections \ref{s321} to \ref{s323} finally encompass
the suitable transformations with the
inverse square root of the metric tensor \((\hat{G}_{\mbox{{\scz Osp}}/ \mbox{{\scz U}}})^{-1/2}\)
of \(\mbox{Osp}(S,S|2L)/ \mbox{U}(L|S)\) in order to replace the nontrivial coset integration
measure by Euclidean path integration measures of the independent fields.
In section \ref{s33} diagonal elements of coset matrices as in
\(\hat{T}^{-1}(\vec{x},t_{p})\;(\,\pp\hat{T}(\vec{x},t_{p})\,)\)
are related to the diagonal elements of the new transformed field variables for
anomalous field degrees of freedom having Euclidean path integration measures.
Furthermore, we describe the problem for the '{\it path-dependent}'
phase values of the coset order-parameter matrix
where one has to include nonlocal time contour dependent histories for the transformation
to Euclidean fields. Section \ref{s34} comprises the variations of the effective actions
for classical field equations with the Euclidean path integration variables.
In section \ref{s41} a brief summary is included how the transformations to
Euclidean coherent state path fields effect the observables following from
differentiation with respect to the source fields. We also point out again for
the possible extensions of the few classical integrable systems to chaotic cases
which may be classified in terms of r-s matrices and symmetry breaking extensions of quantum groups
\cite{abl1}-\cite{kerf1}.

\subsection{Finite versus infinite order gradient expansion of determinants} \lb{e1}

There always appears the problem whether the restriction to a finite order gradient expansion
is sufficient for considering a functional determinant in the
'\(\det(\,\hat{\mscr{O}}\,)=\){\it exponential$\{$trace logarithm($\hat{\mscr{O}}$$\}$}' kind.
As one takes only terms with derivatives for
stable, static energy configurations in 3(+1) spatial dimensions, one has to expand from
second up to fourth order gradients so that one cannot scale the particular configuration
to arbitrary small or large sizes in the three dimensional coordinate space integrations
over the static Hamiltonian density ('Derrick's theorem' \cite{naka}). The spatially two-dimensional
case is expected to contribute to the Goldstone modes in a SSB with second order gradients
as a lowest order approximation. Since we reduce the expansion up to second order gradients
in the present paper, we have only extracted the Goldstone modes of the SSB
\(\mbox{Osp}(S,S|2L)\,/\,\mbox{U}(L|S)\otimes\mbox{U}(L|S)\). Following Ref. \cite{mies1},
one can straightforwardly continue with an expansion to higher order gradients according to
the rules and principles of chapter 4 in \cite{mies1}. However, one has to decide which
transport coefficients, composed of the background field, should remain from the
gradient expansion as the unsaturated operators act to the right or left with
commutator relations of the Green functions. The ambiguity, caused by possible partial
integrations between background field coefficients and coset matrices, can be
diminished by applying Ward identities for gauge transformations of the coset generator;
however, there does not remain a unique Lagrangian of finite order gradients because the Ward
identities may also be used under various approximations. In the present paper we
concentrate on the nontrivial coset integration measure which is transformed
to Euclidean path integration variables for straightforward
classical approximations in the lowest, finite order gradient expansions.

These lowest, finite order gradient expansions of the coset matrices have the particular
property to be related to the Sine(h)-Gordon-like equations which allow for various
integrable, classical solutions. Nevertheless, we point out in Refs. \cite{inf_g1,bmdis2} how
to circumvent the problem of finite order gradients by using the integral representation
(\ref{e1_1}) for the logarithm of an operator which is also similarly applied for the
inverse of an operator (compare e.g. \cite{eng})
\beq\lb{e1_1}
\big(\ln\hat{\mscr{O}}\big) &=&\bigg(\int_{0}^{+\infty}dv\;
\frac{\exp\{-v\:\hat{1}\}-\exp\{-v\:\hat{\mscr{O}}\}}{v}\bigg)\;; \\ \lb{e1_2}
\big(\hat{\mscr{O}}^{-1}\big) &=&
\bigg(\int_{0}^{+\infty}dv\;\;
\exp\big\{-v\:\hat{\mscr{O}}\big\}\bigg)\;.
\eeq
As we use the particular form (\ref{e1_3}) for the gradient operator
\((\hat{1}+\delta\hat{\mscr{H}}(\hat{T}^{-1},\hat{T})\;\langle\hat{\mscr{H}}\rangle^{-1})\) with mean field
approximation \(\langle\ldots\rangle\) for the anomalous doubled, one-particle operator
\(\hat{\mscr{H}}\), we obtain the relation \(\hat{T}^{-1}\:\langle\hat{\mscr{H}}\rangle\:\hat{T}\:
\langle\hat{\mscr{H}}\rangle^{-1}\) which results for vanishing source term \(\hat{\mscr{J}}\equiv0\)
into relation (\ref{e1_4}) for the super-determinant
\beq \lb{e1_3}
\delta\hat{\mscr{H}}(\hat{T}^{-1},\hat{T}) &=&\hat{T}^{-1}\:\langle\hat{\mscr{H}}\rangle\:\hat{T}-
\langle\hat{\mscr{H}}\rangle  \;;  \\
\hat{\mscr{O}} &=&\bigg(\hat{1}+\Big(\delta\hat{\mscr{H}}(\hat{T}^{-1},\hat{T})+
\wt{\mscr{J}}(\hat{T}^{-1},\hat{T})\Big)\:
\langle\hat{\mscr{H}}\rangle^{-1}\bigg)   \\ \no &=&
\hat{T}^{-1}\;\langle\hat{\mscr{H}}\rangle\;\hat{T}\;\langle\hat{\mscr{H}}\rangle^{-1}+
\wt{\mscr{J}}(\hat{T}^{-1},\hat{T})\;\langle\hat{\mscr{H}}\rangle^{-1}  \;;  \\ \lb{e1_4}
\mscr{A}_{SDET}[\hat{T},\langle\hat{\mscr{H}}\rangle;\hat{\mscr{J}}\equiv0] &=&
\frac{1}{2}\;\;\mbox{tr}\;\mbox{STR}\ln
\Big[{T}^{-1}\;\langle\hat{\mscr{H}}\rangle\;\hat{T}\;\langle\hat{\mscr{H}}\rangle^{-1}\Big] \\ \no
&=&\frac{1}{2}\int_{0}^{+\infty}dv\;\;
\mbox{tr}\;\mbox{STR}\bigg[\frac{\exp\{-v\:\hat{1}\}-
\exp\{-v\:\hat{T}^{-1}\;\langle\hat{\mscr{H}}\rangle\;\hat{T}\;\langle\hat{\mscr{H}}\rangle^{-1}\}}{v}\bigg]\;.
\eeq
If one assumes slowly varying finite order gradients of
\(\hat{T}^{-1}\,\langle\hat{\mscr{H}}\rangle\,\hat{T}\), one will also
obtain unintented, extraordinary large spatial and time-like variations with
\(\hat{T}\,\langle\hat{\mscr{H}}\rangle^{-1}\,\hat{T}^{-1}\)
according to the additional trace operation on the logarithm.
In order to circumvent this problem, we use in Refs. \cite{inf_g1,bmdis2}
the particular integral representation (\ref{e1_1},\ref{e1_2}) for the logarithm of an operator
(and similarly for the inverse) which gives a simple representation of the logarithm with the (coset matrix weighted)
combination of \(\langle\hat{\mscr{H}}\rangle\) and its inverse
\(\langle\hat{\mscr{H}}\rangle^{-1}\) in an exponential.
One can emphasize this point by a gauge transformation of the coset decomposition so that the mean field
operator \(\langle\hat{\mscr{H}}\rangle\) is simplified to pure spatial gradient operators (compare
Ref. \cite{inf_g1}).

If we suppose finite, positive eigenvalues for the total operator
\(\hat{\mscr{O}}=\hat{T}^{-1}\,\langle\hat{\mscr{H}}\rangle\,\hat{T}\,
\langle\hat{\mscr{H}}\rangle^{-1}\), the inverse factorials \(1/n!\)
of \(\exp\{-v\:\hat{T}^{-1}\,\langle\hat{\mscr{H}}
\rangle\,\hat{T}\,\langle\hat{\mscr{H}}\rangle^{-1}\,\}\)
cause a meaningful expansion and convergence instead of a pure logarithm
\(\ln(\,\hat{T}^{-1}\,\langle\hat{\mscr{H}}\rangle\,\hat{T}\,
\langle\hat{\mscr{H}}\rangle^{-1}\,)\)
with reciprocal integer numbers in the expansion.
Therefore, one can also rely on the integral representations
(\ref{e1_1},\ref{e1_2}) for the logarithm and for the inverse
of an operator \(\hat{\mscr{O}}\) and can apply these relations for reducing
the path integral part with coset matrices
\(\hat{T}(\vec{x},t_{p})=\exp\{-\hat{Y}(\vec{x},t_{p})\,\}\) and coset generator
\(\hat{Y}(\vec{x},t_{p})\) for molecular and BCS-pair condensates.
We can even choose the eigenbasis of the mean field approximated, one-particle operator \(\langle\hat{H}\rangle\)
or of its anomalous doubled version \(\langle\hat{\mscr{H}}\rangle\)
instead of the d+1 dimensional coordinate representation.
This particular matrix representation for \(\hat{T}\)
in terms of the eigenbasis of \(\langle\hat{\mscr{H}}\rangle\) allows to calculate
observables as correlation functions of anomalous super-field combinations
\(\langle\psi_{\vec{x},\alpha}(t_{p})\:\psi_{\vec{x}\ppr,\beta}(t_{p}\ppr)\rangle\),
density terms \(\langle\psi_{\vec{x},\alpha}^{*}(t_{p})\:\psi_{\vec{x}\ppr,\beta}(t_{p})\rangle\)
and eigenvalue correlations of molecular and BCS-terms.

\subsection{Bosonic and fermionic operators with their coherent state field representations}\lb{s12}

In this section we briefly define the super-fields from their corresponding
bosonic and fermionic operators. We introduce the complex conjugation,
super-transposition, super-trace and super-determinant in analogy to the
operations on ordinary matrices \cite{cor}-\cite{alice1}. The basic constituents are determined by
super-fields \(\psi_{\vec{x},\alpha}(t)\) which have \(L(=2l+1)\) odd-numbered bosonic
and \(S(=2s+1)\) even-numbered fermionic angular momentum degrees of freedom. The summations
over these bosonic and fermionic angular momentum degrees of freedom are abbreviated
by the first greek letters \(\alpha,\:\beta,\:\gamma\ldots\) in bilinear or quartic
relations of the super-fields. We specify these \(N=L+S\) component super-fields
in (\ref{s1_1}) with internal bosonic vector \(\vec{b}_{\vec{x}}(t)=\{b_{\vec{x},m}(t)\}\)
and internal Grassmann-valued, fermionic vector
\(\vec{\alpha}_{\vec{x}}(t)=\{\alpha_{\vec{x},r}(t)\}\)
\beq \lb{s1_1}
\alpha\;,\;\beta\;,\ldots&=&\underbrace{-l,\ldots,+l}_{\mbox{L bosons}}\;;\;
\underbrace{-s,\ldots,+s}_{\mbox{S fermions}}\;;\hspace*{0.5cm}
l=0,1,2,\ldots\;;\hspace*{0.5cm}s=\frac{1}{2},\frac{3}{2},\frac{5}{2},\ldots\;;
 \\ \no N&=&L+S\;; \hspace*{1.0cm}L=2l+1\;;\hspace*{1.0cm}S=2s+1\;; \\ \no
\psi_{\vec{x},\alpha}(t)&=&\left( \bea{c}
\vec{b}_{\vec{x}}(t) \\
\vec{\alpha}_{\vec{x}}(t) \eea\right)\;;\hspace*{0.5cm} \bea{rcl}
\vec{b}_{\vec{x}}(t)&=&\big\{b_{\vec{x},m}(t)\big\}=\Big\{b_{\vec{x},-l}(t)\;,\;
\ldots\;,\;b_{\vec{x},+l}(t)\Big\}^{T}; \\
\vec{\alpha}_{\vec{x}}(t)&=&\big\{\alpha_{\vec{x},r}(t)\big\}=\Big\{\alpha_{\vec{x},-s}(t)\;,\;
\ldots\;,\;\alpha_{\vec{x},+s}(t)\Big\}^{T};\eea \\ \no
\psi_{\vec{x},\alpha}^{+}(t)&=&\Big(b_{\vec{x},-l}^{*}(t)\;,\;\ldots\;,\;
b_{\vec{x},+l}^{*}(t)\;;\;\alpha_{\vec{x},-s}^{*}(t)\;,\; \ldots\;,\;\alpha_{\vec{x},+s}^{*}(t)\Big)\;.
\eeq
These coherent super-fields are applied on the non-equilibrium time contour to define
the coherent state path integral following from the Hamilton operator (\ref{s1_3}) with
combined Bose- and Fermi-operators in (\ref{s1_2}) \footnote{The spatial sum \(\sum_{\vec{x}}\ldots\)
is dimensionless and is scaled with the system volume so that \(\sum_{\vec{x}}\) is equivalent to
\(\int_{L^{d}}\;(d^{d}x/L^{d})\ldots\) .}
\be \lb{s1_2}
\hat{\psi}_{\vec{x},\alpha}=\left\{\hat{\vec{b}}_{\vec{x}}\;,\; \hat{\vec{\alpha}}_{\vec{x}}\right\}^{T}\;;\hspace*{1.0cm}
\hat{\psi}_{\vec{x},\alpha}^{+}=\left\{\hat{\vec{b}}_{\vec{x}}^{+}\;,\; \hat{\vec{\alpha}}_{\vec{x}}^{+}\right\}\;;
\ee
\beq \lb{s1_3}
\lefteqn{\hat{H}(\hat{\psi}^{+},\hat{\psi},t)= \sum_{\vec{x}}\sum_{\alpha}
\hat{\psi}^{+}_{\vec{x},\alpha}\;\; \hat{h}(\vec{x})\;\; \hat{\psi}_{\vec{x},\alpha}+
\sum_{\vec{x},\vec{x}\ppr}\sum_{\alpha,\beta}
\hat{\psi}^{+}_{\vec{x}\ppr,\beta} \hat{\psi}^{+}_{\vec{x},\alpha}\;V_{|\vec{x}\ppr-\vec{x}|}\;
\hat{\psi}_{\vec{x},\alpha} \hat{\psi}_{\vec{x}\ppr,\beta} + } \\ \no &+& \sum_{\vec{x},\alpha}
\Big(j_{\psi;\alpha}^{*}(\vec{x},t)\;\hat{\psi}_{\vec{x},\alpha} +
\hat{\psi}_{\vec{x},\alpha}^{+}(t)\;j_{\psi;\alpha}(\vec{x},t)\Big) +
\\ \no &+&\frac{1}{2}
\sum_{\vec{x}}\strab\Bigg[\;\wt{j}_{\psi\psi;N\times N}^{+}(\vec{x},t)\; \left( \bea{cc}
\hat{c}_{\vec{x},L\times L} & \hat{\eta}^{T}_{\vec{x},L\times S} \\
\hat{\eta}_{\vec{x},S\times L} & \hat{f}_{\vec{x},S\times S} \eea\right) + \left( \bea{cc}
\hat{c}^{+}_{\vec{x},L\times L} & \hat{\eta}^{+}_{\vec{x},L\times S} \\
\hat{\eta}^{*}_{\vec{x},S\times L} & \hat{f}^{+}_{\vec{x},S\times S}
\eea\right)\;\wt{j}_{\psi\psi;N\times N}(\vec{x},t)\; \Bigg]_{\mbox{.}}
\eeq
Aside from the one-particle operator \(\hat{h}(\vec{x})\) (\ref{s1_4}) with kinetic energy,
trap- and chemical potential \(u(\vec{x})\), \(\mu_{0}\), we include
a short-ranged quartic interaction potential \(V_{|\vec{x}\ppr-\vec{x}|}\)
with super-symmetry between Bose- and Fermi-particles. These two operator terms obey a
global super-unitary invariance \(\mbox{U}(L|S)\) so that a super-symmetry results between
bosonic and fermionic angular momentum degrees of freedom. We assume that this super-symmetry
may be achieved by appropriate tuning of Feshbach resonances with similar effective masses
and similar properties concerning the trap potential \cite{hulet2}-\cite{bm1}
\beq \lb{s1_4}
\hat{h}(\vec{x})&=&
\frac{\hat{\vec{p}}^{\;2}}{2m}+u(\vec{x})-\mu_{0} \;; \\  \lb{s1_5}
j_{\psi;N}(\vec{x},t)&=&\Big\{j_{\psi;B,L}(\vec{x},t)\;;\;
j_{\psi;F,S}(\vec{x},t)\Big\}^{T}_{\mbox{.}}
\eeq
Apart from the \(\mbox{U}(L|S)\) symmetry breaking source field \(j_{\psi;\alpha}(\vec{x},t)\)
(\ref{s1_5}) for a coherent BEC wavefunction, we specialize on the investigation of
super-symmetric pair condensates which are created by the \(N\times N=(L+S)\times (L+S)\)
super-symmetric source matrix \(\wt{j}_{\psi\psi;N\times N}(\vec{x},t)\) in (\ref{s1_3}).
The boson-boson pair condensate terms are denoted by a \(L\times L\) symmetric operator
matrix \(\hat{c}_{\vec{x},L\times L}\) (\ref{s1_6}) and the fermion-fermion
pair condensates (acting as a boson in its entity) are marked by the anti-symmetric
operator matrix \(\hat{f}_{\vec{x},S\times S}\) (\ref{s1_7}). The fermion-boson
mixed operator (\ref{s1_8}) and its transpose (\ref{s1_9}) are abbreviated by
\(\hat{\eta}_{\vec{x},S\times L}\) and \(\hat{\eta}_{\vec{x},L\times S}^{T}\) and
have fermionic properties as an entity, due to the composition of a boson and
fermion operator
\beq \lb{s1_6}
\hat{c}_{\vec{x},L\times L}&=&\big\{\hat{c}_{\vec{x},mn}\big\}=\big\{\hat{b}_{\vec{x},m}\;
\hat{b}_{\vec{x},n}\big\}\;;\hspace*{0.5cm}\hat{c}_{\vec{x},mn}^{T}=\hat{c}_{\vec{x},mn}\;;
\hspace*{0.5cm}m,n=-l,\ldots,+l\;; \\   \lb{s1_7}
\hat{f}_{\vec{x},S\times S}&=&\big\{\hat{f}_{\vec{x},rr\ppr}\big\}=\big\{ \hat{\alpha}_{\vec{x},r}\;
\hat{\alpha}_{\vec{x},r\ppr}\big\}\;;\hspace*{0.5cm}\hat{f}_{\vec{x},rr\ppr}^{T}=-\hat{f}_{\vec{x},rr\ppr}\;;
\hspace*{0.5cm}r,r\ppr= -s,\ldots,+s \;; \\   \lb{s1_8}
\hat{\eta}_{\vec{x},S\times L}&=&\big\{\hat{\eta}_{\vec{x},rm}\big\}=\big\{
\hat{\alpha}_{\vec{x},r}\;\hat{b}_{\vec{x},m}\big\}\;;\hspace*{0.5cm}r=-s,\ldots,+s\;\;;\hspace*{0.25cm} m=-l,\ldots,+l \;;
\\  \lb{s1_9} \hat{\eta}_{\vec{x},L\times S}^{T}&=&\big\{\hat{\eta}_{\vec{x},mr}^{T}\big\}=\big\{
\hat{b}_{\vec{x},m}\;\hat{\alpha}_{\vec{x},r}\big\}\;;\hspace*{0.5cm}m=-l,\ldots,+l\;\;;\hspace*{0.37cm} r=-s,\ldots,+s\;.
\eeq
The \(N\times N\) source matrix \(\wt{j}_{\psi\psi;N\times N}(\vec{x},t)\) (\ref{s1_10}) has to respect
the symmetry properties of the super-symmetric, paired terms in (\ref{s1_6}-\ref{s1_9})
and therefore has a symmetric even sub-matrix \(\hat{j}_{B;L\times L}(\vec{x},t)\)
for the boson-boson pair condensates (\ref{s1_11}) and an anti-symmetric even sub-matrix \(\hat{j}_{F;S\times S}(\vec{x},t)\)
for the corresponding fermion-fermion paired terms (\ref{s1_12}).
Furthermore, Grassmann or anti-commuting fields \(\hat{j}_{\eta;S\times L}(\vec{x},t)\),
\(\hat{j}_{\eta;L\times S}^{T}(\vec{x},t)\) (\ref{s1_13}) generate
the boson-fermion \(\hat{\eta}_{\vec{x},L\times S}^{T}\) (\ref{s1_9})
or fermion-boson \(\hat{\eta}_{\vec{x},S\times L}\) (\ref{s1_8}) pair condensates
in a super-trace relation. Appropriate signs have to be taken into account due to
the property of a super-trace in the fermion-fermion part of a super-matrix
\beq \lb{s1_10}
\wt{j}_{\psi\psi;N\times N}(\vec{x},t)&=& \left( \bea{cc}
\hat{j}_{B;L\times L}(\vec{x},t) & -\hat{j}^{T}_{\eta;L\times S}(\vec{x},t) \\
\hat{j}_{\eta;S\times L}(\vec{x},t) & -\hat{j}_{F;S\times S}(\vec{x},t) \eea\right)\;; \\ \lb{s1_11}
\hat{j}_{B;L\times L}(\vec{x},t)&\in& \mbox{\sf C}_{even}\;;\hspace*{0.5cm}
\hat{j}_{B;mn}^{T}(\vec{x},t)=\hat{j}_{B;mn}(\vec{x},t)\;; \\ \lb{s1_12}
\hat{j}_{F;S\times S}(\vec{x},t)&\in&\mbox{\sf C}_{even}\;;\hspace*{0.5cm}
\hat{j}_{F;rr\ppr}^{T}(\vec{x},t)=-\hat{j}_{F;rr\ppr}(\vec{x},t)\;;\hspace*{0.5cm}
\hat{j}_{F;rr}(\vec{x},t)=0 \;; \\ \lb{s1_13}
\hat{j}_{\eta;S\times L}(\vec{x},t)&=&\{j_{\eta;rm}(\vec{x},t)\}\in\mscr{C}_{odd}\;;\hspace*{0.5cm}m=-l,\ldots,+l\;; \\ \no
\hat{j}_{\eta;L\times S}^{+}(\vec{x},t) &=& \{j_{\eta;mr}^{+}(\vec{x},t)\}\in\mscr{C}_{odd}\;; \hspace*{0.5cm}
r=-s,\ldots,+s\;.
\eeq
In the following we define the complex conjugation (\ref{s1_14}) of Grassmann variables
and the super-transposition (\ref{s1_16},\ref{s1_17}), the super-trace (\ref{s1_18}),
the super-hermitian conjugation (\ref{s1_19}) and the super-determinant (\ref{s1_20}) of
graded- or super-matrices as \(\hat{N}_{1}\), \(\hat{N}_{2}\) (\ref{s1_15}) \cite{cor}-\cite{alice1}.
The complex conjugation of a product \((\xi_{1}\ldots\xi_{i}\ldots\xi_{n})^{*}\)
of anti-commuting variables \(\xi_{i}\) changes these n-factors to its reversed
order with complex conjugated, odd numbers \(\xi_{i}^{*}\). This definition
provides the combination \(\xi_{i}^{*}\xi_{i}\) of a Grassmann number \(\xi_{i}\)
and its complex conjugate \(\xi_{i}^{*}\) with the property of an even, real
(but nilpotent) variable
\be \lb{s1_14}
\left(\xi_{1}\ldots\xi_{i}\ldots\xi_{n}\right)^{*}=\xi_{n}^{*}\ldots\xi_{i}^{*}\ldots\xi_{1}^{*}\;;
\hspace*{0.5cm}(\xi_{i}^{*})^{*}=\xi_{i}\;;\hspace*{0.5cm}
(\xi_{i}^{*}\xi_{i})^{*}=\xi_{i}^{*}(\xi_{i}^{*})^{*}=\xi_{i}^{*}\xi_{i}\;.
\ee
Super-matrices as \(\hat{N}_{1}\), \(\hat{N}_{2}\) (\ref{s1_15}) consist of the even
boson-boson blocks \(\hat{c}_{1}\), \(\hat{c}_{2}\) and the even fermion-fermion
blocks \(\hat{f}_{1}\), \(\hat{f}_{2}\). Sub-matrices of anti-commuting variables
are placed in the non-diagonal fermion-boson blocks \(\hat{\chi}_{1}\), \(\hat{\chi}_{2}\)
and boson-fermion blocks \(\hat{\eta}_{1}^{T}\), \(\hat{\eta}_{2}^{T}\).
Under a super-transposition '$\mbox{st}$' of the super-matrices \(\hat{N}_{1}\), \(\hat{N}_{2}\)
(\ref{s1_16}), the even parts \(\hat{c}_{1}\), \(\hat{c}_{2}\) and
\(\hat{f}_{1}\), \(\hat{f}_{2}\) are transposed in the manner of ordinary matrices
whereas the fermion-boson blocks \(\hat{\chi}_{1}\), \(\hat{\chi}_{2}\)
and boson-fermion blocks \(\hat{\eta}_{1}^{T}\), \(\hat{\eta}_{2}^{T}\) are
exchanged with transposition and with the inclusion of an additional minus sign
in the resulting fermion-boson blocks \(-\hat{\eta}_{1}\), \(-\hat{\eta}_{2}\) (\ref{s1_16}).
This definition of super-transposition preserves the property of ordinary matrices to
be reversed under transposition in a product of matrices (\ref{s1_17})
\beq \lb{s1_15}
\hat{N}_{1}&=&\left( \bea{cc}
\hat{c}_{1} & \hat{\eta}^{T}_{1} \\
\hat{\chi}_{1} & \hat{f}_{1} \eea\right)\;;
\hspace*{0.28cm}
\hat{N}_{2}=\left( \bea{cc}
\hat{c}_{2} & \hat{\eta}^{T}_{2} \\
\hat{\chi}_{2} & \hat{f}_{2} \eea\right)\;;  \\ \lb{s1_16}
\hat{N}^{st}_{1}&=&\left( \bea{cc}
\hat{c}^{T}_{1} & \hat{\chi}^{T}_{1} \\
-\hat{\eta}_{1} & \hat{f}^{T}_{1}
\eea\right)\;;\hspace*{0.28cm}
\hat{N}^{st}_{2}=\left( \bea{cc}
\hat{c}^{T}_{2} & \hat{\chi}^{T}_{2} \\
-\hat{\eta}_{2} & \hat{f}^{T}_{2}
\eea\right) \;;    \\   \lb{s1_17}
(\hat{N}_{1}\;\cdot\;\hat{N}_{2})^{st}&=&\hat{N}_{2}^{st}\;\cdot\;\hat{N}_{1}^{st}\;.
\eeq
The super-trace '$\mbox{str}$' of a super-matrix \(\hat{N}\) comprises the traces of the
even boson-boson part \(\hat{c}\) and the even fermion-fermion part \(\hat{f}\) (\ref{s1_18}).
However, an additional minus sign has to be included in the trace of the fermion-fermion
part so that the cyclic invariance of a product of super-matrices is maintained in a
super-trace relation as in a trace with the product of several ordinary matrices
\be   \lb{s1_18}
\mbox{str}[\hat{N}]=\mbox{str}\left(
\bea{cc}
\hat{c} & \hat{\eta}^{T} \\
\hat{\chi} & \hat{f}
\eea\right)=\mbox{tr}[\hat{c}]-\mbox{tr}[\hat{f}]\;; \hspace*{0.5cm}
\mbox{str}[\hat{N}_{1}\;\hat{N}_{2}]=\mbox{str}[\hat{N}_{2}\;\hat{N}_{1}] \;.
\ee
The super-hermitian conjugation (\ref{s1_19}) of super-matrices
\(\hat{N}_{1}\), \(\hat{N}_{2}\) (\ref{s1_15}) does not involve
additional minus signs as the super-transposition (\ref{s1_16},\ref{s1_17}). In comparison
to (\ref{s1_17}), the property \((\hat{N}_{1}\;\hat{N}_{2})^{+}=\hat{N}_{2}^{+}\;\hat{N}_{1}^{+}\)
(reversal of a product under super-hermitian conjugation) is already contained
without additional minus signs because the
complex conjugation (\ref{s1_14}) of a product of anti-commuting numbers
is defined with an exchange of the factors to its reversed order
\be\lb{s1_19}
\hat{N}_{1}^{+}=\left( \bea{cc}
\hat{c}_{1}^{+} & \hat{\chi}_{1}^{+} \\
\hat{\eta}_{1}^{*} & \hat{f}_{1}^{+} \eea\right)\;;\hspace*{0.5cm}
\hat{N}_{2}^{+}=\left( \bea{cc}
\hat{c}_{2}^{+} & \hat{\chi}_{2}^{+} \\
\hat{\eta}_{2}^{*} & \hat{f}_{2}^{+} \eea\right)\;;\hspace*{0.5cm}
(\hat{N}_{1}\;\hat{N}_{2})^{+}=\hat{N}_{2}^{+}\;\hat{N}_{1}^{+}\;.
\ee
The definition of a super-determinant '$\mbox{sdet}$'
for a super-matrix \(\hat{N}\) is generalized from the relation
\(\mbox{det}(\hat{M})=\exp\{\mbox{tr}\ln(\hat{M})\}\) of ordinary
matrices \(\hat{M}\). The ordinary trace relation '\(\mbox{tr}\)'
in the exponent is generalized with the super-trace '\(\mbox{str}\)'
(\ref{s1_18}) for super-matrices \(\hat{N}\), consisting of the even
boson-boson, fermion-fermion blocks \(\hat{c}_{L\times L}\), \(\hat{f}_{S\times S}\)
and the odd fermion-boson, boson-fermion blocks
\(\hat{\chi}_{S\times L}\), \(\hat{\eta}_{L\times S}^{T}\). Using the properties
of a super-trace '\(\mbox{str}\)' (as cyclic invariance), one can transform
the generalized relation \(\mbox{sdet}(\hat{N}_{N\times N})=
\exp\{\mbox{str}\ln(\hat{N}_{N\times N})\}\) (\ref{s1_20}) to ordinary \(L\times L\) and
\(S\times S\) determinants where the determinant \(\mbox{det}(\hat{f}_{S\times S})\)
of the even fermion-fermion section appears in the denominator because of the
additional negative sign in the fermion-fermion section of a super-trace
\beq \lb{s1_20}
\mbox{sdet}\big(\hat{N}_{N\times N}\big)&\stackrel{!}{=}&\exp\left\{\mbox{str}\ln\left(
\bea{cc}
\hat{c}_{L\times L} & \hat{\eta}_{L\times S}^{T} \\
\hat{\chi}_{S\times L} & \hat{f}_{S\times S}
\eea\right)\right\}  \\ \no &=& \exp\left\{\mbox{str}\ln\left(
\bea{cc}
\hat{c}_{L\times L} & 0 \\
 0 & \hat{f}_{S\times S}
\eea\right)\left(
\bea{cc}
\hat{1}_{L\times L} & \hat{c}_{L\times L}^{-1}\;\hat{\eta}_{L\times S}^{T} \\
\hat{f}_{S\times S}^{-1}\;\hat{\chi}_{S\times L} & \hat{1}_{S\times S}
\eea\right)\right\}      \\ \no &=&
\frac{\det\big(\hat{c}_{L\times L}-
\hat{\eta}_{L\times S}^{T}\hat{f}_{S\times S}^{-1}\;\hat{\chi}_{S\times L}\big)}{\det\big(\hat{f}_{S\times S}\big)}\;.
\eeq
In the case of a product of super-matrices, the property \(\mbox{sdet}(\hat{N}_{1}\;\hat{N}_{2})=
\mbox{sdet}(\hat{N}_{1})\;\mbox{sdet}(\hat{N}_{2})\) for the factorization of
the super-determinant holds in a similar manner
as in the case with ordinary matrices because of the cyclic invariance
in the super-trace (\ref{s1_18}).

\subsection{The super-symmetric coherent state path integral}\lb{s13}

In the remainder we consider the time contour integral (\ref{s1_21}) with the
time variable $t_{p}$ on the two branches \(p=\pm\) for the time development
of the Hamiltonian (\ref{s1_3}) in forward \(\int_{-\infty}^{\infty}dt_{+}\ldots\)
and backward \(\int_{\infty}^{-\infty}dt_{-}\ldots\) direction \cite{ke}-\cite{gil1}.
The negative sign of the backward propagation
\(\int_{\infty}^{-\infty}dt_{-}\ldots=-\int_{-\infty}^{\infty}dt_{-}\ldots\)
will be frequently taken into account by the time contour metric-symbol
\(\eta_{p=\pm}=p=\pm\)
\beq \lb{s1_21}
\int_{C}d t_{p}\ldots &=&\int_{-\infty}^{+\infty}d
t_{+}\ldots+ \int_{+\infty}^{-\infty}d t_{-}\ldots= \int_{-\infty}^{+\infty}d t_{+}\ldots-
\int_{-\infty}^{+\infty}d t_{-}\ldots \\ \no &=&
\sum_{p=\pm}\int_{-\infty}^{\infty}d t_{p}\;\;\eta_{p}\;\;\ldots\;;\hspace*{0.5cm}
(\eta_{p=\pm}=\pm)\;.
\eeq
The corresponding coherent state path integral
\(Z[\hat{\mscr{J}},j_{\psi},\wt{j}_{\psi\psi}]\) \cite{ke}-\cite{gil1},\cite{mies1}
of the Hamiltonian (\ref{s1_3}) with its symmetry breaking source
fields \(j_{\psi;N}(\vec{x},t)\) (\ref{s1_5}) and
\(\wt{j}_{\psi\psi;N\times N}(\vec{x},t)\) (\ref{s1_10}-\ref{s1_13}) is given in
relation (\ref{s1_22}) with inclusion of the time contour integrals (\ref{s1_21})
in the exponentials
\beq \lb{s1_22}
\lefteqn{Z[\hat{\mscr{J}},j_{\psi},\wt{j}_{\psi\psi}]=
\int d[\psi_{\vec{x},\alpha}(t_{p})] \;\;
\exp\bigg\{-\frac{\im}{\hbar}\int_{C}d t_{p}\sum_{\vec{x}}\sum_{\alpha}
\psi_{\vec{x},\alpha}^{*}(t_{p})\;\;\hat{H}_{p}(\vec{x},t_{p})\;\;
\psi_{\vec{x},\alpha}(t_{p})\bigg\} } \\ \no &\times& \exp\bigg\{-\frac{\im}{\hbar}
\int_{C}dt_{p}\sum_{\vec{x},\vec{x}\ppr}\sum_{\alpha,\beta}
\psi_{\vec{x}\ppr,\beta}^{*}(t_{p})\;\psi_{\vec{x},\alpha}^{*}(t_{p})\;
V_{|\vec{x}\ppr-\vec{x}|}\;\psi_{\vec{x},\alpha}(t_{p})\;\psi_{\vec{x}\ppr,\beta}(t_{p})
\bigg\} \\ \no &\times &
\exp\bigg\{-\frac{\im}{\hbar}\int_{C}d t_{p}\sum_{\vec{x}}\sum_{\alpha}
\Big(j_{\psi;\alpha}^{*}(\vec{x},t_{p})\;\psi_{\vec{x},\alpha}(t_{p})+
\psi_{\vec{x},\alpha}^{*}(t_{p})\;j_{\psi;\alpha}(\vec{x},t_{p})\Big)\bigg\}
\\ \no &\times&
\exp\Bigg\{-\frac{\im}{2\hbar}\int_{C}d t_{p} \sum_{\vec{x}}\strab\Bigg[ \left( \bea{cc}
\hat{j}_{B;L\times L}^{+}(\vec{x},t_{p}) & \hat{j}_{\eta;L\times S}^{+}(\vec{x},t_{p}) \\
-\hat{j}_{\eta;S\times L}^{*}(\vec{x},t_{p}) & -\hat{j}_{F;S\times S}^{+}(\vec{x},t_{p}) \eea\right)\left( \bea{cc}
\hat{c}_{L\times L}(\vec{x},t_{p}) & \hat{\eta}_{L\times S}^{T}(\vec{x},t_{p}) \\
\hat{\eta}_{S\times L}(\vec{x},t_{p}) & \hat{f}_{S\times S}(\vec{x},t_{p}) \eea\right)+ \\ \no &+&\left( \bea{cc}
\hat{c}_{L\times L}^{+}(\vec{x},t_{p}) & \hat{\eta}_{L\times S}^{+}(\vec{x},t_{p}) \\
\hat{\eta}_{S\times L}^{*}(\vec{x},t_{p}) & \hat{f}_{S\times S}^{+}(\vec{x},t_{p}) \eea\right)\left( \bea{cc}
\hat{j}_{B;L\times L}(\vec{x},t_{p}) & -\hat{j}^{T}_{\eta;L\times S}(\vec{x},t_{p}) \\
\hat{j}_{\eta;S\times L}(\vec{x},t_{p}) & -\hat{j}_{F;S\times S}(\vec{x},t_{p}) \eea\right)\Bigg]\Bigg\} \\ \no &\times&
\exp\bigg\{-\frac{\im}{2\hbar}\int_{C}d t_{p_{1}}^{(1)}\; d t_{p_{2}}^{(2)}
\sum_{\vec{x},\vec{x}\ppr}\sum_{\alpha,\beta} \Psi_{\vec{x}\ppr,\beta}^{+b}(t_{p_{2}}^{(2)})\;
\hat{\mscr{J}}_{\vec{x}\ppr,\beta;\vec{x},\alpha}^{ba}(t_{p_{2}}^{(2)};
t_{p_{1}}^{(1)})\;\Psi_{\vec{x},\alpha}^{a}(t_{p_{1}}^{(1)})\bigg\}_{\mbox{.}}
\eeq
Due to a missing potential for disorder with an ensemble average, the super-symmetric
coherent state fields \(\psi_{\vec{x},\alpha}(t_{p})\),  \(\psi_{\vec{x},\alpha}^{*}(t_{p})\)
only couple on a single specific branch of the time contour without any combinations
between forward '+' and backward '-' propagation (compare with the Refs.\ \cite{bmdis1}-\cite{bmdis2}
in the case of disorder). The one-particle operator \(\hat{h}(\vec{x})\) (\ref{s1_4})
is completed to \(\hat{H}_{p}(\vec{x},t_{p})\) with the time contour derivative
\(\hat{E}_{p}=\im\hbar\;\pp/\pp t_{p}\) and the imaginary time contour increment
\(\im\:\ve_{p=\pm}=(\pm)\:\im\;\ve\), (\(\ve>0_{+}\)) for appropriate convergence
properties of Green functions with propagation according to
suitable time directions
\be \lb{s1_23}
\hat{H}_{p}(\vec{x},t_{p}) = -\hat{E}_{p}-\im\;\ve_{p}+\hat{h}(\vec{x}) =
-\im\hbar\frac{\pp}{\pp t_{p}}-\im\;\ve_{p}+
\frac{\hat{\vec{p}}^{\;2}}{2m}+u(\vec{x})-\mu_{0}\;.
\ee
A further
source matrix \(\hat{\mscr{J}}_{\vec{x}\ppr,\beta;\vec{x},\alpha}^{ba}(t_{p_{2}}^{(2)},t_{p_{1}}^{(1)})\)
is incorporated in the coherent state path integral
\(Z[\hat{\mscr{J}},j_{\psi},\wt{j}_{\psi\psi}]\) (\ref{s1_22})
because it combines 'Nambu' doubled coherent state fields
\(\Psi_{\vec{x},\alpha}^{a(=1/2)}(t_{p})=\{\psi_{\vec{x},\alpha}(t_{p})\:;\:
\psi_{\vec{x},\alpha}^{*}(t_{p})\}^{T}\) \cite{gold,nambu}. Therefore, it is possible to generate
anomalous terms as \(\langle\psi_{\vec{x},\beta}(t_{p})\;\psi_{\vec{x},\alpha}(t_{p})\rangle\)
by a single differentiation of \(Z[\hat{\mscr{J}},j_{\psi},\wt{j}_{\psi\psi}]\) (\ref{s1_22})
with respect to \(\hat{\mscr{J}}_{\vec{x},\beta;\vec{x},\alpha}^{21}(t_{p},t_{p})\).
Furthermore, one has to distinguish between source fields
\(j_{\psi;\alpha}(\vec{x},t_{p})\), \(\wt{j}_{\psi\psi;\alpha\beta}(\vec{x},t_{p})\) with
a dependence on the time contour branch '\(p=\pm\)' for generating observables by
differentiation of (\ref{s1_22}) and the corresponding 'condensate seed' fields
\(j_{\psi;\alpha}(\vec{x},t)\), \(\wt{j}_{\psi\psi;\alpha\beta}(\vec{x},t)\)
for SSB \cite{lipp}-\cite{dick}. The latter 'condensate seeds' follow by setting
the corresponding time branches to equivalent finite values (\ref{s1_24},\ref{s1_25});
this has to be performed at the final end of calculations with
\(Z[\hat{\mscr{J}},j_{\psi},\wt{j}_{\psi\psi}]\) (\ref{s1_22}) after the prevailing
observables have been determined by differentiation with respect to
\(j_{\psi;\alpha}(\vec{x},t_{p})\), \(\wt{j}_{\psi\psi;\alpha\beta}(\vec{x},t_{p})\) or
with respect to \(\hat{\mscr{J}}_{\vec{x}\ppr,\beta;\vec{x},\alpha}^{ba}(t_{p_{2}}^{(2)},t_{p_{1}}^{(1)})\).
The last source matrix
\(\hat{\mscr{J}}_{\vec{x}\ppr,\beta;\vec{x},\alpha}^{ba}(t_{p_{2}}^{(2)},t_{p_{1}}^{(1)})\) has then
to be set to zero with remaining {\it finite} 'condensate seeds'
\(j_{\psi;\alpha}(\vec{x},t)\), \(\wt{j}_{\psi\psi;\alpha\beta}(\vec{x},t)\)
for the creation of a coherent BEC-wavefunction and for the creation of
pair condensates with super-symmetry on the coset space \(\mbox{Osp}(S,S|2L)/ \mbox{U}(L|S)\)
\beq\lb{s1_24}
j_{\psi;\alpha}(\vec{x},t_{p}) &:=&j_{\psi;\alpha}(\vec{x},t)\neq 0\;;\hspace*{1.0cm}
\mbox{'condensate seed' for } \langle\psi_{\vec{x},\alpha}(t_{p})\rangle\;; \\ \lb{s1_25}
\wt{j}_{\psi\psi;\alpha\beta}(\vec{x},t_{p})&:=&
\wt{j}_{\psi\psi;\alpha\beta}(\vec{x},t)\neq 0\;;\hspace*{0.5cm}\mbox{'condensate seed' for }
\langle\psi_{\vec{x},\beta}(t_{p})\;\;\psi_{\vec{x},\alpha}(t_{p})\rangle\;.
\eeq
The even coherent state fields \(\hat{c}_{L\times L}(\vec{x},t_{p})\),
\(\hat{f}_{S\times S}(\vec{x},t_{p})\) and odd coherent fields
\(\hat{\eta}_{S\times L}(\vec{x},t_{p})\), \(\hat{\eta}_{L\times S}^{T}(\vec{x},t_{p})\) (\ref{s1_22}),
corresponding to the operators \(\hat{c}_{\vec{x},L\times L}\), \(\hat{f}_{\vec{x},S\times S}\)
and \(\hat{\eta}_{\vec{x},S\times L}\), \(\hat{\eta}_{\vec{x},L\times S}^{T}\) (\ref{s1_6}-\ref{s1_9}),
involve super-symmetric combinations of anomalous terms as
'\(\psi_{\vec{x},\beta}(t_{p})\;\psi_{\vec{x},\alpha}(t_{p})\)' so that a 'Nambu'-doubling
(\ref{s1_26}) \(\Psi_{\vec{x},\alpha}^{a(=1/2)}(t_{p})=\{\psi_{\vec{x},\alpha}(t_{p})\:;\:
\psi_{\vec{x},\alpha}^{*}(t_{p})\}^{T}\) of coherent state fields has to be taken into
account in transformations of \(Z[\hat{\mscr{J}},j_{\psi},\wt{j}_{\psi\psi}]\) (\ref{s1_22})
\be\lb{s1_26}
\Psi_{\vec{x},\alpha}^{a(=1/2)}(t_{p})=\left(
\bea{c}
\psi_{\vec{x},\alpha}(t_{p}) \\
\psi_{\vec{x},\alpha}^{*}(t_{p})
\eea\right)=
\bigg\{\underbrace{\vec{b}_{\vec{x}}(t_{p})\;,\;\vec{\alpha}_{\vec{x}}(t_{p})}_{a=1}\;;\;
\underbrace{\vec{b}_{\vec{x}}^{*}(t_{p})\;,\;\vec{\alpha}_{\vec{x}}^{*}(t_{p})}_{a=2}\bigg\}^{T}\;.
\ee
According to the presence of anomalous terms, an order-parameter
\(\hat{\Phi}_{\vec{x},\alpha;\vec{x}\ppr,\beta}^{ab}(t_{p})\) has to respect the symmetries of
the dyadic product of 'Nambu' doubled super-fields (\ref{s1_27}) with 'Nambu' indices \(a,b=1,2\).
Apart from the density
terms \(\hat{\Phi}_{\vec{x},\alpha;\vec{x}\ppr,\beta}^{11}(t_{p})\),
\(\hat{\Phi}_{\vec{x},\alpha;\vec{x}\ppr,\beta}^{22}(t_{p})\), this guarantees the inclusion
of pair condensate terms in the off-diagonal blocks with super-matrices
\(\hat{\Phi}_{\vec{x},\alpha;\vec{x}\ppr,\beta}^{12}(t_{p})\),
\(\hat{\Phi}_{\vec{x},\alpha;\vec{x}\ppr,\beta}^{21}(t_{p})\) so that the appropriate
super-symmetries allow for a coset decomposition \(\mbox{Osp}(S,S|2L)/ \mbox{U}(L|S)\otimes \mbox{U}(L|S)\)
of the ortho-symplectic super-group \(\mbox{Osp}(S,S|2L)\) with the super-unitary subgroup \(\mbox{U}(L|S)\)
\beq\lb{s1_27}
\hat{\Phi}_{\vec{x},\alpha;\vec{x}\ppr,\beta}^{ab}(t_{p})&=&\Psi_{\vec{x},\alpha}^{a}(t_{p})\;\otimes\;
\Psi_{\vec{x}\ppr,\beta}^{+b}(t_{p}) = \left( \bea{c}
\psi_{\vec{x},\alpha}(t_{p}) \\
\psi_{\vec{x},\alpha}^{*}(t_{p}) \eea\right)^{a}\otimes
\Big(\psi_{\vec{x}\ppr,\beta}^{*}(t_{p})\;;\;\psi_{\vec{x}\ppr,\beta}(t_{p})\Big)^{b}
\\ \no &=&\left( \bea{cc}
\langle\psi_{\vec{x},\alpha}(t_{p})\;\psi_{\vec{x}\ppr,\beta}^{*}(t_{p})\rangle &
\langle\psi_{\vec{x},\alpha}(t_{p})\;\psi_{\vec{x}\ppr,\beta}(t_{p})\rangle \\
\langle\psi_{\vec{x},\alpha}^{*}(t_{p})\;\psi_{\vec{x}\ppr,\beta}^{*}(t_{p})\rangle &
\langle\psi_{\vec{x},\alpha}^{*}(t_{p})\;\psi_{\vec{x}\ppr,\beta}(t_{p})\rangle \eea\right)^{ab} = \left( \bea{cc}
\hat{\Phi}_{\vec{x},\alpha;\vec{x}\ppr,\beta}^{11}(t_{p}) &
\hat{\Phi}_{\vec{x},\alpha;\vec{x}\ppr,\beta}^{12}(t_{p}) \\
\hat{\Phi}_{\vec{x},\alpha;\vec{x}\ppr,\beta}^{21}(t_{p}) &
\hat{\Phi}_{\vec{x},\alpha;\vec{x}\ppr,\beta}^{22}(t_{p}) \eea\right)^{ab}_{\mbox{.}}
\eeq
In comparison to the \(N\times N=(L+S)\times(L+S)\) super-matrices
\(\hat{N}_{1}\), \(\hat{N}_{2}\) in Eqs.\ (\ref{s1_15}-\ref{s1_20}), we have therefore to
consider \(2N\times2N\) super-matrices \(\hat{\Phi}_{\alpha\beta}^{ab}\) consisting of
four \(N\times N=(L+S)\times(L+S)\) sub-super-matrices \(\hat{\Phi}_{\alpha\beta}^{11}\),
\(\hat{\Phi}_{\alpha\beta}^{22}\) and \(\hat{\Phi}_{\alpha\beta}^{12}\),
\(\hat{\Phi}_{\alpha\beta}^{21}\). Each of the four sub-super-matrices is composed of an
even \(L\times L\) (\(S\times S\)) boson-boson (fermion-fermion) block and the two
odd parts in the \(S\times L\) fermion-boson and \(L\times S\) boson-fermion blocks.
Consequently, we have to generalize operations as the super-transposition '$\mbox{st}$' of
\(N\times N=(L+S)\times(L+S)\) super-matrices to those of \(2N\times2N\) super-matrices
being partitioned into four \(N\times N\) sub-super-matrices. The super-transposition
'$\mbox{st}$' and super-trace '$\mbox{str}$' are straightforwardly extended to the super-transposition
'$\mbox{ST}$' (\ref{s1_28}) and super-trace '$\mbox{STR}$' (\ref{s1_29})
of \(2N\times 2N\) super-matrices.
The total, even, \(2L\times 2L\) boson-boson, \(2S\times 2S\) fermion-fermion sections and the total,
odd, \(2S\times 2L\) fermion-boson, \(2L\times 2S\) boson-fermion sections
are consistently split into four parts, respectively, and are distributed to
four \(N\times N\) sub-super-matrices. The super-hermitian conjugation (\ref{s1_30})
of \(2N\times 2N\) super-matrices \(\hat{\Phi}_{\alpha\beta}^{ab}\) follows by taking
the super-hermitian conjugate of the block diagonal parts \(\hat{\Phi}_{\alpha\beta}^{11}\),
\(\hat{\Phi}_{\alpha\beta}^{22}\) as in (\ref{s1_19}) and also of
\(\hat{\Phi}_{\alpha\beta}^{12}\), \(\hat{\Phi}_{\alpha\beta}^{21}\); in addition the latter
super-hermitian conjugated, off-diagonal \(N\times N\) blocks have to exchange their
places
\beq\lb{s1_28}
\Big(\hat{\Phi}_{\alpha\beta}^{ab}\Big)^{ST}&=&\left( \bea{cc}
\hat{\Phi}_{\alpha\beta}^{11} & \hat{\Phi}_{\alpha\beta}^{12} \\
\hat{\Phi}_{\alpha\beta}^{21} &  \hat{\Phi}_{\alpha\beta}^{22} \eea\right)^{ST}= \left( \bea{cc}
\big(\hat{\Phi}_{\alpha\beta}^{11}\big)^{st} & \big(\hat{\Phi}_{\alpha\beta}^{21}\big)^{st} \\
\big(\hat{\Phi}_{\alpha\beta}^{12}\big)^{st} &  \big(\hat{\Phi}_{\alpha\beta}^{22}\big)^{st}
\eea\right)\;; \\  \lb{s1_29}
\STRAB\Big[\hat{\Phi}_{\alpha\beta}^{ab}\Big]&=&
\strab\Big[\hat{\Phi}_{\alpha\beta}^{11}\Big]+\strab\Big[\hat{\Phi}_{\alpha\beta}^{22}\Big]
= \sum_{m=-l}^{m=+l}\hat{\Phi}_{mm}^{11}-\sum_{r=-s}^{r=+s}\hat{\Phi}_{rr}^{11}+
\sum_{m=-l}^{m=+l}\hat{\Phi}_{mm}^{22}-\sum_{r=-s}^{r=+s}\hat{\Phi}_{rr}^{22} \;; \\  \lb{s1_30}
\Big(\hat{\Phi}_{\alpha\beta}^{ab}\Big)^{+}&=&\left( \bea{cc}
\hat{\Phi}_{\alpha\beta}^{11} & \hat{\Phi}_{\alpha\beta}^{12} \\
\hat{\Phi}_{\alpha\beta}^{21} &  \hat{\Phi}_{\alpha\beta}^{22} \eea\right)^{+}= \left( \bea{cc}
\big(\hat{\Phi}_{\alpha\beta}^{11}\big)^{+} & \big(\hat{\Phi}_{\alpha\beta}^{21}\big)^{+} \\
\big(\hat{\Phi}_{\alpha\beta}^{12}\big)^{+} &  \big(\hat{\Phi}_{\alpha\beta}^{22}\big)^{+}
\eea\right)_{\mbox{.}}
\eeq
In a similar manner the super-determinant '\(\mbox{sdet}(\hat{N})\)' is extended
to a super-determinant '\(\mbox{SDET}(\hat{\Phi}_{\alpha\beta}^{ab})\)' (\ref{s1_31}) of
\(2N\times2N\) super-matrices by substituting the super-trace '\(\mbox{str}\)' (\ref{s1_18})
in relation (\ref{s1_20}) with the super-trace '\(\mbox{STR}\)' (\ref{s1_29}) of
\(2N\times2N\) super-matrices, having symmetries as the dyadic product (\ref{s1_27}) of
'Nambu' doubled coherent state fields
\be\lb{s1_31}
\mbox{SDET}\Big(\hat{\Phi}_{\alpha\beta}^{ab}\Big)=
\exp\Big\{\STRAB\ln\big(\hat{\Phi}_{\alpha\beta}^{ab}\big)\Big\}\;.
\ee
In Ref.\ \cite{mies1} we describe in detail how to transform the coherent
state path integral \(Z[\hat{\mscr{J}},j_{\psi},\wt{j}_{\psi\psi}]\) (\ref{s1_22})
with 'Nambu' doubled super-fields, doubled one-particle and interaction parts
to doubled self-energies using Hubbard-Stratonovich transformations (HST) \cite{st}.
The properties of a \(\mbox{Osp}(S,S|2L)/ \mbox{U}(L|S)\otimes \mbox{U}(L|S)\) coset decomposition are
analyzed in general and require anti-hermitian anomalous terms in
the self-energy matrix \(\delta\wt{\Sigma}_{\alpha\beta}^{ab}(\vec{x},t_{p})\)
for an appropriate parametrization
\footnote{In the remainder the tilde
'$\wt{\ph{\Sigma}}$' of $\delta\wt{\Sigma}_{2N\times 2N}$ refers to a self-energy with
anti-hermitian anomalous terms $\im\;\delta\hat{\Sigma}_{N\times N}^{12}$,
$\im\;\delta\hat{\Sigma}_{N\times N}^{21}$ in comparison to $\delta\hat{\Sigma}_{2N\times 2N}$
with hermitian pair condensates $\delta\hat{\Sigma}_{N\times N}^{12}$, $\delta\hat{\Sigma}_{N\times N}^{21}$;
\(\delta\hat{\Sigma}_{N\times N}^{21}=\big(\delta\hat{\Sigma}_{N\times N}^{12}\big)^{+}\).
We mark the 'Nambu' doubled self-energy \(\delta\wt{\Sigma}_{\alpha\beta}^{ab}(\vec{x},t_{p})\;\wt{K}\) (\ref{s1_37})
with a '\(\delta\)' in order to distinguish from the total sum
\(\wt{\Sigma}_{\alpha\beta}^{ab}(\vec{x},t_{p})\;\wt{K}\) (\ref{s2_1}) with the background field \(\sigma_{D}^{(0)}(\vec{x},t_{p})\)
as the dominant contribution. The metric \(\wt{K}\) (\ref{s1_38}) has to be added to the self-energy for taking values within
the ortho-symplectic super-algebra \(\mbox{osp}(S,S|2L)\).}.
Additionally we have to incorporate a real, scalar background field \(\sigma_{D}^{(0)}(\vec{x},t_{p})\)
as a self-energy density term for \(\sum_{\alpha=1}^{N=L+S}\psi_{\vec{x},\alpha}^{*}(t_{p})\;
\psi_{\vec{x},\alpha}(t_{p})\) in the HST transformations. The \(\mbox{U}(L|S)\)
density terms, as subgroup of \(\mbox{Osp}(S,S|2L)\) in \(\delta\hat{\Sigma}_{\alpha\beta}^{aa}(\vec{x},t_{p})\;\wt{K}\),
only contribute as 'hinge'-fields in the spontaneous symmetry breaking according to the
coset decomposition \(\mbox{Osp}(S,S|2L)/ \mbox{U}(L|S)\otimes \mbox{U}(L|S)\). After a complete 'Nambu' doubling
and suitable HST's of \(Z[\hat{\mscr{J}},j_{\psi},\wt{j}_{\psi\psi}]\) (\ref{s1_22}), we
obtain in Ref.\ \cite{mies1} a coherent state path integral
\(Z[\hat{\mscr{J}},J_{\psi},\im\hat{J}_{\psi\psi}]\) (\ref{s1_32}) which depends
on the real, scalar self-energy density \(\sigma_{D}^{(0)}(\vec{x},t_{p})\) as background
field and on the \(2N\times 2N\) super-symmetric self-energy
\(\delta\wt{\Sigma}_{\alpha\beta}^{ab}(\vec{x},t_{p})\;\wt{K}\) (\ref{s1_37}) with anti-hermitian
anomalous terms \((\:\delta\wt{\Sigma}_{\alpha\beta}^{a\neq b}(\vec{x},t_{p})\:)^{+}=-
\delta\wt{\Sigma}_{\alpha\beta}^{b\neq a}(\vec{x},t_{p})\)
in the off-diagonals (\(a\neq b\)). Corresponding to the short-ranged
interaction potential \(V_{|\vec{x}\ppr-\vec{x}|}\), the spatially nonlocal
self-energies, resulting from the HST's, are approximated to their local form with
an effective, constant interaction parameter $V_{0}$. This approximation is justified by
the assumption that the strong oscillations
lead to a cancellation of phases for exceeding interaction range.
Introducing the \(2N\times 2N\) 'Nambu' doubled super-matrix
\(\wt{\mscr{M}}_{\vec{x},\alpha;\vec{x}\ppr,\beta}^{ab}(t_{p},t_{q}\ppr)\) (\ref{s1_33}),
we achieve in Ref.\ \cite{mies1} the coherent state path integral
\(Z[\hat{\mscr{J}},J_{\psi},\im\hat{J}_{\psi\psi}]\) (\ref{s1_32}) where
the source fields \(j_{\psi;\alpha}(\vec{x},t_{p})\),
\(\wt{j}_{\psi\psi;\alpha\beta}(\vec{x},t_{p})\) are converted to their
'Nambu' doubled form \(J_{\psi;\alpha}^{a(=1/2)}(\vec{x},t_{p})=
\{j_{\psi;\alpha}(\vec{x},t_{p})\;;\;j_{\psi;\alpha}^{*}(\vec{x},t_{p})\}\) (\ref{s1_34})
and to \(\im\;\hat{J}_{\psi\psi;\alpha\beta}^{a\neq b}(\vec{x},t_{p})\) (\ref{s1_35},\ref{s1_36}).
We have also to include various 'Nambu' metric tensors
\(\hat{K}^{a}=\boldsymbol{\{}\,(\hat{1}_{L\times L}\,,\,\hat{1}_{S\times S})^{a=1}\;;\;
(\hat{1}_{L\times L}\,,\,-\hat{1}_{S\times S})^{a=2}\,\boldsymbol{\}}\),
\(\wt{K}^{a}=\boldsymbol{\{}\,(\hat{1}_{L\times L}\,,\,\hat{1}_{S\times S})^{a=1}\;;\;
(-\hat{1}_{L\times L}\,,\,\hat{1}_{S\times S})^{a=2}\,\boldsymbol{\}}\) and
\(\hat{I}^{a}=\boldsymbol{\{}\,(\hat{1}_{L\times L}\,,\,\hat{1}_{S\times S})^{a=1}\;;\;
(\hat{\im}_{L\times L}\,,\,\hat{\im}_{S\times S})^{a=2}\,\boldsymbol{\}}\) so that
the parametrization and propagation of the self-energy fields in the exponentials are confined to the ortho-symplectic
super-group \(\mbox{Osp}(S,S|2L)\)
\beq \no
\lefteqn{\hspace*{-0.6cm}Z[\hat{\mscr{J}},J_{\psi},\im\hat{J}_{\psi\psi}]=
\int\;\;d[\sigma_{D}^{(0)}(\vec{x},t_{p})]\;\; \exp\bigg\{\frac{\im}{2\hbar}\frac{1}{V_{0}}
\int_{C}d t_{p}\sum_{\vec{x}}
\sigma_{D}^{(0)}(\vec{x},t_{p})\;\;\sigma_{D}^{(0)}(\vec{x},t_{p})\bigg\} \times
\int d[\delta\wt{\Sigma}(\vec{x},t_{p})\;\wt{K}]} \\ \no &\times&
\exp\bigg\{\frac{\im}{4\hbar}\frac{1}{V_{0}}
\int_{C}d t_{p}\sum_{\vec{x}}\STRAB\Big[
\Big(\delta\wt{\Sigma}(\vec{x},t_{p})-\im\;\hat{J}_{\psi\psi}(\vec{x},t_{p})\Big)\;
\wt{K}\;\Big(\delta\wt{\Sigma}(\vec{x},t_{p})-\im\;\hat{J}_{\psi\psi}(\vec{x},t_{p})\Big)\;\wt{K}\Big]\bigg\}
\\ \lb{s1_32}  &\times &
\Bigg\{\mbox{SDET}\bigg[\wt{\mscr{M}}_{\vec{x},\alpha;\vec{x}\ppr,\beta}^{ab}(t_{p},t_{q}\ppr)\bigg]
\Bigg\}^{\mathbf{-1/2}} \; \times  \\ \no &\times&
\exp\bigg\{\frac{\im}{2\hbar}\Omega\int_{C}d t_{p}\;d t\ppr_{q}\sum_{\vec{x},\vec{x}\ppr}\mscr{N}_{x}\;
J_{\psi;\beta}^{+b}(\vec{x}\ppr,t_{q}\ppr)\;\hat{I}\wt{K}\;
\wt{\mscr{M}}_{\vec{x}\ppr,\beta;\vec{x},\alpha}^{\mathbf{-1};ba}(t_{q}\ppr,t_{p})\;\hat{I}\;
J_{\psi;\alpha}^{a}(\vec{x},t_{p})\bigg\}\;;
\eeq
\beq \no
\lefteqn{\hspace*{-0.6cm}
\wt{\mscr{M}}_{\vec{x},\alpha;\vec{x}\ppr,\beta}^{ab}(t_{p},t_{q}\ppr)=
\delta_{\vec{x},\vec{x}\ppr}\;\eta_{p}\;\delta_{p,q}\;\delta_{t_{p},t_{q}\ppr}\Bigg[
\left(\bea{cc} \hat{H}_{p}(\vec{x},t_{p})+\sigma_{D}^{(0)}(\vec{x},t_{p}) &  \\
 & \hat{H}_{p}^{T}(\vec{x},t_{p})+\sigma_{D}^{(0)}(\vec{x},t_{p})
\eea\right)+ } \\ \lb{s1_33} &+&
\left(\bea{cc}
\delta\hat{\Sigma}_{\alpha\beta}^{11}(\vec{x},t_{p}) & \im\;
\delta\hat{\Sigma}_{\alpha\beta}^{12}(\vec{x},t_{p}) \\
\im\;\delta\hat{\Sigma}_{\alpha\beta}^{21}(\vec{x},t_{p}) &
\delta\hat{\Sigma}_{\alpha\beta}^{22}(\vec{x},t_{p})
\eea\right)\;\wt{K}\Bigg]_{\alpha\beta}^{ab}+
\underbrace{\hat{I}\;\hat{K}\;\eta_{p}\;
\frac{\hat{\mscr{J}}_{\vec{x},\alpha;\vec{x}\ppr,\beta}^{ab}(t_{p},t_{q}\ppr)}{\Omega\;\mscr{N}_{x}}\;
\eta_{q}\;\hat{K}\;\hat{I}\;\wt{K}
}_{\wt{\mscr{J}}_{\vec{x},\alpha;\vec{x}\ppr,\beta}^{ab}(t_{p},t_{q}\ppr)}\;\;;
\eeq
\beq\lb{s1_34}
J_{\psi;\alpha}^{a(=1/2)}(\vec{x},t_{p})&=&
\Big\{\underbrace{j_{\psi;\alpha}(\vec{x},t_{p})}_{a=1}\;;\;
\underbrace{j_{\psi;\alpha}^{*}(\vec{x},t_{p})}_{a=2}\Big\}^{T} \;; \\ \lb{s1_35}
\hat{J}_{\psi\psi;\alpha\beta}^{a\neq b}(\vec{x},t_{p}) &=&
\left(\bea{cc} 0 & \hat{j}_{\psi\psi;\alpha\beta}(\vec{x},t_{p})  \\
\hat{j}_{\psi\psi;\alpha\beta}^{+}(\vec{x},t_{p}) & 0 \eea\right)  \;; \\ \lb{s1_36}
\hat{j}_{\psi\psi;\alpha\beta}(\vec{x},t_{p}) &=&
\left(\bea{cc} \hat{j}_{B;L\times L}(\vec{x},t_{p}) & \hat{j}_{\eta;L\times S}^{T}(\vec{x},t_{p}) \\
\hat{j}_{\eta;S\times L}(\vec{x},t_{p})  & \hat{j}_{F;S\times S}(\vec{x},t_{p}) \eea\right) \;; \\ \no
\hat{j}_{B;L\times L}^{T}(\vec{x},t_{p})=\hat{j}_{B;L\times L}(\vec{x},t_{p}) &;&
\hat{j}_{F;S\times S}^{T}(\vec{x},t_{p})=-\hat{j}_{F;S\times S}(\vec{x},t_{p})\;.
\eeq
Apart from the self-energy density \(\sigma_{D}^{(0)}(\vec{x},t_{p})\) as
background field in (\ref{s1_32},\ref{s1_33}), the self-energy super-matrix
\(\delta\wt{\Sigma}_{\alpha\beta}^{ab}(\vec{x},t_{p})\;\wt{K}\) (\ref{s1_37}) only enters into
the coherent state path integral (\ref{s1_32},\ref{s1_33}) with independent field degrees of freedom confined
to the parameters of the ortho-symplectic \(\mbox{osp}(S,S|2L)\) super-algebra. It has to
be noted that the self-energy super-matrix \(\delta\wt{\Sigma}_{\alpha\beta}^{ab}(\vec{x},t_{p})\)
has to include the appropriate metric \(\wt{K}\) (\ref{s1_38}) in order to become
an exact element of \(\mbox{osp}(S,S|2L)\)
\beq \lb{s1_37}
\delta\wt{\Sigma}_{\alpha\beta}^{ab}(\vec{x},t_{p})\;\;\wt{K} &=&
\left(\bea{cc}
\delta\hat{\Sigma}_{\alpha\beta}^{11}(\vec{x},t_{p}) & \im\;
\delta\hat{\Sigma}_{\alpha\beta}^{12}(\vec{x},t_{p}) \\
\im\;\delta\hat{\Sigma}_{\alpha\beta}^{21}(\vec{x},t_{p}) &
\delta\hat{\Sigma}_{\alpha\beta}^{22}(\vec{x},t_{p})
\eea\right)_{\alpha\beta}^{ab}\;\wt{K} \;; \\ \lb{s1_38}
\wt{K}&=&\Big\{\underbrace{\hat{1}_{L\times L}\,,\,\hat{1}_{S\times S}}_{a=1}\;;\;
\underbrace{\overbrace{-\hat{1}_{L\times L}\,,\,\hat{1}_{S\times S}}^{\wt{\kappa}_{N\times N}}}_{a=2}\Big\}
\;;\hspace*{0.5cm} \wt{\kappa}_{N\times N} =
\Big\{\underbrace{-\hat{1}_{L\times L}}_{BB}\,,\,\underbrace{\hat{1}_{S\times S}}_{FF}\Big\}\;.
\eeq
The density terms \(\delta\hat{\Sigma}_{\alpha\beta}^{11}(\vec{x},t_{p})\),
\(\delta\hat{\Sigma}_{\alpha\beta}^{22}(\vec{x},t_{p})\), referring to the super-unitary
\(\mbox{U}(L|S)\) group, are eliminated in combination of the coset decomposition with the
gradient expansion and have the effect of 'hinge' functions in a SSB.
According to the symmetry examination in Ref.\ \cite{mies1}, the
coset decomposition \(\mbox{Osp}(S,S|2L)/ \mbox{U}(L|S)\otimes \mbox{U}(L|S)\) requires anti-hermitian anomalous
terms \(\delta\wt{\Sigma}_{\alpha\beta}^{12}(\vec{x},t_{p})=\im\:
\delta\hat{\Sigma}_{\alpha\beta}^{12}(\vec{x},t_{p})\),
\(\delta\wt{\Sigma}_{\alpha\beta}^{21}(\vec{x},t_{p})=\im\:
\delta\hat{\Sigma}_{\alpha\beta}^{21}(\vec{x},t_{p})\),
\(\delta\hat{\Sigma}_{\alpha\beta}^{21}(\vec{x},t_{p})=(\,
\delta\hat{\Sigma}_{\alpha\beta}^{12}(\vec{x},t_{p})\,)^{+}\) (\ref{s1_37},\ref{s1_38}) in order to obtain
the correct number of independent field degrees of freedom as the independent parameters
of \(\mbox{Osp}(S,S|2L)\). The gradient expansion of the super-matrix
\(\wt{\mscr{M}}_{\vec{x},\alpha;\vec{x}\ppr,\beta}^{ab}(t_{p},t_{q}\ppr)\) (\ref{s1_33})
with the coset fields in \(\hat{T}(\vec{x},t_{p})\) results in effective actions
\(\mscr{A}_{\mscr{N}^{-1}}\ppr\big[\hat{T};J_{\psi}\big]\),
\(\mscr{A}_{\mscr{N}^{0}}\ppr\big[\hat{T};J_{\psi}\big]\),
\(\mscr{A}_{\mscr{N}^{+1}}\ppr\big[\hat{T}\big]\) (\ref{s1_39}) which can be classified
according to a parameter \(\mscr{N}=\hbar\,\Omega\;\mscr{N}_{x}\)
(\(\Omega=1/\Delta t\), \(\mscr{N}_{x}=(L/\Delta x)^{d}\)),
denoting the total number of spatial points on an underlying grid and
specifying the maximum possible energy \(\hbar\Omega\) corresponding to the
discrete time steps \(\Delta t\). The nontrivial coset integration measure
is indicated by \(d[\hat{T}^{-1}(\vec{x},t_{p})\:d\hat{T}(\vec{x},t_{p})]\)
in \(Z[\hat{\mscr{J}},J_{\psi},\im \hat{J}_{\psi\psi}]\) (\ref{s1_39}) (see section \ref{s22}).
The source action \(\mscr{A}_{\hat{J}_{\psi\psi}}\big[\hat{T}\big]\),
following from \(\im\;\hat{J}_{\psi\psi;\alpha\beta}^{a\neq b}(\vec{x},t_{p})\)
(\ref{s1_35},\ref{s1_36}), is independent from gradients and the background field
\(\sigma_{D}^{(0)}(\vec{x},t_{p})\) and can be simplified by using
properties of Vandermonde matrices \cite{mehta}. The action, resulting for the
additional source matrix \(\hat{\mscr{J}}_{\vec{x}\ppr,\beta;\vec{x},\alpha}^{ba}(t_{q}\ppr,t_{p})\)
within up to second order of the gradient expansion, is denoted by \(\mscr{A}\ppr\big[\hat{T};\hat{\mscr{J}}\big]\)
and is further investigated in section \ref{s41}
\beq \lb{s1_39}
Z[\hat{\mscr{J}},J_{\psi},\im \hat{J}_{\psi\psi}]&=&
\int d\big[\hat{T}^{-1}(\vec{x},t_{p})\;d\hat{T}(\vec{x},t_{p})\big]\;\;
\exp\Big\{\im\;\mscr{A}_{\hat{J}_{\psi\psi}}\big[\hat{T}\big]\Big\}  \\ \no &\times&
\exp\Big\{-\mscr{A}_{\mscr{N}^{-1}}\ppr\big[\hat{T};J_{\psi}\big]-
\mscr{A}_{\mscr{N}^{0}}\ppr\big[\hat{T};J_{\psi}\big]-
\mscr{A}_{\mscr{N}^{+1}}\ppr\big[\hat{T}\big]\Big\} \;\times\;
\exp\Big\{-\mscr{A}\ppr\big[\hat{T};\hat{\mscr{J}}\big]\Big\}_{\mbox{.}}
\eeq
It remains to identify the various classical actions in (\ref{s1_39}) with
the nontrivial coset integration measure which has to be replaced by Euclidean path integration fields.
This is accomplished in sections \ref{s22} and \ref{s23}; however, we have in advance to
describe the precise parameters of anomalous fields \(\hat{T}(\vec{x},t_{p})\) following
from the total self-energy matrix \(\delta\wt{\Sigma}_{\alpha\beta}^{ab}(\vec{x},t_{p})\;\wt{K}\)
(\ref{s1_37},\ref{s1_38}) in the coset decomposition \(\mbox{Osp}(S,S|2L)/ \mbox{U}(L|S)\otimes \mbox{U}(L|S)\) (section \ref{s21}).

Instead of the gradient expansion for effective actions in
\(Z[\hat{\mscr{J}},J_{\psi},\im \hat{J}_{\psi\psi}]\) (\ref{s1_39}), we remark an
alternative solution (\ref{s1_40}) of the original coherent state path integral (\ref{s1_32})
with super-matrix \(\wt{\mscr{M}}_{\vec{x},\alpha;\vec{x}\ppr,\beta}^{ab}(t_{p},t_{q}\ppr)\) (\ref{s1_33}).
This solution follows from the functional variation of (\ref{s1_32}) with respect to
the self-energy super-matrix \(\delta\wt{\Sigma}_{\alpha\beta}^{ab}(\vec{x},t_{p})\;\wt{K}\) (\ref{s1_37})
as a \(\mbox{osp}(S,S|2L)\) super-generator
\beq\lb{s1_40}
\lefteqn{0\equiv\frac{\Delta \Big(Z[\hat{\mscr{J}},J_{\psi},\im \hat{J}_{\psi\psi}]\Big)}{\Delta
\Big(\delta\wt{\Sigma}(\vec{x},t_{p})\;\wt{K}\Big)_{\alpha\beta}^{ab}}=\bigg\langle
\frac{1}{\mscr{N}}\frac{\im}{2}\frac{\eta_{p}}{V_{0}}\;
\Big[\Big(\delta\wt{\Sigma}(\vec{x},t_{p})-\im\;\hat{J}_{\psi\psi}(\vec{x},t_{p})\Big)\;\wt{K}\Big]_{\beta\alpha}^{ba} + } \\ \no &-&
\frac{\eta_{p}}{2}\;
\wt{\mscr{M}}_{\vec{x},\beta;\vec{x},\alpha}^{\mathbf{-1};ba}(t_{p}+\delta t_{p},t_{p}+\delta t_{p}\ppr) -
\frac{\im}{2}\frac{\Omega}{\hbar}\int_{C}dt_{q_{1}}^{(1)}\;dt_{q_{2}}^{(2)}\sum_{\vec{y}_{1},\vec{y}_{2}}\mscr{N}_{x}\;
\;\eta_{p}\;\wt{\mscr{M}}_{\vec{x},\beta;\vec{y}_{1},\alpha_{1}}^{\mathbf{-1};ba_{1}}(t_{p}+\delta t_{p},t_{q_{1}}^{(1)})
\;\times \\ \no &\times&
\Big(\hat{I}\;J_{\psi;\alpha_{1}}^{a_{1}}(\vec{y}_{1},t_{q_{1}}^{(1)})
\;\;\otimes\;\;
J_{\psi;\beta_{2}}^{+b_{2}}(\vec{y}_{2},t_{q_{2}}^{(2)})\;\hat{I}\;\wt{K}\Big)\;
\wt{\mscr{M}}_{\vec{y}_{2},\beta_{2};\vec{x},\alpha}^{\mathbf{-1};b_{2}a}(t_{q_{2}}^{(2)},t_{p}+\delta t_{p}\ppr)
\bigg\rangle_{Z[\hat{\mscr{J}},J_{\psi},\im \hat{J}_{\psi\psi}]} \;.
\eeq
One can apply continued fractions of \(\delta\wt{\Sigma}_{\alpha\beta}^{ab}(\vec{x},t_{p})\;\wt{K}\) (\ref{s1_37})
for solving the mean field equation (\ref{s1_40}) (compare Ref.\ \cite{bmdis1}). This process considerably
simplifies in case of spatial symmetries and under restriction to stationary solutions. We note
that the matrix \(\wt{\mscr{M}}_{\vec{x},\alpha;\vec{x}\ppr,\beta}^{ab}(t_{p},t_{q}\ppr)\) (\ref{s1_33})
is not only of central importance for the gradient expansion with the anomalous terms, but also
for the saddle point equation (\ref{s1_40}) because it consists of the background field
\(\sigma_{D}^{(0)}(\vec{x},t_{p})\) apart from the self-energy super-matrix
\(\delta\wt{\Sigma}_{\alpha\beta}^{ab}(\vec{x},t_{p})\;\wt{K}\) (\ref{s1_37}).
The exact mean field equation (\ref{s1_40}) can be approximated by averaging of the inverse
\(\wt{\mscr{M}}_{\vec{x},\alpha;\vec{x}\ppr,\beta}^{\mathbf{-1};ab}(t_{p},t_{q}\ppr)\) of the
super-matrix (\ref{s1_33}) as one-point Green function with the background field \(\sigma_{D}^{(0)}(\vec{x},t_{p})\).
This averaging process of \(\wt{\mscr{M}}_{\vec{x},\alpha;\vec{x}\ppr,\beta}^{\mathbf{-1};ab}(t_{p},t_{q}\ppr)\)
by \(\sigma_{D}^{(0)}(\vec{x},t_{p})\) becomes itself more accessible by taking a saddle point
solution for the background density field in order to approximate (\ref{s1_40}).

\section{The classical Lagrangian for anomalous terms} \lb{s2}

\subsection{The independent field variables for the super-symmetric pair condensates}\lb{s21}

According to super-group properties of \(\mbox{Osp}(S,S|2L)\), the sum of self-energies
\(\sigma_{D}^{(0)}\,\hat{1}_{2N\times 2N}\) and \(\delta\wt{\Sigma}_{2N\times 2N}\:\wt{K}\)
is factorized into density terms \(\delta\hat{\Sigma}_{D;N\times N}^{11}\),
\(\delta\hat{\Sigma}_{D;N\times N}^{22}\) (\ref{s2_6},\ref{s2_7}) and super-matrices
\(\hat{T}(\vec{x},t_{p})\) (\ref{s2_2}-\ref{s2_5}) for pair condensates within the
coset decomposition \(\mbox{Osp}(S,S|2L)/ \mbox{U}(L|S)\otimes \mbox{U}(L|S)\) (\ref{s2_1}). The self-energy density
matrices \(\delta\hat{\Sigma}_{D;N\times N}^{11}\),
\(\delta\hat{\Sigma}_{D;N\times N}^{22}\;\wt{\kappa}\) are related to the super-unitary group \(\mbox{U}(L|S)\)
and only act as 'hinge' fields for the SSB and the gradient expansion. The background
self-energy density \(\sigma_{D}^{(0)}(\vec{x},t_{p})\) is invariant under \(\mbox{U}(L|S)\)
subgroup transformations and has therefore to be considered as the invariant ground or
vacuum state in the SSB of \(\mbox{Osp}(S,S|2L)\) with \(\mbox{U}(L|S)\) as subgroup
\beq\lb{s2_1}
\lefteqn{\wt{\Sigma}_{2N\times 2N}(\vec{x},t_{p})\;\wt{K} =
\sigma_{D}^{(0)}(\vec{x},t_{p})\;\hat{1}_{2N\times 2N}+\delta\wt{\Sigma}_{2N\times 2N}(\vec{x},t_{p})\;\wt{K} = }
\\ \no &=&\hat{T}(\vec{x},t_{p})\;\left( \bea{cc}
\sigma_{D}^{(0)}(\vec{x},t_{p})\;\hat{1}_{N\times N}+\delta\hat{\Sigma}_{D;N\times N}^{11}(\vec{x},t_{p}) & 0 \\
0 & \sigma_{D}^{(0)}(\vec{x},t_{p})\;\wt{\kappa}+\delta\hat{\Sigma}_{D;N\times N}^{22}(\vec{x},t_{p}) \eea\right)\;\wt{K}\;\;
\hat{T}^{-1}(\vec{x},t_{p}) \\ \no
&=&\sigma_{D}^{(0)}(\vec{x},t_{p})\;\hat{1}_{2N\times 2N}+
\hat{T}_{2N\times 2N}(\vec{x},t_{p})\;\;\underbrace{\left(\bea{cc}
\delta\hat{\Sigma}_{D;N\times N}^{11}(\vec{x},t_{p}) & 0 \\
0 & \delta\hat{\Sigma}_{D;N\times N}^{22}(\vec{x},t_{p})\;\wt{\kappa}
\eea\right)}_{\delta\hat{\Sigma}_{D;2N\times 2N}(\vec{x},t_{p})\;\wt{K}}\;\;\hat{T}_{2N\times 2N}^{-1}(\vec{x},t_{p}) \;.
\eeq
The independent field degrees of freedom for pair condensates, originally defined by
\(\im\;\delta\hat{\Sigma}_{\alpha\beta}^{12}\), \(\im\;\delta\hat{\Sigma}_{\alpha\beta}^{21}\)
in (\ref{s1_37}), are described by the \(\mbox{osp}(S,S|2L)/ \mbox{u}(L|S)\) super-generator
\(\hat{Y}_{2N\times 2N}(\vec{x},t_{p})\) (\ref{s2_3})
or its exponential form \(\hat{T}_{2N\times 2N}(\vec{x},t_{p})=\exp\{-\hat{Y}_{2N\times 2N}(\vec{x},t_{p})\}\)
(\ref{s2_2}) for the coset manifold \(\mbox{Osp}(S,S|2L)/ \mbox{U}(L|S)\). The super-generator \(\hat{Y}_{2N\times 2N}(\vec{x},t_{p})\)
(\ref{s2_3}) consists of the sub-generators \(\hat{X}_{N\times N}(\vec{x},t_{p})\) (\ref{s2_4}) and its super-hermitian conjugate
\(\wt{\kappa}\;\hat{X}_{N\times N}^{+}\) in the off-diagonal blocks \((a\neq b)\). They are itself
composed of the even complex symmetric matrix \(\hat{c}_{D;L\times L}(\vec{x},t_{p})\) for molecular condensates
and the even complex anti-symmetric matrix \(\hat{f}_{D;S\times S}(\vec{x},t_{p})\) for BCS terms (\ref{s2_5}). The Grassmann
valued field degrees of freedom for anomalous terms are given by the matrix \(\hat{\eta}_{D;S\times L}\)
and its transpose  \(\hat{\eta}_{D;L\times S}^{T}\) in the fermion-boson and boson-fermion blocks
\beq\lb{s2_2}
\hat{T}_{\alpha\beta}^{ab}(\vec{x},t_{p})&=&\exp\Bigg\{\im\left( \bea{cc}
0 & \im\;\hat{X}_{N\times N}(\vec{x},t_{p}) \\
\im\;\wt{\kappa}\;\hat{X}_{N\times N}^{+}(\vec{x},t_{p}) & 0
\eea\right)\Bigg\}   \\ \no &=&\exp\Big\{-\hat{Y}_{2N\times 2N}(\vec{x},t_{p})\Big\}\;; \\ \lb{s2_3}
\hat{Y}_{\alpha\beta}^{ab}(\vec{x},t_{p})&=&\left( \bea{cc}
0 & \hat{X}_{\alpha\beta}(\vec{x},t_{p}) \\
\wt{\kappa}_{N\times N}\;\hat{X}_{\alpha\beta}^{+}(\vec{x},t_{p}) & 0 \eea\right)^{ab}\;;
\\ \lb{s2_4} \hat{X}_{\alpha\beta}(\vec{x},t_{p})&=&\left( \bea{cc}
-\hat{c}_{D;L\times L}(\vec{x},t_{p}) & \hat{\eta}_{D;L\times S}^{T}(\vec{x},t_{p}) \\
-\hat{\eta}_{D;S\times L}(\vec{x},t_{p}) & \hat{f}_{D;S\times S}(\vec{x},t_{p}) \eea\right)\;; \\ \lb{s2_5}
\hat{c}_{D;L\times L}^{T}(\vec{x},t_{p})&=&\hat{c}_{D;L\times L}(\vec{x},t_{p})\;;\hspace*{0.5cm}
\hat{f}_{D;S\times S}^{T}(\vec{x},t_{p})=-\hat{f}_{D;S\times S}(\vec{x},t_{p}) \;.
\eeq
This parametrization with \(\hat{Y}_{2N\times 2N}(\vec{x},t_{p})\) (\ref{s2_3}) takes into account
the exact structure of symmetry breaking source terms for super-symmetric pair condensates
which have been introduced into the original Hamiltonian (\ref{s1_3}) for the coherent state
path integrals. The \(\mbox{U}(L|S)\) self-energy density matrices \(\delta\hat{\Sigma}_{D;N\times N}^{11}(\vec{x},t_{p})\)
(\ref{s2_6}) and its super-transposed copy \(\delta\hat{\Sigma}_{D;N\times N}^{22}(\vec{x},t_{p})\;\wt{\kappa}\)
(\ref{s2_7},\ref{s2_9}) contain the independent field degrees following from the dyadic product density parts
\(\psi_{\vec{x},\alpha}(t_{p})\otimes\psi_{\vec{x},\beta}^{*}(t_{p})\) and
\(\psi_{\vec{x},\alpha}^{*}(t_{p})\otimes\psi_{\vec{x},\beta}(t_{p})\) within the HST transformations
\beq\lb{s2_6}
\delta\hat{\Sigma}_{D;\alpha\beta}^{11}(\vec{x},t_{p})&=&\left( \bea{cc}
\delta \hat{B}_{D;L\times L}(\vec{x},t_{p}) & \delta\hat{\chi}_{D;L\times S}^{+}(\vec{x},t_{p}) \\
\delta\hat{\chi}_{D;S\times L}(\vec{x},t_{p}) & \delta \hat{F}_{D;S\times S}(\vec{x},t_{p}) \eea\right) \;; \\  \lb{s2_7}
\delta\hat{\Sigma}_{D;\alpha\beta}^{22}(\vec{x},t_{p})&=&\left( \bea{cc}
\delta \hat{B}_{D;L\times L}^{T}(\vec{x},t_{p}) & \delta\hat{\chi}_{D;L\times S}^{T}(\vec{x},t_{p}) \\
\delta\hat{\chi}_{D;S\times L}^{*}(\vec{x},t_{p}) & -\delta \hat{F}_{D;S\times S}^{T}(\vec{x},t_{p}) \eea\right)\;;
\\ \lb{s2_8} \delta\hat{B}_{D;L\times L}^{+}(\vec{x},t_{p})&=&
\delta\hat{B}_{D;L\times L}(\vec{x},t_{p})\;;\hspace*{1.0cm}
\delta\hat{F}_{D;S\times S}^{+}(\vec{x},t_{p})=\delta\hat{F}_{D;S\times S}(\vec{x},t_{p}) \;;
\\  \lb{s2_9} \delta\hat{\Sigma}^{11}_{D;N\times N}(\vec{x},t_{p}) &=&
-\Big(\delta\hat{\Sigma}^{22}_{D;N\times N}(\vec{x},t_{p})\;\;\wt{\kappa}\Big)^{st}\;.
\eeq
The even density terms of the boson-boson, fermion-fermion blocks (\ref{s2_8}) are given by
hermitian matrices
\(\delta\hat{B}_{D;L\times L}(\vec{x},t_{p})\), \(\delta\hat{F}_{D;S\times S}(\vec{x},t_{p})\).
The odd density terms in the fermion-boson, boson-fermion sections are determined by
\(\delta\hat{\chi}_{D;S\times L}(\vec{x},t_{p})\) and its super-hermitian conjugate
\(\delta\hat{\chi}_{D;L\times S}^{+}(\vec{x},t_{p})\) (\ref{s2_6},\ref{s2_7}).
These self-energy densities
\(\delta\hat{\Sigma}_{D;N\times N}^{11}(\vec{x},t_{p})\),
\(\delta\hat{\Sigma}_{D;N\times N}^{22}(\vec{x},t_{p})\;\wt{\kappa}\) or
\(\delta\hat{\Sigma}_{D;2N\times 2N}^{aa}(\vec{x},t_{p})\;\wt{K}\) act as 'hinge' fields in the SSB
and can be factorized to real \(N=L+S\) eigenvalues \(\delta\hat{\lambda}_{N\times N}\) or
its 'Nambu' doubled form \(\delta\hat{\Lambda}_{2N\times 2N}\) which comprise the maximal abelian
Cartan subalgebra of rank \(N\) for the \(\mbox{U}(L|S)\) or \(\mbox{Osp}(S,S|2L)\) super-group (\ref{s2_10}-\ref{s2_12}).
The remaining field degrees of freedom \(\hat{Q}_{N\times N}^{11}(\vec{x},t_{p})\),
\(\hat{Q}_{N\times N}^{22}(\vec{x},t_{p})\) (\ref{s2_13}-\ref{s2_15}) for the self-energy density matrices
have their parameters within the ladder operators of the super-unitary algebra \(\mbox{u}(L|S)\). Since
\(N=L+S\) real parameter fields are already contained in the eigenvalues
\(\delta\hat{\lambda}_{\alpha}(\vec{x},t_{p})\) (\ref{s2_12}), the diagonal values of
\(\hat{\mscr{B}}_{D;mm}=0\;\;(m=1,\ldots,L)\) and \(\hat{\mscr{F}}_{D;ii}=0\;\;(i=1,\ldots,S)\)
have to vanish in the generators of \(\hat{Q}^{11}_{N\times N}(\vec{x},t_{p})\) (\ref{s2_14}),
\(\hat{Q}^{22}_{N\times N}(\vec{x},t_{p})\) (\ref{s2_15}). In consequence the ladder operators with their independent fields
only remain from the super-unitary \(\mbox{U}(L|S)\) algebra within the eigenvector matrices \(\hat{Q}^{11}_{N\times N}\),
\(\hat{Q}^{22}_{N\times N}\) of the block diagonal self-energy densities \(\delta\hat{\Sigma}_{D;2N\times 2N}^{aa}\;\wt{K}\)
\beq \lb{s2_10}\hspace*{-0.7cm}
\delta\hat{\Sigma}_{D;2N\times 2N}(\vec{x},t_{p})\;\wt{K}&=&\hat{Q}_{2N\times 2N}^{-1}(\vec{x},t_{p})\;\;
\delta\hat{\Lambda}_{2N\times 2N}(\vec{x},t_{p})\;\;\hat{Q}_{2N\times 2N}(\vec{x},t_{p})\;; \\ \lb{s2_11}
\delta\hat{\Lambda}_{2N\times 2N}(\vec{x},t_{p})&=&\delta\hat{\Lambda}_{\alpha}^{a}(\vec{x},t_{p})=
\mbox{diag}\Big\{\delta\hat{\lambda}_{N\times N}(\vec{x},t_{p})\;;\;
-\delta\hat{\lambda}_{N\times N}(\vec{x},t_{p})\Big\}\;; \\ \lb{s2_12}
\delta\hat{\lambda}_{N\times N}(\vec{x},t_{p})&=&
\delta\hat{\lambda}_{\alpha}(\vec{x},t_{p})=\Big\{\delta\hat{\lambda}_{B;1},\ldots,\delta\hat{\lambda}_{B;m},\ldots,
\delta\hat{\lambda}_{B;L}\;;\;\delta\hat{\lambda}_{F;1},\ldots,\delta\hat{\lambda}_{F;i},\ldots,
\delta\hat{\lambda}_{F;S}\Big\}\;; \\ \lb{s2_13}
\hat{Q}_{2N\times 2N}(\vec{x},t_{p}) &=&\left(\bea{cc}
\hat{Q}_{N\times N}^{11} & 0 \\
0 & \hat{Q}_{N\times N}^{22}
\eea\right)\;;\hspace*{0.5cm}\big(\hat{Q}_{N\times N}^{22}\big)^{st}=
\hat{Q}_{N\times N}^{11,+}=\hat{Q}_{N\times N}^{11,-1}\;; \\ \lb{s2_14}
\hat{Q}^{11}_{N\times N}(\vec{x},t_{p})&=&\exp\left\{\im\left( \bea{cc}
\hat{\mscr{B}}_{D;L\times L} & \hat{\omega}_{D;L\times S}^{+} \\
\hat{\omega}_{D;S\times L} & \hat{\mscr{F}}_{D;S\times S}
\eea\right)\right\}\;;  \hspace*{0.5cm}\hat{\mscr{B}}_{D;L\times L}^{+}=\hat{\mscr{B}}_{D;L\times L}\;;
\\  \lb{s2_15}\hat{Q}^{22}_{N\times N}(\vec{x},t_{p})&=&\exp\left\{\im\left( \bea{cc}
-\hat{\mscr{B}}_{D;L\times L}^{T} & \hat{\omega}_{D;L\times S}^{T} \\
-\hat{\omega}_{D;S\times L}^{*} & -\hat{\mscr{F}}_{D;S\times S}^{T} \eea\right)\right\}\;;
\hspace*{0.5cm}\hat{\mscr{F}}_{D;S\times S}^{+}=\hat{\mscr{F}}_{D;S\times S} \;; \\  \no &&
\hat{\mscr{B}}_{D;mm}=0\;\;(m=1,\ldots,L) \;;\hspace*{0.5cm}
\hat{\mscr{F}}_{D;ii}=0\;\;(i=1,\ldots,S)\;.
\eeq
In analogy one can diagonalize the off-diagonal blocks
\(\hat{X}_{N\times N}(\vec{x},t_{p})\), \(\wt{\kappa}\;\hat{X}_{N\times N}^{+}(\vec{x},t_{p})\) (\ref{s2_2}-\ref{s2_5})
of the super-generator \(\hat{Y}_{2N\times 2N}(\vec{x},t_{p})\) for the pair condensates (\ref{s2_16}-\ref{s2_20}).
The diagonal blocks of \(\hat{Y}_{DD;2N\times 2N}\) in the anomalous sectors
(\ref{s2_16},\ref{s2_17}) consist of the diagonal matrices
\(\hat{X}_{DD;N\times N}\), \(\wt{\kappa}\;\hat{X}_{DD;N\times N}^{+}\) (\ref{s2_18}-\ref{s2_20}) which are rotated
with the parameters \(\hat{\mscr{C}}_{D;L\times L}\), \(\hat{\mscr{G}}_{D;S\times S}\),
\(\hat{\xi}_{D;S\times L}\), \(\hat{\xi}_{D;S\times L}^{*}\) by the matrices
\(\hat{P}_{N\times N}^{11}\), \(\hat{P}_{N\times N}^{22}\) (or \(\hat{P}_{2N\times 2N}^{aa}\)) (\ref{s2_21}-\ref{s2_23})
to the generators
\(\hat{X}_{N\times N}(\vec{x},t_{p})\), \(\wt{\kappa}\;\hat{X}_{N\times N}^{+}(\vec{x},t_{p})\) (\ref{s2_3},\ref{s2_4}).
The complex \(L\) parameters $\ovv{c}_{m}$ within the diagonal matrix \(\hat{\ovv{c}}_{L\times L}\)
describe the anomalous molecular terms and are factorized into its modulus
\(|\ovv{c}_{m}|\) and phase \(\varphi_{m}\) (\ref{s2_19}).
The parameters \(\hat{\ovv{f}}_{S\times S}(\vec{x},t_{p})\) (\ref{s2_20}) of the fermionic degrees
have to consider the anti-symmetric form \(\hat{f}_{D;S\times S}\) of BCS terms (\ref{s2_5}) so that one has to
introduce the quaternion algebra (\ref{s2_25}) with anti-symmetric Pauli matrix \((\tau_{2})_{\mu\nu}\) and
complex field variables \(\ovv{f}_{r}\) as corresponding anti-symmetric eigenvalues for the BCS terms (\ref{s2_20}).
The rotation matrices \(\hat{P}_{2N\times 2N}^{aa}\) (\ref{s2_21}-\ref{s2_23}) include the remaining field degrees of
freedom of \(\hat{X}_{N\times N}(\vec{x},t_{p})\), \(\wt{\kappa}\;\hat{X}_{N\times N}^{+}(\vec{x},t_{p})\) (\ref{s2_4})
with even hermitian matrices \(\hat{\mscr{C}}_{D;L\times L}\),
\(\hat{\mscr{G}}_{D;S\times S}\) (\ref{s2_24},\ref{s2_25}) for the boson-boson, fermion-fermion parts
where $L$ diagonal real parameters (or $S/2$ complex parameters with quaternion \((\tau_{2})_{\mu\nu}\))
have to be excluded from the ladder operators. They are already contained within the $L$ complex
eigenvalues \(\ovv{c}_{m}\) or $S/2$ complex anti-symmetric quaternion eigenvalues
\((\tau_{2})_{\mu\nu}\;\ovv{f}_{r}\)
\footnote{The range of indices for the angular momentum degrees of freedom for the
fermions and bosons is adapted from \(-s,\ldots,+s\) and \(-l,\ldots,+l\) to the range
\(1,\ldots,S=2s+1\) and \(1,\ldots,L=2l+1\). Furthermore, two notations for the index
range of the fermions are used in parallel in the remainder : (i) The first notation
labels the angular momentum degrees of freedom from \(i,j=1,\ldots,S=2s+1\), especially in the density parts.
(ii) The second one regards the quaternionic structure of the fermion-fermion parts concerning the anomalous sectors
and has a \(2\times 2\) block matrix structure with \(\mu,\nu=1,2\) and
\(r,r\ppr=1,\ldots,S/2\) so that e.g. \(\delta\lambda_{F;r\mu}\) corresponds to
\(\delta\lambda_{F;i=2(r-1)+\mu}\).\lb{ft2}}
\beq \lb{s2_16}
\hat{Y}_{2N\times 2N}(\vec{x},t_{p})&=&\hat{P}_{2N\times 2N}^{-1}(\vec{x},t_{p})\;\;
\hat{Y}_{DD;2N\times 2N}(\vec{x},t_{p})\;\;\hat{P}_{2N\times 2N}(\vec{x},t_{p})  \\ \no &=&
\left(\bea{cc} 0 & \hat{P}^{11,-1}\;\hat{X}_{DD}\;\hat{P}^{22}  \\
\hat{P}^{22,-1}\;\wt{\kappa}\;\hat{X}_{DD}^{+}\;\hat{P}^{11} & 0
\eea\right)^{ab}  \;;   \\ \lb{s2_17}
\hat{Y}_{DD;2N\times 2N}(\vec{x},t_{p}) &=&\left(\bea{cc}
0 & \hat{X}_{DD;N\times N}(\vec{x},t_{p}) \\
\wt{\kappa}\;\hat{X}_{DD;N\times N}^{+}(\vec{x},t_{p}) & 0
\eea\right)\;; \\  \lb{s2_18}
\hat{X}_{DD;N\times N}(\vec{x},t_{p}) &=&\left(\bea{cc}
-\hat{\ovv{c}}_{L\times L}(\vec{x},t_{p}) & 0 \\
0 & \hat{\ovv{f}}_{S\times S}(\vec{x},t_{p})
\eea\right)\;; \\  \lb{s2_19}
\hat{\ovv{c}}_{L\times L}(\vec{x},t_{p})&=&\mbox{diag}\Big\{\ovv{c}_{1}(\vec{x},t_{p}),\ldots,
\ovv{c}_{m}(\vec{x},t_{p}),\ldots,\ovv{c}_{L}(\vec{x},t_{p})\Big\}\;;  \\ \no &&
\ovv{c}_{m}(\vec{x},t_{p})=|\ovv{c}_{m}(\vec{x},t_{p})|\;\exp\{\im\;\varphi_{m}(\vec{x},t_{p})\}\;;\;\;
(\ovv{c}_{m}(\vec{x},t_{p})\in\mbox{\sf C}_{even})\;; \\  \lb{s2_20}
\hat{\ovv{f}}_{S\times S}(\vec{x},t_{p})&=&\mbox{diag}\Big\{\big(\tau_{2}\big)_{\mu\nu}\;\ovv{f}_{1}(\vec{x},t_{p}),\ldots,
\big(\tau_{2}\big)_{\mu\nu}\;\ovv{f}_{r}(\vec{x},t_{p}),\ldots,
\big(\tau_{2}\big)_{\mu\nu}\;\ovv{f}_{S/2}(\vec{x},t_{p})\Big\}\;; \\ \no &&
\ovv{f}_{r}(\vec{x},t_{p})=|\ovv{f}_{r}(\vec{x},t_{p})|\;\exp\{\im\;\phi_{r}(\vec{x},t_{p})\}\;; \\ \no &&
(\ovv{f}_{r}(\vec{x},t_{p})\in\mbox{\sf C}_{even})\;;\;\;(r=1,\ldots,S/2),(\mu,\nu=1,2)\;; \\  \lb{s2_21}
\hat{P}_{2N\times 2N}(\vec{x},t_{p}) &=&\left(\bea{cc}
\hat{P}_{N\times N}^{11}(\vec{x},t_{p}) & 0 \\
0 & \hat{P}_{N\times N}^{22}(\vec{x},t_{p})
\eea\right)\;;\hspace*{0.5cm}\big(\hat{P}_{N\times N}^{22}\big)^{st}=
\hat{P}_{N\times N}^{11,+}=\hat{P}_{N\times N}^{11,-1}\;; \\  \lb{s2_22}
\hat{P}^{11}_{N\times N}(\vec{x},t_{p})&=&\exp\left\{\im\left( \bea{cc}
\hat{\mscr{C}}_{D;L\times L}(\vec{x},t_{p}) & \hat{\xi}_{D;L\times S}^{+}(\vec{x},t_{p}) \\
\hat{\xi}_{D;S\times L}(\vec{x},t_{p}) & \hat{\mscr{G}}_{D;S\times S}(\vec{x},t_{p})
\eea\right)\right\}\;;   \\ \lb{s2_23}
\hat{P}^{22}_{N\times N}(\vec{x},t_{p})&=&\exp\left\{\im\left( \bea{cc}
-\hat{\mscr{C}}_{D;L\times L}^{T}(\vec{x},t_{p}) & \hat{\xi}_{D;L\times S}^{T}(\vec{x},t_{p}) \\
-\hat{\xi}_{D;S\times L}^{*}(\vec{x},t_{p}) & -\hat{\mscr{G}}_{D;S\times S}^{T}(\vec{x},t_{p}) \eea\right)\right\}\;;
\\ \lb{s2_24}
\hat{\mscr{C}}_{D;L\times L}^{+}(\vec{x},t_{p})&=&\hat{\mscr{C}}_{D;L\times L}(\vec{x},t_{p}) \;;\hspace*{0.5cm}
\hat{\mscr{C}}_{D;mm}(\vec{x},t_{p})=0\;;\;\;(m=1,\ldots,L)\;;  \\ \lb{s2_25}
\hat{\mscr{G}}_{D;S\times S}^{+}(\vec{x},t_{p})&=&\hat{\mscr{G}}_{D;S\times S}(\vec{x},t_{p})\;;\hspace*{0.5cm}
\hat{\mscr{G}}_{D;r\mu,r\nu}(\vec{x},t_{p})=0\;;\;\;(r=1,\ldots,S/2),(\mu,\nu=1,2)\;;  \\ \no
\hat{\mscr{G}}_{D;S\times S}(\vec{x},t_{p}) &=&\hat{\mscr{G}}_{D;r\mu,r\ppr\nu}(\vec{x},t_{p})=
\sum_{k=0}^{3}(\tau_{k})_{\mu\nu}\;\hat{\mscr{G}}_{D;rr\ppr}^{(k)}(\vec{x},t_{p})\;;\hspace*{0.5cm}
\big(\,\hat{\mscr{G}}_{D;rr\ppr}^{(k)}(\vec{x},t_{p})\,\big)^{+}=
\hat{\mscr{G}}_{D;rr\ppr}^{(k)}(\vec{x},t_{p}) \;.
\eeq
In section \ref{s34} we have to require that the diagonal matrix elements (the quaternionic diagonal, anti-symmetric
matrix elements) of the boson-boson part (fermion-fermion part) have to vanish in the gauge combination
\((\,\pp \hat{P}_{N\times N}^{aa}(\vec{x},t_{p})\,)\;\hat{P}_{N\times N}^{-1;aa}(\vec{x},t_{p})\) of the
block diagonal matrices \(\hat{P}_{N\times N}^{aa}(\vec{x},t_{p})\) with their derivatives (\ref{s2_26},\ref{s2_27})
\beq\lb{s2_26}
0&\stackrel{!}{=}& \Big[\big(\pp \hat{P}_{N\times N}^{aa}(\vec{x},t_{p})\big)\;
\hat{P}_{N\times N}^{-1;aa}(\vec{x},t_{p})\Big]_{BB;mm}\;; \\ \lb{s2_27}
0&\stackrel{!}{=}& \Big[\big(\pp \hat{P}_{N\times N}^{aa}(\vec{x},t_{p})\big)\;
\hat{P}_{N\times N}^{-1;aa}(\vec{x},t_{p})\Big]_{FF;r\mu,r\nu}\;.
\eeq
This can be accomplished by a gauge transformation (\ref{s2_28}) of \(\hat{P}_{N\times N}^{aa}(\vec{x},t_{p})\) with
a diagonal (quaternion diagonal) matrix \(\hat{P}_{DD;N\times N}^{aa}(\vec{x},t_{p})\) which has only
non-vanishing matrix elements \(\hat{\mscr{C}}_{D;mm}\neq 0\), \(\hat{\mscr{G}}_{D;r\mu,r\nu}\neq 0\)
along the diagonals, just in opposite to \(\hat{P}_{N\times N}^{aa}(\vec{x},t_{p})\) (\ref{s2_24},\ref{s2_25}).
These diagonal, real \(\hat{\mscr{C}}_{D;mm}\) and hermitian \(2\times 2\) elements  \(\hat{\mscr{G}}_{D;r\mu,r\nu}\)
(\ref{s2_31},\ref{s2_32}) have to depend on the
off-diagonal parameters of the ladder operators in \(\hat{P}_{N\times N}^{11}\), \(\hat{P}_{N\times N}^{22}\)
and have to be chosen with suitable dependence in such a manner that the block diagonal, gauge transformed
super-matrices \(\hat{\mscr{P}}_{N\times N}^{aa}=\hat{P}_{DD;N\times N}^{aa}\;\hat{P}_{N\times N}^{aa}\)
(\ref{s2_28}) fulfill the property of
\(\sum_{\beta=1}^{N=L+S}(\pp\hat{\mscr{P}}_{\alpha\beta}^{aa})\;\hat{\mscr{P}}_{\beta\alpha}^{-1;aa}\equiv 0\) (\ref{s2_33}-\ref{s2_35}).
One has to take into account the quaternion algebra for the fermion-fermion parts in order to achieve
\(\sum_{\beta=1}^{N=L+S}(\pp\hat{\mscr{P}}_{\alpha\beta}^{aa})\;\hat{\mscr{P}}_{\beta\alpha}^{-1;aa}\equiv 0\) for
diagonal elements \(\alpha\) (with '\(\alpha\)' denoting a quaternion matrix element in the fermion-fermion section !)
\beq \lb{s2_28}
\hat{P}_{N\times N}^{aa}(\vec{x},t_{p}) &\rightarrow &
\hat{\mscr{P}}_{N\times N}^{aa}(\vec{x},t_{p})=\hat{P}_{DD;N\times N}^{aa}(\vec{x},t_{p})\;\;
\hat{P}_{N\times N}^{aa}(\vec{x},t_{p}) \;;  \\  \lb{s2_29}
\hat{P}^{11}_{DD;N\times N}(\vec{x},t_{p})&=&\exp\left\{\im\left( \bea{cc}
\hat{\mscr{C}}_{D;mm}(\vec{x},t_{p}) & 0 \\
0 & \hat{\mscr{G}}_{D;r\mu,r\nu}(\vec{x},t_{p})
\eea\right)\right\}\;;   \\  \lb{s2_30}
\hat{P}^{22}_{DD;N\times N}(\vec{x},t_{p})&=&\exp\left\{-\im\left( \bea{cc}
\hat{\mscr{C}}_{D;mm}(\vec{x},t_{p}) & 0 \\
0 & \hat{\mscr{G}}_{D;r\mu,r\nu}^{T}(\vec{x},t_{p}) \eea\right)\right\}\;; \\  \lb{s2_31}
\hat{\mscr{C}}_{D;mm}(\vec{x},t_{p}) &=&\hat{\mscr{C}}_{D;mm}\Big(\hat{\mscr{C}}_{D;m\neq n};
\hat{\mscr{G}}_{D;r\mu,r\ppr\nu},\,(r\neq r\ppr);\hat{\xi}_{D;S\times L};\hat{\xi}_{D;S\times L}^{*}\Big) \;; \\  \lb{s2_32}
\hat{\mscr{G}}_{D;r\mu,r\nu}(\vec{x},t_{p})&=&\hat{\mscr{G}}_{D;r\mu,r\nu}\Big(\hat{\mscr{C}}_{D;m\neq n};
\hat{\mscr{G}}_{D;r\mu,r\ppr\nu},\,(r\neq r\ppr);\hat{\xi}_{D;S\times L};\hat{\xi}_{D;S\times L}^{*}\Big) \;; \\ \no
\big(\pp \hat{\mscr{P}}_{N\times N}^{aa}(\vec{x},t_{p})\big)\;\hat{\mscr{P}}_{N\times N}^{-1;aa}(\vec{x},t_{p}) &=&
\hat{P}_{DD;N\times N}^{aa}(\vec{x},t_{p})\;\;
\big(\pp \hat{P}_{N\times N}^{aa}(\vec{x},t_{p})\big)\;\hat{P}_{N\times N}^{-1;aa}(\vec{x},t_{p})\;\;
\hat{P}_{DD;N\times N}^{-1;aa}(\vec{x},t_{p}) \\ \lb{s2_33} &+&
\big(\pp \hat{P}_{DD;N\times N}^{aa}(\vec{x},t_{p})\big)\;\hat{P}_{DD;N\times N}^{-1;aa}(\vec{x},t_{p})\;;
\eeq
\beq \lb{s2_34}
\hat{P}_{DD;mm}^{-1;aa}(\vec{x},t_{p})\;\big(\pp \hat{P}_{DD;mm}^{aa}(\vec{x},t_{p})\big) &=&- \sum_{\alpha=1}^{N=L+S}
\big(\pp \hat{P}_{m,\alpha}^{aa}(\vec{x},t_{p})\big)\;\hat{P}_{\alpha,m}^{-1;aa}(\vec{x},t_{p}) \;; \\ \lb{s2_35} \sum_{\lambda=1,2}
\hat{P}_{DD;r\mu,r\lambda}^{-1;aa}(\vec{x},t_{p})\;\big(\pp \hat{P}_{DD;r\lambda,r\nu}^{aa}(\vec{x},t_{p})\big) &=&- \sum_{\alpha=1}^{N=L+S}
\big(\pp \hat{P}_{r\mu,\alpha}^{aa}(\vec{x},t_{p})\big)\;\hat{P}_{\alpha,r\nu}^{-1;aa}(\vec{x},t_{p}) \;.
\eeq

\subsection{The coset integration measure of $\mbox{Osp}(S,S|2L)/ \mbox{U}(L|S)\otimes \mbox{U}(L|S)$}\lb{s22}

The coset decomposition \(\mbox{Osp}(S,S|2L)/ \mbox{U}(L|S)\otimes \mbox{U}(L|S)\), as described in section
\ref{s21} for the self-energy super-matrix \(\delta\wt{\Sigma}_{\alpha\beta}^{ab}(\vec{x},t_{p})\;\wt{K}\),
involves a nontrivial integration measure. The corresponding super-Jacobi matrix of
this transformation follows from the square root \(\hat{G}_{\mbox{{\scz Osp}}/ \mbox{{\scz U}}}^{1/2}\) of the
\(\mbox{Osp}(S,S|2L)/ \mbox{U}(L|S)\) metric tensor \(\hat{G}_{\mbox{{\scz Osp}}/ \mbox{{\scz U}}}\) for the
invariant length \((ds_{\mbox{{\scz Osp}}/ \mbox{{\scz U}}}(\delta\hat{\lambda}))^{2}\) (\ref{s2_36}).
We neglect anomalies, which are caused by the anticommuting variables, and introduce the super-determinant
\(\mbox{SDET}(\hat{G}_{\mbox{{\scz Osp}}/ \mbox{{\scz U}}}^{1/2})=(\mbox{SDET}(\hat{G}_{\mbox{{\scz Osp}}/ \mbox{{\scz U}}}))^{1/2}\)
of this super-Jacobi-matrix \(\hat{G}_{\mbox{{\scz Osp}}/ \mbox{{\scz U}}}^{1/2}\) for the change of integration measure from
\(d[\sigma_{D}^{(0)}(t_{p})]\) \(d[\delta\wt{\Sigma}_{2N\times 2N}(t_{p})\;\wt{K}]\) to
\(d[\sigma_{D}^{(0)}(t_{p})]\) \(d[d\hat{Q}\:\hat{Q}^{-1};\delta\hat{\lambda}]\)
\(d[\hat{T}^{-1}\:d\hat{T};\delta\hat{\lambda}]\). This change of integration measure can be
particularly obtained by diagonalizing the metric tensor \(\hat{G}_{\mbox{{\scz Osp}}/ \mbox{{\scz U}}}\)
with the eigenvalues \(\hat{\ovv{c}}_{L\times L}(\vec{x},t_{p})\),
\(\hat{\ovv{f}}_{S\times S}(\vec{x},t_{p})\) (\(\ovv{c}_{m}(\vec{x},t_{p})\), \(\ovv{f}_{r}(\vec{x},t_{p})\))
(\ref{s2_16}-\ref{s2_20}) and eigenvector matrices \(\hat{\mscr{P}}_{N\times N}^{aa}(\vec{x},t_{p})\)
(\ref{s2_21}-\ref{s2_25}, \ref{s2_28}-\ref{s2_35}) of the
coset matrices \(\hat{X}_{N\times N}(\vec{x},t_{p})\), \(\wt{\kappa}\;\hat{X}_{N\times N}^{+}(\vec{x},t_{p})\)
(\ref{s2_2}-\ref{s2_5}) for the independent anomalous terms
\beq \lb{s2_36}
\Big(ds_{\mbox{{\scz Osp}}/ \mbox{{\scz U}}}(\delta\hat{\lambda})\Big)^{2}&=&-2\;\strab\Big[
\big(\wt{T}_{0}^{-1}\;d\wt{T}_{0}\big)_{\alpha\beta}^{12}\;\;
\big(\wt{T}_{0}^{-1}\;d\wt{T}_{0}\big)_{\beta\alpha}^{21}\;
\big(\delta\wt{\lambda}_{\beta}+\delta\wt{\lambda}_{\alpha}\big)^{2}\Big] \;;  \\ \lb{s2_37}
\big(\wt{T}_{0}^{-1}\;d\wt{T}_{0}\big)_{\alpha\beta}^{ab} &=&
\big(\hat{\mscr{P}}\;\hat{T}^{-1}\;d\hat{T}\;\hat{\mscr{P}}^{-1}\big)_{\alpha\beta}^{ab}-
\big(\hat{\mscr{P}}\;\hat{Q}^{-1}\;d\hat{Q}\;\hat{\mscr{P}}^{-1}\big)_{\alpha\beta}^{ab} \;; \\ \lb{s2_38}
\lefteqn{\hspace*{-1.6cm}d\big[\sigma_{D}^{(0)}(t_{p})\big]\;\;
d\big[\delta\wt{\Sigma}_{2N\times 2N}(t_{p})\;\wt{K}\big] = } \\ \no &=&
d\big[\sigma_{D}^{(0)}(t_{p})\big]\;\;
d\big[d\hat{Q}(t_{p})\;\hat{Q}^{-1}(t_{p});\delta\hat{\lambda}(t_{p})\big]\;\;
d\big[\hat{T}^{-1}(t_{p})\;d\hat{T}(t_{p});\delta\hat{\lambda}(t_{p})\big] \;.
\eeq
Repeated application of Eq.\ (\ref{s2_39}) (for the variation \(\delta\hat{B}\) of general
generators \(\hat{B}\) in the exponent) allows to simplify the combination
\((\hat{\mscr{P}}\;\hat{T}^{-1}\;d\hat{T}\;\hat{\mscr{P}}^{-1})_{\alpha\beta}^{ab}\)
of coset matrices to a relation (\ref{s2_40}) with their eigenvalues
\(\hat{X}_{DD}\), \(\wt{\kappa}\;\hat{X}_{DD}^{+}\), \(\hat{Y}_{DD}\) for the
pair condensates by an additional integration over a parameter \(v\in[0,1]\) with \(\hat{Y}_{DD}\)
\cite{eng,mies1}
\beq \lb{s2_39}
\exp\big\{\hat{B}\big\}\;\delta\Big(\exp\big\{-\hat{B}\big\}\Big) &=&-\int_{0}^{1}dv\;
\exp\big\{v\;\hat{B}\big\}\;\;\delta\hat{B}\;\;\exp\big\{-v\;\hat{B}\big\} \;; \\  \lb{s2_40}
\big(\hat{\mscr{P}}\;\hat{T}^{-1}\;d\hat{T}\;\hat{\mscr{P}}^{-1}\big)_{\alpha\beta}^{ab} &=&
\Big(\hat{\mscr{P}}\;\exp\big\{\hat{Y}\big\}\;\;
d\Big(\exp\big\{-\hat{Y}\big\}\Big)\;\;\hat{\mscr{P}}^{-1}\Big)_{\alpha\beta}^{ab}  \\ \no &=&
-\int_{0}^{1}dv\;\;\Big(\exp\big\{v\;\hat{Y}_{DD}\big\}\;\;d\hat{Y}\ppr\;\;
\exp\big\{-v\;\hat{Y}_{DD}\big\}\Big)_{\alpha\beta}^{ab} \;; \\   \lb{s2_41}
d\hat{Y}\ppr&=&\hat{\mscr{P}}\;\;d\hat{Y}\;\;\hat{\mscr{P}}^{-1}\;;\hspace*{0.5cm}
\mbox{str}\big[d\hat{Y}\ppr\;d\hat{Y}\ppr\big]=\mbox{str}\big[d\hat{Y}\;d\hat{Y}\big]\;.
\eeq
The integration measure of the \(\hat{\mscr{P}}\), \(\hat{\mscr{P}}^{-1}\)
rotated, independent coset elements \(d\hat{Y}\ppr\) (\ref{s2_41}) is
equivalent to the original anomalous terms within matrix \(d\hat{Y}\)
\footnote{In the remainder \(\hat{\mscr{P}}\), \(\hat{\mscr{P}}^{-1}\) transformed coset elements are marked
by an additional prime " ' " , as e.\ g.\ for \(d\hat{Y}\ppr=\hat{\mscr{P}}\;\;d\hat{Y}\;\;\hat{\mscr{P}}^{-1}\).}.
Therefore, one can perform the integrations over the parameter \(v\in [0,1]\)
with the eigenvalues \(\hat{Y}_{DD}\) of \(\hat{Y}\) straightforwardly to obtain
the metric tensor \(\hat{G}_{\mbox{{\scz Osp}}/ \mbox{{\scz U}}}\) for
\((ds_{\mbox{{\scz Osp}}/ \mbox{{\scz U}}}(\delta\hat{\lambda}))^{2}\) (\ref{s2_36}-\ref{s2_38}). In the following we list
the results for the integration measure in terms of the diagonal coset metric
tensor, determined by \(\ovv{c}_{m}\), \(\ovv{f}_{r}\), and specify in relations (\ref{s2_42}-\ref{s2_45})
the integration measure for the block diagonal self-energy densities (\ref{s2_6}-\ref{s2_15})
\(\delta\hat{\Sigma}_{D;2N\times 2N}\;\wt{K}\) in the coset decomposition
\beq \lb{s2_42}
\delta\hat{\Sigma}_{D;2N\times 2N}\;\wt{K}&=&\hat{Q}_{2N\times 2N}^{-1}\;\;
\delta\hat{\Lambda}_{2N\times 2N}\;\;\hat{Q}_{2N\times 2N} \;; \\  \lb{s2_43}
d\big[d\hat{Q}\;\hat{Q}^{-1};\delta\hat{\lambda}\big] &=&
d\big[\delta\hat{\Sigma}_{D}\;\wt{K}\big] \;;
\eeq
\beq\lb{s2_44}
\lefteqn{d\big[d\hat{Q}\;\hat{Q}^{-1},\delta\hat{\Lambda}\big]=
d\big[\delta\hat{\Sigma}_{D}\;\wt{K}\big]=
\prod_{\{\vec{x},t_{p}\}}\Bigg[\bigg\{2^{(L+S)/2}\;
\bigg(\prod_{m=1}^{L}d\big(\delta\lambda_{B;m}\big)\bigg)
\bigg(\prod_{i=1}^{S}d\big(\delta\lambda_{F;i}\big)\bigg)\bigg\}\times } \\ \no &\times&
\bigg\{\prod_{m=1}^{L}\prod_{n=m+1}^{L}\bigg(4\;
\frac{\big(d\hat{Q}^{11}\;\hat{Q}^{11,-1}\big)_{BB;mn}\wedge
\big(d\hat{Q}^{11}\;\hat{Q}^{11,-1}\big)_{BB;nm}}{2\;\im}\;
\big(\delta\hat{\lambda}_{B;n}-\delta\hat{\lambda}_{B;m}\big)^{2}\bigg)\bigg\}
\\ \no &\times&
\bigg\{\prod_{i=1}^{S}\prod_{i\ppr=i+1}^{S}\bigg(4\;
\frac{\big(d\hat{Q}^{11}\;\hat{Q}^{11,-1}\big)_{FF;ii\ppr}\wedge
\big(d\hat{Q}^{11}\;\hat{Q}^{11,-1}\big)_{FF;i\ppr i}}{2\;\im}\;
\big(\delta\hat{\lambda}_{F;i\ppr}-\delta\hat{\lambda}_{F;i}\big)^{2}\bigg)\bigg\}
\\ \no &\times&
\bigg\{\prod_{m=1}^{L}\prod_{i\ppr=1}^{S}\bigg(\frac{1}{4}\;
\big(d\hat{Q}^{11}\;\hat{Q}^{11,-1}\big)_{BF;mi\ppr}\;
\big(d\hat{Q}^{11}\;\hat{Q}^{11,-1}\big)_{FB;i\ppr m}\;
\big(\delta\hat{\lambda}_{F;i\ppr}-\delta\hat{\lambda}_{B;m}\big)^{\mbox{\boldmath{$^{-2}$}}}\bigg)\bigg\}\Bigg]\;;
\eeq
\beq\lb{s2_45}
\lefteqn{d\big[\delta\hat{\Sigma}_{D}\;\wt{K}\big]=
d\big[d\hat{Q}\;\hat{Q}^{-1},\delta\hat{\Lambda}\big]=
\prod_{\{\vec{x},t_{p}\}}\Bigg[\bigg\{2^{(L+S)/2}\;
\bigg(\prod_{m=1}^{L}d\big(\delta\hat{B}_{D;mm}\big)\bigg)
\bigg(\prod_{i=1}^{S}d\big(\delta\hat{F}_{D;ii}\big)\bigg)\bigg\} \times } \\ \no &\times&
\bigg\{\prod_{m=1}^{L}\prod_{n=m+1}^{L}\bigg(4\;\;
\frac{d\big(\delta\hat{B}_{D;mn}^{*}\big)\;\wedge\;
d\big(\delta\hat{B}_{D;mn}\big)}{2\;\im}\bigg)\bigg\}\times
\bigg\{\prod_{i=1}^{S}\prod_{i\ppr=i+1}^{S}\bigg(4\;\;
\frac{d\big(\delta\hat{F}_{D;ii\ppr}^{*}\big)\;\wedge\;
d\big(\delta\hat{F}_{D;ii\ppr}\big)}{2\;\im}\bigg)\bigg\} \times  \\ \no &\times&
\bigg\{\prod_{m=1}^{L}\prod_{i\ppr=1}^{S}\bigg(\frac{1}{4}\;\;
d\big(\delta\hat{\chi}_{D;i\ppr m}^{*}\big)\;\;
d\big(\delta\hat{\chi}_{D;i\ppr m}\big)\bigg)\bigg\}\Bigg] \;.
\eeq
One has to consider that the original invariant length \((ds_{\mbox{{\scz Osp}}/ \mbox{{\scz U}}}(\delta\hat{\lambda}))^{2}\)
(\ref{s2_36}) of the coset integration measure \(d[\hat{T}^{-1}\;d\hat{T};\delta\hat{\lambda}]\)
(\ref{s2_38},\ref{s2_46}) also incorporates the eigenvalues \(\delta\lambda_{\alpha}\) of the
density terms. However, the eigenvalues \(\delta\lambda_{\alpha}\) of the densities factorize into a
polynomial \(\mfrak{P}(\delta\hat{\lambda})\) (\ref{s2_47}) and
separate from the coset integration \(d[\hat{T}^{-1}\;d\hat{T}]\) which
solely depends on the field variables of \(\hat{X}_{N\times N}\),
\(\wt{\kappa}\;\hat{X}_{N\times N}^{+}\) weighted by functions of their eigenvalues \(\ovv{c}_{m}\),
\(\ovv{f}_{r}\)
\beq \lb{s2_46}
d\big[\hat{T}^{-1}\;d\hat{T};\delta\hat{\lambda}\big] &=&\mfrak{P}\big(\delta\hat{\lambda}\big)\;\;
d\big[\hat{T}^{-1}\;d\hat{T}\big] \;;
\eeq
\beq \lb{s2_47}
\lefteqn{\mfrak{P}\big(\delta\hat{\lambda}\big)=\prod_{\{\vec{x},t_{p}\}}\Bigg[\bigg\{
\bigg(\prod_{m=1}^{L}\big(\delta\hat{\lambda}_{B;m}\big)^{2}\bigg)
\bigg(\prod_{r=1}^{S/2}\big(\delta\hat{\lambda}_{F;r1}+\delta\hat{\lambda}_{F;r2}\big)^{2}\bigg)\bigg\} \times\bigg\{
\prod_{m=1}^{L}\prod_{n=m+1}^{L}\bigg(
\big(\delta\hat{\lambda}_{B;n}+\delta\hat{\lambda}_{B;m}\big)^{2}\bigg)\bigg\} }
\\ \no &\times&
\bigg\{\prod_{r=1}^{S/2}\prod_{r\ppr=r+1}^{S/2}\bigg(
\big(\delta\hat{\lambda}_{F;r1}+\delta\hat{\lambda}_{F;r\ppr 1}\big)^{2}\;\;
\big(\delta\hat{\lambda}_{F;r2}+\delta\hat{\lambda}_{F;r\ppr 2}\big)^{2} \times
\big(\delta\hat{\lambda}_{F;r2}+\delta\hat{\lambda}_{F;r\ppr 1}\big)^{2}\;\;
\big(\delta\hat{\lambda}_{F;r1}+\delta\hat{\lambda}_{F;r\ppr 2}\big)^{2} \bigg)\bigg\} \\ \no &\times&
\bigg\{\prod_{m=1}^{L}\prod_{r\ppr=1}^{S/2}\bigg(
\big(\delta\hat{\lambda}_{F;r\ppr 1}+\delta\hat{\lambda}_{B;m}\big)^{2}\;\;
\big(\delta\hat{\lambda}_{F;r\ppr 2}+\delta\hat{\lambda}_{B;m}\big)^{2} \bigg)^{\mbox{\boldmath{$-1$}}}\bigg\}\Bigg]\;.
\eeq
The actual coset integration measure \(d[\hat{T}^{-1}\;d\hat{T}]\) is listed in relation (\ref{s2_48})
where the polynomial \(\mfrak{P}(\delta\hat{\lambda})\) (\ref{s2_47})
has been isolated and been shifted to the action terms,
which are determined by integrations over the
self-energy densities \(\delta\hat{\Sigma}_{D;2N\times 2N}\;\wt{K}\) (\ref{s2_42}-\ref{s2_45}).
After their removal by integration, these self-energy densities or 'hinge' fields of the SSB
yield the action term \(\mscr{A}_{\hat{J}_{\psi\psi}}[\hat{T}]\) of the 'condensate seeds' with the
source matrix \(\im\;\hat{J}_{\psi\psi;\alpha\beta}^{a\neq b}(\vec{x},t)\) for the pair condensates
\beq \lb{s2_48}
\lefteqn{d\big[\hat{T}^{-1}\;d\hat{T}\big]=
\prod_{\{\vec{x},t_{p}\}}\Bigg[\Bigg\{\prod_{m=1}^{L}\bigg(
\frac{d\hat{c}_{D;mm}^{*}\wedge d\hat{c}_{D;mm}}{2\;\im}\;\;2\;
\left|\frac{\sin\big(2\;|\ovv{c}_{m}|\big)}{|\ovv{c}_{m}|}\right|\bigg)\Bigg\} } \\ \no &\times&
\Bigg\{\prod_{m=1}^{L}\prod_{n=m+1}^{L}\bigg(
\frac{d\hat{c}_{D;mn}^{*}\wedge d\hat{c}_{D;mn}}{2\;\im}\;\;2\;
\left|\frac{\sin\big(|\ovv{c}_{m}|+|\ovv{c}_{n}|\big)}{|\ovv{c}_{m}|+|\ovv{c}_{n}|}\right|\;\;
\left|\frac{\sin\big(|\ovv{c}_{m}|-|\ovv{c}_{n}|\big)}{|\ovv{c}_{m}|-|\ovv{c}_{n}|}\right| \bigg)\Bigg\}
\\ \no &\times& \Bigg\{\prod_{r=1}^{S/2}\bigg(
\frac{d\hat{f}_{D;rr}^{(2)*}\wedge d\hat{f}_{D;rr}^{(2)}}{2\;\im}\;\;
\frac{\sinh\big(2\;|\ovv{f}_{r}|\big)}{|\ovv{f}_{r}|}\bigg)\Bigg\} \\ \no &\times&
\Bigg\{\prod_{r=1}^{S/2}\prod_{r\ppr=r+1}^{S/2}\prod_{k=0}^{3}\bigg(
\frac{d\hat{f}_{D;rr\ppr}^{(k)*}\wedge d\hat{f}_{D;rr\ppr}^{(k)}}{2\;\im}\;4\;
\left|\frac{\sinh\big(|\ovv{f}_{r}|+|\ovv{f}_{r\ppr}|\big)}{
|\ovv{f}_{r}|+|\ovv{f}_{r\ppr}|}\right|\;
\left|\frac{\sinh\big(|\ovv{f}_{r}|-|\ovv{f}_{r\ppr}|\big)}{
|\ovv{f}_{r}|-|\ovv{f}_{r\ppr}|}\right|\bigg)\Bigg\}   \\ \no &\times&
\Bigg\{\prod_{m=1}^{L}\prod_{r\ppr=1}^{S/2}
\frac{d\hat{\eta}_{D;r\ppr 1,m}^{*}\; d\hat{\eta}_{D;r\ppr 1,m}\;\;
d\hat{\eta}_{D;r\ppr 2,m}^{*}\; d\hat{\eta}_{D;r\ppr 2,m}}{ {\ds
\left(2\;\left|\frac{\sinh\big(|\ovv{f}_{r\ppr}|+\im\;|\ovv{c}_{m}|\big)}{|\ovv{f}_{r\ppr}|+
\im\;|\ovv{c}_{m}|}\right|\;\;
\left|\frac{\sinh\big(|\ovv{f}_{r\ppr}|-\im\;|\ovv{c}_{m}|\big)}{|\ovv{f}_{r\ppr}|-\im\;|\ovv{c}_{m}|}\right|
\right)^{\mbox{\boldmath{$2$}}}  }}   \Bigg\}\Bigg]_{\mbox{.}}
\eeq

\subsection{Effective action for pair condensates with
coupling coefficients of the background field} \lb{s23}

The effective actions (\ref{s2_49}) of coset matrices \(\hat{T}(\vec{x},t_{p})\) (\ref{s2_2}-\ref{s2_5})
with their independent fields for anomalous terms in the super-generator \(\hat{Y}(\vec{x},t_{p})\)
follow from a gradient expansion of the super-matrix (\ref{s1_33})
\(\wt{\mscr{M}}_{\vec{x},\alpha;\vec{x}\ppr,\beta}^{ab}(t_{p},t_{q}\ppr)\)
in the coherent state path integral \(Z[\hat{\mscr{J}},J_{\psi},\im\;\hat{J}_{\psi\psi}]\)
(\ref{s1_32}). It is of central importance that the coset decomposition allows a factorization
of the integration measure into density terms and fields for the pair condensates. In the following
we give the result of the gradient expansion combined with the coset decomposition and classify
the effective, remaining actions for anomalous fields
\(\hat{Y}(\vec{x},t_{p})\) (\ref{s2_2}-\ref{s2_5},\ref{s2_16}-\ref{s2_25}) according to the
parameter \(\mscr{N}=\hbar\Omega\;\mscr{N}_{x}\) (\(\Omega=1/\Delta t\), \(\mscr{N}_{x}=(L/\Delta x)^{d}\)).
This parameter \(\mscr{N}\) arises e.\ g.\ in the course of the gradient expansion of the super-determinant
when one has to introduce an appropriate integration for the discrete spatial and time-like points
on an underlying grid with intervals \(\Delta x\) and \(\Delta t\)
\beq \lb{s2_49}
Z[\hat{\mscr{J}},J_{\psi},\im \hat{J}_{\psi\psi}]&=&
\int d\big[\hat{T}^{-1}(\vec{x},t_{p})\;d\hat{T}(\vec{x},t_{p})\big]\;\;
\exp\Big\{\im\;\mscr{A}_{\hat{J}_{\psi\psi}}\big[\hat{T}\big]\Big\}  \\ \no &\times&
\exp\Big\{-\mscr{A}_{\mscr{N}^{-1}}\ppr\big[\hat{T};J_{\psi}\big]-
\mscr{A}_{\mscr{N}^{0}}\ppr\big[\hat{T};J_{\psi}\big]-
\mscr{A}_{\mscr{N}^{+1}}\ppr\big[\hat{T}\big]\Big\} \;\times\;
\exp\Big\{-\mscr{A}\ppr\big[\hat{T};\hat{\mscr{J}}\big]\Big\}_{\mbox{.}}
\eeq
The effective action \(\mscr{A}_{\mscr{N}^{-1}}\ppr\big[\hat{T};J_{\psi}\big]\)
(\ref{s2_49},\ref{s2_54}) of order \(\mscr{N}^{-1}\)
is composed of the gradients (\ref{s2_50}) with super-matrices
\(\hat{Z}=\hat{T}\;\hat{S}\;\hat{T}^{-1}\) (\ref{s2_51},\ref{s2_52}), following from
the expansion of the super-determinant, and the gradients \((\wt{\pp}_{i}\hat{T})\;\hat{T}^{-1}\), resulting
from the expansion of the inverse super-matrix \(\wt{\mscr{M}}_{\vec{x},\alpha;\vec{x}\ppr,\beta}^{-1;ab}(t_{p},t_{q}\ppr)\)
(\ref{s1_33}) with the 'Nambu' doubled source
fields \(J_{\psi;\alpha}^{+a}(\vec{x},t_{p})\,\ldots\, J_{\psi;\alpha}^{a}(\vec{x},t_{p})\).
The combination of coset matrices \(\hat{T}\), \(\hat{T}^{-1}\) to the super-matrix \(\hat{Z}=\hat{T}\;\hat{S}\;\hat{T}^{-1}\)
with metric \(\hat{S}=\{\hat{S}^{aa}\;\delta_{\alpha\beta}\}\)
\((\hat{S}^{11}=+1\,\);\(\,\hat{S}^{22}=-1)\) (\ref{s2_52}) completely restricts
the 'Nambu' doubled super-trace '$\mbox{STR}$' (\ref{s1_29}) to terms of the pair condensates in the off-diagonal
blocks of \((\hat{T}^{-1}\;(\wt{\pp}_{i}\hat{T}))_{\alpha\beta}^{a\neq b}\) (\ref{s2_53})
with super-trace '$\mbox{str}$' (\ref{s1_18})
\beq \lb{s2_50}
\wt{\pp}_{i} &:=& \frac{\hbar}{\sqrt{2\, m}}\frac{\pp}{\pp x^{i}}\;;
\\  \lb{s2_51}
\hat{Z}(\vec{x},t_{p}) &=&\hat{T}(\vec{x},t_{p})\;\hat{S}\;\hat{T}(\vec{x},t_{p}) \;; \\  \lb{s2_52}
\hat{S}&=& \big\{\hat{S}^{a}\;\delta_{ab}\;\delta_{\alpha\beta}\big\}=
\Big\{\underbrace{+\hat{1}_{N\times N}}_{a=1}\;;\;\underbrace{-\hat{1}_{N\times N}}_{a=2}\Big\} \;;
\eeq
\beq \lb{s2_53} \lefteqn{
\STRAB\Big[\Big(\wt{\pp}_{i}\hat{Z}(\vec{x},t_{p})\Big)\;\Big(\wt{\pp}_{j}\hat{Z}(\vec{x},t_{p})\Big)\Big] = } \\ \no &=&
-4\sum_{a,b=1,2}^{a\neq b}\strab\Big\{\Big[\hat{T}^{-1}(\vec{x},t_{p})\;
\big(\wt{\pp}_{i}\hat{T}(\vec{x},t_{p})\big)\Big]_{\alpha\beta}^{a\neq b}\;
\Big[\hat{T}^{-1}(\vec{x},t_{p})\;
\big(\wt{\pp}_{j}\hat{T}(\vec{x},t_{p})\big)\Big]_{\beta\alpha}^{b\neq a}\Big\} \;.
\eeq
The effective coupling functions \(c^{ij}(\vec{x},t_{p})\) (\ref{s2_55}-\ref{s2_58})
for gradients of \(\hat{Z}\) and \(d^{ij}(\vec{x},t_{p})\) (\ref{s2_59}) for
\((\wt{\pp}_{i}\hat{T})\;\hat{T}^{-1}\) are achieved from the action of '{\it unsaturated}' gradient operators
'$\boldsymbol{\wt{\pp}_{i}}$' onto Green functions of the background field \(\sigma_{D}^{(0)}(\vec{x},t_{p})\)
and by the average \(\langle\ldots\rangle_{\hat{\sigma}_{D}^{(0)}}\) with the corresponding
background functional \(Z[j_{\psi};\hat{\sigma}_{D}^{(0)}]\) (see Eqs.\ (\ref{s2_74}-\ref{s2_76}) and Ref.\ \cite{mies1} with
chapter 4 and appendix B)
\beq\lb{s2_54}
\lefteqn{
\mscr{A}_{\mscr{N}^{-1}}\ppr\big[\hat{T};J_{\psi}\big]= \frac{1}{4}\frac{1}{\mscr{N}}\int_{C}\frac{d
t_{p}}{\hbar}\sum_{\vec{x}} c^{ij}(\vec{x},t_{p})\;\;
\mbox{STR}\Big[\Big(\wt{\pp}_{i}\hat{Z}(\vec{x},t_{p})\Big)\;
\Big(\wt{\pp}_{j}\hat{Z}(\vec{x},t_{p})\Big)\Big] + } \\ \no & - & \frac{\im}{\mscr{N}} \int_{C}\frac{d
t_{p}}{\hbar}\sum_{\vec{x}}\sum_{a,b=1,2}
\sum_{\alpha,\beta=1}^{N=L+S} d^{ij}(\vec{x},t_{p})\; \times \\ \no &\times&
\frac{J_{\psi;\beta}^{+,b}(\vec{x},t_{p})}{\mscr{N}}\bigg(\hat{I}\;\wt{K}\;
\Big(\wt{\pp}_{i}\hat{T}(\vec{x},t_{p})\Big)\;\hat{T}^{-1}(\vec{x},t_{p})\;
\Big(\wt{\pp}_{j}\hat{T}(\vec{x},t_{p})\Big)\;\hat{T}^{-1}(\vec{x},t_{p})\;
\hat{I}\bigg)_{\beta\alpha}^{ba}\;\frac{J_{\psi;\alpha}^{a}(\vec{x},t_{p})}{\mscr{N}}\;;
\eeq
\beq\lb{s2_55}
c^{ij}(\vec{x},t_{p})&=&c^{(1),ij}(\vec{x},t_{p})+c^{(2),ij}(\vec{x},t_{p})\;; \\  \lb{s2_56}
\breve{v}(\vec{x},t_{p})&=&\breve{u}(\vec{x})+\breve{\sigma}_{D}^{(0)}(\vec{x},t_{p})=
\frac{u(\vec{x})+\sigma_{D}^{(0)}(\vec{x},t_{p})}{\mscr{N}} \;;  \\   \lb{s2_57}
c^{(1),ij}(\vec{x},t_{p})&=&
-2\bigg\langle\Big(\wt{\pp}_{i}\wt{\pp}_{j}\breve{v}(\vec{x},t_{p})\Big)
\bigg\rangle_{\hat{\sigma}_{D}^{(0)}} - \delta_{ij}\;\sum_{k=1}^{d}
\bigg\langle\Big(\wt{\pp}_{k}\wt{\pp}_{k}\breve{v}(\vec{x},t_{p})\Big)
\bigg\rangle_{\hat{\sigma}_{D}^{(0)}} \;; \\            \lb{s2_58}
c^{(2),ij}(\vec{x},t_{p})&=& 2\bigg\langle\Big(\wt{\pp}_{i}\breve{v}(\vec{x},t_{p})\Big)\;
\Big(\wt{\pp}_{j}\breve{v}(\vec{x},t_{p})\Big)\bigg\rangle_{\hat{\sigma}_{D}^{(0)}} -
\delta_{ij}\;\sum_{k=1}^{d} \bigg\langle\Big(\wt{\pp}_{k}\breve{v}(\vec{x},t_{p})\Big)^{2}
\bigg\rangle_{\hat{\sigma}_{D}^{(0)}}  \;;    \\    \lb{s2_59}
d^{ij}(\vec{x},t_{p}) &=& 2\;\bigg\langle 3\;\Big(\wt{\pp}_{i}\breve{v}(\vec{x},t_{p})\Big)\;
\Big(\wt{\pp}_{j}\breve{v}(\vec{x},t_{p})\Big) -
\Big(\wt{\pp}_{i}\wt{\pp}_{j}\breve{v}(\vec{x},t_{p})\Big)
\bigg\rangle_{\hat{\sigma}_{D}^{(0)}}\;.
\eeq
The action \(\mscr{A}_{\mscr{N}^{0}}\ppr\big[\hat{T};J_{\psi}\big]\) (\ref{s2_60})
of order \(\mscr{N}^{0}\) does not contain coupling parameters as
\(c^{ij}(\vec{x},t_{p})\), \(d^{ij}(\vec{x},t_{p})\) of the background
field apart from the average of \(\langle\sigma_{D}^{(0)}(\vec{x},t_{p})\rangle_{\hat{\sigma}_{D}^{(0)}}\)
for an effective potential which modifies the trap potential \(u(\vec{x})\). It is also composed of a part
following from the gradient expansion of the super-determinant and a part for the coherent BEC wavefunction
with the 'Nambu' doubled bilinear source fields \(J_{\psi;\alpha}^{+a}(\vec{x},t_{p})\,\ldots\, J_{\psi;\alpha}^{a}(\vec{x},t_{p})\)
\beq \no
\mscr{A}_{\mscr{N}^{0}}\ppr\big[\hat{T};J_{\psi}\big]
&=& -\frac{1}{2}\int_{C}\frac{d t_{p}}{\hbar}\sum_{\vec{x}}\Bigg\{
\mbox{STR}\bigg[\hat{T}^{-1}(\vec{x},t_{p})\;\hat{S}\;\Big(\hat{E}_{p}\hat{T}(\vec{x},t_{p})\Big)+
\hat{T}^{-1}(\vec{x},t_{p})\;\Big(\wt{\pp}_{i}\wt{\pp}_{i}\hat{T}(\vec{x},t_{p})\Big)\bigg]+ \\ \lb{s2_60} &+&
\Big(u(\vec{x})-\mu_{0}-\im\;\ve_{p}+\big\langle\sigma_{D}^{(0)}(\vec{x},t_{p})\big\rangle_{\hat{\sigma}_{D}^{(0)}}
\Big)\;\mbox{STR}\bigg[\Big(\hat{T}^{-1}(\vec{x},t_{p})\Big)^{2}-\hat{1}_{2N\times 2N}\bigg]\Bigg\} + \\ \no &-&
\frac{\im}{2}\int_{C}\frac{d t_{p}}{\hbar}\sum_{\vec{x}}\sum_{a,b=1,2}\sum_{\alpha,\beta=1}^{N=L+S}
\frac{J_{\psi;\beta}^{+,b}(\vec{x},t_{p})}{\mscr{N}}\bigg[\hat{I}\;\wt{K}\;
\bigg(\Big(\wt{\pp}_{i}\wt{\pp}_{i}\hat{T}(\vec{x},t_{p})\Big)\;\hat{T}^{-1}(\vec{x},t_{p})+ \\ \no &+&
\hat{T}(\vec{x},t_{p})\;\hat{S}\;\hat{T}^{-1}(\vec{x},t_{p})\;\Big(\hat{E}_{p}\hat{T}(\vec{x},t_{p})\Big)\;
\hat{T}^{-1}(\vec{x},t_{p}) + \\ \no &-&2\;
\Big(\wt{\pp}_{i}\hat{T}(\vec{x},t_{p})\Big)\;\hat{T}^{-1}(\vec{x},t_{p})\;
\Big(\wt{\pp}_{i}\hat{T}(\vec{x},t_{p})\Big)\;\hat{T}^{-1}(\vec{x},t_{p})\bigg)
\hat{I}\bigg]_{\beta\alpha}^{ba}\;\frac{J_{\psi;\alpha}^{a}(\vec{x},t_{p})}{\mscr{N}}\;.
\eeq
The action \(\mscr{A}_{\mscr{N}^{+1}}\ppr\big[\hat{T}\big]\) (\ref{s2_61})
of order \(\mscr{N}^{+1}\) does not involve any gradients and has completely
different properties for the variation of classical field solutions,
due to the additional metric \(\eta_{p}\) in the time contour integral
\beq\lb{s2_61}
\mscr{A}_{\mscr{N}^{+1}}\ppr\big[\hat{T}\big]&=& \frac{\mscr{N}}{2}\int_{C}\frac{d
t_{p}}{\hbar}\eta_{p}\sum_{\vec{x}} \mbox{STR}\bigg[\Big(\hat{T}^{-1}(\vec{x},t_{p})\Big)^{2}-
\hat{1}_{2N\times 2N}\bigg]\;.
\eeq
According to the additional contour metric \(\eta_{p}\) in (\ref{s2_61}), the two branches of the time contour
integral in \(\mscr{A}_{\mscr{N}^{+1}}\ppr\big[\hat{T}\big]\) are added whereas the two branches
of time contour integrals in \(\mscr{A}_{\mscr{N}^{-1}}\ppr\big[\hat{T};J_{\psi}\big]\) (\ref{s2_54}),
\(\mscr{A}_{\mscr{N}^{0}}\ppr\big[\hat{T};J_{\psi}\big]\) (\ref{s2_60}) are subtracted. Therefore, the variation
\(\delta\hat{Y}(\vec{x},t_{p})\) (\ref{s2_62}-\ref{s2_64}) of classical fields
in \(\mscr{A}_{\mscr{N}^{+1}}\ppr\big[\hat{T}\big]\) (\ref{s2_61})
has its first contribution in the second order variation with \(\delta\hat{Y}(\vec{x},t_{p})\)
for the independent, anomalous fields \(y_{\kappa}(\vec{x},t_{p})\) in \(\hat{Y}(\vec{x},t_{p})=
y_{\kappa}(\vec{x},t_{p})\;\hat{Y}^{(\kappa)}\) with coset super-generators \(\hat{Y}^{(\kappa)}\)
(concerning variation of actions with contour time integrals in coherent
state path integrals see Refs.\ \cite{bmlv1,bmlv2})
\beq\lb{s2_62}
\hat{Y}(\vec{x},t_{p=\pm})&=&\hat{Y}(\vec{x},t)+\delta\hat{Y}(\vec{x},t_{p=\pm})=
\hat{Y}(\vec{x},t)\pm\frac{1}{2}\;\delta\hat{Y}(\vec{x},t) \;; \\  \lb{s2_63}
\hat{Y}(\vec{x},t_{p=\pm})&=&y_{\kappa}(\vec{x},t_{p=\pm})\;\;\hat{Y}^{(\kappa)} \;; \\   \lb{s2_64}
y_{\kappa}(\vec{x},t_{p=\pm})&=&y_{\kappa}(\vec{x},t)+\delta y_{\kappa}(\vec{x},t_{p=\pm}) =
y_{\kappa}(\vec{x},t)\pm\frac{1}{2}\;\delta y_{\kappa}(\vec{x},t)\;.
\eeq
The variations of \(\mscr{A}_{\mscr{N}^{-1}}\ppr\big[\hat{T};J_{\psi}\big]\) (\ref{s2_54}),
\(\mscr{A}_{\mscr{N}^{0}}\ppr\big[\hat{T};J_{\psi}\big]\) (\ref{s2_60}) already contribute in first order of
\(\delta\hat{Y}(\vec{x},t_{p=\pm})\) (\ref{s2_62}-\ref{s2_64}) and allow for classical field solutions following from
first order variations to a stationary phase in the coherent state path integral (\ref{s2_49}). The second
and all higher even order variations of \(\mscr{A}_{\mscr{N}^{+1}}\ppr\big[\hat{T}\big]\) (\ref{s2_61}) modify
these classical, first order variational solutions of \(\mscr{A}_{\mscr{N}^{-1}}\ppr\big[\hat{T};J_{\psi}\big]\),
\(\mscr{A}_{\mscr{N}^{0}}\ppr\big[\hat{T};J_{\psi}\big]\)  and can be regarded as general fluctuation terms with
{\it universal properties}, entirely determined by the symmetries of the coset decomposition for the anomalous fields.
This property of contributing only from second and higher even order variations with \(\delta\hat{Y}(\vec{x},t_{p=\pm})\)
also holds for the coset integration measure (\ref{s2_48}) and causes the inconsistent treatment in comparison
to the other main actions \(\mscr{A}_{\mscr{N}^{-1}}\ppr\big[\hat{T};J_{\psi}\big]\) (\ref{s2_54}),
\(\mscr{A}_{\mscr{N}^{0}}\ppr\big[\hat{T};J_{\psi}\big]\) (\ref{s2_60}). The transformation with the inverse square root
\(\hat{G}_{\mbox{{\scz Osp}}/ \mbox{{\scz U}}}^{-1/2}\) of the coset metric tensor removes
this {\it artificial} problem and yields Euclidean path integration measures
for the independent, anomalous fields. However, one obtains a different dependence of the pair condensate fields
in the actions \(\mscr{A}_{\mscr{N}^{-1}}\ppr\big[\hat{T};J_{\psi}\big]\) (\ref{s2_54}),
\(\mscr{A}_{\mscr{N}^{0}}\ppr\big[\hat{T};J_{\psi}\big]\) (\ref{s2_60}), according to the transformation with the
super-Jacobi matrix \(\hat{G}_{\mbox{{\scz Osp}}/ \mbox{{\scz U}}}^{-1/2}\). The functional dependence of anomalous fields
is also changed by this transformation in the action term
\(\mscr{A}_{\mscr{N}^{+1}}\ppr\big[\hat{T}\big]\) (\ref{s2_61}) which, however, cannot be eliminated as the coset integration measure
(\ref{s2_48}). In spite of the transformation with the inverse square root of the coset metric tensor, the action term
\(\mscr{A}_{\mscr{N}^{+1}}\ppr\big[\hat{T}\big]\) (\ref{s2_61}) only allows
non-vanishing variations with \(\delta\hat{Y}(\vec{x},t_{p=\pm})\) in second and higher even orders and modifies
the classical solutions of the first order variations of \(\mscr{A}_{\mscr{N}^{-1}}\ppr\big[\hat{T};J_{\psi}\big]\),
\(\mscr{A}_{\mscr{N}^{0}}\ppr\big[\hat{T};J_{\psi}\big]\) by {\it universal fluctuations}.
These {\it universal fluctuations} are solely determined by the coset decomposition
\(\mbox{Osp}(S,S|2L)/ \mbox{U}(L|S)\otimes \mbox{U}(L|S)\), due to the absence of any background
field dependencies.

The pair condensate action term \(\mscr{A}_{\hat{J}_{\psi\psi}}[\hat{T}]\) (\ref{s2_65}) in
\(Z[\hat{\mscr{J}},J_{\psi},\im\;\hat{J}_{\psi\psi}]\) (\ref{s2_49}) arises from the integration
of the quadratic self-energy density action
\(\mscr{A}_{2}\big[\hat{T},\delta\hat{\Sigma}_{D};\im\hat{J}_{\psi\psi}\big]\) (\ref{s2_66})
over the independent density fields of
\(\delta\hat{\Sigma}_{D}\;\wt{K}\) (\ref{s2_45})
with inclusion of the polynomial \(\mfrak{P}(\delta\hat{\lambda})\) (\ref{s2_47})
of the eigenvalues \(\delta\hat{\lambda}\) (\ref{s2_11},\ref{s2_12})
for the density super-matrices \(\delta\hat{\Sigma}_{D}^{11}\) or
\(\delta\hat{\Sigma}_{D}^{22}\;\wt{\kappa}\) (\ref{s2_6}-\ref{s2_9}).
The eigenvalues \(\delta\lambda_{\alpha}\) can be discerned as the
parameters or variables of the \(\mbox{U}(L|S)\) related density terms \(\delta\hat{\Sigma}_{D}^{11}\),
\(\delta\hat{\Sigma}_{D}^{22}\;\wt{\kappa}\) (\ref{s2_6}-\ref{s2_15})
within the characteristic eigenvalue equations of the super-determinants (\ref{s2_67},\ref{s2_68})
so that the integration measure \(d[\delta\hat{\Sigma}_{D}\;\wt{K}]\) (\ref{s2_45}) with additional polynomial
\(\mfrak{P}(\delta\hat{\lambda}(\delta\hat{\Sigma}_{D}\;\wt{K}))\) (\ref{s2_47}) can be used to specify the effective
pair 'condensate seed' action \(\mscr{A}_{\hat{J}_{\psi\psi}}\big[\hat{T}\big]\) (\ref{s2_65}).
Alternatively, the factorization of \(\delta\hat{\Sigma}_{D}\;\wt{K}\) into eigenvalues \(\delta\hat{\lambda}_{\alpha}\)
and eigenvectors \(\hat{Q}_{\alpha\beta}^{aa}\), \(\hat{Q}_{\alpha\beta}^{-1;aa}\) (\ref{s2_6}-\ref{s2_15})
can be applied to determine the action
\(\mscr{A}_{\hat{J}_{\psi\psi}}\big[\hat{T}\big]\) (\ref{s2_65}) after integration of
\(\mscr{A}_{2}\big[\hat{T},\hat{Q}^{-1}\;\delta\hat{\Lambda}\;\hat{Q};\im\hat{J}_{\psi\psi}\big]\) (\ref{s2_69})
over the eigenvalues and eigenvectors with inclusion of \(\mfrak{P}(\delta\hat{\lambda})\) (\ref{s2_47}).
The latter method of factorization with eigenvalues allows to disentangle the integrations with properties of
Vandermonde matrices and Gaussian weights for orthogonal Hermite- (or related Laguerre) polynomials \cite{mehta}
\beq\lb{s2_65}
\lefteqn{
\exp\Big\{\im\;\mscr{A}_{\hat{J}_{\psi\psi}}\big[\hat{T}\big]\Big\}  =
\int d\big[\delta\hat{\Sigma}_{D}(\vec{x},t_{p})\;\wt{K}\big]\;\;
\mfrak{P}\big(\delta\hat{\lambda}(\vec{x},t_{p})\big)\;\;
\exp\Big\{\im\;\mscr{A}_{2}\big[\hat{T},\delta\hat{\Sigma}_{D};\im\hat{J}_{\psi\psi}\big]\Big\} = }
\\ \no &=&
\int d\big[d\hat{Q}(\vec{x},t_{p})\;\hat{Q}^{-1}(\vec{x},t_{p});
\delta\hat{\lambda}(\vec{x},t_{p})\big]\;\;
\mfrak{P}\big(\delta\hat{\lambda}(\vec{x},t_{p})\big)\;\;
\exp\Big\{\im\;\mscr{A}_{2}\big[\hat{T},\hat{Q}^{-1}\;\delta\hat{\Lambda}\;\hat{Q};
\im\hat{J}_{\psi\psi}\big]\Big\}\;;
\eeq
\beq\lb{s2_66}
\lefteqn{\hspace*{-1.8cm}\mscr{A}_{2}\big[\hat{T},\delta\hat{\Sigma}_{D};\im\hat{J}_{\psi\psi}\big] =
\frac{1}{4\hbar\;V_{0}}\int_{C}d t_{p}\sum_{\vec{x}} \bigg\{
\mbox{STR}\Big[\delta\hat{\Sigma}_{D;2N\times 2N}(\vec{x},t_{p})\;
\wt{K}\;\delta\hat{\Sigma}_{D;2N\times 2N}(\vec{x},t_{p})\;
\wt{K}\Big]+  } \\ \no &-&
2\;\mbox{STR}\Big[\im\;\hat{J}_{\psi\psi}(\vec{x},t_{p})\;\wt{K}\;
\hat{T}(\vec{x},t_{p})\;\delta\hat{\Sigma}_{D;2N\times 2N}(\vec{x},t_{p})\;\wt{K}\;
\hat{T}^{-1}(\vec{x},t_{p}) \Big] +
\\ \no &+&\mbox{STR}\Big[\im\;\hat{J}_{\psi\psi}(\vec{x},t_{p})\;\wt{K}\;\im\;
\hat{J}_{\psi\psi}(\vec{x},t_{p})\;\wt{K}\Big]\bigg\} \;;
\eeq
\beq\lb{s2_67}
\mbox{sdet}\Big\{\delta\hat{\Sigma}_{D;\alpha\beta}^{11}-\delta\lambda\;\;\delta_{\alpha\beta}\Big\}
&=&0\;;\hspace*{1.5cm}
\mbox{sdet}\Big\{\delta\hat{\Sigma}_{D;\alpha\beta}^{22}\;\wt{\kappa}-
\big(-\delta\lambda\big)\;\;\delta_{\alpha\beta}\Big\}=0 \;; \\   \lb{s2_68}
\delta\hat{\Sigma}_{D;N\times N}^{11}(\vec{x},t_{p})&=&-
\Big(\delta\hat{\Sigma}_{D;N\times N}^{22}(\vec{x},t_{p})\;\;\wt{\kappa}\Big)^{st}\;;
\eeq
\beq\lb{s2_69}
\lefteqn{\mscr{A}_{2}\big[\hat{T},\hat{Q}^{-1}\;\delta\hat{\Lambda}\;\hat{Q};
\im\hat{J}_{\psi\psi}\big]  = }  \\ \no &=&
\frac{1}{4\hbar\;V_{0}}\int_{C}d t_{p}\sum_{\vec{x}}
\mbox{STR}\Big[\big(\delta\wt{\Sigma}(\vec{x},t_{p})-\im\;\hat{J}_{\psi\psi}(\vec{x},t_{p})\big)\;
\wt{K}\;\big(\delta\wt{\Sigma}(\vec{x},t_{p})-\im\;\hat{J}_{\psi\psi}(\vec{x},t_{p})\big)\;
\wt{K}\Big]
\\ \no  &=&\frac{1}{4\hbar\;V_{0}}\int_{C}d t_{p}\sum_{\vec{x}}\bigg\{
2\;\mbox{str}\Big[\big(\delta\hat{\lambda}_{N\times N}(\vec{x},t_{p})\big)^{2}\Big]+ \\ \no &-&
2\;\mbox{STR}\Big[\im\;\hat{J}_{\psi\psi}(\vec{x},t_{p})\;\wt{K}\;
\hat{T}(\vec{x},t_{p})\;\hat{Q}^{-1}(\vec{x},t_{p})\;
\delta\hat{\Lambda}(\vec{x},t_{p})\;
\hat{Q}(\vec{x},t_{p})\;\hat{T}^{-1}(\vec{x},t_{p})
\Big] + \\ \no &+&\mbox{STR}\Big[\im\;\hat{J}_{\psi\psi}(\vec{x},t_{p})\;\wt{K}\;\im\;
\hat{J}_{\psi\psi}(\vec{x},t_{p})\;\wt{K}\Big]\bigg\}\;.
\eeq
It remains to outline the averaging procedure \(\langle\ldots\rangle_{\hat{\sigma}_{D}^{(0)}}\)
of the coupling coefficients \(c^{ij}(\vec{x},t_{p})\), \(d^{ij}(\vec{x},t_{p})\) with the
generating functional \(Z[j_{\psi};\hat{\sigma}_{D}^{(0)}]\) (\ref{s2_74})
of the background field \(\sigma_{D}^{(0)}(\vec{x},t_{p})\).
Aside from the quadratic term of \(\sigma_{D}^{(0)}(\vec{x},t_{p})\) following from the HST's,
the Hamilton operator \(\hat{\mscr{H}}[\hat{\sigma}_{D}^{(0)}]\) (\ref{s2_70}) specifies the
determinant and the coherent BEC-wavefunction parts with the source fields \(j_{\psi;\alpha}(\vec{x},t_{p})\)
in \(Z[j_{\psi};\hat{\sigma}_{D}^{(0)}]\) (\ref{s2_74}). If the number \(L=2l+1\), \((l=0,1,2,\ldots)\) of bosonic
angular momentum degrees of freedom exceeds those of fermionic angular momentum degrees of freedom
\((L>S=2s+1)\), \((s=\frac{1}{2},\frac{3}{2},\frac{5}{2},\ldots)\), the determinant of the operator
\(\hat{\mscr{H}}[\hat{\sigma}_{D}^{(0)}]\) (\ref{s2_70}) appears in the denominator
\((\,\det(\hat{\mscr{H}}[\hat{\sigma}_{D}^{(0)}])\,)^{-(L-S)}\) with a power \((L-S)>0\).
In this extraordinary case \((L-S)>0\) combined with attractive interactions \(V_{0}<0\), the
background generating functional \(Z[j_{\psi};\hat{\sigma}_{D}^{(0)}]\) (\ref{s2_74})
may describe the experimentally observable,
considerable increase of the coherent bosonic BEC-wavefunctions towards the collapse, due to the appearance of
effective zero eigenvalues of \(\hat{\mscr{H}}[\hat{\sigma}_{D}^{(0)}]\) (\ref{s2_70}) in the
propagation with \((\,\det(\hat{\mscr{H}}[\hat{\sigma}_{D}^{(0)}])\,)^{-(L-S)}\)
in \(Z[j_{\psi};\hat{\sigma}_{D}^{(0)}]\) \cite{hulet4,hulet5}. We list in relations (\ref{s2_70}-\ref{s2_73})
the Hamilton operator \(\hat{\mscr{H}}[\hat{\sigma}_{D}^{(0)}]\) (\ref{s2_70}), its
corresponding Green function \(\hat{g}^{(0)}[\hat{\sigma}_{D}^{(0)}]\) (\ref{s2_71}) and the definitions
of the trace '\(\mbox{tr}\)' (\ref{s2_72}) and unit operator '\(\hat{1}\) (\ref{s2_73}) in the considered Hilbert space
with the complete set of states concerning spatial points and the contour time (compare with Ref.\ \cite{mies1},
chapter 4.2)
\beq \lb{s2_70}
\hat{\mscr{H}}[\hat{\sigma}_{D}^{(0)}]&=&
\bigg(\hat{\eta}\big(-\hat{E}_{p}
\underbrace{-\im\;\ve_{p}+\frac{\hat{\vec{p}}^{\;2}}{2m}+u(\hat{\vec{x}})-\mu_{0}}_{\hat{h}_{p}}\big)
+\hat{\sigma}_{D}^{(0)} \bigg)\;;  \\ \lb{s2_71}
\hat{g}^{(0)}[\hat{\sigma}_{D}^{(0)}]&=&\Big(\hat{\mscr{H}}[\hat{\sigma}_{D}^{(0)}]\Big)^{-1}=
\bigg(\hat{\eta}\big(-\hat{E}_{p}
\underbrace{-\im\;\ve_{p}+\frac{\hat{\vec{p}}^{\;2}}{2m}+u(\hat{\vec{x}})-\mu_{0}}_{\hat{h}_{p}}\big)
+\hat{\sigma}_{D}^{(0)} \bigg)^{-1}\;; \\    \lb{s2_72}
\mbox{tr}\Big[\ldots\Big]&=&\int_{C}\frac{dt_{p}}{\hbar}\eta_{p}\sum_{\vec{x}}\mscr{N}\;
\langle\vec{x},t_{p}|\ldots|\vec{x},t_{p}\rangle \\ \no &=&
\int_{-\infty}^{\infty}\frac{dt_{+}}{\hbar}\sum_{\vec{x}}\mscr{N}\;
\langle\vec{x},t_{+}|\ldots|\vec{x},t_{+}\rangle +
\int_{-\infty}^{\infty}\frac{dt_{-}}{\hbar}\sum_{\vec{x}}\mscr{N}\;
\langle\vec{x},t_{-}|\ldots|\vec{x},t_{-}\rangle \;; \\  \lb{s2_73}
\hat{1} &=&\int_{C}\frac{dt_{p}}{\hbar}\eta_{p}\sum_{\vec{x}}\mscr{N}\;
|\vec{x},t_{p}\rangle\langle\vec{x},t_{p}| \\ \no &=&
\int_{-\infty}^{\infty}\frac{dt_{+}}{\hbar}\sum_{\vec{x}}\mscr{N}\;
|\vec{x},t_{+}\rangle \langle\vec{x},t_{+}|+
\int_{-\infty}^{\infty}\frac{dt_{-}}{\hbar}\sum_{\vec{x}}\mscr{N}\;
|\vec{x},t_{-}\rangle\langle\vec{x},t_{-}|\;;  \\   \lb{s2_74}
Z[j_{\psi};\hat{\sigma}_{D}^{(0)}] &=& \int d[\hat{\sigma}_{D}^{(0)}(\vec{x},t_{p})]\;\;
\exp\bigg\{\frac{\im}{2\hbar}\frac{1}{V_{0}}\int_{C}d t_{p}\sum_{\vec{x}}
\sigma_{D}^{(0)}(\vec{x},t_{p})\;\sigma_{D}^{(0)}(\vec{x},t_{p})\bigg\}    \\ \no &\times &
\exp\bigg\{-(L-S)\;\mbox{tr}\bigg[\ln\bigg(\hat{\eta}\big(-\hat{E}_{p}
\underbrace{-\im\;\ve_{p}+\frac{\hat{\vec{p}}^{\;2}}{2m}+u(\hat{\vec{x}})-\mu_{0}}_{\hat{h}_{p}}\big)
+\hat{\sigma}_{D}^{(0)} \bigg)\bigg]\bigg\}  \\ \no &\times&
\exp\bigg\{\im\sum_{\alpha=1}^{N=L+S}\frac{1}{\mscr{N}}\;
\langle j_{\psi;\alpha}|\hat{\eta}\;\hat{g}^{(0)}[\hat{\sigma}_{D}^{(0)}]\;\hat{\eta}|
j_{\psi;\alpha}\rangle\bigg\}_{\mbox{.}}
\eeq
The averaging procedure \(\langle\ldots\rangle_{\hat{\sigma}_{D}^{(0)}}\) (\ref{s2_75})
for the coupling coefficients \(c^{ij}(\vec{x},t_{p})\), \(d^{ij}(\vec{x},t_{p})\) has therefore to be performed
with the generating function \(Z[j_{\psi};\hat{\sigma}_{D}^{(0)}]\) (\ref{s2_74})
of the background field \(\sigma_{D}^{(0)}(\vec{x},t_{p})\) according to the following relation
\beq\lb{s2_75}
\lefteqn{
\bigg\langle\bigg(\mbox{functional of }\sigma_{D}^{(0)}(\vec{x},t_{p})\mbox{ with gradient terms}\bigg)
\bigg\rangle_{\hat{\sigma}_{D}^{(0)}}= } \\ \no &=&
\int d[\hat{\sigma}_{D}^{(0)}(\vec{x},t_{p})] \;\;
\exp\bigg\{\frac{\im}{2\hbar}\frac{1}{V_{0}}\int_{C}d t_{p}\sum_{\vec{x}}
\sigma_{D}^{(0)}(\vec{x},t_{p})\;\sigma_{D}^{(0)}(\vec{x},t_{p})\bigg\}  \\ \no &\times&
\exp\bigg\{-(L-S)\;\mbox{tr}\ln\bigg(\hat{\eta}\big(-\hat{E}_{p}
\underbrace{-\im\;\ve_{p}+\frac{\hat{\vec{p}}^{2}}{2m}+u(\hat{\vec{x}})-\mu_{0}}_{\hat{h}_{p}}\big)+
\sigma_{D}^{(0)}\bigg)\bigg\}  \\ \no &\times&
\exp\bigg\{\im\sum_{\alpha=1}^{N=L+S}\frac{1}{\mscr{N}}
\langle j_{\psi;\alpha}|\hat{\eta}\;\hat{g}^{(0)}[\hat{\sigma}_{D}^{(0)}]\;\hat{\eta}
|j_{\psi;\alpha}\rangle\bigg\}
\;\times\;\bigg(\mbox{functional of }\sigma_{D}^{(0)}(\vec{x},t_{p})\mbox{ with gradient terms}\bigg)_{\mbox{.}}
\eeq
Instead of functional averaging by \(Z[j_{\psi};\hat{\sigma}_{D}^{(0)}]\) according to Eq. (\ref{s2_75}),
one can also apply a saddle point equation or first order variation with the background field in order
to obtain a mean field solution for \(\sigma_{D}^{(0)}(\vec{x},t_{p})\) (\ref{s2_76}). This mean field solution
can then be substituted for the functional dependence of the coupling coefficients
\(c^{ij}(\vec{x},t_{p})\), \(d^{ij}(\vec{x},t_{p})\) on the background field according to
the defining relations (\ref{s2_55}-\ref{s2_59}). One can expect a good approximation by this mean field solution
because the generating function \(Z[j_{\psi};\hat{\sigma}_{D}^{(0)}]\) is only determined
by the background field \(\sigma_{D}^{(0)}(\vec{x},t_{p})\) which is itself related to
the difference of boson-boson and fermion-fermion densities due to the \(\mbox{U}(1)\) symmetries
in \(Z[j_{\psi};\hat{\sigma}_{D}^{(0)}]\)
\beq\no
\lefteqn{\hspace*{-1.4cm}0\equiv \frac{\im}{V_{0}}\frac{1}{\mscr{N}}\;\sigma_{D}^{(0)}(\vec{x},t_{p})-\big(L-S\big)\;
\bigg[-\hat{E}_{p}-\im\;\ve_{p}+\frac{\hat{\vec{p}}^{2}}{2m}+u(\hat{\vec{x}})-\mu_{0}+\sigma_{D}^{(0)}(\vec{x},t_{p})
\bigg]_{\vec{x},\vec{x}}^{\mathbf{-1}}\!\!\!(t_{p}+\delta t_{p},t_{p}+\delta t_{p}\ppr) + } \\  \lb{s2_76} &-&\im\;
\int_{C}\frac{dt_{q_{1}}^{(1)}}{\hbar}\frac{dt_{q_{2}}^{(2)}}{\hbar}\sum_{\vec{y}_{1},\vec{y}_{2}}\mscr{N}
\sum_{\alpha=1}^{N=L+S} \;\;\times \\ \no &\times&
j_{\psi;\alpha}^{+}(\vec{y}_{2},t_{q_{2}}^{(2)})\;\;
\bigg[-\hat{E}_{p}-\im\;\ve_{p}+\frac{\hat{\vec{p}}^{2}}{2m}+u(\hat{\vec{x}})-\mu_{0}+\sigma_{D}^{(0)}(\vec{x},t_{p})
\bigg]_{\vec{y}_{2},\vec{x}}^{\mathbf{-1}}\!\!\!\!(t_{q_{2}}^{(2)},t_{p}+\delta t_{p}\ppr)   \;\times \\ \no &\times&
\bigg[-\hat{E}_{p}-\im\;\ve_{p}+\frac{\hat{\vec{p}}^{2}}{2m}+u(\hat{\vec{x}})-\mu_{0}+\sigma_{D}^{(0)}(\vec{x},t_{p})
\bigg]_{\vec{x},\vec{y}_{1}}^{\mathbf{-1}}\!\!\!\!(t_{p}+\delta t_{p},t_{q_{1}}^{(1)}) \;\;\;
j_{\psi;\alpha}(\vec{y}_{1},t_{q_{1}}^{(1)})  \;.
\eeq

\subsection{Scaling of physical parameters and quantities to dimensionless values and fields}\lb{s24}

A typical property of classical equations concerns the re-scaling of physical parameters and
quantities to dimensionless values and fields, as e.g.\ in the Gross-Pitaevskii or nonlinear
Schr\"{o}dinger equation. In the considered, effective coherent state path integral
\(Z[\hat{\mscr{J}},J_{\psi},\im\hat{J}_{\psi\psi}]\) (\ref{s2_49}), we therefore scale
the actions \(\mscr{A}_{\mscr{N}^{-1}}\ppr[\hat{T};J_{\psi}]\) (\ref{s2_54}),
\(\mscr{A}_{\mscr{N}^{0}}\ppr[\hat{T};J_{\psi}]\) (\ref{s2_60}) and
\(\mscr{A}_{\mscr{N}^{+1}}\ppr[\hat{T}]\) (\ref{s2_61}) with the pair condensate fields in
dependence on discrete spatial and time-like coordinates to dimensionless values.
In consequence, one can perform the first and higher order variations of the actions
(under the presumed Euclidean path integration fields) in classical correspondence
to the coherent state path integral which represents many-body quantum mechanics. We list
in Eq.\ (\ref{s2_77}) the parameters \(\Omega\) and \(\mscr{N}_{x}\), which indicate
the maximum energy \(\hbar\;\Omega\) and number of discrete spatial points, and combine
them to the parameter \(\mscr{N}\) which inevitably appears with the space and time
contour integrals. Furthermore, we have to scale all energy parameters and potentials
to dimensionless quantities with the parameter \(\mscr{N}\)
\beq \lb{s2_77}\hspace*{-0.6cm}
\Omega=1/\Delta t\;,\;\;\;\mscr{N}_{x}=(L/\Delta x)^{d} &\rightarrow&\mscr{N}=\hbar\Omega\;\mscr{N}_{x}
\;; \\  \lb{s2_78}
\ve_{p}\;,\;\mu_{0}\;,\;V_{0}\;,\;u(\vec{x}) &\rightarrow&
\breve{\ve}_{p}=\ve_{p}/\mscr{N}\;,\;\breve{\mu}_{0}=\mu_{0}/\mscr{N}\;,\;\breve{V}_{0}=V_{0}/\mscr{N}\;,\;
\breve{u}(\vec{x})=u(\vec{x})/\mscr{N} \;;  \\  \lb{s2_79}
\sigma_{D}^{(0)}(\vec{x},t_{p}) &\rightarrow& \breve{\sigma}_{D}^{(0)}(\vec{x},t_{p})=
\sigma_{D}^{(0)}(\vec{x},t_{p})/\mscr{N}\;.
\eeq
The dimensionless, scaled quantities are denoted by the additional symbol
'\(\breve{\ph{V_{0}}}\)' above the corresponding, original physical parameter
or physical quantity symbol. Similarly, the contour time \(t_{p}=t_{0}\;\breve{t}_{p}\),
the contour time derivative \(\hat{E}_{p}=\mscr{N}\;\breve{E}_{p}\) and the spatial
coordinates \(\vec{x}=x_{0}\;\breve{\vec{x}}\) with their gradients \(\pp_{i}=\breve{\pp}_{i}/x_{0}\)
are scaled by the parameters \(t_{0}=\hbar/\mscr{N}\), \(x_{0}=t_{0}\cdot 1/(2\frac{m}{\mscr{N}})^{1/2}\)
to dimensionless quantities
\beq \lb{s2_80}
\hat{E}_{p}=\im\;\hbar\;\frac{\pp}{\pp t_{p}}\;;\;\;\hbar\;\omega_{p} &\rightarrow&
\im\;\frac{\pp}{\pp \breve{t}_{p}}=\hat{E}_{p}/\mscr{N}=\breve{E}_{p}=\im\;\breve{\pp}_{\breve{t}_{p}}
\;;\;\;\breve{\omega}_{p}=\hbar\;\omega_{p}/\mscr{N} \;; \\ \lb{s2_81} &&
t_{0}=\hbar/\mscr{N}\;;\;\;\;d\breve{t}_{p}=\mscr{N}\;dt_{p}/\hbar=dt_{p}/t_{0} \;; \\  \lb{s2_82}
 \frac{\hbar^{2}}{2m}\;\frac{\pp}{\pp\vec{x}}\cdot\frac{\pp}{\pp\vec{x}} &\rightarrow&
\bigg(\frac{\hbar^{2}\;m^{-1}}{2\;\mscr{N}}\bigg)\;\frac{\pp}{\pp\vec{x}}\cdot\frac{\pp}{\pp\vec{x}} =
x_{0}^{2}\;\frac{\pp}{\pp\vec{x}}\cdot\frac{\pp}{\pp\vec{x}}=
\frac{\pp}{\pp\breve{\vec{x}}}\cdot\frac{\pp}{\pp\breve{\vec{x}}} \;;\;\;\;
\breve{\vec{x}}=\vec{x}/x_{0} \;; \\   \lb{s2_83}
d^{d}x/L^{d} &\rightarrow&(x_{0}/L)^{d}\;\;d^{d}\breve{x}\;;\;\;
x_{0}=\Big(\hbar^{2}\;m^{-1}\Big/\big(2\mscr{N}\big)\Big)^{1/2}=
t_{0}\Big(1\Big/{\ts \big(2\frac{m}{\mscr{N}}\big)}\Big)^{1/2}\;.
\eeq
Application of (\ref{s2_77}-\ref{s2_83}) for the re-scaling of
\(\mscr{A}_{\mscr{N}^{-1}}\ppr[\hat{T};J_{\psi}]\) (\ref{s2_54}),
\(\mscr{A}_{\mscr{N}^{0}}\ppr[\hat{T};J_{\psi}]\) (\ref{s2_60}),
\(\mscr{A}_{\mscr{N}^{+1}}\ppr[\hat{T}]\) (\ref{s2_61}) yields the action
\(\mscr{A}^{(d)}[\hat{Z},\hat{T};\breve{J}_{\psi}]\) (\ref{s2_84}) with Lagrangian
\(\mscr{L}^{(d)}[\hat{Z},\hat{T};\breve{J}_{\psi}]\) (\ref{s2_85}) for the anomalous
fields in the coset super-generator \(\hat{Y}(\breve{\vec{x}},\breve{t}_{p})\).
The action term \(\mscr{A}_{\hat{J}_{\psi\psi}}[\hat{T}]\) in (\ref{s2_84}) creates these anomalous fields
from the vacuum state through
the 'condensate seed matrix' \(\hat{J}_{\psi\psi;\alpha\beta}^{a\neq b}(\vec{x},t_{p})\).
However, instead of a detailed creation process by \(\hat{J}_{\psi\psi;\alpha\beta}^{a\neq b}(\vec{x},t_{p})\),
we simply assume suitable, initial conditions of the pair condensate fields in
\(\hat{Y}(\breve{\vec{x}},\breve{t}_{p})\) whose dynamics are determined by the action
\(\mscr{A}^{(d)}[\hat{Z},\hat{T};\breve{J}_{\psi}]\) (\ref{s2_84}) or
\(\mscr{L}^{(d)}[\hat{Z},\hat{T};\breve{J}_{\psi}]\) (\ref{s2_85})
\beq \lb{s2_84}
Z[\hat{\mscr{J}},J_{\psi},\im \hat{J}_{\psi\psi}]&=&
\int d\big[\hat{T}^{-1}(\breve{\vec{x}},\breve{t}_{p})\;d\hat{T}(\breve{\vec{x}},\breve{t}_{p})\big]\;\;
\exp\Big\{\im\;\mscr{A}_{\hat{J}_{\psi\psi}}\big[\hat{T}\big]\Big\}\;\times\;
\exp\Big\{-\mscr{A}\ppr\big[\hat{T};\hat{\mscr{J}}\big]\Big\} \\ \no &\times&
\exp\Big\{-\mscr{A}^{(d)}\big[\hat{Z},\hat{T};\breve{J}_{\psi}\big]\Big\}\;; \\ \lb{s2_85}
\mscr{A}^{(d)}\big[\hat{Z},\hat{T};\breve{J}_{\psi}\big] &=&
\int_{C}d\breve{t}_{p}\int d^{d}\breve{x}\;\bigg(\frac{x_{0}}{L}\bigg)^{d}\;\;\;
\mscr{L}^{(d)}\big[\hat{Z},\hat{T};\breve{J}_{\psi}\big] \;.
\eeq
The action \(\mscr{A}\ppr[\hat{T};\hat{\mscr{J}}]\) in \(Z[\hat{\mscr{J}},J_{\psi},\im \hat{J}_{\psi\psi}]\) (\ref{s2_84})
specifies the observable quantities
by differentiation with respect to \(\hat{\mscr{J}}_{\vec{x},\alpha;\vec{x}\ppr,\beta}^{ab}(t_{p},t_{q}\ppr)\)
(compare section \ref{s41}) which has afterwards to be set to zero. Therefore,
\(\mscr{A}\ppr[\hat{T};\hat{\mscr{J}}]\) cannot effect the dynamics of the pair condensate fields
as the action \(\mscr{A}^{(d)}[\hat{Z},\hat{T};\breve{J}_{\psi}]\) (\ref{s2_85}).
Relation (\ref{s2_86}) finally contains the complete, re-scaled Lagrangian
\(\mscr{L}^{(d)}[\hat{Z},\hat{T};\breve{J}_{\psi}]\) whose dependencies on pair condensates
have to be modified by the transformation of the super-Jacobi matrix \(\hat{G}_{\mbox{{\scz Osp}}/ \mbox{{\scz U}}}^{-1/2}\)
from the coset metric tensor; in consequence,  the nontrivial coset integration measure
\(d[\hat{T}^{-1}(\vec{x},t_{p})\;d\hat{T}(\vec{x},t_{p})]\) (\ref{s2_48}) in (\ref{s2_84})
is eliminated for Euclidean path integration fields as the new, independent anomalous field variables
in \(Z[\hat{\mscr{J}},J_{\psi},\im \hat{J}_{\psi\psi}]\) (\ref{s2_84},\ref{s2_85}) (see following section \ref{s3})
\beq \lb{s2_86}
\lefteqn{\mscr{L}^{(d)}\big[\hat{Z},\hat{T};\breve{J}_{\psi}\big] =
\frac{\breve{c}^{ij}(\breve{\vec{x}},\breve{t}_{p})}{4}\;\;
\mbox{STR}\Big[\Big(\breve{\pp}_{i}\hat{Z}(\breve{\vec{x}},\breve{t}_{p})\Big)\;
\Big(\breve{\pp}_{j}\hat{Z}(\breve{\vec{x}},\breve{t}_{p})\Big)\Big]+  }  \\ \no &-&\frac{1}{2}\;
\mbox{STR}\Big[\hat{T}^{-1}(\breve{\vec{x}},\breve{t}_{p})\;\hat{S}\;
\Big(\breve{E}_{p}\hat{T}(\breve{\vec{x}},\breve{t}_{p})\Big)+
\hat{T}^{-1}(\breve{\vec{x}},\breve{t}_{p})\;\Big(\breve{\pp}_{i}\hat{T}(\breve{\vec{x}},\breve{t}_{p})\Big)\;
\hat{T}^{-1}(\breve{\vec{x}},\breve{t}_{p})\;\Big(\breve{\pp}_{i}\hat{T}(\breve{\vec{x}},\breve{t}_{p})\Big)\Big]+
\\ \no &-&\frac{1}{2}\Big(\breve{u}(\breve{\vec{x}})-\breve{\mu}_{0}-\im\;\breve{\ve}_{p}+
\big\langle\breve{\sigma}_{D}^{(0)}(\breve{\vec{x}},\breve{t}_{p})\big\rangle_{\hat{\sigma}_{D}^{(0)}}\Big)\;
\mbox{STR}\Big[\Big(\hat{T}^{-1}(\breve{\vec{x}},\breve{t}_{p})\Big)^{2}-\hat{1}_{2N\times 2N}\Big]+
\\ \no &-&\im\;\Big(\breve{d}^{ij}(\breve{\vec{x}},\breve{t}_{p})-\frac{1}{2}\;\delta^{ij}\Big)\;\;\times
\\ \no &\times& \breve{J}_{\psi;\beta}^{+,b}(\breve{\vec{x}},\breve{t}_{p})\;
\bigg[\hat{I}\;\wt{K}\;\Big(\breve{\pp}_{i}\hat{T}(\breve{\vec{x}},\breve{t}_{p})\Big)\;
\hat{T}^{-1}(\breve{\vec{x}},\breve{t}_{p})\;\Big(\breve{\pp}_{j}\hat{T}(\breve{\vec{x}},\breve{t}_{p})\Big)\;
\hat{T}^{-1}(\breve{\vec{x}},\breve{t}_{p})\;\hat{I}\bigg]_{\beta\alpha}^{ba}\;
\breve{J}_{\psi;\alpha}^{a}(\breve{\vec{x}},\breve{t}_{p})+  \\  \no &-&\frac{\im}{2}\;
\breve{J}_{\psi;\beta}^{+,b}(\breve{\vec{x}},\breve{t}_{p})\;
\bigg[\hat{I}\;\wt{K}\;\hat{T}(\breve{\vec{x}},\breve{t}_{p})\;\hat{S}\;\hat{T}^{-1}(\breve{\vec{x}},\breve{t}_{p})\;
\Big(\breve{E}_{p}\hat{T}(\breve{\vec{x}},\breve{t}_{p})\Big)\;\hat{T}^{-1}(\breve{\vec{x}},\breve{t}_{p})\;
\hat{I}\bigg]_{\beta\alpha}^{ba}\;
\breve{J}_{\psi;\alpha}^{a}(\breve{\vec{x}},\breve{t}_{p})+ \\  \no &+&\frac{\eta_{p}}{2}\;
\mbox{STR}\bigg[\Big(\hat{T}^{-1}(\breve{\vec{x}},\breve{t}_{p})\Big)^{2}-\hat{1}_{2N\times 2N}\bigg]_{\mbox{.}}
\eeq

\section{Classical field equations with Euclidean path integration variables} \lb{s3}

\subsection{General symmetry considerations for the transformation to Euclidean variables}\lb{s31}

It is the aim of this section to transform the '\(\hat{\mscr{P}}\), \(\hat{\mscr{P}}^{-1}\)' rotated
derivative \(\hat{\mscr{P}}\;\hat{T}^{-1}\;(\pp\hat{T})\;\hat{\mscr{P}}^{-1}\) involving sine- (sinh-) functions
of eigenvalues to a Euclidean form \((\pp\hat{\mscr{Z}}_{\alpha\beta}^{ab})\) (\ref{s3_1})
(compare appendix A and C in Ref. \cite{mies1}).
The general symbolic derivative '\(\pp\)', appearing in this section, representatively replaces
a partial, spatial gradient '\(\breve{\pp}_{i}\)' or time contour derivative
'\(\breve{\pp}_{\breve{t}_{p}}\)', a variation symbol '\(\delta\)' for stationary phases
or a total derivative '\(d\)'. The transformed 'Nambu'
doubled super-matrix \((\pp\hat{\mscr{Z}}_{\alpha\beta}^{ab})\) (\ref{s3_1}) of Euclidean form is composed
of four sub-super-matrices with densities \((\pp\hat{\mscr{Y}}_{\alpha\beta}^{11})\),
\((\pp\hat{\mscr{Y}}_{\alpha\beta}^{22})\)
and anomalous terms \((\pp\hat{\mscr{X}}_{\alpha\beta})\), \(\wt{\kappa}\;(\pp\hat{\mscr{X}}_{\alpha\beta}^{+})\)
\beq \lb{s3_1}
\big(\pp\hat{\mscr{Z}}_{\alpha\beta}^{ab}\big) &=&
-\Big(\hat{\mscr{P}}\;\hat{T}^{-1}\;\big(\pp\hat{T}\big)\;\hat{\mscr{P}}^{-1}\Big)_{\alpha\beta}^{ab}=
\left(\bea{cc} \big(\pp\hat{\mscr{Y}}^{11}_{\alpha\beta}\big) & \big(\pp\hat{\mscr{X}}_{\alpha\beta}\big) \\
\wt{\kappa}\;\big(\pp\hat{\mscr{X}}_{\alpha\beta}^{+}\big) & \big(\pp\hat{\mscr{Y}}^{22}_{\alpha\beta}\big)
\eea\right)_{\mbox{.}}^{ab}
\eeq
Apart from the dependence of the densities \((\pp\hat{\mscr{Y}}_{\alpha\beta}^{11})\),
\((\pp\hat{\mscr{Y}}_{\alpha\beta}^{22})\)
on the pair condensate fields in \((\pp\hat{\mscr{X}}_{\alpha\beta})\),
\(\wt{\kappa}\;(\pp\hat{\mscr{X}}_{\alpha\beta}^{+})\),
the 'Nambu' doubled super-matrix \((\pp\hat{\mscr{Z}}_{\alpha\beta}^{ab})\) (\ref{s3_1})
 has similar symmetries between matrix entries as the self-energy
\(\delta\wt{\Sigma}_{\alpha\beta}^{ab}(\vec{x},t_{p})\;\wt{K}\) (\ref{s1_37},\ref{s1_38})
and is also confined to taking values within
the ortho-symplectic super-algebra \(\mbox{osp}(S,S|2L)\). The super-matrix \((\pp\hat{\mscr{X}}_{\alpha\beta})\) (\ref{s3_2}) and
its super-hermitian conjugate \(\wt{\kappa}\;(\pp\hat{\mscr{X}}_{\alpha\beta}^{+})\) (\ref{s3_3})
consist of the symmetric, complex, even matrices \((\pp\hat{b}_{mn})\), \((\pp\hat{b}_{mn}^{+})\) for the molecular pair condensates and the anti-symmetric, complex, even matrices \((\pp\hat{a}_{r\mu,r\ppr\nu})\),
\((\pp\hat{a}_{r\mu,r\ppr\nu}^{+})\) for the BCS pair condensate terms (\ref{s3_4}).
This is in accordance with the \(N\times N\) super-matrices
\(\hat{X}_{N\times N}\), \(\wt{\kappa}\;\hat{X}_{N\times N}^{+}\)
(\ref{s2_3},\ref{s2_4}) and their symmetric, complex, even boson-boson parts
\(\hat{c}_{D;L\times L}\), \(\hat{c}_{D;L\times L}^{+}\)
and anti-symmetric, complex, even fermion-fermion parts \(\hat{f}_{D;S\times S}\), \(\hat{f}_{D;S\times S}^{+}\) (\ref{s2_5}).
The odd parts \(\hat{\eta}_{D;S\times L}\), \(\hat{\eta}_{D;L\times S}^{T}\) (\ref{s2_4}) (respectively their
complex conjugates \(\hat{\eta}_{D;S\times L}^{*}\), \(\hat{\eta}_{D;L\times S}^{+}\))
are substituted by \(\hat{\zeta}_{r\mu,n}\), \(\hat{\zeta}_{m,r\ppr\nu}^{T}\) (and
\(\hat{\zeta}_{r\mu,n}^{*}\), \(\hat{\zeta}_{m,r\ppr\nu}^{+}\)) in
\((\pp\hat{\mscr{X}}_{\alpha\beta})\), \(\wt{\kappa}\;(\pp\hat{\mscr{X}}_{\alpha\beta}^{+})\)
(compare footnote \ref{ft2} for the indexing and numbering of the anomalous fermion-fermion parts
by \(2\times 2\) quaternion elements)
\beq\lb{s3_2}
\big(\pp\hat{\mscr{X}}_{\alpha\beta}\big)&=&
-\Big(\hat{\mscr{P}}\;\hat{T}^{-1}\;\big(\pp\hat{T}\big)\;\hat{\mscr{P}}^{-1}\Big)_{\alpha\beta}^{12}=
\left(\bea{cc} -\big(\pp\hat{b}_{mn}\big) & \big(\pp\hat{\zeta}_{m,r\ppr\nu}^{T}\big) \\
-\big(\pp\hat{\zeta}_{r\mu,n}\big) & \big(\pp\hat{a}_{r\mu,r\ppr\nu}\big) \eea\right)_{\alpha\beta} \;; \\  \lb{s3_3}
\wt{\kappa}\;\big(\pp\hat{\mscr{X}}^{+}_{\alpha\beta}\big)&=&
-\Big(\hat{\mscr{P}}\;\hat{T}^{-1}\;\big(\pp\hat{T}\big)\;\hat{\mscr{P}}^{-1}\Big)_{\alpha\beta}^{21}=
\left(\bea{cc} \big(\pp\hat{b}_{mn}^{*}\big) & \big(\pp\hat{\zeta}_{m,r\ppr\nu}^{+}\big) \\
\big(\pp\hat{\zeta}_{r\mu,n}^{*}\big) & \big(\pp\hat{a}_{r\mu,r\ppr\nu}^{+}\big) \eea\right)_{\alpha\beta} \;; \\  \lb{s3_4}
&& \big(\pp\hat{b}_{mn}\big)=\big(\pp\hat{b}_{mn}^{T}\big)\;;
\hspace*{0.5cm}\big(\pp\hat{a}_{r\mu,r\ppr\nu}\big)=-\big(\pp\hat{a}_{r\mu,r\ppr\nu}^{T}\big)
=\sum_{k=0}^{3}\big(\tau_{k}\big)_{\mu\nu}\;\big(\pp\hat{a}_{rr\ppr}^{(k)}\big) \;.
\eeq
The density part \((\pp\hat{\mscr{Y}}_{\alpha\beta}^{11})\) (\ref{s3_5},\ref{s3_1}) has in its entity as a \(N\times N\)
super-matrix anti-hermitian properties, with the even, hermitian boson-boson and fermion-fermion matrices
\((\pp\hat{d}_{mn})\), \((\pp\hat{g}_{r\mu,r\ppr\nu})\) (\ref{s3_6}). The odd parts
of \((\hat{\mscr{P}}\;\hat{T}^{-1}\;(\pp\hat{T})\;\hat{\mscr{P}}^{-1})^{11}\) (similar to
\(\delta\hat{\chi}_{D;S\times L}\), \(\delta\hat{\chi}_{D;L\times S}^{+}\) of
\(\delta\hat{\Sigma}_{D;N\times N}^{11}\)) are represented by \((\pp\hat{\xi}_{r\mu,n})\),
\((\pp\hat{\xi}_{m,r\ppr\nu}^{+})\) so that one obtains an anti-hermitian property of
\((\pp\hat{\mscr{Y}}_{\alpha\beta}^{11})\) because of the total imaginary factor
\beq\lb{s3_5}
\big(\pp\hat{\mscr{Y}}^{11}_{\alpha\beta}\big)&=&
-\Big(\hat{\mscr{P}}\;\hat{T}^{-1}\;\big(\pp\hat{T}\big)\;\hat{\mscr{P}}^{-1}\Big)_{\alpha\beta}^{11}=\im\;
\left(\bea{cc} \big(\pp\hat{d}_{mn}\big) & \big(\pp\hat{\xi}_{m,r\ppr\nu}^{+}\big) \\
\big(\pp\hat{\xi}_{r\mu,n}\big) & \big(\pp\hat{g}_{r\mu,r\ppr\nu}\big) \eea\right)_{\alpha\beta}^{11} \;; \\  \lb{s3_6}
\big(\pp\hat{\mscr{Y}}^{11}_{\alpha\beta}\big)^{+}&=&-\big(\pp\hat{\mscr{Y}}^{11}_{\alpha\beta}\big)\;;\hspace*{0.5cm}
\big(\pp\hat{d}_{mn}^{+}\big)=\big(\pp\hat{d}_{mn}\big)\;;\hspace*{0.5cm}
\big(\pp\hat{g}_{r\mu,r\ppr\nu}^{+}\big)=\big(\pp\hat{g}_{r\mu,r\ppr\nu}\big)\;.
\eeq
The '22' \(N\times N\) super-matrix \((\pp\hat{\mscr{Y}}_{\alpha\beta}^{22})\) (\ref{s3_7},\ref{s3_1})
is related by super-transposition 'st' (\ref{s1_15}-\ref{s1_17})
to the '11' \(N\times N\) density super-matrix \((\pp\hat{\mscr{Y}}_{\alpha\beta}^{11})\) (\ref{s3_5},\ref{s3_6})
with inclusion of an additional minus sign. It is composed of the same variables with \((\pp\hat{d}_{L\times L}^{T})\),
\((\pp\hat{g}_{S\times S}^{T})\) and also contains the same odd parts
\((\pp\hat{\xi}_{S\times L}^{*})\), \((\pp\hat{\xi}_{L\times S}^{T})\)
apart from the additional super-transposition
\beq \lb{s3_7}
\big(\pp\hat{\mscr{Y}}^{22}_{\alpha\beta}\big)&=&
-\Big(\hat{\mscr{P}}\;\hat{T}^{-1}\;\big(\pp\hat{T}\big)\;\hat{\mscr{P}}^{-1}\Big)_{\alpha\beta}^{22}=-\im\;
\left(\bea{cc} \big(\pp\hat{d}_{mn}^{T}\big) & -\big(\pp\hat{\xi}_{m,r\ppr\nu}^{T}\big) \\
\big(\pp\hat{\xi}_{r\mu,n}^{*}\big) & \big(\pp\hat{g}_{r\mu,r\ppr\nu}^{T}\big)
\eea\right)_{\alpha\beta}^{22} \;; \\ \lb{s3_8}
\big(\pp\hat{\mscr{Y}}^{22}_{\alpha\beta}\big)^{st}&=&-\big(\pp\hat{\mscr{Y}}^{11}_{\alpha\beta}\big)=
-\Big(\hat{\mscr{P}}\;\hat{T}^{-1}\;\big(\pp\hat{T}\big)\;\hat{\mscr{P}}^{-1}\Big)_{\alpha\beta}^{22,st}=
\Big(\hat{\mscr{P}}\;\hat{T}^{-1}\;\big(\pp\hat{T}\big)\;\hat{\mscr{P}}^{-1}\Big)_{\alpha\beta}^{11}\;.
\eeq
According to the coset decomposition, both parts, density (\ref{s3_5}-\ref{s3_8})
and anomalous terms (\ref{s3_2}-\ref{s3_4})
of \((\pp\hat{\mscr{Z}}_{\alpha\beta}^{ab})\) (\ref{s3_1}), depend on the
original, independent anomalous fields in \(\hat{X}_{N\times N}\), \(\wt{\kappa}\;\hat{X}_{N\times N}^{+}\)
(\ref{s2_2}-\ref{s2_5}) and on their eigenvalues \(\ovv{c}_{m}\), \(\ovv{f}_{r}\) (\ref{s2_16}-\ref{s2_25}).
In consequence there exists a cross-dependence of the density fields
\((\pp\hat{\mscr{Z}}_{\alpha\beta}^{11})=(\pp\hat{\mscr{Y}}_{\alpha\beta}^{11})\),
\((\pp\hat{\mscr{Z}}_{\alpha\beta}^{22})=(\pp\hat{\mscr{Y}}_{\alpha\beta}^{22})\)
over the original matrices \(\hat{X}_{N\times N}\), \(\wt{\kappa}\;\hat{X}_{N\times N}^{+}\) of the coset decomposition
to the Euclidean pair condensate integration variables in
\((\pp\hat{\mscr{X}}_{\alpha\beta})\), \(\wt{\kappa}\;(\pp\hat{\mscr{X}}_{\alpha\beta}^{+})\).

\subsection{Removal of the coset integration measure and transformation to
Euclidean integration variables} \lb{s32}

In appendix C of Ref. \cite{mies1} we have explicitly computed the general derivative
\(\big(\hat{\mscr{P}}\;\hat{T}^{-1}\;\big(\pp\hat{T}\big)\;\hat{\mscr{P}}^{-1}\big)_{\alpha\beta}^{ab}\)
in terms of the eigenvalues \(\hat{Y}_{DD}\), \(\hat{X}_{DD}\), \(\wt{\kappa}\;\hat{X}_{DD}^{+}\) (\ref{s2_16}-\ref{s2_25})
and the rotated derivative \(\hat{\mscr{P}}\;(\pp\hat{Y})\;\hat{\mscr{P}}^{-1}\) (\ref{s3_11}) of the independent anomalous
fields \((\pp\hat{c}_{D;L\times L})\), \((\pp\hat{f}_{D;S\times S})\), \((\pp\hat{\eta}_{D;S\times L})\),
\((\pp\hat{\eta}_{D;L\times S}^{+})\) (\ref{s2_2}-\ref{s2_5}).
Using relation (\ref{s3_9}) for the derivative of an exponential of a matrix \cite{eng,mies1},
it remains to multiply the rotated derivative
\((\pp\hat{Y}\ppr)=\hat{\mscr{P}}\;(\pp\hat{Y})\;\hat{\mscr{P}}^{-1}\) (\ref{s3_11}) by the
diagonal anomalous matrices in the various block parts of \(\exp\{\pm v\;\hat{Y}_{DD}\}\)
with eigenvalues \(\ovv{c}_{m}\)
and quaternion eigenvalues \((\tau_{2})_{\mu\nu}\;\ovv{f}_{r}\) (\ref{s2_16}-\ref{s2_25})
\beq \lb{s3_9}
\exp\big\{\hat{B}\big\}\;\delta\Big(\exp\big\{-\hat{B}\big\}\Big) &=&-\int_{0}^{1}dv\;
\exp\big\{v\;\hat{B}\big\}\;\;\delta\hat{B}\;\;\exp\big\{-v\;\hat{B}\big\} \;; \\  \lb{s3_10}
-\big(\pp\hat{\mscr{Z}}_{\alpha\beta}^{ab}\big)=
\Big(\hat{\mscr{P}}\;\hat{T}^{-1}\;\big(\pp\hat{T}\big)\;\hat{\mscr{P}}^{-1}\Big)_{\alpha\beta}^{ab} &=&
\Big(\hat{\mscr{P}}\;\exp\big\{\hat{Y}\big\}\;\;
\Big(\pp\exp\big\{-\hat{Y}\big\}\Big)\;\;\hat{\mscr{P}}^{-1}\Big)_{\alpha\beta}^{ab}  \\ \no &=&
\Big(\hat{\mscr{P}}\;\exp\big\{\hat{\mscr{P}}^{-1}\;\hat{Y}_{DD}\;\hat{\mscr{P}}\big\}\;\;
\Big(\pp\exp\big\{-\hat{\mscr{P}}^{-1}\;\hat{Y}_{DD}\;\hat{\mscr{P}}\big\}\Big)\;\;
\hat{\mscr{P}}^{-1}\Big)_{\alpha\beta}^{ab}  \\ \no &=&
-\int_{0}^{1}dv\;\;\Big(\exp\big\{v\;\hat{Y}_{DD}\big\}\;\;\big(\pp\hat{Y}\ppr\big)\;\;
\exp\big\{-v\;\hat{Y}_{DD}\big\}\Big)_{\alpha\beta}^{ab} \;; \\  \lb{s3_11}
\big(\pp\hat{Y}\ppr\big)=\hat{\mscr{P}}\;\;\big(\pp\hat{Y}\big)\;\;\hat{\mscr{P}}^{-1} &;&
\mbox{STR}\Big[\big(\pp\hat{Y}\ppr\big)\;\big(\pp\hat{Y}\ppr\big)\Big]=
\mbox{STR}\Big[\big(\pp\hat{Y}\big)\;\big(\pp\hat{Y}\big)\Big]\;.
\eeq
After straightforward integration (\ref{s3_10})
of (hyperbolic) trigonometric functions with parameter \(v\in[0,1]\), one acquires the dependence
of \((\pp\hat{\mscr{X}}_{\alpha\beta})\), \(\wt{\kappa}\;(\pp\hat{\mscr{X}}_{\alpha\beta}^{+})\) (\ref{s3_2}-\ref{s3_4})
on the independent rotated variables within \((\pp\hat{c}_{D;L\times L}\ppr)\),
\((\pp\hat{f}_{D;S\times S}\ppr)\), \((\pp\hat{\eta}_{D;S\times L}\ppr)\),
\((\pp\hat{\eta}_{D;L\times S}^{\bprime,+})\) (\ref{s3_11},\ref{s2_2}-\ref{s2_5})
of \((\pp\hat{X}_{N\times N})\), \(\wt{\kappa}\;(\pp\hat{X}_{N\times N}^{+})\) and with corresponding
eigenvalues \(\ovv{c}_{m}\), \(\ovv{f}_{r}\) (\ref{s2_16}-\ref{s2_25}). Similarly one achieves the relation between the
densities \((\pp\hat{\mscr{Y}}_{\alpha\beta}^{11})\), \((\pp\hat{\mscr{Y}}_{\alpha\beta}^{22})\)
(\ref{s3_5}-\ref{s3_8}) and the original anomalous fields \((\pp\hat{X}_{N\times N})\),
\(\wt{\kappa}\;(\pp\hat{X}_{N\times N}^{+})\) (\ref{s2_2}-\ref{s2_5})
of the coset decomposition. Therefore, the densities \((\pp\hat{\mscr{Y}}_{\alpha\beta}^{11})\),
\((\pp\hat{\mscr{Y}}_{\alpha\beta}^{22})\) can also be related to the Euclidean, independent anomalous
integration variables of \((\pp\hat{\mscr{X}}_{\alpha\beta})\), \(\wt{\kappa}\;(\pp\hat{\mscr{X}}_{\alpha\beta}^{+})\).
In the following subsections \ref{s321}, \ref{s322} and \ref{s323}, we apply the results
for \(\big(\hat{\mscr{P}}\;\hat{T}^{-1}\;\big(\pp\hat{T}\big)\;\hat{\mscr{P}}^{-1}\big)_{\alpha\beta}^{ab}\)
of Ref.\ \cite{mies1} with appendix C  in order to transform to Euclidean fields. These are separated
into the split even boson-boson , fermion-fermion and odd fermion-boson, boson-fermion parts.

\subsubsection{Boson-boson part of the transformation to
Euclidean integration variables of pair condensate fields}\lb{s321}

In this subsection we record the transformation of
\((\hat{\mscr{P}}\;\hat{T}^{-1}\;(\pp\hat{T})\;\hat{\mscr{P}}^{-1})_{BB;mn}^{ab}\) to
\((\pp\hat{b}_{mn})\), \((\pp\hat{b}_{mn}^{*})\) \((a\neq b)\) and corresponding density parts
\((\pp\hat{d}_{mn})\) \((a=b)\) and have furthermore to distinguish between the diagonal \((m=n)\)
and off-diagonal \((m\neq n)\) matrix elements of transformations which are restricted to the total, even
boson-boson part
\beq \lb{s3_12}
-\Big(\hat{\mscr{P}}\;\hat{T}^{-1}\;\big(\pp\hat{T}\big)\;\hat{\mscr{P}}^{-1}\Big)_{\alpha\beta}^{12}&=&
\big(\pp\hat{\mscr{X}}_{\alpha\beta}\big)=
\left(\bea{cc} -\big(\pp\hat{b}_{mn}\big) & \big(\pp\hat{\zeta}_{m,r\ppr\nu}^{T}\big)  \\
-\big(\pp\hat{\zeta}_{r\mu,n}\big) & \big(\pp\hat{a}_{r\mu,r\ppr\nu}\big)
\eea\right)_{\alpha\beta} \;; \\   \lb{s3_13}
\big(\pp\hat{b}_{mn}\big)=\big(\pp\hat{b}_{mn}^{T}\big) &;&
\big(\pp\hat{a}_{r\mu,r\ppr\nu}\big)=-\big(\pp\hat{a}_{r\mu,r\ppr\nu}^{T}\big) \;; \\  \lb{s3_14}
-\Big(\hat{\mscr{P}}\;\hat{T}^{-1}\;\big(\pp\hat{T}\big)\;\hat{\mscr{P}}^{-1}\Big)_{\alpha\beta}^{11}
&=&\big(\pp\hat{\mscr{Y}}^{11}_{\alpha\beta}\big)=\im\;
\left(\bea{cc} \big(\pp\hat{d}_{mn}\big) & \big(\pp\hat{\xi}_{m,r\ppr\nu}^{+}\big) \\
\big(\pp\hat{\xi}_{r\mu,n}\big) & \big(\pp\hat{g}_{r\mu,r\ppr\nu}\big) \eea\right)_{\alpha\beta}^{11} \;; \\  \lb{s3_15}
\big(\pp\hat{d}_{mn}^{+}\big)=\big(\pp\hat{d}_{mn}\big) &;&
\big(\pp\hat{g}_{r\mu,r\ppr\nu}^{+}\big)=\big(\pp\hat{g}_{r\mu,r\ppr\nu}\big)\;.
\eeq
In relations (\ref{s3_16}-\ref{s3_19}) we give the detailed transformation (\ref{s3_16},\ref{s3_17})
from the diagonal, rotated, anomalous molecular condensates
\((\pp\hat{c}_{D;mm}\ppr)\), \((\pp\hat{c}_{D;mm}^{\bprime *})\) to Euclidean fields
\((\pp\hat{b}_{mm})\), \((\pp\hat{b}_{mm}^{*})\) and also determine the reverse transformations (\ref{s3_18})
from \((\pp\hat{b}_{mm})\), \((\pp\hat{b}_{mm}^{*})\) to the original fields
\((\pp\hat{c}_{D;mm}\ppr)\), \((\pp\hat{c}_{D;mm}^{\bprime *})\) in the coset decomposition.
The back transformation (\ref{s3_18}) allows to calculate the change of integration measure from
\(d\hat{c}_{D;mm}\ppr\wedge d\hat{c}_{D;mm}^{\bprime *}\) or in equivalence from the un-rotated fields
\(d\hat{c}_{D;mm}\wedge d\hat{c}_{D;mm}^{*}\) to the diagonal, Euclidean elements
\(d\hat{b}_{mm}\wedge d\hat{b}_{mm}^{*}\)
\beq \lb{s3_16} \lefteqn{
-\Big(\hat{\mscr{P}}\;\hat{T}^{-1}\;\big(\pp\hat{T}\big)\;\hat{\mscr{P}}^{-1}\Big)_{BB;mm}^{12}=
\Big(\hat{\mscr{P}}\;\hat{T}^{-1}\;\big(\pp\hat{T}\big)\;\hat{\mscr{P}}^{-1}\Big)_{BB;mm}^{21,*}=
-\big(\pp\hat{b}_{mm}\big)= }  \\ \no &=&
-\bigg(\frac{1}{2}+\frac{\sin\big(2\;|\ovv{c}_{m}|\big)}{4\;|\ovv{c}_{m}|}\bigg)\;
\big(\pp\hat{c}_{D;mm}\ppr\big)-
\bigg(\frac{1}{2}-\frac{\sin\big(2\;|\ovv{c}_{m}|\big)}{4\;|\ovv{c}_{m}|}\bigg)\;e^{\im\;2\varphi_{m}}\;
\big(\pp\hat{c}_{D;mm}^{\bprime *}\big) \;;
\eeq
\beq \lb{s3_17}
\left(\bea{c} \big(\pp\hat{b}_{mm}\big) \\  \big(\pp\hat{b}_{mm}^{*}\big) \eea\right) &=&
\left(\bea{cc}
\bigg(\frac{1}{2}+\frac{\sin\big(2\;|\ovv{c}_{m}|\big)}{4\;|\ovv{c}_{m}|}\bigg)  &
e^{\im\;2\varphi_{m}}\;\bigg(\frac{1}{2}-\frac{\sin\big(2\;|\ovv{c}_{m}|\big)}{4\;|\ovv{c}_{m}|}\bigg)
\\ e^{-\im\;2\varphi_{m}}\;\bigg(\frac{1}{2}-\frac{\sin\big(2\;|\ovv{c}_{m}|\big)}{4\;|\ovv{c}_{m}|}\bigg) &
\bigg(\frac{1}{2}+\frac{\sin\big(2\;|\ovv{c}_{m}|\big)}{4\;|\ovv{c}_{m}|}\bigg) \eea\right)
\left(\bea{c} \big(\pp\hat{c}_{D;mm}\ppr\big) \\
\big(\pp\hat{c}_{D;mm}^{\bprime *}\big) \eea\right)_{\mbox{;}}   \\ \lb{s3_18}
\left(\bea{c} \big(\pp\hat{c}_{D;mm}\ppr\big) \\
\big(\pp\hat{c}_{D;mm}^{\bprime *}\big) \eea\right)  &=&
\left(\bea{cc}
\bigg(\frac{1}{2}+\frac{|\ovv{c}_{m}|}{\sin\big(2\;|\ovv{c}_{m}|\big)}\bigg)  &
e^{\im\;2\varphi_{m}}\;\bigg(\frac{1}{2}-\frac{|\ovv{c}_{m}|}{\sin\big(2\;|\ovv{c}_{m}|\big)}\bigg)
\\ e^{-\im\;2\varphi_{m}}\;\bigg(\frac{1}{2}-\frac{|\ovv{c}_{m}|}{\sin\big(2\;|\ovv{c}_{m}|\big)}\bigg) &
\bigg(\frac{1}{2}+\frac{|\ovv{c}_{m}|}{\sin\big(2\;|\ovv{c}_{m}|\big)}\bigg) \eea\right)
\left(\bea{c} \big(\pp\hat{b}_{mm}\big) \\  \big(\pp\hat{b}_{mm}^{*}\big) \eea\right)_{\mbox{;}}   \\  \lb{s3_19}
d\hat{c}_{D;mm}\ppr\wedge d\hat{c}_{D;mm}^{\bprime *}  &=&
d\hat{c}_{D;mm}\wedge d\hat{c}_{D;mm}^{*}=
d\hat{b}_{mm}\wedge d\hat{b}_{mm}^{*}\;\;\;
\frac{2\;|\ovv{c}_{m}|}{\sin\big(2\;|\ovv{c}_{m}|\big)} \;.
\eeq
The matrix in (\ref{s3_17}) with eigenvalues \(\ovv{c}_{m}\) (\ref{s2_19}) represents the square root of the coset metric tensor
\(\hat{G}_{\mbox{{\scz Osp}}/ \mbox{{\scz U}}}^{1/2}\) concerning the boson-boson part, however, in its diagonalized form
with eigenvalues \(\ovv{c}_{m}\). The inverse transformation (\ref{s3_18}) of Euclidean fields
\((\pp\hat{b}_{mm})\), \((\pp\hat{b}_{mm}^{*})\) to
\((\pp\hat{c}_{D;mm}\ppr)\), \((\pp\hat{c}_{D;mm}^{\bprime *})\) therefore contains
the inverse square root of the coset metric tensor \(\hat{G}_{\mbox{{\scz Osp}}/ \mbox{{\scz U}}}^{-1/2}\)  as
already mentioned in the introduction. Both kinds of metric tensors,
\(\hat{G}_{\mbox{{\scz Osp}}/ \mbox{{\scz U}}}^{1/2}\) and \(\hat{G}_{\mbox{{\scz Osp}}/ \mbox{{\scz U}}}^{-1/2}\), are considerably simplified
because of the inclusion of the transformation from \(\hat{X}_{N\times N}\),
\(\wt{\kappa}\;\hat{X}_{N\times N}^{+}\) to their eigenvalues in
\(\hat{X}_{DD}\), \(\wt{\kappa}\;\hat{X}_{N\times N}^{+}\)
with eigenvectors '\(\hat{\mscr{P}}_{\alpha\beta}^{aa}\), \(\hat{\mscr{P}}_{\alpha\beta}^{-1;aa}\)' (\ref{s2_16}-\ref{s2_35}).
Since the transformation of \((\pp\hat{b}_{mm})\), \((\pp\hat{b}_{mm}^{*})\) (\ref{s3_18})
is composed of the inverse square root of the metric tensor \(\hat{G}_{\mbox{{\scz Osp}}/ \mbox{{\scz U}}}^{-1/2}\), the
corresponding integration measure
\(\mbox{SDET}(\hat{G}_{\mbox{{\scz Osp}}/ \mbox{{\scz U}}}^{-1/2})=(\,\mbox{SDET}(\hat{G}_{\mbox{{\scz Osp}}/ \mbox{{\scz U}}})\,)^{-1/2}\)
cancels the nontrivial integration measure originally introduced for the coset decomposition of
\(\delta\wt{\Sigma}_{\alpha\beta}^{ab}(\vec{x},t_{p})\;\wt{K}\) to
\(\hat{T}\;\delta\hat{\Sigma}_{D;2N\times 2N}\;\wt{K}\;\hat{T}^{-1}\) (compare section \ref{s22}).
According to appendix C of Ref.\ \cite{mies1}, we can also give
the transformation of the diagonal elements of the boson-boson density part
\((\,\hat{\mscr{P}}\;\hat{T}^{-1}\;\big(\pp\hat{T}\big)\;\hat{\mscr{P}}^{-1}\,)_{BB;mm}^{11}\). It yields with the sub-metric tensor
\((\hat{G}_{\mbox{{\scz Osp}}/ \mbox{{\scz U}}}^{-1/2})_{BB;mm}\) (\ref{s3_18}) for the relations between
\((\pp\hat{b}_{mm})\), \((\pp\hat{b}_{mm}^{*})\) and
\((\pp\hat{c}_{D;mm}\ppr)\), \((\pp\hat{c}_{D;mm}^{\bprime *})\), the diagonal density elements
\(\im\,(\pp\hat{d}_{mm})\) (\ref{s3_20}) in terms of
\(\tan(|\ovv{c}_{m}|)\) of the eigenvalues \(\ovv{c}_{m}\) (\ref{s2_19}) and in terms of the diagonal, Euclidean anomalous fields
\((\pp\hat{b}_{mm})\), \((\pp\hat{b}_{mm}^{*})\)
\beq \lb{s3_20}
\lefteqn{\hspace*{-1.9cm}-\Big(\hat{\mscr{P}}\;\hat{T}^{-1}\;\big(\pp\hat{T}\big)\;\hat{\mscr{P}}^{-1}\Big)_{BB;mm}^{11} =
\Big(\hat{\mscr{P}}\;\hat{T}^{-1}\;\big(\pp\hat{T}\big)\;\hat{\mscr{P}}^{-1}\Big)_{BB;mm}^{22} =
\im\;\big(\pp\hat{d}_{mm}\big) =}  \\  \no &=&
-\frac{\Big(\sin\big(|\ovv{c}_{m}|\big)\Big)^{2}}{2\;|\ovv{c}_{m}|}\;
\Big(\big(\pp\hat{c}_{D;mm}^{\bprime *}\big)\;e^{\im\;\varphi_{m}}-
\big(\pp\hat{c}_{D;mm}^{\bprime}\big)\;e^{-\im\;\varphi_{m}}\Big) \\ \no & = &
\frac{1}{2}\;\tan\big(|\ovv{c}_{m}|\big)\;
\Big(\big(\pp\hat{b}_{mm}\big)\;e^{-\im\;\varphi_{m}}-
\big(\pp\hat{b}_{mm}^{*}\big)\;e^{\im\;\varphi_{m}}\Big)\;.
\eeq
We transfer the results of the transformation to Euclidean fields from the diagonal boson-boson parts to
the case of off-diagonal matrix elements \((m\neq n)\) (\ref{s3_21}) and
introduce the coefficients \(A_{BB}\), \(B_{BB}\) (\ref{s3_22},\ref{s3_23}),
depending on the modulus of eigenvalues \(|\ovv{c}_{m}|\), \(|\ovv{c}_{n}|\) for specification of
\((\pp\hat{b}_{m\neq n})\), \((\pp\hat{b}_{m\neq n}^{*})\). The transformation (\ref{s3_21},\ref{s3_24})
determines the off-diagonal matrix elements of the boson-boson part of \((\hat{G}_{\mbox{{\scz Osp}}/ \mbox{{\scz U}}}^{1/2})_{BB;m\neq n}\)
whereas relation (\ref{s3_25}) describes the back transformation from
\((\pp\hat{b}_{m\neq n})\), \((\pp\hat{b}_{m\neq n}^{*})\) to the original,
'\(\hat{\mscr{P}}_{\alpha\beta}^{aa}\), \(\hat{\mscr{P}}_{\alpha\beta}^{-1;aa}\)' rotated fields
\((\pp\hat{c}_{D;m\neq n}\ppr)\), \((\pp\hat{c}_{D;m\neq n}^{\bprime *})\)
of the coset decomposition. The diagonalized forms of
\(\hat{G}_{\mbox{{\scz Osp}}/ \mbox{{\scz U}}}^{1/2}\), \(\hat{G}_{\mbox{{\scz Osp}}/ \mbox{{\scz U}}}^{-1/2}\)
(\ref{s3_24},\ref{s3_25}) are simplified by the
coefficients \(A_{BB}\), \(B_{BB}\) which are defined by the relations
(\ref{s3_22},\ref{s3_23}) of the eigenvalues \(|\ovv{c}_{m}|\), \(|\ovv{c}_{n}|\).
Note that the limit process \(|\ovv{c}_{n}|\rightarrow|\ovv{c}_{m}|\)
reproduces the results (\ref{s3_16}-\ref{s3_19}) of diagonal
elements \((m=n)\) for the boson-boson part of the metric tensor. This holds in particular for the case of the
integration measure (\ref{s3_28}) which attains the identical form of (\ref{s3_19}) in case of the limit process
\(|\ovv{c}_{n}|\rightarrow|\ovv{c}_{m}|\). The integration measure (\ref{s3_28}) with eigenvalues
\(|\ovv{c}_{m}|\), \(|\ovv{c}_{n}|\) for \(d\hat{b}_{m\neq n}\), \(d\hat{b}_{m\neq n}^{*}\) results into the
Euclidean form after substitution into the original coset integration (\ref{s2_48})
for \(d\hat{c}_{D;m\neq n}\ppr\), \(d\hat{c}_{D;m\neq n}^{\bprime *}\) of the coset decomposition
\(\mbox{Osp}(S,S|2L)/ \mbox{U}(L|S)\otimes \mbox{U}(L|S)\)
\beq \lb{s3_21} \lefteqn{
-\Big(\hat{\mscr{P}}\;\hat{T}^{-1}\;\big(\pp\hat{T}\big)\;\hat{\mscr{P}}^{-1}\Big)_{BB;mn}^{12}
\stackrel{m\neq n}{=}
\Big(\hat{\mscr{P}}\;\hat{T}^{-1}\;\big(\pp\hat{T}\big)\;\hat{\mscr{P}}^{-1}\Big)_{BB;mn}^{21,*}
\stackrel{m\neq n}{=} } \\ \no &=&
-\big(\pp\hat{b}_{mn}\big)=-A_{BB}\;\big(\pp\hat{c}_{D;mn}\ppr\big)-
e^{\im(\varphi_{m}+\varphi_{n})}\;B_{BB}\;\big(\pp\hat{c}_{D;mn}^{\bprime *}\big)\;;
\eeq
\beq\lb{s3_22}
A_{BB} &=& \frac{|\ovv{c}_{m}|\;\cos\big(|\ovv{c}_{n}|\big)\;\sin\big(|\ovv{c}_{m}|\big)-
|\ovv{c}_{n}|\;\cos\big(|\ovv{c}_{m}|\big)\;\sin\big(|\ovv{c}_{n}|\big)}
{|\ovv{c}_{m}|^{2}-|\ovv{c}_{n}|^{2}}  \;;   \\   \lb{s3_23}
B_{BB} &=& \frac{|\ovv{c}_{n}|\;\cos\big(|\ovv{c}_{n}|\big)\;\sin\big(|\ovv{c}_{m}|\big)-
|\ovv{c}_{m}|\;\cos\big(|\ovv{c}_{m}|\big)\;\sin\big(|\ovv{c}_{n}|\big)}
{|\ovv{c}_{m}|^{2}-|\ovv{c}_{n}|^{2}}  \;;  \\  \lb{s3_24}
\left(\bea{c} \big(\pp\hat{b}_{mn}\big) \\  \big(\pp\hat{b}_{mn}^{*}\big) \eea\right) &=&
\left(\bea{cc}
A_{BB}  &  e^{\im(\varphi_{m}+\varphi_{n})}\;B_{BB} \\  e^{-\im(\varphi_{m}+\varphi_{n})}\;B_{BB} & A_{BB} \eea\right)
\left(\bea{c} \big(\pp\hat{c}_{D;mn}\ppr\big) \\
\big(\pp\hat{c}_{D;mn}^{\bprime *}\big) \eea\right)_{\mbox{;}}   \\ \lb{s3_25}
\left(\bea{c} \big(\pp\hat{c}_{D;mn}\ppr\big) \\
\big(\pp\hat{c}_{D;mn}^{\bprime *}\big) \eea\right)  &=&\frac{1}{A_{BB}^{2}-B_{BB}^{2}}
\left(\bea{cc}
A_{BB}  &  -e^{\im(\varphi_{m}+\varphi_{n})}\;B_{BB} \\
 -e^{-\im(\varphi_{m}+\varphi_{n})}\;B_{BB} & A_{BB} \eea\right)
\left(\bea{c} \big(\pp\hat{b}_{mn}\big) \\
\big(\pp\hat{b}_{mn}^{*}\big) \eea\right)_{\mbox{;}}  \\ \lb{s3_26}
\frac{A_{BB}}{A_{BB}^{2}-B_{BB}^{2}}&=&\frac{1}{2}\bigg(
\frac{|\ovv{c}_{m}|-|\ovv{c}_{n}|}{\sin\big(|\ovv{c}_{m}|-|\ovv{c}_{n}|\big)}+
\frac{|\ovv{c}_{m}|+|\ovv{c}_{n}|}{\sin\big(|\ovv{c}_{m}|+|\ovv{c}_{n}|\big)}\bigg) \;;  \\   \lb{s3_27}
\frac{B_{BB}}{A_{BB}^{2}-B_{BB}^{2}}&=&\frac{1}{2}\bigg(
\frac{|\ovv{c}_{m}|+|\ovv{c}_{n}|}{\sin\big(|\ovv{c}_{m}|+|\ovv{c}_{n}|\big)}-
\frac{|\ovv{c}_{m}|-|\ovv{c}_{n}|}{\sin\big(|\ovv{c}_{m}|-|\ovv{c}_{n}|\big)}\bigg)  \;;  \\  \lb{s3_28}
d\hat{c}_{D;mn}\ppr\wedge d\hat{c}_{D;mn}^{\bprime *}&=&
d\hat{c}_{D;mn}\wedge d\hat{c}_{D;mn}^{*}=
d\hat{b}_{mn}\wedge d\hat{b}_{mn}^{*}\;\;
\frac{|\ovv{c}_{m}|+|\ovv{c}_{n}|}{\sin\big(|\ovv{c}_{m}|+|\ovv{c}_{n}|\big)}\;\;
\frac{|\ovv{c}_{m}|-|\ovv{c}_{n}|}{\sin\big(|\ovv{c}_{m}|-|\ovv{c}_{n}|\big)}\;.
\eeq
According to appendix C of Ref.\ \cite{mies1}, we sate the corresponding boson-boson density part
for off-diagonal matrix elements \(\im\,(\pp\hat{d}_{mn})\) (\ref{s3_29})
by using coefficient functions \(C_{BB}\), \(D_{BB}\) (\ref{s3_30}-\ref{s3_32})
of the eigenvalues \(|\ovv{c}_{m}|\), \(|\ovv{c}_{n}|\) (\ref{s2_19}).
Combining the transformation (\ref{s3_25}) with (\ref{s3_29}),
the coefficients \(A_{BB}\), \(B_{BB}\) and \(C_{BB}\), \(D_{BB}\) allow to reduce the dependence
of \((\,\hat{\mscr{P}}\;\hat{T}^{-1}\;(\pp\hat{T})\;\hat{\mscr{P}}^{-1}\,)_{BB;m\neq n}^{11}\) onto
the Euclidean variables \((\pp\hat{b}_{m\neq n})\), \((\pp\hat{b}_{m\neq n}^{*})\)
of anomalous terms for final relation (\ref{s3_33})
\beq \lb{s3_29}\hspace*{-1.0cm}
-\Big(\hat{\mscr{P}}\;\hat{T}^{-1}\;\big(\pp\hat{T}\big)\;\hat{T}^{-1}\Big)_{BB;mn}^{11}
&\stackrel{m\neq n}{=}&\Big(\hat{\mscr{P}}\;\hat{T}^{-1}\;\big(\pp\hat{T}\big)\;
\hat{T}^{-1}\Big)_{BB;mn}^{22,T}
\stackrel{m\neq n}{=} \im\;\big(\pp\hat{d}_{mn}\big)=  \\ \no &=&
-e^{\im\;\varphi_{m}}\;D_{BB}\;\big(\pp\hat{c}_{D;mn}^{\bprime *}\big)+
e^{-\im\;\varphi_{n}}\;C_{BB}\;\big(\pp\hat{c}_{D;mn}\ppr\big) \;;   \\   \lb{s3_30}
C_{BB}&=&\frac{|\ovv{c}_{n}|\;-|\ovv{c}_{n}|\;\cos\big(|\ovv{c}_{n}|\big)\;\cos\big(|\ovv{c}_{m}|\big)-
|\ovv{c}_{m}|\;\sin\big(|\ovv{c}_{n}|\big)\;\sin\big(|\ovv{c}_{m}|\big)}{|\ovv{c}_{n}|^{2}-|\ovv{c}_{m}|^{2}}
 \;;  \\  \lb{s3_31}
D_{BB}&=&\frac{|\ovv{c}_{m}|\;-|\ovv{c}_{m}|\;\cos\big(|\ovv{c}_{m}|\big)\;\cos\big(|\ovv{c}_{n}|\big)-
|\ovv{c}_{n}|\;\sin\big(|\ovv{c}_{m}|\big)\;\sin\big(|\ovv{c}_{n}|\big)}{|\ovv{c}_{m}|^{2}-|\ovv{c}_{n}|^{2}}
 \;; \\ \lb{s3_32}  &&    D_{BB}\big(|\ovv{c}_{m}|,|\ovv{c}_{n}|\big)=C_{BB}\big(|\ovv{c}_{n}|,|\ovv{c}_{m}|\big)
  \;;  \\   \lb{s3_33}\hspace*{-1.0cm}
-\Big(\hat{\mscr{P}}\;\hat{T}^{-1}\;\big(\pp\hat{T}\big)\;\hat{T}^{-1}\Big)_{BB;mn}^{11}
&\stackrel{m\neq n}{=}&\Big(\hat{\mscr{P}}\;\hat{T}^{-1}\;\big(\pp\hat{T}\big)\;\hat{T}^{-1}\Big)_{BB;mn}^{22,T}
\stackrel{m\neq n}{=} \im\;\big(\pp\hat{d}_{mn}\big)=  \\ \no &=&
\frac{e^{-\im\;\varphi_{n}}\;\sin\big(|\ovv{c}_{n}|\big)\;\big(\pp\hat{b}_{mn}\big)-
e^{\im\;\varphi_{m}}\;\sin\big(|\ovv{c}_{m}|\big)\;\big(\pp\hat{b}_{mn}^{*}\big)}{\cos\big(|\ovv{c}_{m}|\big)+
\cos\big(|\ovv{c}_{n}|\big)} = \\ \no &=&
\frac{1}{2}\;\tan\bigg(\frac{|\ovv{c}_{m}|+|\ovv{c}_{n}|}{2}\bigg)\;
\Big[e^{-\im\;\varphi_{n}}\;\big(\pp\hat{b}_{mn}\big)-
e^{\im\;\varphi_{m}}\;\big(\pp\hat{b}_{mn}^{*}\big)\Big] +  \\ \no &-&
\frac{1}{2}\;\tan\bigg(\frac{|\ovv{c}_{m}|-|\ovv{c}_{n}|}{2}\bigg)\;
\Big[e^{-\im\;\varphi_{n}}\;\big(\pp\hat{b}_{mn}\big)+
e^{\im\;\varphi_{m}}\;\big(\pp\hat{b}_{mn}^{*}\big)\Big]_{\mbox{.}}
\eeq
In case of a limit process \(|\ovv{c}_{n}|\rightarrow|\ovv{c}_{m}|\), we obtain from Eq.\ (\ref{s3_33}) the result (\ref{s3_20})
for the diagonal density elements \((m=n)\).

\subsubsection{Fermion-fermion part of the transformation to
Euclidean integration variables of pair condensate fields}\lb{s322}

In analogy to the boson-boson part, we list corresponding results of the transformation to Euclidean field
variables for the fermion-fermion parts (appendix C in Ref.\ \cite{mies1}).
However, one has to exchange the ordinary matrix elements of the
boson-boson parts by matrix elements of the quaternion algebra with standard \(2\times 2\) Pauli matrices
and the \(2\times 2\) unit matrix (\ref{s3_34}). In the case of quaternionic, diagonal elements of the pair condensates,
one has to restrict to the quaternion element with Pauli matrix \((\tau_{2})_{\mu\nu}\;(\pp\hat{a}_{rr}^{(2)})\) (\ref{s3_34}),
due to the anti-symmetry of \((\,\hat{\mscr{P}}\;\hat{T}^{-1}\;(\pp\hat{T})\;\hat{\mscr{P}}^{-1}\,)_{FF;r\mu,r\nu}^{12}\)
and of \((\pp\hat{a}_{r\mu,r\nu})=(\tau_{2})_{\mu\nu}\;(\pp\hat{a}_{rr}^{(2)})\) for the BCS pair condensate terms
\beq \lb{s3_34}
\lefteqn{\hspace*{-1.5cm}-\Big(\hat{\mscr{P}}\;\hat{T}^{-1}\;\big(\pp\hat{T}\big)\;
\hat{\mscr{P}}^{-1}\Big)_{FF;r\mu,r\nu}^{12} =
-\Big(\hat{\mscr{P}}\;\hat{T}^{-1}\;\big(\pp\hat{T}\big)\;\hat{\mscr{P}}^{-1}\Big)_{FF;r\mu,r\nu}^{21,+}=
\big(\pp\hat{a}_{r\mu,r\nu}\big)= \big(\tau_{2}\big)_{\mu\nu}\;\big(\pp\hat{a}_{rr}^{(2)}\big)=  }
  \\ \no &=&\big(\tau_{2}\big)_{\mu\nu}\;
\bigg[\bigg(\frac{1}{2}+\frac{\sinh\big(2\;|\ovv{f}_{r}|\big)}{4\;|\ovv{f}_{r}|}\bigg)\;
\big(\pp\hat{f}_{D;rr}^{\bprime(2)}\big)+
\bigg(\frac{1}{2}-\frac{\sinh\big(2\;|\ovv{f}_{r}|\big)}{4\;|\ovv{f}_{r}|}\bigg)\;e^{\im\;2\phi_{r}}\;
\big(\pp\hat{f}_{D;rr}^{\bprime(2)*}\big)\bigg]_{\mbox{.}}
\eeq
The complementary diagonal forms of \(\hat{G}_{\mbox{{\scz Osp}}/ \mbox{{\scz U}}}^{1/2}\), \(\hat{G}_{\mbox{{\scz Osp}}/ \mbox{{\scz U}}}^{-1/2}\)
with diagonal elements of the fermion-fermion section follow from the transformations of
\((\pp\hat{f}_{D;rr}^{\bprime(2)})\), \((\pp\hat{f}_{D;rr}^{\bprime(2)*})\) (\ref{s2_2}-\ref{s2_5}) to
\((\pp\hat{a}_{rr}^{(2)})\), \((\pp\hat{a}_{rr}^{(2)*})\) and vice versa (\ref{s3_35},\ref{s3_36}). The change of integration
measure is given in (\ref{s3_37}) and yields with the original coset integration measure of
\(\mbox{Osp}(S,S|2L)/ \mbox{U}(L|S)\otimes \mbox{U}(L|S)\) (\ref{s2_48}) Euclidean integration variables
\(d\hat{a}_{rr}^{(2)}\wedge d\hat{a}_{rr}^{(2)*}\)
\beq\lb{s3_35}
\left(\bea{c} \big(\pp\hat{a}_{rr}^{(2)}\big) \\  \big(\pp\hat{a}_{rr}^{(2)*}\big) \eea\right) &=&
\left(\bea{cc}
\bigg(\frac{1}{2}+\frac{\sinh\big(2\;|\ovv{f}_{r}|\big)}{4\;|\ovv{f}_{r}|}\bigg)  &
e^{\im\;2\phi_{r}}\;\bigg(\frac{1}{2}-\frac{\sinh\big(2\;|\ovv{f}_{r}|\big)}{4\;|\ovv{f}_{r}|}\bigg)
\\ e^{-\im\;2\phi_{r}}\;\bigg(\frac{1}{2}-\frac{\sinh\big(2\;|\ovv{f}_{r}|\big)}{4\;|\ovv{f}_{r}|}\bigg) &
\bigg(\frac{1}{2}+\frac{\sinh\big(2\;|\ovv{f}_{r}|\big)}{4\;|\ovv{f}_{r}|}\bigg) \eea\right)
\left(\bea{c} \big(\pp\hat{f}_{D;rr}^{\bprime(2)}\big) \\
\big(\pp\hat{f}_{D;rr}^{\bprime(2)*}\big) \eea\right)_{\mbox{;}}   \\ \lb{s3_36}
\left(\bea{c} \big(\pp\hat{f}_{D;rr}^{\bprime(2)}\big) \\
\big(\pp\hat{f}_{D;rr}^{\bprime(2)*}\big) \eea\right)  &=&
\left(\bea{cc}
\bigg(\frac{1}{2}+\frac{|\ovv{f}_{r}|}{\sinh\big(2\;|\ovv{f}_{r}|\big)}\bigg)  &
e^{\im\;2\phi_{r}}\;\bigg(\frac{1}{2}-\frac{|\ovv{f}_{r}|}{\sinh\big(2\;|\ovv{f}_{r}|\big)}\bigg)
\\ e^{-\im\;2\phi_{r}}\;\bigg(\frac{1}{2}-\frac{|\ovv{f}_{r}|}{\sinh\big(2\;|\ovv{f}_{r}|\big)}\bigg) &
\bigg(\frac{1}{2}+\frac{|\ovv{f}_{r}|}{\sinh\big(2\;|\ovv{f}_{r}|\big)}\bigg) \eea\right)
\left(\bea{c} \big(\pp\hat{a}_{rr}^{(2)}\big) \\  \big(\pp\hat{a}_{rr}^{(2)*}\big) \eea\right)_{\mbox{;}}  \\ \lb{s3_37}
d\hat{f}_{D;rr}^{\bprime(2)}\wedge d\hat{f}_{D;rr}^{\bprime(2)*}&=&
d\hat{f}_{D;rr}^{(2)}\wedge d\hat{f}_{D;rr}^{(2)*}=
d\hat{a}_{rr}^{(2)}\wedge d\hat{a}_{rr}^{(2)*}\;\;
\frac{2\;|\ovv{f}_{r}|}{\sinh\big(2\;|\ovv{f}_{r}|\big)}\;.
\eeq
We quote the result (\ref{s3_38}) for the quaternionic, diagonal densities
\((\,\hat{\mscr{P}}\;\hat{T}^{-1}\;(\pp\hat{T})\;\hat{\mscr{P}}^{-1}\,)_{FF;r\mu,r\nu}^{11}\) in terms of the
coset fields \((\pp\hat{f}_{D;rr}^{\bprime(2)})\), \((\pp\hat{f}_{D;rr}^{\bprime(2)*})\) (\ref{s2_2}-\ref{s2_5})
according to appendix C in Ref.\ \cite{mies1}. Incorporating the
transformation (\ref{s3_36}) with diagonal sub-metric tensor \(\hat{G}_{\mbox{{\scz Osp}}/ \mbox{{\scz U}}}^{-1/2}\), we obtain the diagonal
density elements \(\im\,(\pp\hat{g}_{r\mu,r\nu})\)
of the fermion-fermion part in dependence on \(\tanh(|\ovv{f}_{r}|)\) and the
Euclidean fermion-fermion pair condensate fields \((\pp\hat{a}_{rr}^{(2)})\), \((\pp\hat{a}_{rr}^{(2)*})\)
\beq\lb{s3_38}
\lefteqn{\hspace*{-2.0cm}
-\Big(\hat{\mscr{P}}\;\hat{T}^{-1}\;\big(\pp\hat{T}\big)\;\hat{\mscr{P}}^{-1}\Big)_{FF;r\mu,r\nu}^{11} =
\Big(\hat{\mscr{P}}\;\hat{T}^{-1}\;\big(\pp\hat{T}\big)\;\hat{\mscr{P}}^{-1}\Big)_{FF;r\mu,r\nu}^{22,T} =
\im\;\big(\pp\hat{g}_{r\mu,r\nu}\big) = } \\  \no &=&
\delta_{\mu\nu}\;\frac{\Big(\sinh\big(|\ovv{f}_{r}|\big)\Big)^{2}}{2\;|\ovv{f}_{r}|}\;
\Big(\big(\pp\hat{f}_{D;rr}^{\bprime(2)*}\big)\;e^{\im\;\phi_{r}}-
\big(\pp\hat{f}_{D;rr}^{\bprime(2)}\big)\;e^{-\im\;\phi_{r}}\Big)  \\ \no &=&-\delta_{\mu\nu}\;
\frac{1}{2}\;\tanh\big(|\ovv{f}_{r}|\big)\;
\Big(\big(\pp\hat{a}_{rr}^{(2)}\big)\;e^{-\im\;\phi_{r}}-
\big(\pp\hat{a}_{rr}^{(2)*}\big)\;e^{\im\;\phi_{r}}\Big)\;.
\eeq
Concerning the off-diagonal, anomalous fields of the fermion-fermion sections, one has to apply all four quaternion
elements \((\pp\hat{a}_{rr\ppr}^{(k)})\) with \((\tau_{0})_{\mu\nu}\), \((\tau_{1})_{\mu\nu}\),
\((\tau_{2})_{\mu\nu}\), \((\tau_{3})_{\mu\nu}\) and \((\pp\hat{a}_{rr\ppr}^{(k)})=-(\pp\hat{a}_{r\ppr r}^{(k)})\)
being anti-symmetric for \(k=0,1,3\) and being symmetric \((\pp\hat{a}_{rr\ppr}^{(2)})=(\pp\hat{a}_{r\ppr r}^{(2)})\) for \(k=2\)
\be \lb{s3_39}
-\Big(\hat{\mscr{P}}\;\hat{T}^{-1}\;\big(\pp\hat{T}\big)\;\hat{\mscr{P}}^{-1}\Big)_{FF;r\mu,r\ppr\nu}^{12}
\stackrel{r\neq r\ppr}{=}
-\Big(\hat{\mscr{P}}\;\hat{T}^{-1}\;\big(\pp\hat{T}\big)\;\hat{\mscr{P}}^{-1}\Big)_{FF;r\mu,r\ppr\nu}^{21,+}
\stackrel{r\neq r\ppr}{=}  \sum_{k=0}^{3}\big(\tau_{k}\big)_{\mu\nu}\;\big(\pp\hat{a}_{rr\ppr}^{(k)}\big)\;.
\ee
We abbreviate various terms of eigenvalues \(|\ovv{f}_{r}|\), \(|\ovv{f}_{r\ppr}|\) (\ref{s2_20}) by the coefficients
\(A_{FF}\) (\ref{s3_40}), \(B_{FF}\) (\ref{s3_41}) and have to
distinguish between quaternion elements of \((\tau_{0})_{\mu\nu}\)
and \((\tau_{k})_{\mu\nu}\) (k=1,2,3) for the transformation of the original, anomalous coset fields
\((\pp\hat{f}_{D;rr\ppr}^{\bprime (k)})\), \((\pp\hat{f}_{D;rr\ppr}^{\bprime(k)*})\) (\ref{s2_2}-\ref{s2_5})
to their Euclidean correspondents
\beq \lb{s3_40}
A_{FF} &=& \frac{|\ovv{f}_{r}|\;\cosh\big(|\ovv{f}_{r\ppr}|\big)\;\sinh\big(|\ovv{f}_{r}|\big)-
|\ovv{f}_{r\ppr}|\;\cosh\big(|\ovv{f}_{r}|\big)\;\sinh\big(|\ovv{f}_{r\ppr}|\big)}
{|\ovv{f}_{r}|^{2}-|\ovv{f}_{r\ppr}|^{2}} \;; \\  \lb{s3_41}
B_{FF} &=& \frac{|\ovv{f}_{r}|\;\cosh\big(|\ovv{f}_{r}|\big)\;\sinh\big(|\ovv{f}_{r\ppr}|\big)-
|\ovv{f}_{r\ppr}|\;\cosh\big(|\ovv{f}_{r\ppr}|\big)\;\sinh\big(|\ovv{f}_{r}|\big)}
{|\ovv{f}_{r}|^{2}-|\ovv{f}_{r\ppr}|^{2}}   \;;   \\  \lb{s3_42}
\left(\bea{c} \big(\pp\hat{a}_{rr\ppr}^{(0)}\big) \\ \big(\pp\hat{a}_{rr\ppr}^{(0)*}\big)\eea\right) &=&
\left(\bea{cc}
A_{FF}  &  e^{\im(\phi_{r}+\phi_{r\ppr})}\;B_{FF} \\  e^{-\im(\phi_{r}+\phi_{r\ppr})}\;B_{FF} & A_{FF} \eea\right)
\left(\bea{c} \big(\pp\hat{f}_{D;rr\ppr}^{\bprime(0)}\big) \\
\big(\pp\hat{f}_{D;rr\ppr}^{\bprime(0)*}\big) \eea\right)_{\mbox{;}}  \\ \lb{s3_43}
\left(\bea{c} \big(\pp\hat{f}_{D;rr\ppr}^{\bprime(0)}\big) \\
\big(\pp\hat{f}_{D;rr\ppr}^{\bprime(0)*}\big) \eea\right)  &=& \frac{1}{A_{FF}^{2}-B_{FF}^{2}}
\left(\bea{cc}
A_{FF}  &  -e^{\im(\phi_{r}+\phi_{r\ppr})}\;B_{FF} \\  -e^{-\im(\phi_{r}+\phi_{r\ppr})}\;B_{FF} & A_{FF} \eea\right)
\left(\bea{c} \big(\pp\hat{a}_{rr\ppr}^{(0)}\big) \\ \big(\pp\hat{a}_{rr\ppr}^{(0)*}\big) \eea\right)_{\mbox{;}}  \\ \lb{s3_44}
\left(\bea{c} \big(\pp\hat{a}_{rr\ppr}^{(k)}\big) \\ \big(\pp\hat{a}_{rr\ppr}^{(k)*}\big)\eea\right)
&\stackrel{\boldsymbol{k=1,2,3}}{=}   & \left(\bea{cc}
A_{FF}  &  -e^{\im(\phi_{r}+\phi_{r\ppr})}\;B_{FF} \\  -e^{-\im(\phi_{r}+\phi_{r\ppr})}\;B_{FF} & A_{FF} \eea\right)
\left(\bea{c} \big(\pp\hat{f}_{D;rr\ppr}^{\bprime(k)}\big) \\
\big(\pp\hat{f}_{D;rr\ppr}^{\bprime(k)*}\big) \eea\right)_{\mbox{;}}   \\ \lb{s3_45}
\left(\bea{c} \big(\pp\hat{f}_{D;rr\ppr}^{\bprime(k)}\big) \\
\big(\pp\hat{f}_{D;rr\ppr}^{\bprime(k)*}\big) \eea\right)  &\stackrel{\boldsymbol{k=1,2,3}}{=} &
\frac{1}{A_{FF}^{2}-B_{FF}^{2}} \left(\bea{cc}
A_{FF}  &  e^{\im(\phi_{r}+\phi_{r\ppr})}\;B_{FF} \\  e^{-\im(\phi_{r}+\phi_{r\ppr})}\;B_{FF} & A_{FF} \eea\right)
\left(\bea{c} \big(\pp\hat{a}_{rr\ppr}^{(k)}\big) \\ \big(\pp\hat{a}_{rr\ppr}^{(k)*}\big) \eea\right)_{\mbox{;}}  \\  \lb{s3_46}
\frac{A_{FF}}{A_{FF}^{2}-B_{FF}^{2}}&=&\frac{1}{2}\bigg(
\frac{|\ovv{f}_{r}|-|\ovv{f}_{r\ppr}|}{\sinh\big(|\ovv{f}_{r}|-|\ovv{f}_{r\ppr}|\big)}+
\frac{|\ovv{f}_{r}|+|\ovv{f}_{r\ppr}|}{\sinh\big(|\ovv{f}_{r}|+|\ovv{f}_{r\ppr}|\big)}\bigg) \;;  \\  \lb{s3_47}
\frac{B_{FF}}{A_{FF}^{2}-B_{FF}^{2}}&=&-\frac{1}{2}\bigg(
\frac{|\ovv{f}_{r}|+|\ovv{f}_{r\ppr}|}{\sinh\big(|\ovv{f}_{r}|+|\ovv{f}_{r\ppr}|\big)}-
\frac{|\ovv{f}_{r}|-|\ovv{f}_{r\ppr}|}{\sinh\big(|\ovv{f}_{r}|-|\ovv{f}_{r\ppr}|\big)}\bigg)_{\mbox{.}}
\eeq
Taking the determinants of the \(\hat{G}_{\mbox{{\scz Osp}}/ \mbox{{\scz U}}}^{-1/2}\) coset sub-metric transformations (\ref{s3_43},\ref{s3_45}),
we acquire the integration measure \(\mbox{SDET}(\hat{G}_{\mbox{{\scz Osp}}/ \mbox{{\scz U}}}^{-1/2})\) of the particular,
diagonalized fermion-fermion parts (\ref{s3_48}) which are eliminated with the sub-metric tensor parts
\((\,\mbox{SDET}(\hat{G}_{\mbox{{\scz Osp}}/ \mbox{{\scz U}}})\,)^{1/2}\) (\ref{s2_48}) of the original coset decomposition to
Euclidean integration variables \(d\hat{a}_{rr\ppr}^{(k)}\wedge d\hat{a}_{rr\ppr}^{(k)*}\), \((k=0,1,2,3)\)
\be \lb{s3_48}
d\hat{f}_{D;rr\ppr}^{\bprime(k)}\wedge d\hat{f}_{D;rr\ppr}^{\bprime(k)*}=
d\hat{f}_{D;rr\ppr}^{(k)}\wedge d\hat{f}_{D;rr\ppr}^{(k)*}=
d\hat{a}_{rr\ppr}^{(k)}\wedge d\hat{a}_{rr\ppr}^{(k)*}\;\;
\frac{|\ovv{f}_{r}|+|\ovv{f}_{r\ppr}|}{\sinh\big(|\ovv{f}_{r}|+|\ovv{f}_{r\ppr}|\big)}\;\;
\frac{|\ovv{f}_{r}|-|\ovv{f}_{r\ppr}|}{\sinh\big(|\ovv{f}_{r}|-|\ovv{f}_{r\ppr}|\big)}\;.
\ee
The off-diagonal density parts
\(-(\hat{\mscr{P}}\;\hat{T}^{-1}\;(\pp\hat{T})\;\hat{\mscr{P}}^{-1})_{FF;r\mu,r\ppr\nu}^{11}=
\im\,(\pp\hat{g}_{r\mu,r\ppr\nu})\) (\ref{s3_49})
are given as quaternion matrix elements and by coefficients \(C_{FF}\) (\ref{s3_51}), \(D_{FF}\) (\ref{s3_52})
according to appendix C in Ref.\ \cite{mies1}
\beq \lb{s3_49}
\lefteqn{\hspace*{-1.5cm}
-\Big(\hat{\mscr{P}}\;\hat{T}^{-1}\;\big(\pp\hat{T}\big)\;\hat{\mscr{P}}^{-1}\Big)_{FF;r\mu,r\ppr\nu}^{11}
 \stackrel{r\neq r\ppr}{=}  \Big(\hat{\mscr{P}}\;\hat{T}^{-1}\;\big(\pp\hat{T}\big)\;
\hat{\mscr{P}}^{-1}\Big)_{FF;r\mu,r\ppr\nu}^{22,T}
\stackrel{r\neq r\ppr}{=} \im\;\big(\pp\hat{g}_{r\mu,r\ppr\nu}\big)  = } \\ \no &=& -\big(\tau_{2}\big)_{\mu\nu}\;
\Big[e^{-\im\;\phi_{r\ppr}}\;C_{FF}\;\big(\pp\hat{f}_{D;rr\ppr}^{\bprime(0)}\big)+
e^{\im\;\phi_{r}}\;D_{FF}\;\big(\pp\hat{f}_{D;rr\ppr}^{\bprime(0)*}\big)\Big]+ \\ \no &-&
\sum_{k=1,2,3}\big(\hat{m}_{k}\big)_{\mu\nu}\;
\Big[e^{-\im\;\phi_{r\ppr}}\;C_{FF}\;\big(\pp\hat{f}_{D;rr\ppr}^{\bprime(k)}\big)-
e^{\im\;\phi_{r}}\;D_{FF}\;\big(\pp\hat{f}_{D;rr\ppr}^{\bprime(k)*}\big)\Big]\;;  \\ \lb{s3_50} &&
\big(\hat{m}_{1}\big)_{\mu\nu}=\im\;\big(\hat{\tau}_{3}\big)_{\mu\nu}\;;\hspace*{0.5cm}
\big(\hat{m}_{2}\big)_{\mu\nu}=\big(\hat{\tau}_{0}\big)_{\mu\nu}\;;\hspace*{0.5cm}
\big(\hat{m}_{3}\big)_{\mu\nu}=-\im\;\big(\hat{\tau}_{1}\big)_{\mu\nu} \;;
\eeq
\beq  \lb{s3_51}
C_{FF}&=&\frac{-|\ovv{f}_{r\ppr}|+|\ovv{f}_{r\ppr}|\;\cosh\big(|\ovv{f}_{r\ppr}|\big)\;\cosh\big(|\ovv{f}_{r}|\big)-
|\ovv{f}_{r}|\;\sinh\big(|\ovv{f}_{r\ppr}|\big)\;\sinh\big(|\ovv{f}_{r}|\big)}{|\ovv{f}_{r\ppr}|^{2}-|\ovv{f}_{r}|^{2}} \;;  \\ \lb{s3_52}
D_{FF}&=&\frac{-|\ovv{f}_{r}|+|\ovv{f}_{r}|\;\cosh\big(|\ovv{f}_{r}|\big)\;\cosh\big(|\ovv{f}_{r\ppr}|\big)-
|\ovv{f}_{r\ppr}|\;\sinh\big(|\ovv{f}_{r}|\big)\;\sinh\big(|\ovv{f}_{r\ppr}|\big)}{|\ovv{f}_{r}|^{2}-|\ovv{f}_{r\ppr}|^{2}}  \;;
  \\ \lb{s3_53}  &&  C_{FF}\big(|\ovv{f}_{r}|,|\ovv{f}_{r\ppr}|\big)=D_{FF}\big(|\ovv{f}_{r\ppr}|,|\ovv{f}_{r}|\big)\;.
\eeq
Insertion of relations (\ref{s3_43},\ref{s3_45}) with coefficients \(A_{FF}\) (\ref{s3_40}), \(B_{FF}\) (\ref{s3_41})
into (\ref{s3_49}) yields the fermion-fermion
density part in terms of \(\tanh(\,(|\ovv{f}_{r}|\pm|\ovv{f}_{r\ppr}|)/2\,)\) and the Euclidean integration
variables \((\pp\hat{a}_{rr\ppr}^{(k)})\), \((\pp\hat{a}_{rr\ppr}^{(k)*})\) combined with the quaternion,
\(2\times 2\) elements \((\tau_{k})_{\mu\nu}\), \((\tau_{0})_{\mu\nu}\), \((\hat{m}_{k})_{\mu\nu}\) (\ref{s3_50})
\beq \lb{s3_54}
\lefteqn{\hspace*{-1.0cm}
-\Big(\hat{\mscr{P}}\;\hat{T}^{-1}\;\big(\pp\hat{T}\big)\;\hat{T}^{-1}\Big)_{FF;r\mu,r\ppr\nu}^{11}
 \stackrel{r\neq r\ppr}{=}  \Big(\hat{\mscr{P}}\;\hat{T}^{-1}\;\big(\pp\hat{T}\big)\;
\hat{T}^{-1}\Big)_{FF;r\mu,r\ppr\nu}^{22,T}
\stackrel{r\neq r\ppr}{=} \im\;\big(\pp\hat{g}_{r\mu,r\ppr\nu}\big)  = } \\ \no &=&-\big(\tau_{2}\big)_{\mu\nu}\;\bigg[
\frac{e^{-\im\;\phi_{r\ppr}}\;\sinh\big(|\ovv{f}_{r\ppr}|\big)\;\big(\pp\hat{a}_{rr\ppr}^{(0)}\big)+
e^{\im\;\phi_{r}}\;\sinh\big(|\ovv{f}_{r}|\big)\;\big(\pp\hat{a}_{rr\ppr}^{(0)*}\big)}{
\cosh\big(|\ovv{f}_{r}|\big)+\cosh\big(|\ovv{f}_{r\ppr}|\big)} \bigg] + \\ \no &-&\sum_{k=1,2,3}
\big(\hat{m}_{k}\big)_{\mu\nu}\;\bigg[
\frac{e^{-\im\;\phi_{r\ppr}}\;\sinh\big(|\ovv{f}_{r\ppr}|\big)\;\big(\pp\hat{a}_{rr\ppr}^{(k)}\big)-
e^{\im\;\phi_{r}}\;\sinh\big(|\ovv{f}_{r}|\big)\;\big(\pp\hat{a}_{rr\ppr}^{(k)*}\big)}{
\cosh\big(|\ovv{f}_{r}|\big)+\cosh\big(|\ovv{f}_{r\ppr}|\big)} \bigg] \\ \no &=&
-\frac{1}{2}\;\big(\tau_{2}\big)_{\mu\nu}\;\bigg\{\tanh\bigg(\frac{|\ovv{f}_{r}|+|\ovv{f}_{r\ppr}|}{2}\bigg)\;
\Big[e^{-\im\;\phi_{r\ppr}}\;\big(\pp\hat{a}_{rr\ppr}^{(0)}\big)+
e^{\im\;\phi_{r}}\;\big(\pp\hat{a}_{rr\ppr}^{(0)*}\big)\Big] +  \\ \no &-&
\tanh\bigg(\frac{|\ovv{f}_{r}|-|\ovv{f}_{r\ppr}|}{2}\bigg)\;
\Big[e^{-\im\;\phi_{r\ppr}}\;\big(\pp\hat{a}_{rr\ppr}^{(0)}\big)-
e^{\im\;\phi_{r}}\;\big(\pp\hat{a}_{rr\ppr}^{(0)*}\big)\Big]\bigg\} +  \\ \no &-&
\frac{1}{2}\sum_{k=1,2,3}
\big(\hat{m}_{k}\big)_{\mu\nu}\;\Bigg\{\tanh\bigg(\frac{|\ovv{f}_{r}|+|\ovv{f}_{r\ppr}|}{2}\bigg)\;
\Big[e^{-\im\;\phi_{r\ppr}}\;\big(\pp\hat{a}_{rr\ppr}^{(k)}\big)-
e^{\im\;\phi_{r}}\;\big(\pp\hat{a}_{rr\ppr}^{(k)*}\big)\Big] +  \\ \no &-&
\tanh\bigg(\frac{|\ovv{f}_{r}|-|\ovv{f}_{r\ppr}|}{2}\bigg)\;
\Big[e^{-\im\;\phi_{r\ppr}}\;\big(\pp\hat{a}_{rr\ppr}^{(k)}\big)+
e^{\im\;\phi_{r}}\;\big(\pp\hat{a}_{rr\ppr}^{(k)*}\big)\Big] \Bigg\}  \\ \no &=&
-\frac{1}{2} \sum_{k=0}^{3}\big(\tau_{k}\tau_{2}\big)_{\mu\nu} \;
\bigg\{\tanh\bigg(\frac{|\ovv{f}_{r}|+|\ovv{f}_{r\ppr}|}{2}\bigg)\;
\Big[e^{-\im\;\phi_{r\ppr}}\;\big(\pp\hat{a}_{rr\ppr}^{(k)}\big)-\big(-1\big)^{k}\;
e^{\im\;\phi_{r}}\;\big(\pp\hat{a}_{rr\ppr}^{(k)+}\big)\Big] +  \\ \no &-&
\tanh\bigg(\frac{|\ovv{f}_{r}|-|\ovv{f}_{r\ppr}|}{2}\bigg)\;
\Big[e^{-\im\;\phi_{r\ppr}}\;\big(\pp\hat{a}_{rr\ppr}^{(k)}\big)+\big(-1\big)^{k}\;
e^{\im\;\phi_{r}}\;\big(\pp\hat{a}_{rr\ppr}^{(k)+}\big)\Big]\bigg\}_{\mbox{.}}
\eeq

\subsubsection{Fermion-boson and boson-fermion parts of the transformation to
Euclidean integration variables of pair condensate fields}\lb{s323}

In the case of transformations in the odd fermion-boson and boson-fermion sections,
we have to consider the two quaternion elements \((\tau_{0})_{\mu\nu}\) and \((\tau_{2})_{\mu\nu}\)
and cite the result (\ref{s3_55},\ref{s3_56})
for \((\,\hat{\mscr{P}}\;\hat{T}^{-1}\;(\pp\hat{T})\;\hat{\mscr{P}}^{-1}\,)_{FB;r\mu,n}^{12}\)
from appendix C of Ref.\ \cite{mies1}. The coefficients \(A_{FB}\), \(B_{FB}\) abbreviate
the relations (\ref{s3_57}),(\ref{s3_58}) for the eigenvalues
\(|\ovv{c}_{n}|\) (\ref{s2_19}), \(|\ovv{f}_{r}|\) (\ref{s2_20});
(\(\kappa,\,\mu,\,\nu=1,2\)), (\(r,\,r\ppr=1,\ldots,S/2\)), (\(m,n=1,\ldots,L\))
\beq \lb{s3_55}
-\big(\pp\hat{\zeta}_{r\mu,n}\big) &=&
-\Big(\hat{\mscr{P}}\;\hat{T}^{-1}\;\big(\pp\hat{T}\big)\;\hat{\mscr{P}}^{-1}\Big)_{FB;r\mu,n}^{12}=
\Big(\hat{\mscr{P}}\;\hat{T}^{-1}\;\big(\pp\hat{T}\big)\;\hat{\mscr{P}}^{-1}\Big)_{BF;n,r\mu}^{21,+}  \\ \no &=&
-A_{FB}\;\big(\tau_{0}\big)_{\mu\kappa}\;\big(\pp\hat{\eta}_{D;r\kappa,n}\ppr\big)-
B_{FB}\;e^{\im(\phi_{r}+\varphi_{n})}\;\big(\tau_{2}\big)_{\mu\kappa}\;
\big(\pp\hat{\eta}_{D;r\kappa,n}^{\bprime *}\big) \;;  \\ \lb{s3_56}
-\big(\pp\hat{\zeta}_{r\mu,n}^{*}\big)&=&
-A_{FB}\;\big(\tau_{0}\big)_{\mu\kappa}\;\big(\pp\hat{\eta}_{D;r\kappa,n}^{\bprime *}\big)+
B_{FB}\;e^{-\im(\phi_{r}+\varphi_{n})}\;\big(\tau_{2}\big)_{\mu\kappa}\;
\big(\pp\hat{\eta}_{D;r\kappa,n}^{\bprime}\big) \;; \\   \lb{s3_57}
A_{FB}&=&\frac{|\ovv{c}_{n}|\;\cosh\big(|\ovv{f}_{r}|\big)\;\sin\big(|\ovv{c}_{n}|\big)+
|\ovv{f}_{r}|\;\cos\big(|\ovv{c}_{n}|\big)\;\sinh\big(|\ovv{f}_{r}|\big)}{|\ovv{c}_{n}|^{2}+|\ovv{f}_{r}|^{2}} \;; \\  \lb{s3_58}
B_{FB}&=&\frac{|\ovv{c}_{n}|\;\cos\big(|\ovv{c}_{n}|\big)\;\sinh\big(|\ovv{f}_{r}|\big)-
|\ovv{f}_{r}|\;\cosh\big(|\ovv{f}_{r}|\big)\;\sin\big(|\ovv{c}_{n}|\big)}{|\ovv{c}_{n}|^{2}+|\ovv{f}_{r}|^{2}}\;.
\eeq
The diagonalized, \(4\times 4\) sub-metric tensors \(\hat{G}_{\mbox{{\scz Osp}}/ \mbox{{\scz U}}}^{1/2}\),
\(\hat{G}_{\mbox{{\scz Osp}}/ \mbox{{\scz U}}}^{-1/2}\) of the fermion-boson, boson-fermion sections
are described in relations (\ref{s3_59},\ref{s3_60}) by using the coefficients
\(A_{FB}\), \(B_{FB}\) for abbreviating relations (\ref{s3_61},\ref{s3_62})
\beq  \lb{s3_59}
\left(\bea{c} \big(\pp\hat{\zeta}_{r\mu,n}\big) \\  \big(\pp\hat{\zeta}_{r\mu,n}^{*}\big) \eea\right)
&=&\left(\bea{cc}  A_{FB}\;\;\big(\tau_{0}\big)_{\mu\kappa}  &
B_{FB}\;\;e^{\im(\phi_{r}+\varphi_{n})}\;\big(\tau_{2}\big)_{\mu\kappa} \\
-B_{FB}\;\;e^{-\im(\phi_{r}+\varphi_{n})}\;\big(\tau_{2}\big)_{\mu\kappa}  &
 A_{FB}\;\;\big(\tau_{0}\big)_{\mu\kappa}
\eea\right)
\left(\bea{c} \big(\pp\hat{\eta}_{D;r\kappa,n}^{\bprime}\big) \\
\big(\pp\hat{\eta}_{D;r\kappa,n}^{\bprime *}\big) \eea\right)_{\mbox{;}}   \\  \lb{s3_60}
\left(\bea{c} \big(\pp\hat{\eta}_{D;r\kappa,n}^{\bprime}\big) \\
\big(\pp\hat{\eta}_{D;r\kappa,n}^{\bprime *}\big) \eea\right)  &=&
\left(\bea{cc}
\wt{A}_{FB}\;\;\big(\tau_{0}\big)_{\mu\kappa}  &
-\wt{B}_{FB}\;\;e^{\im(\phi_{r}+\varphi_{n})}\;\big(\tau_{2}\big)_{\mu\kappa} \\
\wt{B}_{FB}\;\;e^{-\im(\phi_{r}+\varphi_{n})}\;\big(\tau_{2}\big)_{\mu\kappa}  &
\wt{A}_{FB}\;\;\big(\tau_{0}\big)_{\mu\kappa}
\eea\right)
\left(\bea{c} \big(\pp\hat{\zeta}_{r\kappa,n}\big) \\
\big(\pp\hat{\zeta}_{r\kappa,n}^{*}\big) \eea\right)_{\mbox{;}}  \\    \lb{s3_61}
\wt{A}_{FB}&=&\frac{A_{FB}}{A_{FB}^{2}+B_{FB}^{2}}=\frac{1}{2}\bigg(
\frac{|\ovv{f}_{r}|-\im\;|\ovv{c}_{n}|}{\sinh\big(|\ovv{f}_{r}|-\im\;|\ovv{c}_{n}|\big)}+
\frac{|\ovv{f}_{r}|+\im\;|\ovv{c}_{n}|}{\sinh\big(|\ovv{f}_{r}|+\im\;|\ovv{c}_{n}|\big)}\bigg) \;;  \\  \lb{s3_62}
\wt{B}_{FB}&=&\frac{B_{FB}}{A_{FB}^{2}+B_{FB}^{2}}=\frac{\im}{2}\bigg(
\frac{|\ovv{f}_{r}|-\im\;|\ovv{c}_{n}|}{\sinh\big(|\ovv{f}_{r}|-\im\;|\ovv{c}_{n}|\big)}-
\frac{|\ovv{f}_{r}|+\im\;|\ovv{c}_{n}|}{\sinh\big(|\ovv{f}_{r}|+\im\;|\ovv{c}_{n}|\big)}\bigg)_{\mbox{.}}
\eeq
The integration measure (\ref{s3_63}) follows from the {\it 'inverse'} of the determinant of transformation (\ref{s3_60})
where the eigenvalue \(|\ovv{c}_{n}|\) of the boson-boson part fits into the hyperbolic \(\sinh\)-function
with the eigenvalue \(|\ovv{f}_{r}|\) of the fermion-fermion section by using an imaginary factor.
In consequence, the original, odd, anomalous coset fields
\(d\hat{\eta}_{D;r1,n}^{\bprime *}\), \(d\hat{\eta}_{D;r1,n}^{\bprime}\), \(d\hat{\eta}_{D;r2,n}^{\bprime *}\),
\(d\hat{\eta}_{D;r2,n}^{\bprime}\) (\ref{s2_2}-\ref{s2_5}) are substituted by the odd Euclidean fields
\(d\hat{\zeta}_{r1,n}^{*}\), \(d\hat{\zeta}_{r1,n}\), \(d\hat{\zeta}_{r2,n}^{*}\),
\(d\hat{\zeta}_{r2,n}\) in combination of the coset integration measure (\ref{s2_48})
\beq \lb{s3_63}
\lefteqn{d\hat{\eta}_{D;r1,n}^{\bprime*}\; d\hat{\eta}_{D;r1,n}^{\bprime}\;\;\;
d\hat{\eta}_{D;r2,n}^{\bprime*}\; d\hat{\eta}_{D;r2,n}^{\bprime} = }  \\ \no &=&
\Bigg\{d\hat{\zeta}_{r1,n}^{*}\;d\hat{\zeta}_{r1,n}\;\;
\bigg(\frac{\sinh\big(|\ovv{f}_{r}|+\im\;|\ovv{c}_{n}|\big)}{|\ovv{f}_{r}|+\im\;|\ovv{c}_{n}|}\bigg)\;
\bigg(\frac{\sinh\big(|\ovv{f}_{r}|-\im\;|\ovv{c}_{n}|\big)}{|\ovv{f}_{r}|-\im\;|\ovv{c}_{n}|}\bigg)\Bigg\}\times
\\ \no &\times&
\Bigg\{d\hat{\zeta}_{r2,n}^{*}\;d\hat{\zeta}_{r2,n}\;\;
\bigg(\frac{\sinh\big(|\ovv{f}_{r}|+\im\;|\ovv{c}_{n}|\big)}{|\ovv{f}_{r}|+\im\;|\ovv{c}_{n}|}\bigg)\;
\bigg(\frac{\sinh\big(|\ovv{f}_{r}|-\im\;|\ovv{c}_{n}|\big)}{|\ovv{f}_{r}|-\im\;|\ovv{c}_{n}|}\bigg)\Bigg\}_{\mbox{.}}
\eeq
According to Ref.\ \cite{mies1} with appendix C, we list the odd density part (\ref{s3_64})
for the fermion-boson, boson-fermion sections
by introducing the coefficients \(C_{FB}\) (\ref{s3_65}), \(D_{FB}\) (\ref{s3_66}) as abbreviation
\beq \lb{s3_64}
\lefteqn{\hspace*{-2.8cm}-\Big(\hat{\mscr{P}}\;\hat{T}^{-1}\;\big(\pp\hat{T}\big)\;\hat{\mscr{P}}^{-1}\Big)_{FB;r\mu,n}^{11}  =
-\Big(\hat{\mscr{P}}\;\hat{T}^{-1}\;\big(\pp\hat{T}\big)\;\hat{\mscr{P}}^{-1}\Big)_{BF;n,r\mu}^{22,T}=
\im\;\big(\pp\hat{\xi}_{r\mu,n}\big) =}  \\ \no &=&
-e^{\im\;\phi_{r}}\;C_{FB}\;\big(\tau_{2}\big)_{\mu\kappa}\;
\big(\pp\hat{\eta}_{D;r\kappa,n}^{\bprime*}\big)+
e^{-\im\;\varphi_{n}}\;D_{FB}\;\big(\tau_{0}\big)_{\mu\kappa}\;
\big(\pp\hat{\eta}_{D;r\kappa,n}^{\bprime}\big) \;;  \\   \lb{s3_65}
C_{FB}&=&\frac{|\ovv{f}_{r}|-|\ovv{f}_{r}|\;\cos\big(|\ovv{c}_{n}|\big)\;\cosh\big(|\ovv{f}_{r}|\big)-
|\ovv{c}_{n}|\;\sin\big(|\ovv{c}_{n}|\big)\;\sinh\big(|\ovv{f}_{r}|\big)}{|\ovv{c}_{n}|^{2}+|\ovv{f}_{r}|^{2}}  \;;
  \\  \lb{s3_66} D_{FB}&=&\frac{|\ovv{c}_{n}|-|\ovv{c}_{n}|\;\cos\big(|\ovv{c}_{n}|\big)\;\cosh\big(|\ovv{f}_{r}|\big)+
|\ovv{f}_{r}|\;\sin\big(|\ovv{c}_{n}|\big)\;\sinh\big(|\ovv{f}_{r}|\big)}{|\ovv{c}_{n}|^{2}+|\ovv{f}_{r}|^{2}}\;.
\eeq
We apply (\ref{s3_60}) with the \(A_{FB}\), \(B_{FB}\) coefficients (\ref{s3_57},\ref{s3_58}), together with (\ref{s3_64})
and coefficients \(C_{FB}\), \(D_{FB}\) (\ref{s3_65},\ref{s3_66}), and
finally obtain relation (\ref{s3_67}) for the odd density parts. One achieves a dependence
on the odd, anomalous Euclidean fields
\((\pp\hat{\zeta}_{r1,n}^{*})\), \((\pp\hat{\zeta}_{r1,n})\), \((\pp\hat{\zeta}_{r2,n}^{*})\),
\((\pp\hat{\zeta}_{r2,n})\) for the odd, fermion-boson, boson-fermion density parts (\ref{s3_64}) in combination
of the eigenvalues \(|\ovv{f}_{r}|\), \(|\ovv{c}_{n}|\) (\ref{s2_19},\ref{s2_20}) appearing with
\(\tanh(\,(|\ovv{f}_{r}|\pm\:\im\:|\ovv{c}_{n}|)/2\,)\)
\beq \lb{s3_67}
\lefteqn{-\Big(\hat{\mscr{P}}\;\hat{T}^{-1}\;\big(\pp\hat{T}\big)\;\hat{\mscr{P}}^{-1}\Big)_{FB;r\mu,n}^{11}=
\im\;\big(\pp\hat{\xi}_{r\mu,n}\big)  = }    \\ \no &=&
\frac{e^{-\im\;\varphi_{n}}\;\sin\big(|\ovv{c}_{n}|\big)\;\big(\tau_{0}\big)_{\mu\kappa}\;
\big(\pp\hat{\zeta}_{r\kappa,n}\big)+
e^{\im\;\phi_{r}}\;\sinh\big(|\ovv{f}_{r}|\big)\;\big(\tau_{2}\big)_{\mu\kappa}\;
\big(\pp\hat{\zeta}_{r\kappa,n}^{*}\big)}{\cosh\big(|\ovv{f}_{r}|\big)+\cos\big(|\ovv{c}_{n}|\big)}
\\  \no &=&
\frac{1}{2}\;\tanh\bigg(\frac{|\ovv{f}_{r}|+\im\;|\ovv{c}_{n}|}{2}\bigg)\;\Big[
e^{\im\;\phi_{r}}\;\big(\tau_{2}\big)_{\mu\kappa}\;\big(\pp\hat{\zeta}_{r\kappa,n}^{*}\big)-\im\;
e^{-\im\;\varphi_{n}}\;\big(\tau_{0}\big)_{\mu\kappa}\;\big(\pp\hat{\zeta}_{r\kappa,n}\big)\Big]+ \\  \no &+&
\frac{1}{2}\;\tanh\bigg(\frac{|\ovv{f}_{r}|-\im\;|\ovv{c}_{n}|}{2}\bigg)\;\Big[
e^{\im\;\phi_{r}}\;\big(\tau_{2}\big)_{\mu\kappa}\;\big(\pp\hat{\zeta}_{r\kappa,n}^{*}\big)+\im\;
e^{-\im\;\varphi_{n}}\;\big(\tau_{0}\big)_{\mu\kappa}\;\big(\pp\hat{\zeta}_{r\kappa,n}\big)\Big]_{\mbox{.}}
\eeq

\subsection{Eigenvalues of cosets for anomalous terms and
their transformed, Euclidean correspondents}  \lb{s33}

In section \ref{s32} with subsections \ref{s321}, \ref{s322}, \ref{s323},
we have used the results of appendix C in Ref.\ \cite{mies1}
for relations (\ref{s3_9}-\ref{s3_11}) in order to transform the original, coset fields in \((\pp\hat{X}_{N\times N})\),
\(\wt{\kappa}\;(\pp\hat{X}_{N\times N}^{+})\) (\ref{s2_2}-\ref{s2_5}) and in
\((\,\hat{\mscr{P}}\;\hat{T}^{-1}\;(\pp\hat{T})\;\hat{\mscr{P}}^{-1}\,)_{\alpha\beta}^{a\neq b}\)
to Euclidean integration variables \((\pp\hat{\mscr{Z}}_{\alpha\beta}^{a\neq b})\) (\ref{s3_1}) depending on
\((\pp\hat{\mscr{X}}_{\alpha\beta})\), \(\wt{\kappa}\;(\pp\hat{\mscr{X}}_{\alpha\beta}^{+})\) (\ref{s3_2}-\ref{s3_4}).
(In this section we have to specialize on the total derivative '\(d\)' for the pair condensate
path field variables in place of the general symbolic derivative '\(\pp\)' of section \ref{s32}.
The general symbolic derivative '\(\pp\)' has been applied as abbreviation for partial, spatial or time-contour-like gradients
'\(\breve{\pp}_{i}\)', '\(\breve{\pp}_{\breve{t}_{p}}\)' and '\(\delta\)'-variations of fields for classical equations
or total derivatives '\(d\)' for the independent path fields of the integration measure.)
However, apart from the dependence on Euclidean integration variables \((d\hat{b}_{mn})\), \((d\hat{a}_{r\mu,r\ppr\nu})\),
\((d\hat{\zeta}_{r\mu,n})\), (+c.c.) (\ref{s3_2}-\ref{s3_4}), there also appear the
eigenvalues \(\ovv{c}_{m}\) (\ref{s2_19}), \(\ovv{f}_{r}\) (\ref{s2_20}) of the
original coset decomposition for anomalous fields. Their dependence has to be determined in terms of the new,
independent Euclidean pair condensate fields \((d\hat{b}_{mn})\), \((d\hat{a}_{r\mu,r\ppr\nu})\),
\((d\hat{\zeta}_{r\mu,n})\), (+c.c.). According to section \ref{s32},
we therefore list again relations (\ref{s3_68}-\ref{s3_75}) which have all been calculated in terms of the
anomalous Euclidean fields \((d\hat{\mscr{X}}_{\alpha\beta})\) and their super-hermitian conjugate
\(\wt{\kappa}\;(d\hat{\mscr{X}}_{\alpha\beta}^{+})\)
\beq \lb{s3_68}
\big(d\hat{\mscr{Z}}_{\alpha\beta}^{ab}\big) &=&
-\Big(\hat{\mscr{P}}\;\hat{T}^{-1}\;\big(d\hat{T}\big)\;\hat{\mscr{P}}^{-1}\Big)_{\alpha\beta}^{ab}=
\left(\bea{cc} \big(d\hat{\mscr{Y}}^{11}_{\alpha\beta}\big) & \big(d\hat{\mscr{X}}_{\alpha\beta}\big) \\
\wt{\kappa}\;\big(d\hat{\mscr{X}}_{\alpha\beta}^{+}\big) &
\big(d\hat{\mscr{Y}}^{22}_{\alpha\beta}\big) \eea\right)_{\mbox{;}}  \\ \lb{s3_69}
\big(d\hat{\mscr{X}}_{\alpha\beta}\big)&=&
-\Big(\hat{\mscr{P}}\;\hat{T}^{-1}\;\big(d\hat{T}\big)\;\hat{\mscr{P}}^{-1}\Big)_{\alpha\beta}^{12}=
\left(\bea{cc} -\big(d\hat{b}_{mn}\big) & \big(d\hat{\zeta}_{m,r\ppr\nu}^{T}\big) \\
-\big(d\hat{\zeta}_{r\mu,n}\big) & \big(d\hat{a}_{r\mu,r\ppr\nu}\big) \eea\right)_{\alpha\beta} \;;  \\  \lb{s3_70}
\wt{\kappa}\;\big(d\hat{\mscr{X}}^{+}_{\alpha\beta}\big)&=&
-\Big(\hat{\mscr{P}}\;\hat{T}^{-1}\;\big(d\hat{T}\big)\;\hat{\mscr{P}}^{-1}\Big)_{\alpha\beta}^{21}=
\left(\bea{cc} \big(d\hat{b}_{mn}^{*}\big) & \big(d\hat{\zeta}_{m,r\ppr\nu}^{+}\big) \\
\big(d\hat{\zeta}_{r\mu,n}^{*}\big) & \big(d\hat{a}_{r\mu,r\ppr\nu}^{+}\big) \eea\right)_{\alpha\beta} \;; \\ \lb{s3_71}
&& \big(d\hat{b}_{mn}\big)=\big(d\hat{b}_{mn}^{T}\big)\;;
\hspace*{0.5cm}\big(d\hat{a}_{r\mu,r\ppr\nu}\big)=-\big(d\hat{a}_{r\mu,r\ppr\nu}^{T}\big) \;; \\ \lb{s3_72}
\big(d\hat{\mscr{Y}}^{11}_{\alpha\beta}\big)&=&
-\Big(\hat{\mscr{P}}\;\hat{T}^{-1}\;\big(d\hat{T}\big)\;\hat{\mscr{P}}^{-1}\Big)_{\alpha\beta}^{11}=\im\;
\left(\bea{cc} \big(d\hat{d}_{mn}\big) & \big(d\hat{\xi}_{m,r\ppr\nu}^{+}\big) \\
\big(d\hat{\xi}_{r\mu,n}\big) & \big(d\hat{g}_{r\mu,r\ppr\nu}\big) \eea\right)_{\alpha\beta}^{11} \;; \\  \lb{s3_73}
\big(d\hat{\mscr{Y}}^{11}_{\alpha\beta}\big)^{+}&=&-\big(d\hat{\mscr{Y}}^{11}_{\alpha\beta}\big)\;;\hspace*{0.5cm}
\big(d\hat{d}_{mn}^{+}\big)=\big(d\hat{d}_{mn}\big)\;;\hspace*{0.5cm}
\big(d\hat{g}_{r\mu,r\ppr\nu}^{+}\big)=\big(d\hat{g}_{r\mu,r\ppr\nu}\big) \;; \\  \lb{s3_74}
\big(d\hat{\mscr{Y}}^{22}_{\alpha\beta}\big)&=&
-\Big(\hat{\mscr{P}}\;\hat{T}^{-1}\;\big(d\hat{T}\big)\;\hat{\mscr{P}}^{-1}\Big)_{\alpha\beta}^{22}=-\im\;
\left(\bea{cc} \big(d\hat{d}_{mn}^{T}\big) & -\big(d\hat{\xi}_{m,r\ppr\nu}^{T}\big) \\
\big(d\hat{\xi}_{r\mu,n}^{*}\big) & \big(d\hat{g}_{r\mu,r\ppr\nu}^{T}\big) \eea\right)_{\alpha\beta}^{22}
\;; \\ \lb{s3_75}
\big(d\hat{\mscr{Y}}_{\alpha\beta}^{22}\big)^{st}&=&-\big(d\hat{\mscr{Y}}^{11}_{\alpha\beta}\big)=
-\Big(\hat{\mscr{P}}\;\hat{T}^{-1}\;\big(d\hat{T}\big)\;\hat{\mscr{P}}^{-1}\Big)_{\alpha\beta}^{22,st}=
\Big(\hat{\mscr{P}}\;\hat{T}^{-1}\;\big(d\hat{T}\big)\;\hat{\mscr{P}}^{-1}\Big)_{\alpha\beta}^{11}\;.
\eeq
Since the coset eigenvalues \(\ovv{c}_{m}\) (\ref{s2_19}), \(\ovv{f}_{r}\) (\ref{s2_20}) only take part in
the transformed actions of Euclidean path integration variables with their total values of the moduli
\(|\ovv{c}_{m}|\), \(|\ovv{f}_{r}|\) and phases \(\varphi_{m}\), \(\phi_{r}\), but {\it without any
derivatives '\(\pp\)'}, we have to relate the coset eigenvalues \(\ovv{c}_{m}\), \(\ovv{f}_{r}\)
with the total derivative '\(d\)' to the Euclidean pair condensate variables. The causal structure
of the original time development operators, which result with over-complete sets of states at every time step
into coherent state path integrals as an underlying lattice theory,
also leads to a natural time ordering in our case (\ref{s2_84}-\ref{s2_86}).
The causal structure of (\ref{s2_84}-\ref{s2_86}) is determined by the time contour,
due to the two branches of forward and backward propagation.
At each slice of the time contour propagation along the coherent state path generating function, we choose
independent, Euclidean pair condensate path fields where the independence of the Euclidean, anomalous
path fields at every time slice refers to the spatial distribution and internal indices of the super-matrices.
After having chosen the set of independent spatial fields at every time slice, the exponential phases
with the actions assign a weight to the particular chosen sets of Euclidean fields according to the
quadratic couplings of the kinetic energies and composed densities from anomalous variables.
Therefore, the total derivative '\(d\)' in (\ref{s3_68}-\ref{s3_75}), which relates the coset eigenvalues
\(\ovv{c}_{m}\), \(\ovv{f}_{r}\) to the total values of anomalous, Euclidean fields, only contains the partial
time contour derivative, corresponding to chosen time contour paths for every spatial point of an underlying
lattice field theory
\be \lb{s3_76}
\bea{rclrcl}
\big(d|\ovv{c}_{m}(t_{p})|\big) &=&{\ds\frac{\big(\pp|\ovv{c}_{m}(t_{p})|\big)}{\pp t_{p}}}\;\;dt_{p}\;;\hspace*{1.0cm} &
\big(d\varphi_{m}(t_{p})\big) &=&{\ds\frac{\big(\pp\varphi_{m}(t_{p})\big)}{\pp t_{p}}}\;\;dt_{p}\;; \\
\big(d|\ovv{f}_{r}(t_{p})|\big) &=&{\ds\frac{\big(\pp|\ovv{f}_{r}(t_{p})|\big)}{\pp t_{p}}}\;\;dt_{p}\;; &
\big(d\phi_{r}(t_{p})\big) &=&{\ds\frac{\big(\pp\phi_{r}(t_{p})\big)}{\pp t_{p}}}\;\;dt_{p}\;.
\eea
\ee
The spatial vector \(\vec{x}\) is omitted in relations (\ref{s3_76}) because the Euclidean integration variables
of the generating function (\ref{s2_84}-\ref{s2_86}) are determined by time contour paths with spatially
independent points which could be abbreviated by an additional index '\(_{\vec{x}}\)' apart from the indices
\(m=1,\ldots,L\) or \(r=1,\ldots,S/2\) for the angular momentum degrees of freedom. In consequence, it
suffices to calculate the partial time contour derivatives of relations (\ref{s3_76}) in dependence
on the Euclidean, pair condensate, integration variables of \(\hat{\mscr{X}}_{\alpha\beta}\) and
\(\wt{\kappa}\;\hat{\mscr{X}}_{\alpha\beta}^{+}\). In fact, it turns out that the differential, absolute values
\((d|\ovv{c}_{m}(t_{p})|)\), \((d|\ovv{f}_{r}(t_{p})|)\) can be integrated to their total values
\(|\ovv{c}_{m}(t_{p})|\), \(|\ovv{f}_{r}(t_{p})|\), according to the property of a total derivative
for a state variable, whereas the phase values \(\varphi_{m}(t_{p})\), \(\phi_{r}(t_{p})\) involve
the detailed time contour path or the past time contour history of the Euclidean, pair condensate path variables
in order to perform the time contour integrals of the phases in (\ref{s3_76}).

We consider again the relations (\ref{s3_9}-\ref{s3_11})
for the variation of exponentials of matrices and obtain Eq.\ (\ref{s3_77}).
However, the suitable choice of gauge (\ref{s3_78})
for diagonal elements of \((d\hat{\mscr{P}})\;\hat{\mscr{P}}^{-1}\), described in Eqs. (\ref{s2_26}-\ref{s2_35}),
gives rise to a vanishing of the diagonal commutator matrix elements
\([\hat{Y}_{DD}\,,\,(d\hat{\mscr{P}})\,\hat{\mscr{P}}^{-1}]_{-,\alpha\alpha}^{ab}\)
between \(\hat{Y}_{DD}\) (\ref{s3_79}) (the diagonal, original
coset generator with eigenvalues \(\ovv{c}_{m}\), \(\ovv{f}_{r}\)) and the diagonal
(quaternion diagonal), vanishing elements of \((d\hat{\mscr{P}})\;\hat{\mscr{P}}^{-1}\) (\ref{s3_78}).
In consequence, eigenvalues \(\ovv{c}_{m}\) (quaternion eigenvalues \((\tau_{2})_{\mu\nu}\;\ovv{f}_{r}\))
of \(\hat{Y}_{DD}\) (\ref{s3_79}) are mapped onto the diagonal (quaternion diagonal) anomalous matrix elements
in \((d\hat{\mscr{X}}_{BB;mm})=-(d\hat{b}_{mm})\),
\(\wt{\kappa}\;(d\hat{\mscr{X}}_{BB;mm}^{+})=(d\hat{b}_{mm}^{*})\) and
\((d\hat{\mscr{X}}_{FF;r\mu,r\nu})=(d\hat{a}_{r\mu,r\nu})\),
\(\wt{\kappa}\;(d\hat{\mscr{X}}_{FF;r\mu,r\nu}^{+})=(d\hat{a}_{r\mu,r\nu}^{+})\) (\ref{s3_80}-\ref{s3_85}).
The latter fields are also taken as the independent variables for the various
diagonal elements of the densities (\ref{s3_86}-\ref{s3_89}) and block parts in
\((d\hat{\mscr{Y}}_{BB;mm}^{11})=\im\;(d\hat{d}_{mm})\),
\((d\hat{\mscr{Y}}_{FF;r\mu,r\nu}^{11})=\im\;(d\hat{g}_{rr}^{(0)})\;\delta_{\mu\nu}\)
and correspondingly in \((d\hat{\mscr{Y}}_{N\times N}^{22})\);
(\((d\hat{g}_{r\mu,r\ppr\nu})=\sum_{k=0}^{3}\big(\tau_{k}\big)_{\mu\nu}\;
(d\hat{g}_{rr\ppr}^{(k)})\))
\beq \lb{s3_77}
\lefteqn{\hspace*{-3.7cm}-\big(d\hat{\mscr{Z}}_{\alpha\beta}^{ab}\big)=
\Big(\hat{\mscr{P}}\;\hat{T}^{-1}\;\big(d\hat{T}\big)\;\hat{\mscr{P}}^{-1}\Big)_{\alpha\beta}^{ab}
 = -\int_{0}^{1}dv\;\;e^{v\;\hat{Y}_{DD}}\;\hat{\mscr{P}}\;\Big(d\;\hat{\mscr{P}}^{-1}\;\hat{Y}_{DD}\;
\hat{\mscr{P}}\Big)\;\hat{\mscr{P}}^{-1}\;
e^{-v\;\hat{Y}_{DD}} } \\ \no &=&-\int_{0}^{1}dv\;\;e^{v\;\hat{Y}_{DD}}\;
\bigg(\big(d\hat{Y}_{DD}\big)+\Big[\hat{Y}_{DD}\;,\;\big(d\hat{\mscr{P}}\big)\;\hat{\mscr{P}}^{-1}\Big]_{-}\bigg)\;
e^{-v\;\hat{Y}_{DD}}   \\  \lb{s3_78}
\Big(\big(d\hat{\mscr{P}}\big)\;\hat{\mscr{P}}^{-1}\Big)_{BB;mm} &\equiv&  0\;; \hspace*{0.5cm}
\Big(\big(d\hat{\mscr{P}}\big)\;\hat{\mscr{P}}^{-1}\Big)_{FF;r\mu,r\nu} \equiv  0\;\hspace*{0.5cm}
\bigg(\Big[\hat{Y}_{DD}\:,\:\big(d\hat{\mscr{P}})\,\hat{\mscr{P}}^{-1}\Big]_{-}\bigg)_{\alpha\alpha}^{ab}\equiv 0\;;
\eeq
\beq\lb{s3_79}
\lefteqn{\hspace*{-4.6cm}\big(d\hat{Y}_{DD}\big) = \left(\bea{cc} 0 & \big(d\hat{X}_{DD}\big) \\
\wt{\kappa}\;\big(d\hat{X}_{DD}^{+}\big) & 0 \eea\right)\;;\hspace*{0.5cm}
(d\hat{X}_{DD})=\left(\bea{cc} -\big(d\ovv{c}_{m}\big)\;\delta_{m,n} & 0 \\
0 & (\tau_{2})_{\mu\nu}\;\big(d\ovv{f}_{r}\big)\;\delta_{r,r\ppr} \eea\right)\;;}  \\ \lb{s3_80}
\big(d\hat{\mscr{Z}}_{BB;mm}^{ab}\big)&=&\int_{0}^{1}dv\;\;
\Big(e^{v\;\hat{Y}_{DD}}\;\;\big(d\hat{Y}_{DD}\big)\;\;e^{-v\;\hat{Y}_{DD}}\Big)_{BB;mm}^{ab} \\ \lb{s3_81}
\big(d\hat{\mscr{Z}}_{FF;r\mu,r\nu}^{ab}\big)&=&\int_{0}^{1}dv\;\;
\Big(e^{v\;\hat{Y}_{DD}}\;\;\big(d\hat{Y}_{DD}\big)\;\;e^{-v\;\hat{Y}_{DD}}\Big)_{FF;r\mu,r\nu}^{ab}  \\ \lb{s3_82}
\big(d\hat{\mscr{Z}}^{12}_{\alpha\beta}\big) &=&
\left(\bea{cc} -\big(d\hat{b}_{mm}\big)\;\delta_{mn} & 0 \\
0 & (\tau_{2})_{\mu\nu}\;\big(d\hat{a}_{rr}^{(2)}\big)\;\delta_{rr\ppr}\eea\right)_{\alpha\beta}^{12}  \\ \lb{s3_83}
\big(d\hat{\mscr{Z}}^{21}_{\alpha\beta}\big) &=&
\left(\bea{cc} \big(d\hat{b}_{mm}^{*}\big)\;\delta_{mn} & 0 \\
0 & (\tau_{2})_{\mu\nu}\;\big(d\hat{a}_{rr}^{(2)*}\big)\;\delta_{rr\ppr}\eea\right)_{\alpha\beta}^{21}  \\ \lb{s3_84}
|\ovv{c}_{m}(t_{p})|- |\ovv{c}_{m}(-\infty_{+})| &=&
\int_{-\infty_{+}}^{t_{p}}\hspace*{-0.37cm}dt_{q}\ppr\;\;{\ts\frac{\big(\pp|\ovv{c}_{m}(t_{q}\ppr)|\big)}{\pp t_{q}\ppr}}\;;\hspace*{0.1cm}
\varphi_{m}(t_{p})-\varphi_{m}(-\infty_{+}) =
\int_{-\infty_{+}}^{t_{p}}\hspace*{-0.37cm}dt_{q}\ppr\;\;{\ts\frac{\big(\pp\varphi_{m}(t_{q}\ppr)\big)}{\pp t_{q}\ppr}}\;; \\  \lb{s3_85}
|\ovv{f}_{r}(t_{p})|- |\ovv{f}_{r}(-\infty_{+})| &=&
\int_{-\infty_{+}}^{t_{p}}\hspace*{-0.37cm}dt_{q}\ppr\;\;{\ts\frac{\big(\pp|\ovv{f}_{r}(t_{q}\ppr)|\big)}{\pp t_{q}\ppr}}\;;\hspace*{0.1cm}
\phi_{r}(t_{p})-\phi_{r}(-\infty_{+}) =
\int_{-\infty_{+}}^{t_{p}}\hspace*{-0.37cm}dt_{q}\ppr\;\;{\ts\frac{\big(\pp\phi_{r}(t_{q}\ppr)\big)}{\pp t_{q}\ppr}}\;;  \\  \lb{s3_86}
\big(d\hat{\mscr{Z}}^{11}_{\alpha\beta}\big) &=& \im\;\left(\bea{cc}
\big(d\hat{d}_{mm}\big)\;\delta_{mn} & 0 \\
0 & \delta_{\mu\nu}\;\delta_{rr\ppr}\;\big(d\hat{g}_{rr}^{(0)}\big) \eea\right)_{\alpha\beta}^{11} \;;  \\ \lb{s3_87}
\big(d\hat{\mscr{Z}}^{22}_{\alpha\beta}\big) &=& -\im\;\left(\bea{cc}
\big(d\hat{d}_{mm}\big)\;\delta_{mn} & 0 \\
0 & \delta_{\mu\nu}\;\delta_{rr\ppr}\;\big(d\hat{g}_{rr}^{(0)}\big) \eea\right)_{\alpha\beta}^{22} \;; \\ \lb{s3_88}
\hat{d}_{mm}(t_{p})- \hat{d}_{mm}(-\infty_{+}) &=&
\int_{-\infty_{+}}^{t_{p}}dt_{q}\ppr\;\;{\ts\frac{\big(\pp|\hat{d}_{mm}(t_{q}\ppr)|\big)}{\pp t_{q}\ppr}}\;;
\;\;(m=1,\ldots,L=2l+1)\;;  \\   \lb{s3_89}
\hat{g}_{r\mu,r\nu}(t_{p})- \hat{g}_{r\mu,r\nu}(-\infty_{+}) &=&\delta_{\mu\nu}
\int_{-\infty_{+}}^{t_{p}}dt_{q}\ppr\;\;{\ts\frac{\big(\pp|\hat{g}_{rr}^{(0)}(t_{q}\ppr)|\big)}{\pp t_{q}\ppr}}\;;
\;\;(r=1,\ldots,S/2=s+1/2)\;.
\eeq
According to the relations in section 3.2 and appendix C of Ref.\ \cite{mies1},
we can apply the known transformations (\ref{s3_80}-\ref{s3_83})
between eigenvalues \(\ovv{c}_{m}\), \(\ovv{f}_{r}\) and the new, independent Euclidean elements
\((d\hat{b}_{mm})\), \((d\hat{a}_{rr}^{(2)})\) (\ref{s3_82},\ref{s3_83})
and as well the dependent, diagonal density terms
\((d\hat{d}_{mm})\), \((d\hat{g}_{rr\ppr}^{(0)})\) in order to determine the
functions (\ref{s3_84},\ref{s3_85},\ref{s3_88},\ref{s3_89}) with following relations
(\ref{s3_90}-\ref{s3_93})
\beq \lb{s3_90} \hspace*{-0.9cm}
\big(d\hat{b}_{mm}\big)&=&d\big(|\hat{b}_{mm}|\;e^{\im\;\beta_{m}}\big) =
\big(d\ovv{c}_{m}^{*}\big)\;e^{\im\;2\varphi_{m}}\;
\bigg(\frac{1}{2}-\frac{\sin\big(2\;|\ovv{c}_{m}|\big)}{4\;|\ovv{c}_{m}|}\bigg)+\big(d\ovv{c}_{m}\big)\;
\bigg(\frac{1}{2}+\frac{\sin\big(2\;|\ovv{c}_{m}|\big)}{4\;|\ovv{c}_{m}|}\bigg) \;; \\ \lb{s3_91} \hspace*{-0.9cm}
\big(d\hat{a}_{rr}^{(2)}\big)&=&d\big(|\hat{a}_{rr}^{(2)}|\;e^{\im\;\alpha_{r}}\big) =
\big(d\ovv{f}_{r}^{*}\big)\;e^{\im\;2\phi_{r}}\;
\bigg(\frac{1}{2}-\frac{\sinh\big(2\;|\ovv{f}_{r}|\big)}{4\;|\ovv{f}_{r}|}\bigg)+\big(d\ovv{f}_{r}\big)\;
\bigg(\frac{1}{2}+\frac{\sinh\big(2\;|\ovv{f}_{r}|\big)}{4\;|\ovv{f}_{r}|}\bigg)  \;; \\ \lb{s3_92}\hspace*{-0.9cm}
\im\;\big(d\hat{d}_{mm}\big) &=&\Big[\big(d\ovv{c}_{m}\big)\;e^{-\im\;\varphi_{m}}-
\big(d\ovv{c}_{m}^{*}\big)\;e^{\im\;\varphi_{m}}\Big]\;\;
\frac{\Big(\sin\big(|\ovv{c}_{m}|\big)\Big)^{2}}{2\;|\ovv{c}_{m}|} \;; \hspace*{0.5cm}
\hat{d}_{mm}\in\mbox{\sf R}\;; \\  \lb{s3_93}\hspace*{-0.9cm}
\im\;\big(d\hat{g}_{rr}^{(0)}\big) &=&-\Big[\big(d\ovv{f}_{r}\big)\;e^{-\im\;\phi_{r}}-
\big(d\ovv{f}_{r}^{*}\big)\;e^{\im\;\phi_{r}}\Big]\;\;
\frac{\Big(\sinh\big(|\ovv{f}_{r}|\big)\Big)^{2}}{2\;|\ovv{f}_{r}|}\;;\hspace*{0.5cm}
\hat{g}_{rr}^{(0)}\in\mbox{\sf R}\;.
\eeq
The separation into real and imaginary parts of Eqs. (\ref{s3_90}-\ref{s3_93}) guides us to the
relations (\ref{s3_94}-\ref{s3_97}) where one can also observe the additional negative sign of the fermion-fermion
density element \((d\hat{g}_{rr}^{(0)})\) with respect to the boson-boson density element
\((d\hat{d}_{mm})\). This additional negative sign is caused by the \(\mbox{U}(L|S)\) super-symmetry
of the original super-symmetric density \(\psi_{\vec{x},m}^{*}(t_{p})\;\psi_{\vec{x},m}(t_{p})+
\psi_{\vec{x},r\mu}^{*}(t_{p})\;\psi_{\vec{x},r\mu}(t_{p})\) which corresponds to the {\it difference}
of boson-boson and fermion-fermion densities
\beq \lb{s3_94}
\big(d\hat{b}_{mm}\big)&=&d\big(|\hat{b}_{mm}|\;e^{\im\;\beta_{m}}\big) =
e^{\im\;\varphi_{m}}\;\Big[\big(d|\ovv{c}_{m}|\big)+\im\;\frac{\sin\big(2\;|\ovv{c}_{m}|\big)}{2}\;\;
\big(d\varphi_{m}\big)\Big]  \;;  \\  \lb{s3_95}
\big(d\hat{a}_{rr}^{(2)}\big)&=&d\big(|\hat{a}_{rr}^{(2)}|\;e^{\im\;\alpha_{r}}\big) =
e^{\im\;\phi_{r}}\;\Big[\big(d|\ovv{f}_{r}|\big)+\im\;\frac{\sinh\big(2\;|\ovv{f}_{r}|\big)}{2}\;\;
\big(d\phi_{r}\big)\Big]  \;;  \\   \lb{s3_96}
\big(d\hat{d}_{mm}\big) &=&\Big(\sin\big(|\ovv{c}_{m}|\big)\Big)^{2}\;\;\big(d\varphi_{m}\big)\;;  \\   \lb{s3_97}
\big(d\hat{g}_{rr}^{(0)}\big) &=&-\Big(\sinh\big(|\ovv{f}_{r}||\big)\Big)^{2}\;\;\big(d\phi_{r}\big)\;.
\eeq
We introduce new pair condensate integration variables
\(\wt{b}_{mm}=|\wt{b}_{mm}|\;e^{\im\;\wt{\beta}_{m}}\) (\ref{s3_98}),
\(\wt{a}_{rr}^{(2)}=|\wt{a}_{rr}^{(2)}|\;e^{\im\;\wt{\alpha}_{r}}\) (\ref{s3_99})
and perform integration measure preserving phase rotations with \(e^{-\im\;\varphi_{m}}\), \(e^{-\im\;\wt{\beta}_{m}}\)
and \(e^{-\im\;\phi_{r}}\), \(e^{-\im\;\wt{\alpha}_{r}}\), respectively
\beq  \lb{s3_98}
\wt{b}_{mm} &=&|\wt{b}_{mm}|\;\;e^{\im\;\wt{\beta}_{m}} \;;  \\ \no
e^{-\im\;\wt{\beta}_{m}}\;\;\big(d\wt{b}_{mm}\big)&=&e^{-\im\;\varphi_{m}}\;\;\big(d\hat{b}_{mm}\big)
\;\;\Longrightarrow\;\;
\big(d\wt{b}_{mm}^{*}\big)\wedge\big(d\wt{b}_{mm}\big)=\big(d\hat{b}_{mm}^{*}\big)\wedge\big(d\hat{b}_{mm}\big)
\;; \\  \lb{s3_99}
\wt{a}_{rr}^{(2)} &=& |\wt{a}_{rr}^{(2)}|\;\;e^{\im\;\wt{\alpha}_{r}}\;;  \\  \no
e^{-\im\;\wt{\alpha}_{r}}\;\;\big(d\wt{a}_{rr}^{(2)}\big)&=&e^{-\im\;\phi_{r}}\;\;\big(d\hat{a}_{rr}^{(2)}\big)
\;\;\Longrightarrow\;\;
\big(d\wt{a}_{rr}^{(2)*}\big)\wedge\big(d\wt{a}_{rr}^{(2)}\big)=\big(d\hat{a}_{rr}^{(2)*}\big)\wedge\big(d\hat{a}_{rr}^{(2)}\big)\;.
\eeq
Accordingly, we can replace the diagonal, Euclidean, pair condensate integration variables
of Eqs. (\ref{s3_90}-\ref{s3_93}) by \((d\wt{b}_{mm})\),
\((d\wt{a}_{rr}^{(2)})\), (+c.c.) and obtain new relations between the coset eigenvalues \(|\ovv{c}_{m}|\), \(\varphi_{m}\),
\(|\ovv{f}_{r}|\), \(\phi_{r}\) and the rotated Euclidean, anomalous elements
\(\wt{b}_{mm}=|\wt{b}_{mm}|\;e^{\im\;\wt{\beta}_{m}}\), \(\wt{a}_{rr}^{(2)}=|\wt{a}_{rr}^{(2)}|\;e^{\im\;\wt{\alpha}_{r}}\)
of Eqs. (\ref{s3_98},\ref{s3_99})
\beq  \lb{s3_100}
\big(d|\wt{b}_{mm}|\big)+\im\;|\wt{b}_{mm}|\;\big(d\wt{\beta}_{m}\big) &=&
\big(d|\ovv{c}_{m}|\big) +\im\;\frac{\sin\big(2\;|\ovv{c}_{m}|\big)}{2}\;\big(d\varphi_{m}\big) \;;  \\  \lb{s3_101}
\big(d|\wt{a}_{rr}^{(2)}|\big)+\im\;|\wt{a}_{rr}^{(2)}|\;\big(d\wt{\alpha}_{r}\big) &=&
\big(d|\ovv{f}_{r}|\big) +\im\;\frac{\sinh\big(2\;|\ovv{f}_{r}|\big)}{2}\;\big(d\phi_{r}\big)  \;.
\eeq
In consequence of measure preserving phase rotations, the total derivatives
\((d|\wt{b}_{mm}|)\), \((d|\ovv{c}_{m}|)\) and \((d|\wt{a}_{rr}^{(2)}|)\), \((d|\ovv{f}_{r}|)\) result between the
absolute values of Euclidean, diagonal variables and the coset eigenvalues so that the absolute values of these transformations
are related to {\it path-independent 'state variables'} of thermodynamics in a 'transferred sense'.
(One can even substitute the contour time '\(t_{p}\)' by the inverse temperature '\(\tau\)' and
the contour integrals by the inverse temperature path '\(0\ldots\beta=1/(KT)\)' of
grand canonical statistical operators. In this case, thermodynamical state variables of the
absolute values \(|\wt{b}_{mm}(\tau)|\), \(|\ovv{c}_{m}(\tau)|\) and \(|\wt{a}_{rr}^{(2)}(\tau)|\), \(|\wt{f}_{r}(\tau)|\)
can be identified after similar HST's and coset decompositions of coherent state representations
of grand canonical {\it 'inverse temperature'} development operators so that the analogy becomes exact.)
\beq \lb{s3_102}
|\ovv{c}_{m}(\breve{\vec{x}},\breve{t}_{p})|-\underbrace{|\ovv{c}_{m}(\breve{\vec{x}},-\breve{\infty}_{+})|}_{=0} &=&
|\wt{b}_{mm}(\breve{\vec{x}},\breve{t}_{p})|-\underbrace{|\wt{b}_{mm}(\breve{\vec{x}},-\breve{\infty}_{+})|}_{=0} \;;   \\  \lb{s3_103}
|\ovv{f}_{r}(\breve{\vec{x}},\breve{t}_{p})|-\underbrace{|\ovv{f}_{r}(\breve{\vec{x}},-\breve{\infty}_{+})|}_{=0} &=&
|\wt{a}_{rr}^{(2)}(\breve{\vec{x}},\breve{t}_{p})|-\underbrace{|\wt{a}_{rr}^{(2)}(\breve{\vec{x}},-\breve{\infty}_{+})|}_{=0} \;.
\eeq
The phases \(\varphi_{m}\), \(\phi_{r}\) (\ref{s3_100},\ref{s3_101}) of the complex coset eigenvalues \(\ovv{c}_{m}\), \(\ovv{f}_{r}\)
are path-dependent with respect to the contour time \(t_{p}\) because they are not determined by total
derivatives and therefore correspond to a kind of {\it 'heat'-} or {\it 'work'-variables} of
thermodynamics, also in a transferred sense
\beq \lb{s3_104} \hspace*{-0.9cm}
\varphi_{m}(\breve{\vec{x}},\breve{t}_{p})-\underbrace{\varphi_{m}(\breve{\vec{x}},-\breve{\infty}_{+})}_{=0} &=&
\int_{-\breve{\infty}_{+}}^{\breve{t}_{p}} \hspace*{-0.3cm}d\breve{t}_{q}\ppr\;
\frac{\pp\varphi_{m}(\breve{\vec{x}},\breve{t}_{q}\ppr)}{\pp \breve{t}_{q}\ppr}=
\int_{-\breve{\infty}_{+}}^{\breve{t}_{p}} \hspace*{-0.3cm}d\breve{t}_{q}\ppr\;
\frac{2\;|\wt{b}_{mm}(\breve{\vec{x}},\breve{t}_{q}\ppr)|}{\sin\big(2\;|\wt{b}_{mm}(\breve{\vec{x}},\breve{t}_{q}\ppr)|\big)}\;
\frac{\pp\wt{\beta}_{m}(\breve{\vec{x}},\breve{t}_{q}\ppr)}{\pp \breve{t}_{q}\ppr} \;; \\ \lb{s3_105} \hspace*{-0.9cm}
\phi_{r}(\breve{\vec{x}},\breve{t}_{p})-\underbrace{\phi_{r}(\breve{\vec{x}},-\breve{\infty}_{+})}_{=0} &=&
\int_{-\breve{\infty}_{+}}^{\breve{t}_{p}}\hspace*{-0.3cm}d\breve{t}_{q}\ppr\;
\frac{\pp\phi_{r}(\breve{\vec{x}},\breve{t}_{q}\ppr)}{\pp \breve{t}_{q}\ppr}=
\int_{-\breve{\infty}_{+}}^{\breve{t}_{p}}\hspace*{-0.3cm}d\breve{t}_{q}\ppr\;
\frac{2\;|\wt{a}_{rr}^{(2)}(\breve{\vec{x}},\breve{t}_{q}\ppr)|}{\sinh\big(2\;|\wt{a}_{rr}^{(2)}(\breve{\vec{x}},\breve{t}_{q}\ppr)|\big)}\;
\frac{\pp\wt{\alpha}_{r}(\breve{\vec{x}},\breve{t}_{q}\ppr)}{\pp \breve{t}_{q}\ppr}  \;.
\eeq
The diagonal boson-boson and fermion-fermion densities are also path-dependent, due to the phases
\((d\varphi_{m})\) and \((d\phi_{r})\), and are specified by following Eqs. (\ref{s3_106},\ref{s3_107}),
after substitution of (\ref{s3_100},\ref{s3_101}) into (\ref{s3_96},\ref{s3_97})
\beq \lb{s3_106} \hspace*{-0.5cm}
\hat{d}_{mm}(\breve{\vec{x}},\breve{t}_{p})-\underbrace{\hat{d}_{mm}(\breve{\vec{x}},-\breve{\infty}_{+})}_{=0} &=&
\int_{-\breve{\infty}_{+}}^{\breve{t}_{p}}\hspace*{-0.3cm}d\breve{t}_{q}\ppr\;
\tan\big(|\wt{b}_{mm}(\breve{\vec{x}},\breve{t}_{q}\ppr)|\big)\;|\wt{b}_{mm}(\breve{\vec{x}},\breve{t}_{q}\ppr)|\;
\frac{\pp\wt{\beta}_{m}(\breve{\vec{x}},\breve{t}_{q}\ppr)}{\pp \breve{t}_{q}\ppr} \;;     \\ \lb{s3_107} \hspace*{-0.5cm}
\hat{g}_{rr}^{(0)}(\breve{\vec{x}},\breve{t}_{p})-\underbrace{\hat{g}_{rr}^{(0)}(\breve{\vec{x}},-\breve{\infty}_{+})}_{=0}
&=&-\int_{-\breve{\infty}_{+}}^{\breve{t}_{p}}\hspace*{-0.3cm}d\breve{t}_{q}\ppr\;
\tanh\big(|\wt{a}_{rr}^{(2)}(\breve{\vec{x}},\breve{t}_{q}\ppr)|\big)\;|\wt{a}_{rr}^{(2)}(\breve{\vec{x}},\breve{t}_{q}\ppr)|\;
\frac{\pp\wt{\alpha}_{r}(\breve{\vec{x}},\breve{t}_{q}\ppr)}{\pp \breve{t}_{q}\ppr}  \;.
\eeq
In summary, we have defined new integration variables \(\wt{b}_{mm}(\breve{\vec{x}},\breve{t}_{p})\),
\(\wt{a}_{rr}^{(2)}(\breve{\vec{x}},\breve{t}_{p})\)
for the diagonal matrix elements of Eqs.\ (\ref{s3_82},\ref{s3_83}) and (\ref{s3_86},\ref{s3_87}) so that,
according to the contour time ordering, we choose partial derivatives of phases
\(\pp\wt{\beta}_{m}/\pp t_{q}\ppr\), \(\pp\wt{\alpha}/\pp t_{q}\ppr\) which determine
the path-dependent phases \(\varphi_{m}(\vec{x},t_{p})\), \(\phi_{r}(\vec{x},t_{p})\) (\ref{s3_104},\ref{s3_105}) and also
the path-dependent diagonal density elements
\(\hat{d}_{mm}(\vec{x},t_{p})\), \(\hat{g}_{rr}^{(0)}(\vec{x},t_{p})\) (\ref{s3_106},\ref{s3_107}).
The absolute values of the coset eigenvalues \(|\ovv{c}_{m}|\), \(|\ovv{f}_{r}|\) transform in a path-independent
manner as corresponding {\it 'state variables'} and are equivalent to the absolute values
\(|\wt{b}_{mm}|\), \(|\wt{a}_{rr}^{(2)}|\) of the new, phase-rotated, Euclidean integration variables (\ref{s3_98}-\ref{s3_101}).

It remains to determine the block diagonal \(\hat{\mscr{P}}\), \(\hat{\mscr{P}}^{-1}\) super-matrix
of Eqs. (\ref{s3_77},\ref{s3_108_1}) in terms of the anomalous parts \((d\hat{\mscr{Z}}_{\alpha\beta}^{a\neq b})\).
Since we have accomplished definite relations between coset eigenvalues of \(\hat{Y}_{DD}\) and the diagonal elements
of pair condensates \((\hat{\mscr{Z}}_{\alpha\beta}^{a\neq b})\) (also comprising the diagonal parts of densities),
we can apply relations (\ref{s3_77},\ref{s3_78}) or (\ref{s3_108_1},\ref{s3_109_1})
in order to calculate \((d\hat{\mscr{P}})\;\hat{\mscr{P}}^{-1}\)
in terms of \((d\hat{\mscr{Z}}_{\alpha\beta}^{ab})\)
\beq \lb{s3_108_1}
\lefteqn{\hspace*{-3.7cm}-\big(d\hat{\mscr{Z}}_{\alpha\beta}^{ab}\big)=
\Big(\hat{\mscr{P}}\;\hat{T}^{-1}\;\big(d\hat{T}\big)\;\hat{\mscr{P}}^{-1}\Big)_{\alpha\beta}^{ab}
 = -\int_{0}^{1}dv\;\;e^{v\;\hat{Y}_{DD}}\;\hat{\mscr{P}}\;\Big(d\;\hat{\mscr{P}}^{-1}\;\hat{Y}_{DD}\;
\hat{\mscr{P}}\Big)\;\hat{\mscr{P}}^{-1}\;
e^{-v\;\hat{Y}_{DD}} } \\ \no &=&-\int_{0}^{1}dv\;\;e^{v\;\hat{Y}_{DD}}\;
\bigg(\big(d\hat{Y}_{DD}\big)+\Big[\hat{Y}_{DD}\;,\;\big(d\hat{\mscr{P}}\big)\;\hat{\mscr{P}}^{-1}\Big]_{-}\bigg)\;
e^{-v\;\hat{Y}_{DD}}   \\  \lb{s3_109_1}
\Big(\big(d\hat{\mscr{P}}\big)\;\hat{\mscr{P}}^{-1}\Big)_{BB;mm} &\equiv&  0\;; \hspace*{0.5cm}
\Big(\big(d\hat{\mscr{P}}\big)\;\hat{\mscr{P}}^{-1}\Big)_{FF;r\mu,r\nu} \equiv  0\;\hspace*{0.5cm}
\bigg(\Big[\hat{Y}_{DD}\:,\:\big(d\hat{\mscr{P}})\,\hat{\mscr{P}}^{-1}\Big]_{-}\bigg)_{\alpha\alpha}^{ab}\equiv 0\;.
\eeq
This can be achieved by a separation of the block diagonal \(N\times N\) matrices \(\hat{\mscr{P}}\),
\(\hat{\mscr{P}}^{-1}\) into subsequent multiplications of matrices where each matrix factor only contains
a generator for a single parameter or (single quaternion parameter for the fermion-fermion parts).
As consequence, one has to treat only \(2\times 2\) matrices (or \(2\times 2\) quaternion-valued matrices)
which connect the different parts of the \(N\times N\) ladder generators within the block diagonal
\(N\times N\) super-matrices \(\hat{\mscr{P}}\), \(\hat{\mscr{P}}^{-1}\).
After the factorization of \((d\hat{\mscr{P}})\;\hat{\mscr{P}}^{-1}\) into single group parts
with generators comprising only single parameters, we use again (\ref{s3_108_1}) in order to integrate
over \(v\in [0,1)\) within \(\exp\{\pm v\;\hat{Y}_{DD}\}\) and the commutator
\([\hat{Y}_{DD}\;,\;(d\hat{\mscr{P}})\;\hat{\mscr{P}}^{-1}]\) with the coset eigenvalues.
This is a straightforward procedure, but tedious task for general \(N\times N\) super-matrices;
we have also to point out that the resulting relation between \((d\hat{\mscr{P}})\;\hat{\mscr{P}}^{-1}\)
and \((d\hat{\mscr{Z}}_{\alpha\beta}^{ab})\) strongly depends on the details of the parametrization,
as e.\ g.\ the chosen sequence of factors with generators having only a single parameter.

\section{Classical field equations and observables}

\subsection{Variation for classical field equations with Euclidean integration variables}\lb{s34}

In consequence to the previous section \ref{s33}, we rename the diagonal, Euclidean integration variables
\(\wt{b}_{mm}\), \(\wt{a}_{r\mu,r\nu}=(\tau_{2})_{\mu\nu}\;\wt{a}_{rr}^{(2)}\) (\ref{s3_98},\ref{s3_99}),
which preserve the Euclidean integration measure of anomalous fields,
to their original symbols \(\hat{b}_{mm}=e^{\im\;\beta_{m}}\;|\hat{b}_{mm}|\)
and
\(\hat{a}_{r\mu,r\nu}=(\tau_{2})_{\mu\nu}\;\hat{a}_{rr}^{(2)}=(\tau_{2})_{\mu\nu}\;e^{\im\;\alpha_{r}}\;|\hat{a}_{rr}^{(2)}|\).
The total, Euclidean integration measure therefore consists of the time contour path fields
\(d\hat{b}_{mn}\), \(d\hat{a}_{r\mu,r\ppr\nu}\), \(d\zeta_{r\mu,n}\), (+c.c.) or of the terms of the
pair condensate matrices \((d\hat{\mscr{X}}_{\alpha\beta})=(d\hat{\mscr{Z}}_{\alpha\beta}^{12})\),
\(\wt{\kappa}\;(d\hat{\mscr{X}}_{\alpha\beta}^{+})=(d\hat{\mscr{Z}}_{\alpha\beta}^{21})\)
(compare Eqs. (\ref{s3_1}-\ref{s3_4}))
\beq \lb{s3_108}
\lefteqn{\hspace*{-1.9cm}d\big[(\!\hat{\mscr{Z}}_{\alpha\beta}^{12}),(d\hat{\mscr{Z}}_{\alpha\beta}^{21})\big]=
d\big[(d\hat{\mscr{X}}_{\alpha\beta}),\wt{\kappa}\;(d\hat{\mscr{X}}_{\alpha\beta}^{+})\big]=
\prod_{\{\breve{\vec{x}},\breve{t}_{p}\}} \Bigg[
\bigg\{\prod_{m=1}^{L}\prod_{n=m}^{L}\frac{\big(d\hat{b}_{mn}^{*}\big)\wedge\big(d\hat{b}_{mn}\big)}{2\;\im}\bigg\}\times } \\ \no
&\times& \bigg\{\prod_{r=1}^{S/2}\frac{\big(d\hat{a}_{rr}^{(2)*}\big)\wedge\big(d\hat{a}_{rr}^{(2)}\big)}{2\;\im}\bigg\}\times
\bigg\{\prod_{r=1}^{S/2}\prod_{r\ppr=r+1}^{S/2}\prod_{k=0}^{3}
\frac{\big(d\hat{a}_{rr\ppr}^{(k)*}\big)\wedge\big(d\hat{a}_{rr\ppr}^{(k)}\big)}{2\;\im}\bigg\}\times \\ \no &\times&
\bigg\{\prod_{r=1}^{S/2}\prod_{\mu=1,2}\prod_{n=1}^{L}\big(d\hat{\zeta}_{r\mu,n}^{*}\big)\;\;
\big(d\hat{\zeta}_{r\mu,n}\big)\bigg\}\Bigg]\;.
\eeq
The coherent state path integral of Eqs.\ (\ref{s2_84},\ref{s2_85}) thus takes the form (\ref{s3_109}) where we have
transformed the coset integration measure \(d[\hat{T}^{-1}(\breve{\vec{x}},\breve{t}_{p})\;
(d\hat{T}(\breve{\vec{x}},\breve{t}_{p}))]\) (\ref{s2_48}) in sections \ref{s321}-\ref{s323} to the Euclidean
correspondents of integration variables (\ref{s3_108}) for the anomalous pair condensate fields
\beq\lb{s3_109}
Z\big[\hat{\mscr{J}},\breve{J}_{\psi},\im\breve{J}_{\psi\psi}\big] &=&
\int d\big[(\!\hat{\mscr{Z}}_{\alpha\beta}^{12}),(d\hat{\mscr{Z}}_{\alpha\beta}^{21})\big]\;\;
\exp\Big\{\im\;\mscr{A}_{\breve{J}_{\psi\psi}}\big[\hat{T}\big]\Big\}\times
\exp\Big\{-\mscr{A}\ppr\big[\hat{T};\hat{\mscr{J}}\big]\Big\}\times \\ \no &\times&
\exp\Big\{-\mscr{A}^{(d)}\big[\hat{\mscr{Z}};\breve{J}_{\psi}\big]\Big\}\;.
\eeq
The action \(\mscr{A}_{\breve{J}_{\psi\psi}}[\hat{T}]\) in (\ref{s3_109}) generates the anomalous fields;
however, we neglect the detailed process of generation of pair condensates, which depends on temperature,
the trap potential and further special properties in the experiments, and assume initial conditions
for the Euclidean, anomalous fields \(\hat{\mscr{Z}}_{\alpha\beta}^{12}\), \(\hat{\mscr{Z}}_{\alpha\beta}^{21}\).
The second action \(\mscr{A}\ppr[\hat{T};\hat{\mscr{J}}]\) in (\ref{s3_109}) determines the observables
with the original source field \(\hat{\mscr{J}}_{\vec{x}\ppr,\beta;\vec{x},\alpha}^{ba}(t_{q}\ppr,t_{p})\)
which relates observables, obtained by differentiation, to the original coherent state path integral
(\ref{s1_22}) of super-fields \(\psi_{\vec{x},\alpha}(t_{p})\), \(\psi_{\vec{x},\alpha}^{*}(t_{p})\).
Hence, one can track the original observables (\ref{s1_22}) composed of the super-fields
\(\psi_{\vec{x},\alpha}(t_{p})\), \(\psi_{\vec{x},\alpha}^{*}(t_{p})\) to the transformed
generating function (\ref{s3_109}) with the Euclidean, pair condensate integration variables whose action
\(\mscr{A}^{(d)}[\hat{\mscr{Z}};\breve{J}_{\psi}]\) contains the coupling coefficients
\(\breve{c}^{ij}(\breve{\vec{x}},\breve{t}_{p})\) (\ref{s2_55}-\ref{s2_58}) and
\(\breve{d}^{ij}(\breve{\vec{x}},\breve{t}_{p})\) (\ref{s2_59}) of the background density field.
In terms of the new, Euclidean, pair condensate fields, the action
\(\mscr{A}^{(d)}[\hat{\mscr{Z}};\breve{J}_{\psi}]\) (\ref{s2_85},\ref{s2_86}) is altered to relations
(\ref{s3_110},\ref{s3_111}) which allow variations for classical field equations, avoiding inconsistencies
of nontrivial integration measures
\be \lb{s3_110}
\mscr{A}^{(d)}[\hat{\mscr{Z}};\breve{J}_{\psi}]=\int_{C}d\breve{t}_{p}\int d^{d}\!\breve{x}\;\bigg(\frac{x_{0}}{L}\bigg)^{d}\;\;
\mscr{L}^{(d)}[\hat{\mscr{Z}};\breve{J}_{\psi}]\;;
\ee
\beq  \lb{s3_111}
\lefteqn{\mscr{L}^{(d)}[\hat{\mscr{Z}};\breve{J}_{\psi}] =
-\Big(\breve{c}^{ij}+\frac{1}{2}\;\delta_{ij}\Big)
\sum_{a,b=1,2}^{(a\neq b)}
\strab\Big[\big(\breve{\pp}_{i}\hat{\mscr{Z}}_{\alpha\beta}^{a\neq b}\big)
\big(\breve{\pp}_{j}\hat{\mscr{Z}}_{\beta\alpha}^{b\neq a}\big)\Big] - \frac{1}{2}\sum_{a=1,2}
\strab\Big[\big(\breve{\pp}_{i}\hat{\mscr{Z}}_{\alpha\beta}^{aa}\big)
\big(\breve{\pp}_{i}\hat{\mscr{Z}}_{\beta\alpha}^{aa}\big) \Big] + } \\ \no  &-&\frac{1}{2}
\Big(\breve{u}(\breve{\vec{x}})-\breve{\mu}_{0}-\im\;\breve{\ve}_{p}+
\big\langle\breve{\sigma}_{D}^{(0)}(\breve{\vec{x}},\breve{t}_{p})\big\rangle_{\hat{\sigma}_{D}^{(0)}}\Big)
\bigg(2\sum_{m=1}^{L}\Big[\cos\big(2\,|\hat{b}_{mm}|\big)-1\Big]-4\sum_{r=1}^{S/2}
\Big[\cosh\big(2\,|\hat{a}_{rr}^{(2)}|\big)-1\Big]\bigg)+ \\ \no &-&\frac{\im}{2}
\STRAB\Big[\exp\big\{2\,\hat{Y}_{DD}\big\}\;\hat{S}\;\big(\breve{\pp}_{\breve{t}_{p}}\hat{\mscr{Z}}\big)\Big]-
\im\;\Big(\breve{d}^{ij}-\frac{1}{2}\delta_{ij}\Big)\;
\breve{J}_{\psi}^{+}\;\hat{I}\;\wt{K}\;\hat{\mscr{P}}^{-1}\;\big(\breve{\pp}_{i}\hat{\mscr{Z}}\big)\;
\big(\breve{\pp}_{j}\hat{\mscr{Z}}\big)\;\hat{\mscr{P}}\;\hat{I}\;\breve{J}_{\psi} +  \\ \no &-&\frac{1}{2}\;
\breve{J}_{\psi}^{+}\;\hat{I}\;\wt{K}\;\hat{\mscr{P}}^{-1}\;\exp\big\{-2\;\hat{Y}_{DD}\big\}\;
\big(\breve{\pp}_{\breve{t}_{p}}\hat{\mscr{Z}}\big)\;\hat{S}\;\hat{\mscr{P}}\;\hat{I}\;\breve{J}_{\psi} +  \\  \no &+&
\frac{\eta_{p}}{2}\;\bigg\{2\sum_{m=1}^{L}\Big[\cos\big(2\;|\hat{b}_{mm}(\breve{\vec{x}},\breve{t}_{p})|\big)-1\Big]-
4\sum_{r=1}^{S/2}\Big[\cosh\big(2\;|\hat{a}_{rr}^{(2)}(\breve{\vec{x}},\breve{t}_{p})|\big)-1\Big]\bigg\}\;.
\eeq
After classification of the independent, Euclidean pair condensate integration variables
\((d\hat{\mscr{Z}}_{\alpha\beta}^{12})\), \((d\hat{\mscr{Z}}_{\alpha\beta}^{21})\) in (\ref{s3_108}),
we list again the matrices \((\pp\hat{\mscr{Z}}_{\alpha\beta}^{ab})\), \(\hat{Y}_{DD}\) of (\ref{s3_111})
with the block diagonal \(\mbox{U}(L|S)\) rotation matrices \(\hat{\mscr{P}}_{\alpha\beta}^{aa}\),
\(\hat{\mscr{P}}_{\alpha\beta}^{aa,-1}\), defined in (\ref{s2_21}-\ref{s2_35}). Furthermore, we
apply the transformations of sections \ref{s32} and \ref{s33}, especially for the time contour
path dependent density terms and phases \(\varphi_{m}\), \(\phi_{r}\) of the
coset eigenvalues \(\ovv{c}_{m}\), \(\ovv{f}_{r}\)
\beq \lb{s3_112}
\big(\pp\hat{\mscr{Z}}_{\alpha\beta}^{ab}\big) &=&
-\Big(\hat{\mscr{P}}\;\hat{T}^{-1}\;\big(\pp\hat{T}\big)\;\hat{\mscr{P}}^{-1}\Big)_{\alpha\beta}^{ab}=
\left(\bea{cc} \big(\pp\hat{\mscr{Y}}^{11}_{\alpha\beta}\big) & \big(\pp\hat{\mscr{X}}_{\alpha\beta}\big) \\
\wt{\kappa}\;\big(\pp\hat{\mscr{X}}_{\alpha\beta}^{+}\big) & \big(\pp\hat{\mscr{Y}}^{22}_{\alpha\beta}\big)
\eea\right)^{ab} \;;   \\ \no   && \mbox{\bf Euclidean, pair condensate integration variables}  \\  \lb{s3_113}
\big(\pp\hat{\mscr{X}}_{\alpha\beta}\big)&=&
-\Big(\hat{\mscr{P}}\;\hat{T}^{-1}\;\big(\pp\hat{T}\big)\;\hat{\mscr{P}}^{-1}\Big)_{\alpha\beta}^{12}=
\left(\bea{cc} -\big(\pp\hat{b}_{mn}\big) & \big(\pp\hat{\zeta}_{m,r\ppr\nu}^{T}\big) \\
-\big(\pp\hat{\zeta}_{r\mu,n}\big) & \big(\pp\hat{a}_{r\mu,r\ppr\nu}\big) \eea\right)_{\alpha\beta} \;; \\  \lb{s3_114}
\wt{\kappa}\;\big(\pp\hat{\mscr{X}}^{+}_{\alpha\beta}\big)&=&
-\Big(\hat{\mscr{P}}\;\hat{T}^{-1}\;\big(\pp\hat{T}\big)\;\hat{\mscr{P}}^{-1}\Big)_{\alpha\beta}^{21}=
\left(\bea{cc} \big(\pp\hat{b}_{mn}^{*}\big) & \big(\pp\hat{\zeta}_{m,r\ppr\nu}^{+}\big) \\
\big(\pp\hat{\zeta}_{r\mu,n}^{*}\big) & \big(\pp\hat{a}_{r\mu,r\ppr\nu}^{+}\big) \eea\right)_{\alpha\beta} \;; \\  \lb{s3_115}
&& \big(\pp\hat{b}_{mn}\big)=\big(\pp\hat{b}_{mn}^{T}\big)\;;
\hspace*{0.5cm}\big(\pp\hat{a}_{r\mu,r\ppr\nu}\big)=-\big(\pp\hat{a}_{r\mu,r\ppr\nu}^{T}\big)
=\sum_{k=0}^{3}\big(\tau_{k}\big)_{\mu\nu}\;\big(\pp\hat{a}_{rr\ppr}^{(k)}\big) \;;  \\ \no  &&
\mbox{\bf density terms in dependence on the Euclidean pair condensate fields} \\  \lb{s3_116}
\big(\pp\hat{\mscr{Y}}^{11}_{\alpha\beta}\big)&=&
-\Big(\hat{\mscr{P}}\;\hat{T}^{-1}\;\big(\pp\hat{T}\big)\;\hat{\mscr{P}}^{-1}\Big)_{\alpha\beta}^{11}=\im\;
\left(\bea{cc} \big(\pp\hat{d}_{mn}\big) & \big(\pp\hat{\xi}_{m,r\ppr\nu}^{+}\big) \\
\big(\pp\hat{\xi}_{r\mu,n}\big) & \big(\pp\hat{g}_{r\mu,r\ppr\nu}\big) \eea\right)_{\alpha\beta}^{11} \;; \\  \lb{s3_117}
\big(\pp\hat{\mscr{Y}}^{11}_{\alpha\beta}\big)^{+}&=&-\big(\pp\hat{\mscr{Y}}^{11}_{\alpha\beta}\big)\;;\hspace*{0.5cm}
\big(\pp\hat{d}_{mn}^{+}\big)=\big(\pp\hat{d}_{mn}\big)\;;\hspace*{0.5cm}
\big(\pp\hat{g}_{r\mu,r\ppr\nu}^{+}\big)=\big(\pp\hat{g}_{r\mu,r\ppr\nu}\big)\;; \\ \lb{s3_118}
\big(\pp\hat{\mscr{Y}}^{22}_{\alpha\beta}\big)&=&
-\Big(\hat{\mscr{P}}\;\hat{T}^{-1}\;\big(\pp\hat{T}\big)\;\hat{\mscr{P}}^{-1}\Big)_{\alpha\beta}^{22}=-\im\;
\left(\bea{cc} \big(\pp\hat{d}_{mn}^{T}\big) & -\big(\pp\hat{\xi}_{m,r\ppr\nu}^{T}\big) \\
\big(\pp\hat{\xi}_{r\mu,n}^{*}\big) & \big(\pp\hat{g}_{r\mu,r\ppr\nu}^{T}\big)
\eea\right)_{\alpha\beta}^{22} \;; \\ \lb{s3_119}
\big(\pp\hat{\mscr{Y}}^{22}_{\alpha\beta}\big)^{st}&=&-\big(\pp\hat{\mscr{Y}}^{11}_{\alpha\beta}\big)=
-\Big(\hat{\mscr{P}}\;\hat{T}^{-1}\;\big(\pp\hat{T}\big)\;\hat{\mscr{P}}^{-1}\Big)_{\alpha\beta}^{22,st}=
\Big(\hat{\mscr{P}}\;\hat{T}^{-1}\;\big(\pp\hat{T}\big)\;\hat{\mscr{P}}^{-1}\Big)_{\alpha\beta}^{11}\;.
\eeq
Appropriately to sections \ref{s32} and \ref{s33}, we can also relate the listed density terms (\ref{s3_116}-\ref{s3_119}) to
the anomalous, Euclidean integration variables with time contour path dependent phases
\(\varphi_{m}(\breve{\vec{x}},\breve{t}_{p})\), \(\phi_{r}(\breve{\vec{x}},\breve{t}_{p})\)
\beq \lb{s3_120}
\im\;\big(\pp\hat{d}_{mn}\big)&=&\frac{1}{2}\tan\bigg(\frac{|\hat{b}_{mm}|+|\hat{b}_{nn}|}{2}\bigg)\;
\Big[e^{-\im\;\varphi_{n}}\;\big(\pp\hat{b}_{mn}\big)-e^{\im\;\varphi_{m}}\;\big(\pp\hat{b}_{mn}^{*}\big)\Big]+ \\  \no &-&
\frac{1}{2}\tan\bigg(\frac{|\hat{b}_{mm}|-|\hat{b}_{nn}|}{2}\bigg)\;
\Big[e^{-\im\;\varphi_{n}}\;\big(\pp\hat{b}_{mn}\big)+e^{\im\;\varphi_{m}}\;\big(\pp\hat{b}_{mn}^{*}\big)\Big]  \\ \no
\im\;\big(\pp\hat{g}_{r\mu,r\ppr\nu}\big) &=&
-\frac{1}{2} \sum_{k=0}^{3}\big(\tau_{k}\tau_{2}\big)_{\mu\nu} \;
\bigg\{\tanh\bigg(\frac{|\hat{a}_{rr}^{(2)}|+|\hat{a}_{r\ppr r\ppr}^{(2)}|}{2}\bigg)\;
\Big[e^{-\im\;\phi_{r\ppr}}\;\big(\pp\hat{a}_{rr\ppr}^{(k)}\big)-\big(-1\big)^{k}\;
e^{\im\;\phi_{r}}\;\big(\pp\hat{a}_{rr\ppr}^{(k)+}\big)\Big] +  \\ \lb{s3_121} &-&
\tanh\bigg(\frac{|\hat{a}_{rr}^{(2)}|-|\hat{a}_{r\ppr r\ppr}^{(2)}|}{2}\bigg)\;
\Big[e^{-\im\;\phi_{r\ppr}}\;\big(\pp\hat{a}_{rr\ppr}^{(k)}\big)+\big(-1\big)^{k}\;
e^{\im\;\phi_{r}}\;\big(\pp\hat{a}_{rr\ppr}^{(k)+}\big)\Big]\bigg\}_{\mbox{;}}
\\ \lb{s3_122} \hat{a}_{rr}^{(2)}\neq 0 &;& \hat{a}_{rr}^{(k)}\equiv 0\;\;\mbox{ for : }k=0,1,3   \\ \lb{s3_124}
\im\;\big(\pp\hat{\xi}_{r\mu,n}\big)&=&\frac{1}{2}\tanh\bigg(\frac{|\hat{a}_{rr}^{(2)}|+\im\;|\hat{b}_{nn}|}{2}\bigg)\;
\Big[e^{\im\;\phi_{r}}\;\big(\tau_{2}\big)_{\mu\kappa}\;\big(\pp\hat{\zeta}_{r\kappa,n}^{*}\big)
-\im\;e^{-\im\;\varphi_{n}}\;\big(\tau_{0}\big)_{\mu\kappa}\;\big(\pp\hat{\zeta}_{r\kappa,n}\big)\Big] + \\ \no &+&
\frac{1}{2}\tanh\bigg(\frac{|\hat{a}_{rr}^{(2)}|-\im\;|\hat{b}_{nn}|}{2}\bigg)\;
\Big[e^{\im\;\phi_{r}}\;\big(\tau_{2}\big)_{\mu\kappa}\;\big(\pp\hat{\zeta}_{r\kappa,n}^{*}\big)
+\im\;e^{-\im\;\varphi_{n}}\;\big(\tau_{0}\big)_{\mu\kappa}\;\big(\pp\hat{\zeta}_{r\kappa,n}\big)\Big] \;;
\eeq
\beq \no &&
\mbox{\bf phases $\boldsymbol{\varphi_{m}(\breve{\vec{x}},\breve{t}_{p})}$,
$\boldsymbol{\phi_{r}(\breve{\vec{x}},\breve{t}_{p})}$
of coset eigenvalues in $\boldsymbol{\hat{Y}_{DD}}$, $\boldsymbol{\hat{X}_{DD}}$ :}
\\ \lb{s3_125}
\varphi_{m}(\breve{\vec{x}},\breve{t}_{p}) &=&\int_{-\breve{\infty}_{+}}^{\breve{t}_{p}}\hspace*{-0.3cm}d\breve{t}_{q}\ppr\;
\frac{2\;|\hat{b}_{mm}(\breve{\vec{x}},\breve{t}_{q}\ppr)|}{\sin\big(2\;|\hat{b}_{mm}(\breve{\vec{x}},\breve{t}_{q}\ppr)|\big)}\;
\frac{\pp\beta_{m}(\breve{\vec{x}},\breve{t}_{q}\ppr)}{\pp\breve{t}_{q}\ppr} \;;
\\ \lb{s3_126}
\phi_{r}(\breve{\vec{x}},\breve{t}_{p}) &=&\int_{-\breve{\infty}_{+}}^{\breve{t}_{p}}\hspace*{-0.3cm}d\breve{t}_{q}\ppr\;
\frac{2\;|\hat{a}_{rr}^{(2)}(\breve{\vec{x}},\breve{t}_{q}\ppr)|}{\sinh\big(2\;|\hat{a}_{rr}^{(2)}(\breve{\vec{x}},\breve{t}_{q}\ppr)|\big)}\;
\frac{\pp\alpha_{r}(\breve{\vec{x}},\breve{t}_{q}\ppr)}{\pp\breve{t}_{q}\ppr} \;;
\\ \lb{s3_127}
\hat{Y}_{DD}(\breve{\vec{x}},\breve{t}_{p}) &=&
\left(\bea{cc}  0 & \hat{X}_{DD}(\breve{\vec{x}},\breve{t}_{p}) \\
\wt{\kappa}\;\hat{X}_{DD}^{+}(\breve{\vec{x}},\breve{t}_{p}) & 0
\eea\right) \;;  \\ \lb{s3_128}
\hat{X}_{DD}(\breve{\vec{x}},\breve{t}_{p}) &=&
\left( \bea{cc}  -|\hat{b}_{mm}(\breve{\vec{x}},\breve{t}_{p})|\;\;e^{\im\;\varphi_{m}(\breve{\vec{x}},\breve{t}_{p})} & 0 \\
0 & \big(\tau_{2}\big)_{\mu\nu}\;\;
|\hat{a}_{rr}^{(2)}(\breve{\vec{x}},\breve{t}_{p})|\;\;e^{\im\;\phi_{r}(\breve{\vec{x}},\breve{t}_{p})}
\eea\right)_{\mbox{.}}
\eeq
According to the listed Eqs. (\ref{s3_110}-\ref{s3_128}), the Lagrangian \(\mscr{L}^{(d)}[\hat{\mscr{Z}};\breve{J}_{\psi}]\) (\ref{s3_111})
is outlined for its various boson-boson, fermion-fermion and odd fermion-boson, boson-fermion parts of the super-matrices.
As already mentioned in Ref.\ \cite{mies1}, the fermion-fermion density parts are always accompanied by a phase factor of
\(e^{\pm\im\;\pi}=-1\) relative to the corresponding boson-boson density part of super-matrices
\footnote{Although the given relation (\ref{s3_129}) for \(\mscr{L}^{(d)}[\hat{\mscr{Z}};\breve{J}_{\psi}]\) obscures
the underlying super-symmetries and has a complicated appearance, we represent the Lagrangian in its expanded version
with split boson-boson, fermion-fermion, odd fermion-boson and boson-fermion parts in order to verify the phase
jump between the boson-boson and fermion-fermion densities, respectively.}
\beq\no
\lefteqn{\mscr{L}^{(d)}[\hat{\mscr{Z}};\breve{J}_{\psi}] =
2\Big(\breve{c}^{ij}+\frac{1}{2}\;\delta_{ij}\Big)\;
\mbox{tr}\Big[\big(\breve{\pp}_{i}\hat{b}_{mn}^{+}\big)\;\big(\breve{\pp}_{j}\hat{b}_{nm}\big)+
\big(\breve{\pp}_{i}\hat{a}_{r\mu,r\ppr\nu}^{+}\big)\;\big(\breve{\pp}_{j}\hat{a}_{r\ppr\nu,r\mu}\big)+2\;
\big(\breve{\pp}_{i}\hat{\zeta}_{m,r\ppr\nu}^{+}\big)\;\big(\breve{\pp}_{j}\hat{\zeta}_{r\ppr\nu,m}\big)\Big]+ } \\ \lb{s3_129} &+&
\mbox{tr}\Big[\big(\breve{\pp}_{i}\hat{d}_{mn}\big)\;\big(\breve{\pp}_{i}\hat{d}_{nm}\big)-
\big(\breve{\pp}_{i}\hat{g}_{r\mu,r\ppr\nu}\big)\;\big(\breve{\pp}_{i}\hat{g}_{r\ppr\nu,r\mu}\big)+2\;
\big(\breve{\pp}_{i}\hat{\xi}_{m,r\ppr\nu}^{+}\big)\;\big(\breve{\pp}_{ij}\hat{\xi}_{r\ppr\nu,m}\big)\Big] + \\ \no &-&
\frac{1}{2}\Big(\breve{u}(\breve{\vec{x}})-\breve{\mu}_{0}-\im\;\breve{\ve}_{p}+
\big\langle\breve{\sigma}_{D}^{(0)}(\breve{\vec{x}},\breve{t}_{p})\big\rangle_{\hat{\sigma}_{D}^{(0)}}\Big)
\Big(2\sum_{m=1}^{L}\Big[\cos\big(2\;|\hat{b}_{mm}|\big)-1\Big]-4\sum_{r=1}^{S/2}
\Big[\cosh\big(2\;|\hat{a}_{rr}^{(2)}|\big)-1\Big]\Big) +  \\ \no &+&
\bigg[\sum_{m=1}^{L}\cos\big(2\;|\hat{b}_{mm}|\big)\;\big(\breve{\pp}_{\breve{t}_{p}}\hat{d}_{mm}\big)-
\sum_{r=1}^{S/2}\cosh\big(2\;|\hat{a}_{rr}^{(2)}|\big)\sum_{\mu=1,2}
\big(\breve{\pp}_{\breve{t}_{p}}\hat{g}_{r\mu,r\mu}\big)\bigg] +  \\ \no &-&\frac{\im}{2}\bigg\{
\sum_{m=1}^{L}\sin\big(2\;|\hat{b}_{mm}|\big)\;
\Big[e^{\im\;\varphi_{m}}\;\big(\breve{\pp}_{\breve{t}_{p}}\hat{b}_{mm}^{+}\big)-
e^{-\im\;\varphi_{m}}\;\big(\breve{\pp}_{\breve{t}_{p}}\hat{b}_{mm}\big)\Big]+  \\  \no &+&\sum_{r=1}^{S/2}
\sinh\big(2\;|\hat{a}_{rr}^{(2)}|\big)\;
\Big[e^{\im\;\phi_{r}}\;\mbox{tr}\big[\big(\tau_{2}\big)_{\mu\nu}\;
\big(\breve{\pp}_{\breve{t}_{p}}\hat{a}_{r\nu,r\mu}^{+}\big)\big]-
e^{-\im\;\phi_{r}}\;\mbox{tr}\big[\big(\tau_{2}\big)_{\nu\mu}\;
\big(\breve{\pp}_{\breve{t}_{p}}\hat{a}_{r\mu,r\nu}\big)\big]\Big]\bigg\} +  \\ \no &+&
2\im\;\Big(\breve{d}^{ij}-\frac{1}{2}\delta_{ij}\Big)\;
\bigg\{\breve{j}_{\hat{\mscr{P}}\psi;B}^{+}\Big[\big(\breve{\pp}_{i}\hat{b}\big)\;\big(\breve{\pp}_{j}\hat{b}^{+}\big)+
\big(\breve{\pp}_{i}\hat{d}\big)\;\big(\breve{\pp}_{j}\hat{d}\big)-
\big(\breve{\pp}_{i}\hat{\zeta}^{T}\big)\;\big(\breve{\pp}_{j}\hat{\zeta}^{*}\big)+
\big(\breve{\pp}_{i}\hat{\xi}^{+}\big)\;\big(\breve{\pp}_{j}\hat{\xi}\big)\Big]\breve{j}_{\hat{\mscr{P}}\psi;B}
+ \\  \no &-&
\breve{j}_{\hat{\mscr{P}}\psi;F}^{+}\Big[\big(\breve{\pp}_{i}\hat{a}\big)\;\big(\breve{\pp}_{j}\hat{a}^{+}\big)-
\big(\breve{\pp}_{i}\hat{g}\big)\;\big(\breve{\pp}_{j}\hat{g}\big)-
\big(\breve{\pp}_{i}\hat{\zeta}\big)\;\big(\breve{\pp}_{j}\hat{\zeta}^{+}\big)-
\big(\breve{\pp}_{i}\hat{\xi}\big)\;\big(\breve{\pp}_{j}\hat{\xi}^{+}\big)\Big]\breve{j}_{\hat{\mscr{P}}\psi;F} + \\ \no &+&
\breve{j}_{\hat{\mscr{P}}\psi;F}^{+}\Big[\big(\breve{\pp}_{i}\hat{\xi}\big)\;\big(\breve{\pp}_{j}\hat{d}\big)+
\big(\breve{\pp}_{i}\hat{g}\big)\;\big(\breve{\pp}_{j}\hat{\xi}\big)+
\big(\breve{\pp}_{i}\hat{\zeta}\big)\;\big(\breve{\pp}_{j}\hat{b}^{+}\big)-
\big(\breve{\pp}_{i}\hat{a}\big)\;\big(\breve{\pp}_{j}\hat{\zeta}^{*}\big)\Big]\breve{j}_{\hat{\mscr{P}}\psi;B} + \\ \no &+&
\breve{j}_{\hat{\mscr{P}}\psi;B}^{+}\Big[\big(\breve{\pp}_{i}\hat{d}\big)\;\big(\breve{\pp}_{j}\hat{\xi}^{+}\big)+
\big(\breve{\pp}_{i}\hat{\xi}^{+}\big)\;\big(\breve{\pp}_{j}\hat{g}\big)+
\big(\breve{\pp}_{i}\hat{b}\big)\;\big(\breve{\pp}_{j}\hat{\zeta}^{+}\big)-
\big(\breve{\pp}_{i}\hat{\zeta}^{T}\big)\;\big(\breve{\pp}_{j}\hat{a}^{+}\big)\Big]\breve{j}_{\hat{\mscr{P}}\psi;F}\bigg\} +
\\ \no &-&
\frac{1}{2}\left(\bea{c} \breve{j}_{\hat{\mscr{P}}\psi}^{+} \\ \breve{j}_{\hat{\mscr{P}}\psi}^{T}\;\wt{\kappa}\;\im \eea\right)^{T}
\left(\bea{cc}\big(e^{-2\;\hat{Y}_{DD}}\big)^{11} & \big(e^{-2\;\hat{Y}_{DD}}\big)^{12}  \\
\big(e^{-2\;\hat{Y}_{DD}}\big)^{21} & \big(e^{-2\;\hat{Y}_{DD}}\big)^{22} \eea\right)
\left(\bea{cc} \big(\breve{\pp}_{\breve{t}_{p}}\hat{\mscr{Y}}^{11}\big) &
-\big(\breve{\pp}_{\breve{t}_{p}}\hat{\mscr{X}}\big) \\
\wt{\kappa}\;\big(\breve{\pp}_{\breve{t}_{p}}\hat{\mscr{X}}^{+}\big) &
-\big(\breve{\pp}_{\breve{t}_{p}}\hat{\mscr{Y}}^{22}\big) \eea\right)
\left(\bea{c} \breve{j}_{\hat{\mscr{P}}\psi} \\ \im\;\breve{j}_{\hat{\mscr{P}}\psi}^{*} \eea\right) +  \\ \no &+&
\frac{\eta_{p}}{2}\;\bigg\{2\sum_{m=1}^{L}\Big[\cos\big(2\;|\hat{b}_{mm}(\breve{\vec{x}},\breve{t}_{p})|\big)-1\Big]-
4\sum_{r=1}^{S/2}\Big[\cosh\big(2\;|\hat{a}_{rr}^{(2)}|\big)-1\Big]\bigg\}\;; \\ \lb{s3_130} &&
\breve{j}_{\hat{\mscr{P}}\psi}(\vec{x},t_{p})=\hat{\mscr{P}}^{11}(\vec{x},t_{p})\;\;\breve{j}_{\psi}(\vec{x},t_{p}) \;.
\eeq
It remains to perform the variation of \(\mscr{L}^{(d)}[\hat{\mscr{Z}};\breve{J}_{\psi}]\) with
respect to the independent, Euclidean fields in \(\delta\hat{\mscr{Z}}_{\alpha\beta}^{12}\),
\(\delta\hat{\mscr{Z}}_{\alpha\beta}^{21}\) for classical equations and quadratic fluctuations
and we thus tabulate the variations of the different terms occurring in (\ref{s3_111}) or its
expanded version (\ref{s3_129})
\beq \lb{s3_131}
\hat{\mscr{Z}}_{\alpha\beta}^{ab}(\breve{\vec{x}},\breve{t}_{\boldsymbol{p=\pm}})&=&
\hat{\mscr{Z}}_{\alpha\beta}^{ab}(\breve{\vec{x}},\breve{t})\boldsymbol{\pm}\frac{1}{2}\;
\delta\hat{\mscr{Z}}_{\alpha\beta}^{ab}(\breve{\vec{x}},\breve{t})  \\  \lb{s3_132}
\delta\hat{\mscr{Z}}_{\alpha\beta}^{ab}(\breve{\vec{x}},\breve{t}) &=&
\delta\hat{\mscr{Z}}_{\alpha\beta}^{a\neq b}(\breve{\vec{x}},\breve{t}) +
\delta\hat{\mscr{Z}}_{\alpha\beta}^{a=b}(\breve{\vec{x}},\breve{t})   \\  \lb{s3_133}
\delta\hat{\mscr{Z}}_{\alpha\beta}^{12}(\breve{\vec{x}},\breve{t})  &=&
\left(\bea{cc} -\delta\hat{b}_{mn}(\breve{\vec{x}},\breve{t}) &
\delta\hat{\zeta}_{m,r\ppr\nu}^{T}(\breve{\vec{x}},\breve{t}) \\
-\delta\hat{\zeta}_{r\mu,n}(\breve{\vec{x}},\breve{t})  &
\big(\tau_{k}\big)_{\mu\nu}\,\delta\hat{a}_{rr\ppr}^{(k)}(\breve{\vec{x}},\breve{t})
\eea\right)  \\ \lb{s3_134}
\delta\hat{\mscr{Z}}_{\alpha\beta}^{21}(\breve{\vec{x}},\breve{t})  &=&
\wt{\kappa}\; \big(\delta\hat{\mscr{Z}}_{\alpha\beta}^{12}(\breve{\vec{x}},\breve{t})\big)^{+}=
\left(\bea{cc} \delta\hat{b}_{mn}^{*}(\breve{\vec{x}},\breve{t}) &
\delta\hat{\zeta}_{m,r\ppr\nu}^{+}(\breve{\vec{x}},\breve{t}) \\
\delta\hat{\zeta}_{r\mu,n}^{*}(\breve{\vec{x}},\breve{t})  &
\big(\tau_{k}\big)_{\mu\nu}\,\delta\hat{a}_{rr\ppr}^{(k)+}(\breve{\vec{x}},\breve{t})
\eea\right)  \\ \lb{s3_135}
\delta\hat{\mscr{Z}}_{\alpha\beta}^{11}(\breve{\vec{x}},\breve{t}) &=&\im\;
\left(\bea{cc} \delta\hat{d}_{mn}(\breve{\vec{x}},\breve{t}) &
\delta\hat{\xi}_{m,r\ppr\nu}^{+}(\breve{\vec{x}},\breve{t}) \\
\delta\hat{\xi}_{r\mu,n}(\breve{\vec{x}},\breve{t})  &
\delta\hat{g}_{r\mu,r\ppr\nu}(\breve{\vec{x}},\breve{t})
\eea\right)\;;     \\ \lb{s3_136}
\delta\hat{\mscr{Z}}_{\alpha\beta}^{22}(\breve{\vec{x}},\breve{t}) &=&-\im\;
\left(\bea{cc} \delta\hat{d}_{mn}^{T}(\breve{\vec{x}},\breve{t}) &
-\delta\hat{\xi}_{m,r\ppr\nu}^{T}(\breve{\vec{x}},\breve{t}) \\
\delta\hat{\xi}_{r\mu,n}^{*}(\breve{\vec{x}},\breve{t})  &
\delta\hat{g}_{r\mu,r\ppr\nu}^{T}(\breve{\vec{x}},\breve{t})
\eea\right)\;;\hspace*{0.5cm}
\big(\delta\hat{\mscr{Z}}_{\alpha\beta}^{22}(\breve{\vec{x}},\breve{t})\big)^{st}=
-\delta\hat{\mscr{Z}}_{\alpha\beta}^{11}(\breve{\vec{x}},\breve{t})      \\ \lb{s3_137}
\im\;\big(\delta\hat{d}_{mm}\big)&=&\frac{1}{2}\tan\big(|\hat{b}_{mm}|\big)\;
\Big[e^{-\im\;\varphi_{m}}\;\big(\delta\hat{b}_{mm}\big)-e^{\im\;\varphi_{m}}\;\big(\delta\hat{b}_{mm}^{*}\big)\Big]\;;  \\ \lb{s3_138}
\im\;\big(\delta\hat{d}_{mn}\big)&\stackrel{m\neq n}{=}&\frac{1}{2}\tan\bigg(\frac{|\hat{b}_{mm}|+|\hat{b}_{nn}|}{2}\bigg)\;
\Big[e^{-\im\;\varphi_{n}}\;\big(\delta\hat{b}_{mn}\big)-e^{\im\;\varphi_{m}}\;\big(\delta\hat{b}_{mn}^{*}\big)\Big]+ \\  \no &-&
\frac{1}{2}\tan\bigg(\frac{|\hat{b}_{mm}|-|\hat{b}_{nn}|}{2}\bigg)\;
\Big[e^{-\im\;\varphi_{n}}\;\big(\delta\hat{b}_{mn}\big)+e^{\im\;\varphi_{m}}\;\big(\delta\hat{b}_{mn}^{*}\big)\Big]\;;
\\ \lb{s3_139} \im\big(\delta\hat{g}_{r\mu,r\nu}\big) &=&
\im\;\big(\tau_{0}\big)_{\mu\nu}\;\big(\delta\hat{g}_{rr}^{(0)}\big)= \\ \no &=&
-\frac{1}{2}\;\delta_{\mu\nu} \;
\tanh\big(|\hat{a}_{rr}^{(2)}|\big)\;
\Big[e^{-\im\;\phi_{r}}\;\big(\delta\hat{a}_{rr}^{(2)}\big)-
e^{\im\;\phi_{r}}\;\big(\delta\hat{a}_{rr}^{(2)*}\big)\Big]\;;
\\ \no
\im\;\big(\delta\hat{g}_{r\mu,r\ppr\nu}\big) &\stackrel{r\neq r\ppr}{=}&
-\frac{1}{2} \sum_{k=0}^{3}\big(\tau_{k}\tau_{2}\big)_{\mu\nu} 
\bigg\{\tanh\bigg(\frac{|\hat{a}_{rr}^{(2)}|+|\hat{a}_{r\ppr r\ppr}^{(2)}|}{2}\bigg)
\Big[e^{-\im\;\phi_{r\ppr}}\;\big(\delta\hat{a}_{rr\ppr}^{(k)}\big)-\big(-1\big)^{k}\;
e^{\im\;\phi_{r}}\;\big(\delta\hat{a}_{rr\ppr}^{(k)+}\big)\Big] +  \\ \lb{s3_140} &-&
\tanh\bigg(\frac{|\hat{a}_{rr}^{(2)}|-|\hat{a}_{r\ppr r\ppr}^{(2)}|}{2}\bigg)
\Big[e^{-\im\;\phi_{r\ppr}}\;\big(\delta\hat{a}_{rr\ppr}^{(k)}\big)+\big(-1\big)^{k}\;
e^{\im\;\phi_{r}}\;\big(\delta\hat{a}_{rr\ppr}^{(k)+}\big)\Big]\bigg\}_{\mbox{;}}
\\ \no && \hat{a}_{rr}^{(2)}\neq 0 \;;\hspace*{0.5cm} \hat{a}_{rr}^{(k)}\equiv 0\;\;\mbox{ for : }k=0,1,3 \;;
\\ \lb{s3_141}
\im\;\big(\delta\hat{\xi}_{r\mu,n}\big)&=&\frac{1}{2}\tanh\bigg(\frac{|\hat{a}_{rr}^{(2)}|+\im\;|\hat{b}_{nn}|}{2}\bigg)\;
\Big[e^{\im\;\phi_{r}}\;\big(\tau_{2}\big)_{\mu\kappa}\;\big(\delta\hat{\zeta}_{r\kappa,n}^{*}\big)
-\im\;e^{-\im\;\varphi_{n}}\;\big(\tau_{0}\big)_{\mu\kappa}\;\big(\delta\hat{\zeta}_{r\kappa,n}\big)\Big] + \\ \no &+&
\frac{1}{2}\tanh\bigg(\frac{|\hat{a}_{rr}^{(2)}|-\im\;|\hat{b}_{nn}|}{2}\bigg)\;
\Big[e^{\im\;\phi_{r}}\;\big(\tau_{2}\big)_{\mu\kappa}\;\big(\delta\hat{\zeta}_{r\kappa,n}^{*}\big)
+\im\;e^{-\im\;\varphi_{n}}\;\big(\tau_{0}\big)_{\mu\kappa}\;\big(\delta\hat{\zeta}_{r\kappa,n}\big)\Big] \;.
\eeq
The variation \(\delta(\exp\{2\;\hat{Y}_{DD}\})\) of the coset eigenvalues also involves a variation of
the phases \(\delta\varphi_{m}(\breve{\vec{x}},\breve{t}_{p})\),
\(\delta\phi_{r}(\breve{\vec{x}},\breve{t}_{p})\) with respect to
\(\delta\beta_{m}(\breve{\vec{x}},\breve{t}_{p})\), \(\delta\alpha_{r}(\breve{\vec{x}},\breve{t}_{p})\)
apart from the variation of absolute values
\(\delta|\hat{b}_{mm}(\breve{\vec{x}},\breve{t}_{p})|\), \(\delta|\hat{a}_{rr}^{(2)}(\breve{\vec{x}},\breve{t}_{p})|\).
The first order variation of the last term in (\ref{s3_111},\ref{s3_129}) vanishes completely, due to the
additional contour metric \(\eta_{p}\). This term begins to contribute from second and all higher even order variations
with universal fluctuations which are entirely determined by the coset decomposition
\(\mbox{Osp}(S,S|2L)/\mbox{U}(L|S)\otimes \mbox{U}(L|S)\). In the following relations (\ref{s3_142}-\ref{s3_145}) we arrange the various
diagonal parts of the coset eigenvalue matrix \((\exp\{2\;\hat{Y}_{DD}\})^{ab}_{\alpha\beta}\) and also point out
the non-Markovian, path dependent phases \(\varphi_{m}(\vec{x},t_{p})\), \(\phi_{r}(\vec{x},t_{p})\) determined by the
contour time history of the anomalous, Euclidean fields \(\hat{b}_{mm}(\vec{x},t_{p})\), \(\hat{a}_{rr}^{(2)}(\vec{x},t_{p})\)
\beq\lb{s3_142}
\lefteqn{\Big(\exp\{2\,\hat{Y}_{DD}\}\Big)_{\alpha\beta}^{11}=\Big(\exp\{2\,\hat{Y}_{DD}\}\Big)_{\alpha\beta}^{22}=}  \\ \no  &=&
\left(\bea{cc} \cos\big(2\;|\hat{b}_{mm}(\breve{\vec{x}},\breve{t}_{p})|\big)\;\delta_{mn} & 0  \\ 0  &
\cosh\big(2\,|\hat{a}_{rr}^{(2)}(\breve{\vec{x}},\breve{t}_{p})|\big)\;\delta_{rr\ppr}\;\delta_{\mu\nu} \eea\right)_{\alpha\beta}^{ab} \;; \\ \lb{s3_143}
\lefteqn{\Big(\exp\{2\,\hat{Y}_{DD}\}\Big)_{\alpha\beta}^{12}=  } \\ \no &=&
\left(\bea{cc} -\sin\big(2\;|\hat{b}_{mm}(\breve{\vec{x}},\breve{t}_{p})|\big)\;e^{\im\,\varphi_{m}(\breve{\vec{x}},\breve{t}_{p})} \;
\delta_{mn} & 0  \\ 0  & \big(\tau_{2}\big)_{\mu\nu}\;
\sinh\big(2\,|\hat{a}_{rr}^{(2)}(\breve{\vec{x}},\breve{t}_{p})|\big)\;
e^{\im\,\phi_{r}(\breve{\vec{x}},\breve{t}_{p})}\;\delta_{rr\ppr} \eea\right)_{\alpha\beta}^{12} \;; \\  \lb{s3_144}
\lefteqn{\Big(\exp\{2\,\hat{Y}_{DD}\}\Big)_{\alpha\beta}^{21}= }  \\ \no &=&
\left(\bea{cc} \sin\big(2\;|\hat{b}_{mm}(\breve{\vec{x}},\breve{t}_{p})|\big)\;e^{-\im\,\varphi_{m}(\breve{\vec{x}},\breve{t}_{p})} \;
\delta_{mn} & 0  \\ 0  & \big(\tau_{2}\big)_{\mu\nu}\;
\sinh\big(2\,|\hat{a}_{rr}^{(2)}(\breve{\vec{x}},\breve{t}_{p})|\big)\;
e^{-\im\,\phi_{r}(\breve{\vec{x}},\breve{t}_{p})}\;\delta_{rr\ppr} \eea\right)_{\alpha\beta}^{21} \;; \\   \lb{s3_145}
\varphi_{m}(\breve{\vec{x}},\breve{t}_{p}) \hspace*{-0.25cm}&=&\int_{-\breve{\infty}_{+}}^{\breve{t}_{p}}\hspace*{-0.3cm}d\breve{t}_{q}\ppr\;
{\ts \frac{2\;|\hat{b}_{mm}(\breve{\vec{x}},\breve{t}_{q}\ppr)|}{\sin\big(2\;|\hat{b}_{mm}(\breve{\vec{x}},\breve{t}_{q}\ppr)|\big)}\;
\frac{\pp\beta_{m}(\breve{\vec{x}},\breve{t}_{q}\ppr)}{\pp\breve{t}_{q}\ppr} } \;;\hspace*{0.5cm}
\phi_{r}(\breve{\vec{x}},\breve{t}_{p}) = \int_{-\breve{\infty}_{+}}^{\breve{t}_{p}}\hspace*{-0.3cm}d\breve{t}_{q}\ppr\;
{\ts \frac{2\;|\hat{a}_{rr}^{(2)}(\breve{\vec{x}},\breve{t}_{q}\ppr)|}{\sinh\big(2\;|\hat{a}_{rr}^{(2)}(\breve{\vec{x}},\breve{t}_{q}\ppr)|\big)}\;
\frac{\pp\alpha_{r}(\breve{\vec{x}},\breve{t}_{q}\ppr)}{\pp\breve{t}_{q}\ppr} } \;.
\eeq
After partial integrations in (\ref{s3_111},\ref{s3_129}), we obtain the first order variation
\(\delta\mscr{L}^{(d)}[\hat{\mscr{Z}};\breve{J}_{\psi}]\) of the Lagrangian in terms of the matrices
\(\big(\delta\hat{\mscr{Z}}_{\alpha\beta}^{ab}\big)\) and \(\big(\delta\exp\{2\;\hat{Y}_{DD}\}\big)_{\alpha\beta}^{ab}\)
and also list the part which starts to contribute after second order variations for universal fluctuations
around the classical solutions within the coset decomposition \(\mbox{Osp}(S,S|2L)/\mbox{U}(L|S)\otimes \mbox{U}(L|S)\)
\beq \lb{s3_146}
\lefteqn{\delta\mscr{L}^{(d)}[\hat{\mscr{Z}};\breve{J}_{\psi}]=\STRAB\Bigg[\big(\delta\hat{\mscr{Z}}_{\alpha\beta}^{ab}\big)\;
\bigg[\breve{\pp}_{i}\bigg( 2\;\breve{c}^{ij}(\breve{\vec{x}},\breve{t})\;
\big(\breve{\pp}_{j}\hat{\mscr{Z}}_{\beta\alpha}^{b\neq a}\big)\;\big(1-\delta_{ab}\big) +
\big(\breve{\pp}_{i}\hat{\mscr{Z}}_{\beta\alpha}^{ba}\big)+ } \\  \no &+& \boldsymbol{\bigg\{}\im\;
\Big(\breve{d}^{ij}(\breve{\vec{x}},\breve{t})-\frac{1}{2}\delta_{ij}\Big)\;
\big(\breve{\pp}_{j}\hat{\mscr{Z}}\big)\;\boldsymbol{,}\;
\hat{\mscr{P}}\;\hat{I}\;\breve{J}_{\psi}\otimes\breve{J}_{\psi}^{+}\;\hat{I}\;\wt{K}\;\hat{\mscr{P}}^{-1}\boldsymbol{\bigg\}_{+}}
\bigg)+  \\ \no &+&
\frac{1}{2}\;\breve{\pp}_{\breve{t}}\bigg(\Big(\im\;\hat{1}_{2N\times 2N}+
\hat{\mscr{P}}\;\hat{I}^{-1}\;\breve{J}_{\psi}\otimes\breve{J}_{\psi}^{+}\;\hat{I}\;\hat{K}\;\hat{\mscr{P}}^{-1}\Big)\;
\exp\{2\;\hat{Y}_{DD}\}\;\hat{S}\bigg) + \bigg(\bigg(\frac{\overrightarrow{\pp}\hat{\mscr{P}}}{\overrightarrow{\pp}\hat{\mscr{Z}}}\bigg)
\hat{\mscr{P}}^{-1} \times \\ \no &\times& \boldsymbol{\bigg[}\Big(\big(\breve{\pp}_{i}\hat{\mscr{Z}}\big)
\big(\breve{\pp}_{j}\hat{\mscr{Z}}\big)\im\Big(\breve{d}^{ij}-\frac{1}{2}\;\delta_{ij}\Big)+
\frac{1}{2}\exp\big\{-2\;\hat{Y}_{DD}\big\}\big(\breve{\pp}_{\breve{t}_{p}}\hat{\mscr{Z}}\big)\;\hat{S}\Big)\;\boldsymbol{,}\;
\hat{\mscr{P}}\;\hat{I}\;\breve{J}_{\psi}\otimes\breve{J}_{\psi}^{+}\;\hat{I}\;\wt{K}\;\hat{\mscr{P}}^{-1}\boldsymbol{\bigg]_{-}}\bigg)
\bigg]_{\beta\alpha}^{ba}\Bigg]+
\\ \no   &-&   \frac{1}{2}\STRAB\Bigg[\big(\delta\exp\{2\;\hat{Y}_{DD}\}\big)_{\alpha\beta}^{ab}
\bigg(\Big[\im\;\hat{S}\;\big(\breve{\pp}_{\breve{t}}\hat{\mscr{Z}}\big)+
\hat{1}_{2N\times 2N}\Big(\breve{u}(\breve{\vec{x}})-\breve{\mu}_{0}+
\Re\big(\langle\breve{\sigma}_{D}^{(0)}(\breve{\vec{x}},\breve{t})\rangle_{\hat{\sigma}_{D}^{(0)}}\big)\Big)\Big]+  \\ \no &+&
\Big[\hat{S}\;\big(\breve{\pp}_{\breve{t}}\hat{\mscr{Z}}\big)\;\hat{\mscr{P}}\;\hat{I}^{-1}\;\breve{J}_{\psi}\otimes
\breve{J}_{\psi}^{+}\;\hat{I}\;\hat{K}\;\hat{\mscr{P}}^{-1}\Big]\bigg)_{\beta\alpha}^{ba}
\Bigg] +  \underbrace{\frac{\eta_{p}}{2}\;\STRAB\Big[\big(\delta\exp\{2\;\hat{Y}_{DD}\}\big)\Big]}_{\equiv 0} + \\ \no &+&
\frac{\eta_{p}}{2}\;\Big(1+\im\;\big(\ve_{+}+\Im\big(\langle\breve{\sigma}_{D}^{(0)}(\breve{\vec{x}},\breve{t})\rangle\big)\Big)\;
\bigg(\delta\STRAB\Big[\big(\delta\exp\{2\;\hat{Y}_{DD}\}\big)\Big]\bigg)   \;.
\eeq
The following steps seem to be involved and complicated, but are straightforward in order to attain
the first order variations for the classical equations with the independent, anomalous Euclidean
fields \(\hat{b}_{m\geq n}(\vec{x},t_{p})\), \(\hat{a}_{rr}^{(2)}(\vec{x},t_{p})\), \(\hat{a}_{r>r\ppr}^{(k)}(\vec{x},t_{p})\)
\((k=0,1,2,3)\), \(\hat{\zeta}_{r\mu,n}(\vec{x},t_{p})\), (+c.c.). One has to relate
the variations  of the matrices
\(\big(\delta\hat{\mscr{Z}}_{\alpha\beta}^{ab}\big)\) and \(\big(\delta\exp\{2\;\hat{Y}_{DD}\}\big)_{\alpha\beta}^{ab}\)
in (\ref{s3_146}) to these independent, anomalous, Euclidean fields where each of these finally defines
a classical equation. At first we specify the variations of the coset eigenvalue matrix
\(\big(\delta\exp\{2\;\hat{Y}_{DD}\}\big)_{\alpha\beta}^{ab}\) in terms of \(\hat{b}_{mm}(\vec{x},t_{p})\)
and \(\hat{a}_{rr}^{(2)}(\vec{x},t_{p})\) which introduce the Sine(h)- and Cos(h)-functions of
these diagonal elements into the first order variations to classical field equations
\beq\lb{s3_147}
\big(\delta\exp\{2\;\hat{Y}_{DD}\}\big)_{\alpha\beta}^{aa} &=&
\left(\bea{cc} \big(\delta\exp\{2\;\hat{Y}_{DD}\}\big)_{BB;mm}^{aa}\;\delta_{mn} & 0 \\
0 & \big(\delta\exp\{2\;\hat{Y}_{DD}\}\big)_{FF;r\mu,r\nu}^{aa}\;\delta_{rr\ppr}\;\delta_{\mu\nu}
\eea\right)_{\alpha\beta}^{aa} \!\!\!\!;  \\ \lb{s3_148}
\big(\delta\exp\{2\;\hat{Y}_{DD}\}\big)_{BB;mm}^{aa} &=& -\sin\big(2\;|\hat{b}_{mm}(\breve{\vec{x}},\breve{t})|\big)\;\times
\\   \no &\times&
\Big[e^{-\im\;\beta_{m}(\breve{\vec{x}},\breve{t})}\;\big(\delta\hat{b}_{mm}(\breve{\vec{x}},\breve{t})\big)+
e^{\im\;\beta_{m}(\breve{\vec{x}},\breve{t})}\;\big(\delta\hat{b}_{mm}^{*}(\breve{\vec{x}},\breve{t})\big)\Big]\;;  \\  \lb{s3_149}
\big(\delta\exp\{2\;\hat{Y}_{DD}\}\big)_{FF;r\mu,r\nu}^{aa} &=& \delta_{\mu\nu}\;
\sinh\big(2\;|\hat{a}_{rr}^{(2)}(\breve{\vec{x}},\breve{t})|\big)\;\times    \\ \no &\times&
\Big[e^{-\im\;\alpha_{r}(\breve{\vec{x}},\breve{t})}\;\big(\delta\hat{a}_{rr}^{(2)}(\breve{\vec{x}},\breve{t})\big)+
e^{\im\;\alpha_{r}(\breve{\vec{x}},\breve{t})}\;\big(\delta\hat{a}_{rr}^{(2)*}(\breve{\vec{x}},\breve{t})\big)\Big] \;; \\  \lb{s3_150}
\big(\delta\exp\{2\;\hat{Y}_{DD}\}\big)_{\alpha\beta}^{12} &=&
\left(\bea{cc} \big(\delta\exp\{2\;\hat{Y}_{DD}\}\big)_{BB;mm}^{12}\;\delta_{mn} & 0 \\
0 & \big(\delta\exp\{2\;\hat{Y}_{DD}\}\big)_{FF;r\mu,r\nu}^{12}\;\delta_{rr\ppr}
\eea\right)_{\alpha\beta}^{12} \;;  \\  \lb{s3_151}
\big(\delta\exp\{2\;\hat{Y}_{DD}\}\big)_{\alpha\beta}^{21} &=&
\left(\bea{cc} \big(\delta\exp\{2\;\hat{Y}_{DD}\}\big)_{BB;mm}^{21}\;\delta_{mn} & 0 \\
0 & \big(\delta\exp\{2\;\hat{Y}_{DD}\}\big)_{FF;r\mu,r\nu}^{21}\;\delta_{rr\ppr}
\eea\right)_{\alpha\beta}^{21}  \;; \\    \lb{s3_152}
\big(\delta\exp\{2\;\hat{Y}_{DD}\}\big)_{BB;mm}^{12} &=& -\big(\delta\exp\{2\;\hat{Y}_{DD}\}\big)_{BB;mm}^{21,*} = \\ \no &=&
-\Big(e^{\im\;\varphi_{m}(\breve{\vec{x}},\breve{t})}\;\cos\big(2\;|\hat{b}_{mm}(\breve{\vec{x}},\breve{t})|\big)+1\Big)\;
e^{-\im\;\beta_{m}(\breve{\vec{x}},\breve{t})}\;\big(\delta\hat{b}_{mm}(\breve{\vec{x}},\breve{t})\big) +  \\ \no &-&
\Big(e^{\im\;\varphi_{m}(\breve{\vec{x}},\breve{t})}\;\cos\big(2\;|\hat{b}_{mm}(\breve{\vec{x}},\breve{t})|\big)-1\Big)\;
e^{\im\;\beta_{m}(\breve{\vec{x}},\breve{t})}\;\big(\delta\hat{b}_{mm}^{*}(\breve{\vec{x}},\breve{t})\big) \;; \\   \lb{s3_153}
\big(\delta\exp\{2\;\hat{Y}_{DD}\}\big)_{FF;r\mu,r\nu}^{12} &=& \big(\delta\exp\{2\;\hat{Y}_{DD}\}\big)_{FF;r\mu,r\nu}^{21,+} = \\ \no &=&
\big(\tau_{2}\big)_{\mu\nu}
\bigg[\Big(e^{\im\;\phi_{r}(\breve{\vec{x}},\breve{t})}\;\cosh\big(2\;|\hat{a}_{rr}^{(2)}(\breve{\vec{x}},\breve{t})|\big)+1\Big)\;
e^{-\im\;\alpha_{r}(\breve{\vec{x}},\breve{t})}\;\big(\delta\hat{a}_{rr}^{(2)}(\breve{\vec{x}},\breve{t})\big) +  \\ \no &+&
\Big(e^{\im\;\phi_{r}(\breve{\vec{x}},\breve{t})}\;\cosh\big(2\;|\hat{a}_{rr}^{(2)}(\breve{\vec{x}},\breve{t})|\big)-1\Big)\;
e^{\im\;\alpha_{r}(\breve{\vec{x}},\breve{t})}\;\big(\delta\hat{a}_{rr}^{(2)*}(\breve{\vec{x}},\breve{t})\big)   \;.
\eeq
At first we specialize on the variation with diagonal (quaternion diagonal) matrix elements
\(\delta\hat{b}_{mm}^{*}(\breve{\vec{x}},\breve{t})\), (\(\delta\hat{a}_{rr}^{(2)*}(\breve{\vec{x}},\breve{t})\))
and have to extract these Euclidean fields from the variation within the matrices
\(\big(\delta\hat{\mscr{Z}}_{\alpha\beta}^{ab}\big)\) and \(\big(\delta\exp\{2\;\hat{Y}_{DD}\}\big)_{\alpha\beta}^{ab}\)
which appear in \(\delta\mscr{L}^{(d)}[\hat{\mscr{Z}};\breve{J}_{\psi}]\) (\ref{s3_146}). As these fields
\(\delta\hat{b}_{mm}^{*}(\breve{\vec{x}},\breve{t})\), (\(\delta\hat{a}_{rr}^{(2)*}(\breve{\vec{x}},\breve{t})\))
are separated from the various parts of the variations of the matrices
\(\big(\delta\hat{\mscr{Z}}_{\alpha\beta}^{ab}\big)\) and \(\big(\delta\exp\{2\;\hat{Y}_{DD}\}\big)_{\alpha\beta}^{ab}\),
one has to include coefficients \(\mfrak{B}_{mm}^{a\geq b}(\breve{\vec{x}},\breve{t})\),
\(\mfrak{Y}_{mm}^{a\geq b}(\breve{\vec{x}},\breve{t})\) (derived from \(\big(\delta\hat{\mscr{Z}}_{\alpha\beta}^{ab}\big)\),
\(\big(\delta\exp\{2\;\hat{Y}_{DD}\}\big)_{\alpha\beta}^{ab}\)) into the resulting field equation
\beq \no
&&\mbox{Variation with }\;\big(\delta\hat{b}_{mm}^{*}(\breve{\vec{x}},\breve{t})\big)  \\  \lb{s3_154}
\mfrak{B}_{mm}^{21}(\breve{\vec{x}},\breve{t}) &=& 1\;;  \\   \lb{s3_155}
\mfrak{B}_{mm}^{11}(\breve{\vec{x}},\breve{t}) &=&-\mfrak{B}_{mm}^{22}(\breve{\vec{x}},\breve{t})=
-\frac{1}{2}\;\tan\big(|\hat{b}_{mm}(\breve{\vec{x}},\breve{t})|\big)\;e^{\im\;\varphi_{m}(\breve{\vec{x}},\breve{t})} \;;  \\   \lb{s3_156}
\mfrak{Y}_{mm}^{21}(\breve{\vec{x}},\breve{t}) &=&
\Big(e^{-\im\;\varphi_{m}(\breve{\vec{x}},\breve{t})}\;
\cos\big(2\;|\hat{b}_{mm}(\breve{\vec{x}},\breve{t})|\big)+1\Big)\;e^{\im\;\beta_{m}(\breve{\vec{x}},\breve{t})} \;;  \\   \lb{s3_157}
\mfrak{Y}_{mm}^{11}(\breve{\vec{x}},\breve{t}) &=& \mfrak{Y}_{mm}^{22}(\breve{\vec{x}},\breve{t}) =
-\sin\big(2\;|\hat{b}_{mm}(\breve{\vec{x}},\breve{t})|\big)\;e^{\im\;\beta_{m}(\breve{\vec{x}},\breve{t})}  \;.
\eeq
According to the above coefficients \(\mfrak{B}_{mm}^{a\geq b}(\breve{\vec{x}},\breve{t})\),
\(\mfrak{Y}_{mm}^{a\geq b}(\breve{\vec{x}},\breve{t})\) of \(\big(\delta\hat{\mscr{Z}}_{\alpha\beta}^{ab}\big)\) and
\(\big(\delta\exp\{2\;\hat{Y}_{DD}\}\big)_{\alpha\beta}^{ab}\), we can simplify the resulting equation of the
diagonal pair condensate fields in the boson-boson part
\beq \lb{s3_158}
\lefteqn{0 \equiv \sum_{a,b=1,2}^{(a\geq b)}\mfrak{B}_{mm}^{a\geq b}(\breve{\vec{x}},\breve{t})\;
\bigg[\breve{\pp}_{i}\bigg( 2\;\breve{c}^{ij}(\breve{\vec{x}},\breve{t})\;
\big(\breve{\pp}_{j}\hat{\mscr{Z}}_{\beta\alpha}^{b\neq a}\big)\;\big(1-\delta_{ab}\big) +
\big(\breve{\pp}_{i}\hat{\mscr{Z}}_{\beta\alpha}^{ba}\big)+ } \\  \no &+& \boldsymbol{\bigg\{}\im\;
\Big(\breve{d}^{ij}(\breve{\vec{x}},\breve{t})-\frac{1}{2}\delta_{ij}\Big)\;
\big(\breve{\pp}_{j}\hat{\mscr{Z}}\big)\;\boldsymbol{,}\;\hat{\mscr{P}}\;
\hat{I}\;\breve{J}_{\psi}\otimes\breve{J}_{\psi}^{+}\;\hat{I}\;\wt{K}\;\hat{\mscr{P}}^{-1}
\boldsymbol{\bigg\}_{+}} \bigg)+  \\ \no &+&
\frac{1}{2}\;\breve{\pp}_{\breve{t}}\bigg(\Big(\im\;\hat{1}_{2N\times 2N}+
\hat{\mscr{P}}\;\hat{I}^{-1}\;\breve{J}_{\psi}\otimes\breve{J}_{\psi}^{+}\;\hat{I}\;\hat{K}\;\hat{\mscr{P}}^{-1}\Big)\;
\exp\{2\;\hat{Y}_{DD}\}\;\hat{S}\bigg)
+ \bigg(\bigg(\frac{\overrightarrow{\pp}\hat{\mscr{P}}}{\overrightarrow{\pp}\hat{\mscr{Z}}}\bigg)
\hat{\mscr{P}}^{-1} \times \\ \no &\times& \boldsymbol{\bigg[}\Big(\big(\breve{\pp}_{i}\hat{\mscr{Z}}\big)
\big(\breve{\pp}_{j}\hat{\mscr{Z}}\big)\im\Big(\breve{d}^{ij}-\frac{1}{2}\;\delta_{ij}\Big)+
\frac{1}{2}\exp\big\{-2\;\hat{Y}_{DD}\big\}\big(\breve{\pp}_{\breve{t}_{p}}\hat{\mscr{Z}}\big)\;\hat{S}\Big)\;\boldsymbol{,}\;
\hat{\mscr{P}}\;\hat{I}\;\breve{J}_{\psi}\otimes\breve{J}_{\psi}^{+}\;\hat{I}\;\wt{K}\;\hat{\mscr{P}}^{-1}\boldsymbol{\bigg]}\bigg)
 \bigg]_{BB;mm}^{b\leq a}+
\\ \no   &-&   \frac{1}{2}\sum_{a,b=1,2}^{(a\geq b)}\mfrak{Y}_{mm}^{a\geq b}(\breve{\vec{x}},\breve{t})\;
\bigg(\Big[\im\;\hat{S}\;\big(\breve{\pp}_{\breve{t}}\hat{\mscr{Z}}\big)+
\hat{1}_{2N\times 2N}\Big(\breve{u}(\breve{\vec{x}})-\breve{\mu}_{0}+
\Re\big(\langle\breve{\sigma}_{D}^{(0)}(\breve{\vec{x}},\breve{t})\rangle_{\hat{\sigma}_{D}^{(0)}}\big)\Big)\Big]+  \\ \no &+&
\Big[\hat{S}\;\big(\breve{\pp}_{\breve{t}}\hat{\mscr{Z}}\big)\;\hat{\mscr{P}}\;\hat{I}^{-1}\;\breve{J}_{\psi}\otimes
\breve{J}_{\psi}^{+}\;\hat{I}\;\hat{K}\;\hat{\mscr{P}}^{-1}\Big]\bigg)_{BB;mm}^{b\leq a} \;\;.
\eeq
Similarly, we specify coefficients
\(\mfrak{F}_{r\mu,r\nu}^{a\geq b}(\breve{\vec{x}},\breve{t})\), \(\mfrak{Y}_{r\mu,r\nu}^{a\geq b}(\breve{\vec{x}},\breve{t})\)
from varying \(\big(\delta\hat{\mscr{Z}}_{\alpha\beta}^{ab}\big)\) and
\(\big(\delta\exp\{2\;\hat{Y}_{DD}\}\big)_{\alpha\beta}^{ab}\) for the quaternion diagonal elements of
the BCS pair condensates within the fermion-fermion section. The analogous list of coefficients is defined in relations (\ref{s3_159}-\ref{s3_162})
for the equations following by the variation with respect to \(\delta\hat{a}_{rr}^{(2)*}(\breve{\vec{x}},\breve{t})\)
\beq \no
&&\mbox{Variation with }\;\big(\delta\hat{a}_{rr}^{(2)*}(\breve{\vec{x}},\breve{t})\big)\;\mbox{ or} \\ \no &&
\big(\delta\hat{a}_{r\mu,r\nu}(\breve{\vec{x}},\breve{t})\big)=-\big(\delta\hat{a}_{r\nu,r\mu}(\breve{\vec{x}},\breve{t})\big)=
\big(\tau_{2}\big)_{\mu\nu}\;\big(\delta\hat{a}_{rr}^{(2)*}(\breve{\vec{x}},\breve{t})\big)
  \\  \lb{s3_159}
\mfrak{F}_{r\mu,r\nu}^{21}(\breve{\vec{x}},\breve{t}) &=& -\mfrak{F}_{r\nu,r\mu}^{21}(\breve{\vec{x}},\breve{t}) =
\big(\tau_{2}\big)_{\mu\nu}\;;  \\   \lb{s3_160}
\mfrak{F}_{r\mu,r\nu}^{11}(\breve{\vec{x}},\breve{t}) &=&-\mfrak{F}_{r\nu,r\mu}^{22}(\breve{\vec{x}},\breve{t})=\big(\tau_{0}\big)_{\mu\nu}\;
\frac{1}{2}\;\tanh\big(|\hat{a}_{rr}^{(2)}(\breve{\vec{x}},\breve{t})|\big)\;e^{\im\;\phi_{r}(\breve{\vec{x}},\breve{t})} \;;  \\   \lb{s3_161}
\mfrak{Y}_{r\mu,r\nu}^{21}(\breve{\vec{x}},\breve{t}) &=& \big(\tau_{2}\big)_{\mu\nu}\;
\Big(e^{-\im\;\phi_{r}(\breve{\vec{x}},\breve{t})}\;
\cosh\big(2\;|\hat{a}_{rr}^{(2)}(\breve{\vec{x}},\breve{t})|\big)+1\Big)\;e^{\im\;\alpha_{r}(\breve{\vec{x}},\breve{t})} \;;  \\   \lb{s3_162}
\mfrak{Y}_{r\mu,r\nu}^{11}(\breve{\vec{x}},\breve{t}) &=& \mfrak{Y}_{r\nu,r\mu}^{22}(\breve{\vec{x}},\breve{t}) =\big(\tau_{0}\big)_{\mu\nu}\;
\sinh\big(2\;|\hat{a}_{rr}^{(2)}(\breve{\vec{x}},\breve{t})|\big)\;e^{\im\;\alpha_{r}(\breve{\vec{x}},\breve{t})}  \;.
\eeq
These coefficients contain Sinh-, Cosh- and Tanh-functions instead of their trigonometric correspondents for
the diagonal pair condensates within the boson-boson part and allow to disentangle the resulting
matrix equation for the quaternion diagonal fermion-fermion section. Since one has to consider quaternion elements
in the case of fermion-fermion parts, we have to perform a trace '\(\mbox{tr}_{\mu\nu}\)' over the \(2\times 2\) quaternion matrices
occurring in the coefficients and the other parts of the resulting field equation
\beq \lb{s3_163}
\lefteqn{0 \equiv -\sum_{a,b=1,2}^{(a\geq b)}\trmn\Bigg[\mfrak{F}_{r\mu,r\nu}^{a\geq b}(\breve{\vec{x}},\breve{t})\;
\bigg[\breve{\pp}_{i}\bigg( 2\;\breve{c}^{ij}(\breve{\vec{x}},\breve{t})\;
\big(\breve{\pp}_{j}\hat{\mscr{Z}}_{\beta\alpha}^{b\neq a}\big)\;\big(1-\delta_{ab}\big) +
\big(\breve{\pp}_{i}\hat{\mscr{Z}}_{\beta\alpha}^{ba}\big)+ } \\  \no &+& \boldsymbol{\bigg\{}\im\;
\Big(\breve{d}^{ij}(\breve{\vec{x}},\breve{t})-\frac{1}{2}\delta_{ij}\Big)\;
\big(\breve{\pp}_{j}\hat{\mscr{Z}}\big)\;\boldsymbol{,}\;
\hat{\mscr{P}}\;\hat{I}\;\breve{J}_{\psi}\otimes\breve{J}_{\psi}^{+}\;\hat{I}\;\wt{K}\;\hat{\mscr{P}}^{-1}
\boldsymbol{\bigg\}_{+}} \bigg)+  \\ \no &+&
\frac{1}{2}\;\breve{\pp}_{\breve{t}}\bigg(\Big(\im\;\hat{1}_{2N\times 2N}+
\hat{\mscr{P}}\;\hat{I}^{-1}\;\breve{J}_{\psi}\otimes\breve{J}_{\psi}^{+}\;\hat{I}\;\hat{K}\;\hat{\mscr{P}}^{-1}\Big)\;
\exp\{2\;\hat{Y}_{DD}\}\;\hat{S}\bigg)  +
\bigg(\bigg(\frac{\overrightarrow{\pp}\hat{\mscr{P}}}{\overrightarrow{\pp}\hat{\mscr{Z}}}\bigg)
\hat{\mscr{P}}^{-1} \times \\ \no &\times& \boldsymbol{\bigg[}\Big(\big(\breve{\pp}_{i}\hat{\mscr{Z}}\big)
\big(\breve{\pp}_{j}\hat{\mscr{Z}}\big)\im\Big(\breve{d}^{ij}-\frac{1}{2}\;\delta_{ij}\Big)+
\frac{e^{-2\;\hat{Y}_{DD}}}{2}\big(\breve{\pp}_{\breve{t}_{p}}\hat{\mscr{Z}}\big)\;\hat{S}\Big)\;\boldsymbol{,}\;
\hat{\mscr{P}}\;\hat{I}\;\breve{J}_{\psi}\otimes\breve{J}_{\psi}^{+}\;\hat{I}\;\wt{K}\;\hat{\mscr{P}}^{-1}\boldsymbol{\bigg]_{-}}\bigg)
\bigg]_{FF;r\nu,r\mu}^{b\leq a}\Bigg]+
\\ \no   &+&   \frac{1}{2}\sum_{a,b=1,2}^{(a\geq b)}\trmn\Bigg[\mfrak{Y}_{r\mu,r\nu}^{a\geq b}(\breve{\vec{x}},\breve{t})\;
\bigg(\Big[\im\;\hat{S}\;\big(\breve{\pp}_{\breve{t}}\hat{\mscr{Z}}\big)+
\hat{1}_{2N\times 2N}\Big(\breve{u}(\breve{\vec{x}})-\breve{\mu}_{0}+
\Re\big(\langle\breve{\sigma}_{D}^{(0)}(\breve{\vec{x}},\breve{t})\rangle_{\hat{\sigma}_{D}^{(0)}}\big)\Big)\Big]+  \\ \no &+&
\Big[\hat{S}\;\big(\breve{\pp}_{\breve{t}}\hat{\mscr{Z}}\big)\;\hat{\mscr{P}}\;\hat{I}^{-1}\;\breve{J}_{\psi}\otimes
\breve{J}_{\psi}^{+}\;\hat{I}\;\hat{K}\;\hat{\mscr{P}}^{-1}\Big]\bigg)_{FF;r\nu,r\mu}^{b\leq a}\Bigg]
\eeq
In the case of off-diagonal variations \(\delta\hat{b}_{m\neq n}^{*}(\breve{\vec{x}},\breve{t})\),
\(\delta\hat{a}_{rr\ppr}^{(k)+}(\breve{\vec{x}},\breve{t})\) in the anomalous boson-boson or
fermion-fermion part, we can neglect variations \(\big(\delta\exp\{2\;\hat{Y}_{DD}\}\big)_{\alpha\beta}^{ab}\)
of the coset eigenvalue matrix because these only consist of diagonal (quaternion diagonal) elements
\(\hat{b}_{mm}^{*}(\breve{\vec{x}},\breve{t})\) (\(\hat{a}_{rr}^{(2)*}(\breve{\vec{x}},\breve{t})\))
apart from the non-Markovian phases \(\varphi_{m}(\breve{\vec{x}},\breve{t})\), \(\phi_{r}(\breve{\vec{x}},\breve{t})\).
Therefore, we have only to take into account coefficients \(\mfrak{B}_{mn}^{a\geq b}(\breve{\vec{x}},\breve{t})\)
arising from the variation with the matrix \(\big(\delta\hat{\mscr{Z}}_{\alpha\beta}^{ab}\big)\)
\beq \no
&&\mbox{Variation with }\;\big(\delta\hat{b}_{m\neq n}^{*}(\breve{\vec{x}},\breve{t})\big) \\   \no &&
\mbox{Mind the symmetry : }\;\big(\delta\hat{b}_{m\neq n}^{*}(\breve{\vec{x}},\breve{t})\big)=
\big(\delta\hat{b}_{n\neq m}^{*}(\breve{\vec{x}},\breve{t})\big)  \\  \lb{s3_164}
\mfrak{B}_{mn}^{21}(\breve{\vec{x}},\breve{t}) &=& 1\;;  \\   \lb{s3_165}
\mfrak{B}_{mn}^{11}(\breve{\vec{x}},\breve{t}) &=&-\mfrak{B}_{nm}^{22}(\breve{\vec{x}},\breve{t})=
-\frac{\;e^{\im\;\varphi_{m}(\breve{\vec{x}},\breve{t})}}{2} \;
\bigg[\tan\bigg(\frac{|\hat{b}_{mm}|+|\hat{b}_{nn}|}{2}\bigg)+\tan\bigg(\frac{|\hat{b}_{mm}|-|\hat{b}_{nn}|}{2}\bigg)\bigg] \;.
\eeq
Application of the above coefficients \(\mfrak{B}_{mn}^{a\geq b}(\breve{\vec{x}},\breve{t})\) finally allows
to give the classical field equations for the off-diagonal anomalous, boson-boson part in abbreviated from
which also includes the trigonometric functions
\beq \lb{s3_166}
\lefteqn{0 \equiv \boldsymbol{\Bigg\{}\sum_{a,b=1,2}^{(a\geq b)}\mfrak{B}_{mn}^{a\geq b}(\breve{\vec{x}},\breve{t})\;
\bigg[\breve{\pp}_{i}\bigg( 2\;\breve{c}^{ij}(\breve{\vec{x}},\breve{t})\;
\big(\breve{\pp}_{j}\hat{\mscr{Z}}_{\beta\alpha}^{b\neq a}\big)\;\big(1-\delta_{ab}\big) +
\big(\breve{\pp}_{i}\hat{\mscr{Z}}_{\beta\alpha}^{ba}\big)+ } \\  \no &+& \boldsymbol{\bigg\{}\im\;
\Big(\breve{d}^{ij}(\breve{\vec{x}},\breve{t})-\frac{1}{2}\delta_{ij}\Big)\;
\big(\breve{\pp}_{j}\hat{\mscr{Z}}\big)\;\boldsymbol{,}\;
\hat{\mscr{P}}\;\hat{I}\;\breve{J}_{\psi}\otimes\breve{J}_{\psi}^{+}\;\hat{I}\;\wt{K}\;\hat{\mscr{P}}^{-1}
\boldsymbol{\bigg\}_{+}}  \bigg)+  \\ \no &+&
\frac{1}{2}\;\breve{\pp}_{\breve{t}}\bigg(\Big(\im\;\hat{1}_{2N\times 2N}+
\hat{\mscr{P}}\;\hat{I}^{-1}\;\breve{J}_{\psi}\otimes\breve{J}_{\psi}^{+}\;\hat{I}\;\hat{K}\;\hat{\mscr{P}}^{-1}\Big)\;
\exp\{2\;\hat{Y}_{DD}\}\;\hat{S}\bigg) +
\bigg(\bigg(\frac{\overrightarrow{\pp}\hat{\mscr{P}}}{\overrightarrow{\pp}\hat{\mscr{Z}}}\bigg)
\hat{\mscr{P}}^{-1} \times \\ \no &\times& \boldsymbol{\bigg[}\Big(\big(\breve{\pp}_{i}\hat{\mscr{Z}}\big)
\big(\breve{\pp}_{j}\hat{\mscr{Z}}\big)\im\Big(\breve{d}^{ij}-\frac{1}{2}\;\delta_{ij}\Big)+
\frac{e^{-2\;\hat{Y}_{DD}}}{2}\big(\breve{\pp}_{\breve{t}_{p}}\hat{\mscr{Z}}\big)\;\hat{S}\Big)\;\boldsymbol{,}\;
\hat{\mscr{P}}\;\hat{I}\;\breve{J}_{\psi}\otimes\breve{J}_{\psi}^{+}\;\hat{I}\;\wt{K}\;\hat{\mscr{P}}^{-1}\boldsymbol{\bigg]}\bigg)
\bigg]_{BB;nm}^{b\leq a}\boldsymbol{\Bigg\}}+
\\ \no   &+& \boldsymbol{\Bigg\{}\;\mbox{\bf entire above terms with }\;\boldsymbol{m\leftrightarrow n}\boldsymbol{\Bigg\}} \;.
\eeq
In analogy we catalogue the coefficients \(\mfrak{F}_{r\mu,r\ppr\nu}^{21,(k)}(\breve{\vec{x}},\breve{t})\)
of the variations \(\big(\delta\hat{\mscr{Z}}_{\alpha\beta}^{ab}\big)\)
for the off-diagonal quaternion matrix elements within the fermion-fermion section. Since the variation
with \(\big(\delta\exp\{2\;\hat{Y}_{DD}\}\big)_{\alpha\beta}^{ab}\) is restricted to the
quaternion diagonal elements, we have only to introduce coefficients \(\mfrak{F}_{r\mu,r\ppr\nu}^{21,(k)}(\breve{\vec{x}},\breve{t})\)
for the variations within \(\big(\delta\hat{\mscr{Z}}_{\alpha\beta}^{ab}\big)\)
\beq \no
&&\mbox{Variation with }\;\big(\delta\hat{a}_{rr\ppr}^{(k)+}(\breve{\vec{x}},\breve{t})\big)\;\mbox{ in }\;
\big(\delta\hat{a}_{r\mu,r\ppr\nu}^{+}(\breve{\vec{x}},\breve{t})\big)=
\sum_{k=0}^{3}\big(\tau_{k}\big)_{\mu\nu}\,\big(\delta\hat{a}_{rr\ppr}^{(k)+}(\breve{\vec{x}},\breve{t})\big); \\   \no
\mbox{Mind the symmetries}&:&\big(\delta\hat{a}_{r\mu,r\ppr\nu}^{+}(\breve{\vec{x}},\breve{t})\big)=-
\big(\delta\hat{a}_{r\ppr\nu,r\mu}^{+}(\breve{\vec{x}},\breve{t})\big)\;; \\ \no &&
\big(\delta\hat{a}_{rr\ppr}^{(2)+}\big)=+\big(\delta\hat{a}_{r\ppr r}^{(2)+}\big)\;;\;\;\mbox{ but }
\big(\delta\hat{a}_{rr\ppr}^{(k)+}\big)=-\big(\delta\hat{a}_{r\ppr r}^{(k)+}\big)\;
\mbox{ for }\;k=0,1,3\;;    \\  \lb{s3_167}
\mfrak{F}_{r\mu,r\ppr\nu}^{21,(k)}(\breve{\vec{x}},\breve{t}) &=& \big(\tau_{k}\big)_{\mu\nu}  \;;  \\   \lb{s3_168}
\mfrak{F}_{r\mu,r\ppr\nu}^{11,(k)}(\breve{\vec{x}},\breve{t}) &=&-\mfrak{F}_{r\ppr\nu,r\mu}^{22,(k)}(\breve{\vec{x}},\breve{t}) =
\frac{\big(-1\big)^{k}}{2}\;\big(\tau_{k}\;\tau_{2}\big)_{\mu\nu}\;
e^{\im\;\phi_{r}(\breve{\vec{x}},\breve{t})}\;\times  \\ \no &\times&
\bigg[\tanh\bigg(\frac{|\hat{a}_{rr}^{(2)}|+|\hat{a}_{r\ppr r\ppr}^{(2)}|}{2}\bigg)+
\tanh\bigg(\frac{|\hat{a}_{rr}^{(2)}|-|\hat{a}_{r\ppr r\ppr}|}{2}\bigg)\bigg]
\eeq
The above coefficients with hyperbolic trigonometric functions again reduce the field equations
to a compact form which includes traces over the \(2\times 2\) quaternion elements
for the anomalous, Euclidean field variables in the off-diagonal fermion-fermion parts
\beq \lb{s3_169}
0 &\equiv& \boldsymbol{\Bigg\{} -\sum_{a,b=1,2}^{(a\geq b)}\trmn
\Bigg[\mfrak{F}_{r\mu,r\ppr\nu}^{a\geq b,\boldsymbol{(k)}}(\breve{\vec{x}},\breve{t})\;
\bigg[\breve{\pp}_{i}\bigg( 2\;\breve{c}^{ij}(\breve{\vec{x}},\breve{t})\;
\big(\breve{\pp}_{j}\hat{\mscr{Z}}_{\beta\alpha}^{b\neq a}\big)\;\big(1-\delta_{ab}\big) +
\big(\breve{\pp}_{i}\hat{\mscr{Z}}_{\beta\alpha}^{ba}\big)+  \\  \no &+& \boldsymbol{\bigg\{}\im\;
\Big(\breve{d}^{ij}(\breve{\vec{x}},\breve{t})-\frac{1}{2}\delta_{ij}\Big)\;
\big(\breve{\pp}_{j}\hat{\mscr{Z}}\big)\;\boldsymbol{,}\;
\hat{\mscr{P}}\;\hat{I}\;\breve{J}_{\psi}\otimes\breve{J}_{\psi}^{+}\;\hat{I}\;\wt{K}\;\hat{\mscr{P}}^{-1}
\boldsymbol{\bigg\}_{+}}  \bigg)+  \\ \no &+&
\frac{1}{2}\;\breve{\pp}_{\breve{t}}\bigg(\Big(\im\;\hat{1}_{2N\times 2N}+
\hat{\mscr{P}}\;\hat{I}^{-1}\;\breve{J}_{\psi}\otimes\breve{J}_{\psi}^{+}\;\hat{I}\;\hat{K}\;\hat{\mscr{P}}^{-1}\Big)\;
\exp\{2\;\hat{Y}_{DD}\}\;\hat{S}\bigg)   +
\bigg(\bigg(\frac{\overrightarrow{\pp}\hat{\mscr{P}}}{\overrightarrow{\pp}\hat{\mscr{Z}}}\bigg)
\hat{\mscr{P}}^{-1} \times \\ \no &\times& \boldsymbol{\bigg[}\Big(\big(\breve{\pp}_{i}\hat{\mscr{Z}}\big)
\big(\breve{\pp}_{j}\hat{\mscr{Z}}\big)\im\Big(\breve{d}^{ij}-\frac{1}{2}\;\delta_{ij}\Big)+
\frac{e^{-2\;\hat{Y}_{DD}}}{2}\big(\breve{\pp}_{\breve{t}_{p}}\hat{\mscr{Z}}\big)\;\hat{S}\Big)\;\boldsymbol{,}\;
\hat{\mscr{P}}\;\hat{I}\;\breve{J}_{\psi}\otimes\breve{J}_{\psi}^{+}\;\hat{I}\;\wt{K}\;\hat{\mscr{P}}^{-1}\boldsymbol{\bigg]_{-}}\bigg)
\bigg]_{FF;r\ppr\nu,r\mu}^{b\leq a}\Bigg] \boldsymbol{\Bigg\}}+
\\ \no   &+&\boldsymbol{\Bigg\{}\;\mbox{\bf entire above terms with }\;\boldsymbol{r \leftrightarrow r\ppr}\;
\mbox{\bf for }\;\boldsymbol{k=2}\boldsymbol{\Bigg\}}\;\;\;\mbox{\bf or }
\\ \no   &-&\boldsymbol{\Bigg\{}\;\mbox{\bf entire upper terms with }\;\boldsymbol{r \leftrightarrow r\ppr}\;
\mbox{\bf for }\;\boldsymbol{k=0,1,3}\boldsymbol{\Bigg\}}
\eeq
Finally, we approach the variation with respect to the anti-commuting, anomalous fields
\(\delta\hat{\zeta}_{r\kappa,n}^{*}(\breve{\vec{x}},\breve{t})\) and extract corresponding
coefficients \(\mfrak{Z}_{FB;r\mu,n}^{a\geq b,(\kappa)}(\breve{\vec{x}},\breve{t})\),
\(\mfrak{Z}_{BF;n,r\mu}^{a\geq b,(\kappa)}(\breve{\vec{x}},\breve{t})\) for the
fermion-boson and boson-fermion parts which are derived from the matrix
\(\big(\delta\hat{\mscr{Z}}_{\alpha\beta}^{ab}\big)\)
\beq \no
&&\mbox{Variation with }\;\big(\delta\hat{\zeta}_{r\kappa,n}^{*}(\breve{\vec{x}},\breve{t})\big)\;;
\big(\delta\hat{\zeta}_{n,r\kappa}^{+}(\breve{\vec{x}},\breve{t})\big)  \\   \lb{s3_170}
\mfrak{Z}_{FB;r\mu,n}^{21,(\kappa)}(\breve{\vec{x}},\breve{t}) &=& 1   \;;  \\   \lb{s3_171}
\mfrak{Z}_{FB;r\mu,n}^{11,(\kappa)}(\breve{\vec{x}},\breve{t}) &=&
\frac{e^{\im\;\phi_{r}(\breve{\vec{x}},\breve{t})}}{2}\;
\bigg[\tanh\bigg(\frac{|\hat{a}_{rr}^{(2)}|+\im\;|\hat{b}_{nn}|}{2}\bigg)+
\tanh\bigg(\frac{|\hat{a}_{rr}^{(2)}|-\im\;|\hat{b}_{nn}|}{2}\bigg)\bigg]\;\big(\tau_{2}\big)_{\mu\kappa} \;; \\   \lb{s3_172}
\mfrak{Z}_{FB;r\mu,n}^{22,(\kappa)}(\breve{\vec{x}},\breve{t}) &=&\im\;
\frac{e^{\im\;\varphi_{n}(\breve{\vec{x}},\breve{t})}}{2}\;
\bigg[\tanh\bigg(\frac{|\hat{a}_{rr}^{(2)}|-\im\;|\hat{b}_{nn}|}{2}\bigg)-
\tanh\bigg(\frac{|\hat{a}_{rr}^{(2)}|+\im\;|\hat{b}_{nn}|}{2}\bigg)\bigg]\;\big(\tau_{0}\big)_{\mu\kappa} \;; \\   \lb{s3_173}
\mfrak{Z}_{BF;n,r\mu}^{21,(\kappa)}(\breve{\vec{x}},\breve{t}) &=& 1   \;;   \\   \lb{s3_174}
\mfrak{Z}_{BF;n,r\mu}^{11,(\kappa)}(\breve{\vec{x}},\breve{t}) &=&-\im\;
\frac{e^{\im\;\varphi_{n}(\breve{\vec{x}},\breve{t})}}{2}\;
\bigg[\tanh\bigg(\frac{|\hat{a}_{rr}^{(2)}|-\im\;|\hat{b}_{nn}|}{2}\bigg)-
\tanh\bigg(\frac{|\hat{a}_{rr}^{(2)}|+\im\;|\hat{b}_{nn}|}{2}\bigg)\bigg]\;\big(\tau_{0}\big)_{\kappa\mu} \;; \\   \lb{s3_175}
\mfrak{Z}_{BF;n,r\mu}^{22,(\kappa)}(\breve{\vec{x}},\breve{t}) &=&-
\frac{e^{\im\;\phi_{r}(\breve{\vec{x}},\breve{t})}}{2}\;
\bigg[\tanh\bigg(\frac{|\hat{a}_{rr}^{(2)}|+\im\;|\hat{b}_{nn}|}{2}\bigg)+
\tanh\bigg(\frac{|\hat{a}_{rr}^{(2)}|-\im\;|\hat{b}_{nn}|}{2}\bigg)\bigg]\;\big(\tau_{2}\big)_{\kappa\mu} \;.
\eeq
The coefficients \(\mfrak{Z}_{FB;r\mu,n}^{a\geq b,(\kappa)}(\breve{\vec{x}},\breve{t})\),
\(\mfrak{Z}_{BF;n,r\mu}^{a\geq b,(\kappa)}(\breve{\vec{x}},\breve{t})\) are partially composed
of compact and non-compact (hyperbolic) trigonometric functions and have to be summed over
the two spin degrees of freedom with the rest of the field equation which finally takes values
within the Grassmann sector of the super-symmetric matrices
\beq \lb{s3_176}
\lefteqn{0 \equiv -\sum_{a,b=1,2}^{(a\geq b)} \sum_{\mu=1,2}
\Bigg[\mfrak{Z}_{FB;r\mu,n}^{a\geq b,\boldsymbol{(\kappa)}}(\breve{\vec{x}},\breve{t})\;
\bigg[\breve{\pp}_{i}\bigg( 2\;\breve{c}^{ij}(\breve{\vec{x}},\breve{t})\;
\big(\breve{\pp}_{j}\hat{\mscr{Z}}_{\beta\alpha}^{b\neq a}\big)\;\big(1-\delta_{ab}\big) +
\big(\breve{\pp}_{i}\hat{\mscr{Z}}_{\beta\alpha}^{ba}\big)+ } \\  \no &+& \boldsymbol{\bigg\{}\im\;
\Big(\breve{d}^{ij}(\breve{\vec{x}},\breve{t})-\frac{1}{2}\delta_{ij}\Big)\;
\big(\breve{\pp}_{j}\hat{\mscr{Z}}\big)\;\boldsymbol{,}\;
\hat{\mscr{P}}\;\hat{I}\;\breve{J}_{\psi}\otimes\breve{J}_{\psi}^{+}\;\hat{I}\;\wt{K}\;\hat{\mscr{P}}^{-1}
\boldsymbol{\bigg\}_{+}}  \bigg)+  \\ \no &+&
\frac{1}{2}\;\breve{\pp}_{\breve{t}}\bigg(\Big(\im\;\hat{1}_{2N\times 2N}+
\hat{\mscr{P}}\;\hat{I}^{-1}\;\breve{J}_{\psi}\otimes\breve{J}_{\psi}^{+}\;\hat{I}\;\hat{K}\hat{\mscr{P}}^{-1}\Big)\;
\exp\{2\;\hat{Y}_{DD}\}\;\hat{S}\bigg) +
\bigg(\bigg(\frac{\overrightarrow{\pp}\hat{\mscr{P}}}{\overrightarrow{\pp}\hat{\mscr{Z}}}\bigg)
\hat{\mscr{P}}^{-1} \times \\ \no &\times& \boldsymbol{\bigg[}\Big(\big(\breve{\pp}_{i}\hat{\mscr{Z}}\big)
\big(\breve{\pp}_{j}\hat{\mscr{Z}}\big)\im\Big(\breve{d}^{ij}-\frac{1}{2}\;\delta_{ij}\Big)+
\frac{e^{-2\;\hat{Y}_{DD}}}{2}\big(\breve{\pp}_{\breve{t}_{p}}\hat{\mscr{Z}}\big)\;\hat{S}\Big)\;\boldsymbol{,}\;
\hat{\mscr{P}}\;\hat{I}\;\breve{J}_{\psi}\otimes\breve{J}_{\psi}^{+}\;\hat{I}\;\wt{K}\;\hat{\mscr{P}}^{-1}\boldsymbol{\bigg]_{-}}\bigg)
\bigg]_{BF;n,r\mu}^{b\leq a}\Bigg] + \\ \no &+&
\sum_{a,b=1,2}^{(a\geq b)} \sum_{\mu=1,2}
\Bigg[\mfrak{Z}_{BF;n,r\mu}^{a\geq b,\boldsymbol{(\kappa)}}(\breve{\vec{x}},\breve{t})\;
\bigg[\breve{\pp}_{i}\bigg( 2\;\breve{c}^{ij}(\breve{\vec{x}},\breve{t})\;
\big(\breve{\pp}_{j}\hat{\mscr{Z}}_{\beta\alpha}^{b\neq a}\big)\;\big(1-\delta_{ab}\big) +
\big(\breve{\pp}_{i}\hat{\mscr{Z}}_{\beta\alpha}^{ba}\big)+  \\  \no &+& \boldsymbol{\bigg\{}\im\;
\Big(\breve{d}^{ij}(\breve{\vec{x}},\breve{t})-\frac{1}{2}\delta_{ij}\Big)\;
\big(\breve{\pp}_{j}\hat{\mscr{Z}}\big)\;\boldsymbol{,}\;
\hat{\mscr{P}}\;\hat{I}\;\breve{J}_{\psi}\otimes\breve{J}_{\psi}^{+}\;\hat{I}\;\wt{K}\;\hat{\mscr{P}}^{-1}
\boldsymbol{\bigg\}_{+}}  \bigg)+  \\ \no &+&
\frac{1}{2}\;\breve{\pp}_{\breve{t}}\bigg(\Big(\im\;\hat{1}_{2N\times 2N}+
\hat{\mscr{P}}\;\hat{I}^{-1}\;\breve{J}_{\psi}\otimes\breve{J}_{\psi}^{+}\;\hat{I}\;\hat{K}\;\hat{\mscr{P}}^{-1}\Big)\;
\exp\{2\;\hat{Y}_{DD}\}\;\hat{S}\bigg)+
\bigg(\bigg(\frac{\overrightarrow{\pp}\hat{\mscr{P}}}{\overrightarrow{\pp}\hat{\mscr{Z}}}\bigg)
\hat{\mscr{P}}^{-1} \times \\ \no &\times& \boldsymbol{\bigg[}\Big(\big(\breve{\pp}_{i}\hat{\mscr{Z}}\big)
\big(\breve{\pp}_{j}\hat{\mscr{Z}}\big)\im\Big(\breve{d}^{ij}-\frac{1}{2}\;\delta_{ij}\Big)+
\frac{e^{-2\;\hat{Y}_{DD}}}{2}\big(\breve{\pp}_{\breve{t}_{p}}\hat{\mscr{Z}}\big)\;\hat{S}\Big)\;\boldsymbol{,}\;
\hat{\mscr{P}}\;\hat{I}\;\breve{J}_{\psi}\otimes\breve{J}_{\psi}^{+}\;\hat{I}\;\wt{K}\;\hat{\mscr{P}}^{-1}\boldsymbol{\bigg]_{-}}\bigg)
\bigg]_{FB;r\mu,n}^{b\leq a}\Bigg]
\eeq

We have classified in relations (\ref{s3_154}-\ref{s3_176}) the various classical field equations following from
first order variations of the independent, Euclidean pair condensate fields.
Since we have considered general angular momentum degrees of freedom of the
boson-boson, fermion-fermion and the odd fermion-boson, boson-fermion parts,
various coefficients [cf. Eqs. (\ref{s3_154}-\ref{s3_157}), (\ref{s3_159}-\ref{s3_162}),
(\ref{s3_164}-\ref{s3_165}), (\ref{s3_167}-\ref{s3_168}), (\ref{s3_170}-\ref{s3_175})]
have to be used as part of the variation within
\(\big(\delta\hat{\mscr{Z}}_{\alpha\beta}^{ab}\big)\) or
\(\big(\delta\exp\{2\;\hat{Y}_{DD}\}\big)_{\alpha\beta}^{ab}\).
One achieves coupled super-symmetric matrix equations which are composed
of Sine(h)-, Cos(h)- or Tan(h)-functions of the diagonal,
Euclidean pair condensate fields so that these illustrate modifications
of the well-known, integrable Sine-Gordon equations in 1+1 or 2+1
dimensions. These matrix equations correspond to the Gross-Pitaevskii equation
in a transferred sense if one regards the coherent super-symmetric
pair condensates in analogy to the coherent BEC-wavefunctions.

\subsection{Observable quantities in terms of coset fields and their corresponding,
Euclidean variables} \lb{s41}

Apart from the gradient expansion with (\ref{B29}), we have also to take into account the
generating source field \(\wt{\mscr{J}}(\hat{T}^{-1},\hat{T})\) (\ref{B30}) whose second
order expansion of the effective actions is listed in relation (\ref{B31}). This generating
source field \(\wt{\mscr{J}}(\hat{T}^{-1},\hat{T})\) (\ref{B30}) can be replaced by derivatives with
respect to the pair condensate 'seeds' \(\im\;\hat{J}_{\psi\psi;\alpha\beta}^{a\neq b}(\vec{x},t_{p})\;\wt{K}\)
(\ref{s2_63},\ref{s2_64}) of the action
\(\mscr{A}_{\hat{J}_{\psi\psi}}[\hat{T}]\) (\ref{s2_65}-\ref{s2_69}) for observables
which go beyond the second order expansion of \(\wt{\mscr{J}}(\hat{T}^{-1},\hat{T})\) in relation (\ref{B31}).
However, the pair condensate 'seed' fields \(\im\;\hat{J}_{\psi\psi;\alpha\beta}^{a\neq b}(\vec{x},t_{p})\;\wt{K}\)
(\ref{s2_63},\ref{s2_64}) of the action \(\mscr{A}_{\hat{J}_{\psi\psi}}[\hat{T}]\) (\ref{s2_65}-\ref{s2_69})
do not allow for generating density terms as
the source field \(\wt{\mscr{J}}(\hat{T}^{-1},\hat{T})\) (\ref{B30}). In correspondence to chapter 4 of Ref.\ \cite{mies1},
one can perform the gradient expansion with \(\delta\hat{\mscr{H}}(\hat{T}^{-1},\hat{T})\) (\ref{B29}) and with the
source term \(\wt{\mscr{J}}(\hat{T}^{-1},\hat{T})\) (\ref{B30}) in order to classify the various terms for the
pair condensate observables or density related observables
\be \lb{B29}
\delta\hat{\mscr{H}}(\hat{T}^{-1},\hat{T}) = -\hat{\eta}\Big(\hat{T}^{-1}\;\hat{S}\;\big(E_{p}\hat{T}\big)
+\hat{T}^{-1}\;\big(\wt{\pp}_{i}\wt{\pp}_{i}\hat{T}\big)  +
\big(\hat{T}^{-1}\;\hat{S}\;\hat{T}-\hat{S}\big)\;\mbox{{\boldmath$\hat{E}_{p}$}} +
2\;\hat{T}^{-1}\;\big(\wt{\pp}_{i}\hat{T}\big)\;\mbox{{\boldmath$\wt{\pp}_{i}$}}\Big)
\ee
\beq \lb{B30}
\wt{\mscr{J}}_{\vec{x},\alpha;\vec{x}\ppr,\beta}^{ab}(\hat{T}^{-1}(t_{p}),\hat{T}(t_{q}\ppr)) &=&
\hat{T}_{\alpha\alpha\ppr}^{-1;aa\ppr}(\vec{x},t_{p})\;\;
\hat{I}\;\hat{K}\;\eta_{p}\;\frac{\hat{\mscr{J}}_{\vec{x},\alpha\ppr;
\vec{x}\ppr,\beta\ppr}^{a\ppr b\ppr}(t_{p},t_{q}\ppr)}{\mscr{N}_{x}}\;\eta_{q}\;
\hat{K}\;\hat{I}\;\wt{K}\;\;\hat{T}_{\beta\ppr\beta}^{b\ppr b}(\vec{x}\ppr,t_{q}\ppr)
\eeq
\beq\lb{B31}
\lefteqn{\mscr{A}\ppr\big[\hat{T};\hat{\mscr{J}}\big]=-\frac{1}{4}\bigg\langle
\mbox{Tr}\;\mbox{STR}\bigg[\wt{\mscr{J}}(\hat{T}^{-1},\hat{T})\;\hat{G}^{(0)}[\hat{\sigma}_{D}^{(0)}]\;
\wt{\mscr{J}}(\hat{T}^{-1},\hat{T})\;\hat{G}^{(0)}[\hat{\sigma}_{D}^{(0)}]\bigg]
\bigg\rangle_{\hat{\sigma}_{D}^{(0)}} +}  \\ \no &-&
\frac{\im}{2}\frac{1}{\mscr{N}}\widehat{\langle J_{\psi;\beta}^{b}}|\hat{\eta}\Big(
\hat{I}\;\wt{K}\;\hat{T}\;\hat{G}^{(0)}[\hat{\sigma}_{D}^{(0)}]\;
\Big(\wt{\mscr{J}}(\hat{T}^{-1},\hat{T})\;\hat{G}^{(0)}[\hat{\sigma}_{D}^{(0)}]\Big)^{2}\;
\hat{T}^{-1}\;\hat{I}\Big)_{\beta\alpha}^{ba}\hat{\eta}|\widehat{J_{\psi;\alpha}^{a}\rangle}
\bigg\rangle_{\hat{\sigma}_{D}^{(0)}}  +  \\ \no &-&\frac{1}{2}\bigg\langle
\mbox{Tr}\;\mbox{STR}\bigg[\delta\hat{\mscr{H}}(\hat{T}^{-1},\hat{T})\;
\hat{G}^{(0)}[\hat{\sigma}_{D}^{(0)}]\;\wt{\mscr{J}}(\hat{T}^{-1},\hat{T})\;
\hat{G}^{(0)}[\hat{\sigma}_{D}^{(0)}]\bigg]\bigg\rangle_{\hat{\sigma}_{D}^{(0)}} + \\ \no &-&
\frac{\im}{2}\frac{1}{\mscr{N}}\bigg\langle
\widehat{\langle J_{\psi;\beta}^{b}}|\hat{\eta}\bigg(
\hat{I}\;\wt{K}\;\hat{T}\;\hat{G}^{(0)}[\hat{\sigma}_{D}^{(0)}]\;\Big(
\delta\hat{\mscr{H}}(\hat{T}^{-1},\hat{T})\;
\hat{G}^{(0)}[\hat{\sigma}_{D}^{(0)}]\;\wt{\mscr{J}}(\hat{T}^{-1},\hat{T})+  \\ \no &+&
\wt{\mscr{J}}(\hat{T}^{-1},\hat{T})\;\hat{G}^{(0)}[\hat{\sigma}_{D}^{(0)}]\;
\delta\hat{\mscr{H}}(\hat{T}^{-1},\hat{T})\Big)\;\hat{G}^{(0)}[\hat{\sigma}_{D}^{(0)}]\;
\hat{T}^{-1}\;\hat{I}\bigg)_{\beta\alpha}^{ba}\hat{\eta}|\widehat{J_{\psi;\alpha}^{a}\rangle}
\bigg\rangle_{\hat{\sigma}_{D}^{(0)}} +  \\ \no &+& \frac{1}{2}
\mbox{Tr}\;\mbox{STR}\bigg[\wt{\mscr{J}}(\hat{T}^{-1},\hat{T})\;
\Big\langle\hat{G}^{(0)}[\hat{\sigma}_{D}^{(0)}]\Big\rangle_{\hat{\sigma}_{D}^{(0)}}\bigg]
+  \\ \no &+&
\frac{\im}{2}\frac{1}{\mscr{N}}\bigg\langle
\widehat{\langle J_{\psi;\beta}^{b}}|\hat{\eta}\Big(
\hat{I}\;\wt{K}\;\hat{T}\;\hat{G}^{(0)}[\hat{\sigma}_{D}^{(0)}]\;
\wt{\mscr{J}}(\hat{T}^{-1},\hat{T})\;\hat{G}^{(0)}[\hat{\sigma}_{D}^{(0)}]\;
\hat{T}^{-1}\;\hat{I}\Big)_{\beta\alpha}^{ba}\hat{\eta}|\widehat{J_{\psi;\alpha}^{a}\rangle}
\bigg\rangle_{\hat{\sigma}_{D}^{(0)}} \;\;\;.
\eeq
The action \(\mscr{A}\ppr[\hat{T};\hat{\mscr{J}}]\) also involves the averaging
\(\langle\ldots\rangle_{\hat{\sigma}_{D}^{(0)}}\) over the background density field
\(\sigma_{D}^{(0)}(\vec{x},t_{p})\) with generating function \(Z[j_{\psi};\hat{\sigma}_{D}^{(0)}]\)
(\ref{s2_74}) (compare (\ref{s2_70}-\ref{s2_75})). However, we can simplify this averaging
process by taking the classical field value which results from the saddle point equation
outlined in (\ref{s2_76}).

\subsection{Outlook for relations between chaotic and integrable systems with modified r-s matrices}

A particular property of the nonlinear sigma-model equations (\ref{s3_154}-\ref{s3_176}) is the integrability
for special dimensions, as 1+1 or even 2+1 \cite{abl1}-\cite{kerf1}. These properties of integrability
are determined by r-s matrix properties which can be investigated as quantum groups \cite{ma}-\cite{kerf1}.
However, as already suggested in \cite{mies1}, one can also try to classify chaotic systems as
extensions of these r-s matrix bi-algebras in analogy of extensions of group or
algebraic properties if one adds symmetry breaking generators or group elements
to the classical equations. If one considers the general
BCH-formulas for abstract time- or spatial development operators, one can always
relate the multiplication of two exponentials \(\exp\{\hat{A}\}\), \(\exp\{\hat{B}\}\)
with some operators \(\hat{A}\) and \(\hat{B}\) to the exponential of commutator terms
between these generators. If the commutator algebra between \(\hat{A}\) and
\(\hat{B}\) is closed, one has specific, closed group properties which may be related
to integrable systems. If the commutator algebra of operators \(\hat{A}\) and \(\hat{B}\)
has deviations from the prevailing, closed algebraic structures, it should be possible
to refer the non-closed algebraic structure in the exponential to some chaotic
behavior. According to these suggestions, it might be possible to examine chaotic
systems as extensions of the few integrable classical equations determined
by r-s matrix structures.


\begin{thebibliography}{99}
\bibitem{neg} J.W. Negele and H. Orland, {\it "Quantum Many-Particle Systems"},
(Addison-Wesley, Reading, MA, 1988)
\bibitem{ka} T. Kashiwa, Y. Ohnuki and M. Suzuki, {\it "Path Integral Methods"},
(Oxford Science Publications, Clarendon Press, Oxford 1997)
\bibitem{nag1} N. Nagaosa, {\it "Quantum Field Theory in Condensed Matter Physics"},
(Springer, 'Series: Theoretical and Mathematical Physics', 1999)
\bibitem{nag2} N. Nagaosa, {\it "Quantum Field Theory in Strongly Correlated Electronic Systems"},
(Springer, 'Series: Theoretical and Mathematical Physics', 1999)
\bibitem{Das1} Ashok Das, {\it "Field Theory (A Path Integral Approach)"},
(World Scientific, 'World Scientific Lecture Notes in Physics, Vol. 75', 2nd edition, 2006)
\bibitem{mies1} B. Mieck, {\it "Coherent state path integral and super-symmetry for condensates
composed of bosonic and fermionic atoms"}, Fortschr. Phys. ("Progress of Physics")
\textbf{55} (No. 9-10) (2007), 989-1120; ({\bf cond-mat/0702223})
\bibitem{cor} L. Corwin, {\it "Graded Lie algebras in mahematics and physics(Bose-Fermi symmetry)"},
Rev. Mod. Phys. \textbf{47} (1975), 573-602
\bibitem{fay} P. Fayet and S. Ferrara, {\it "Supersymmetry"}, Phys. Rep. \textbf{32}(5) (1977), 249-334
\bibitem{soh} M.F. Sohnius, {\it "Introducing Supersymmetry"}, Phys. Rep. \textbf{128}(2-3) (1985), 39-204
\bibitem{bere} F.A. Berezin, {\it "Introduction to Superanalysis"}, (D. Reidel Publishing Company, Dordrecht, 1987)
\bibitem{witt} Bryce de Witt, {\it "Supermanifolds"} (2nd ed.), (Cambridge University Press, Cambridge, 1992)
\bibitem{efe} K. Efetov, {\it "Supersymmetry in Disorder and Chaos" (and references therein)},
(Cambridge University Press, Cambridge, 1997)
\bibitem{corn1} J.F. Cornwell, {\it "Group Theory in Physics, Vol. I"},
(Academic Press, "Techniques in Physics", London, Fifth printing 1994)
\bibitem{corn2} J.F. Cornwell, {\it "Group Theory in Physics, Vol. II"},
(Academic Press, "Techniques in Physics", London, Fifth printing 1993)
\bibitem{corn3} J.F. Cornwell, {\it "Group Theory in Physics, Vol. III
(Supersymmetries and Infinite-Dimensional Algebras)"},
(Academic Press, "Techniques in Physics", London, 1989)
\bibitem{luc} L. Frappat, A. Sciarrino and P. Sorba,
{\it "Dictionary on Lie Algebras and Superalgebras"}, (Academic Press, London, 2000)
\bibitem{alice1} A. Rogers, {\it "Supermanifolds (Theory and Applications)"},
(World Scientific, Singapore, 2007)
\bibitem{gold} J. Goldstone, Nuovo Cimento \textbf{19} (1961), 154
\bibitem{nambu} Y. Nambu, Phys. Rev. Lett. \textbf{4} (1960), 380
\bibitem{ke} L.P. Keldysh, Sov. Phys. JETP \textbf{20} (1965), 1018
\bibitem{schw1} J. Schwinger, {\it "Brownian Motion of a Quantum Oscillator"},
J. Math. Phys. \textbf{2} (1961), 407-432
\bibitem{mahan1} P.M. Bakshi and K.T. Mahanthappa, {\it "Expectation Value Formalism in
Quantum Field Theory. I"}, J. Math. Phys. \textbf{4} (1963), 1-11
\bibitem{mahan2} P.M. Bakshi and K.T. Mahanthappa, {\it "Expectation Value Formalism in
Quantum Field Theory. II"}, J. Math. Phys. \textbf{4} (1963), 12-16
\bibitem{klau} J.R. Klauder and B.S. Skagerstam, {\it "Coherent States (Applications in
Physics and Mathematical Physics)"} (World Scientific, Singapore, 1985)
\bibitem{gil1} W.M. Zhang, D.H. Feng and R. Gilmore, {\it "Coherent states: theory and some
applications"}, Rev. Mod. Phys. \textbf{62}(4), (1990), 867-927
\bibitem{lipp} E. Lipparini, {\it "Modern Many-Particle Physics (Atomic Gases, Quantum Dots and
Quantum Fluids)"}, (World Scientific, Singapore, 2003)
\bibitem{wen} Xiao-Gang Wen, {\it "Quantum Field Theory of Many-Body Systems
(From the Origin of Sound to an Origin of Light and Electrons)"},
(Oxford University Press, Oxford, 2004)
\bibitem{bruus} H. Bruus and K. Flensberg, {\it "Many-Body Quantum Theory in Condensed
Matter Physics (An Introduction)"}, (Oxford University Press, Oxford, 2004)
\bibitem{dick} W.H. Dickhoff and D. Van Neck, {\it "Many Body Theory Exposed!
(Propagator Description of Quantum Mechanics in Many-Body Systems)"},
(World Scientific, Singapore, 2005)
\bibitem{haken1} H. Haken, {\it "Laser Theory"}, (Springer, Berlin, 1970)
\bibitem{haken2} H. Haken, Rev. Mod. Phys. \textbf{47}, (1975), 67
\bibitem{haken3} H. Haken, {\it "Light, (Vols. I and II)"}, (North-Holland, Amsterdam, 1981)
\bibitem{haug1} H. Haug, Z. Phys. \textbf{200}, (1967), 57
\bibitem{haug2} H. Haug and H. Haken, Z. Phys. \textbf{204}, (1967), 262
\bibitem{haake1} F. Haake, Z. Phys. \textbf{227}, (1969), 179
\bibitem{grah1} R. Graham and H. Haken, Z. Phys. \textbf{237}, (1970), 31
\bibitem{scully} Marlan O. Scully and M. Suhail Zubairy, {\it "Quantum Optics"},
(Chapters 11 and 12, Cambridge University Press, 1997)
\bibitem{hulet2} K.E. Strecker, G.B. Partridge and R.G. Hulet,
{\it "Conversion of an Atomic Fermi Gas to Long-Lived Molecular Bose Gas"},
Phys. Rev. Lett. \textbf{91}, (2003), 080406
\bibitem{hulet6} W. Zhang, C.A. Sackett and R.G. Hulet,
{\it "Optical detection of a Bardeen-Cooper-Schrieffer phase transition in a trapped gas of
fermionic atoms"}, Phys. Rev. A \textbf{60}, (1999), 504
\bibitem{tim1} E. Timmermans, K. Furuya, P.W. Milonni and A. K. Kerman,
{\it "Prospect of creating a composite Fermi-Bose superfluid"},
Phys. Lett. A \textbf{285} (2001), 228-233
\bibitem{schreck} F. Schreck, {\it "Mixtures of ultracold gases :
Fermi sea and Bose-Einstein condensate of lithium isotopes"},
Ann. Phys. Fr. \textbf{28} (2003) 1-165
\bibitem{bm1} B. Mieck, {\it "Nonlinear sigma model for a condensate composed of fermionic atoms"},
Physica A \textbf{358} (2005), 347-365
\bibitem{bmdis1} B. Mieck, {\it "Ensemble averaged coherent state path integral for disordered bosons
with a repulsive interaction (Derivation of mean field equations)"},
Fortschr. Phys. ("Progress of Physics") \textbf{55} (No. 9-10) (2007), 951-988; ({\bf cond-mat/0611416})
\bibitem{bmrep} B. Mieck, Rep. Math. Phys. \textbf{47} (No. 1) (2000), 139
\bibitem{bmdis2} B. Mieck, {\it "Ensemble averaged coherent state path integral for disordered bosons
with a repulsive interaction (Infinite order gradient expansion of the functional determinant)"}, (in preparation)
\bibitem{st} R.L. Stratonovich, Sov. Phys. Dokl. \textbf{2} (1958), 416
\bibitem{mehta} M.L. Mehta , {\it "Random Matrices"},
(pages 90 and 125 for Vandermonde determinants, Academic Press, revised and enlarged 2nd edition, London, 1991)
\bibitem{eng} B.-G. Englert, {\it "Lectures on Quantum Mechanics (Vol. III: Perturbed Evolution)"},
(chap. 1.4.2 {\it "Insertion : Varying an Exponential function"}, pages 41-43),
(World Scientific, Singapore, 2006)
\bibitem{bmlv1} B. Mieck, {\it "Coherent state path integral and Langevin equations
of interacting bosons"}, Physica A \textbf{294} (2001), 96-110
\bibitem{bmlv2} B. Mieck, {\it "Coherent state path integral and Langevin equation
of interacting fermions"}, Physica A \textbf{312} (2002), 431-446
\bibitem{hulet4} C.A. Sackett, H.T.C. Stoof and R.G. Hulet,
{\it "Growth and Collapse of a Bose-Einstein Condensate with Attractive Interactions"},
Phys. Rev. Lett. \textbf{80}, (1998), 2031
\bibitem{hulet5} C.A. Sackett, J.M. Gerton, M. Welling and R.G. Hulet,
{\it "Measurements of Collective Collapse in a Bose-Einstein Condensate with Attractive Interactions"},
Phys. Rev. Lett. \textbf{82}, (1999), 876
\bibitem{abl1} M.A. Ablowitz and P.A. Clarkson, {\it "Solitons, Nonlinear Evolution Equations and
Inverse Scattering"}, (Cambridge University Press, {\it "London Mathematical Society
Lecture Note Series (No. 149)"}, London, 1991)
\bibitem{abl2} M.J. Ablowitz, B. Prinari and A.D. Trubatch, {\it "Discrete and Continuous Nonlinear
Schr\"{o}dinger Systems"}, (Cambridge University Press, {\it "London Mathematical Society
Lecture Note Series (No. 302)"}, London, 2003)
\bibitem{gra1} B. Mieck and R. Graham, {\it "Bose-Einstein condensate of kicked rotators"},
J. Phys. A : Math. Gen. \textbf{37} No44 (2004), L581-L588
\bibitem{gra2} B. Mieck and R. Graham, {\it "Bose-Einstein condensate of kicked rotators
with time-dependent interaction"},
J. Phys. A : Math. Gen. \textbf{38} No7 (2005), L139-L144
\bibitem{ma} Zhong-Qi Ma , {\it "Yang-Baxter Equation and Quantum Enveloping Algebras"},
(World Scientific, {\it "Advanced Series on Theoretical Physical Science, Vol. 1"}, Singapore, 1993)
\bibitem{mai1} J.-M. Maillet, {\it "New Integrable Canonical Structures in Two-Dimensional
Models"}, Nucl. Phys. B \textbf{269} (1986), 54-76
\bibitem{bo} Bo-Yu Hou, {\it "Differential Geometry for Physicists
(Advanced Series on Theoretical Physics : Vol. 6)"},
(chap. 6.7 on "{\it Nonlinear $\sigma$-models, soliton solutions and their geometric meaning}"),
(World Scientific, Singapore, 1997)
\bibitem{kerf1} G.G.A. B\"{a}uerle and E.A. de Kerf,
{\it "Lie Algebras, Part 1, (Finite and Infinite Dimensional Lie Algebras and Applications in Physics)"},
(chap. 17 with section 17.5 {\it "Current algebras"},
North-Holland, Elsevier Science Publishers, Amsterdam, 1990)
\bibitem{naka} M. Nakahara, {\it "Geometry, Topology and Physics"},
(chap. 1 with problems 1,Graduate student studies in physics,
Institute of Physics Publishing, Bristol and Philadelphia, 1990)
\bibitem{inf_g1} B. Mieck, (in preparation) {\it "Infinite order gradient expansion
for the determinant of Fermi fields in QCD-type, non-Abelian gauge theories with
chiral anomalies" (Derivation for an effective action of BCS-terms with nontrivial
topology)}
\end{thebibliography}
\end{document}